\documentclass[
	12pt, 
	a4paper,
]{LegrandOrangeBook}

\usepackage{etoolbox}
\DeclareCaptionFont{lmss}{\fontfamily{lmss}\selectfont}
\usepackage[labelfont=bf, font=lmss]{caption}
\usepackage{eurosym}

\hypersetup{
	pdftitle={Aging and memory effects in social and economic dynamics}, 
	pdfauthor={David Abella}, 
	pdfsubject={Aging and memory effects in social and economic dynamics}, 
	pdfkeywords={Aging, memory, social, economic, dynamics, complex, networks, stochastic, modeling, housing market, segregation, segmentation, consensus, spread, complex, contagion}, 
	pdfcreator={LaTeX}, 
}

\DeclareUnicodeCharacter{0301}{*************************************}

\makeatletter
\newcommand{\tempmaxup}[1]{\def\blx@maxcitenames{99}#1}
\makeatother

\DeclareCiteCommand{\fullcite}[\tempmaxup]
    {\usebibmacro{prenote}}
    {\usedriver
        {}
        {\thefield{entrytype}}}
    {\multicitedelim}
    {\usebibmacro{postnote}}

\definecolor{ocre}{RGB}{0, 153, 0} 

\chapterimage{} 
\chapterspaceabove{6.5cm} 
\chapterspacebelow{6.75cm} 

\begin{document}

\fontsize{13pt}{13pt}\selectfont\sffamily
\setstretch{1.2} 


\begin{titlepage}
    	\large

        \begin{center}
            \sffamily
            \includegraphics[width=.35\textwidth]{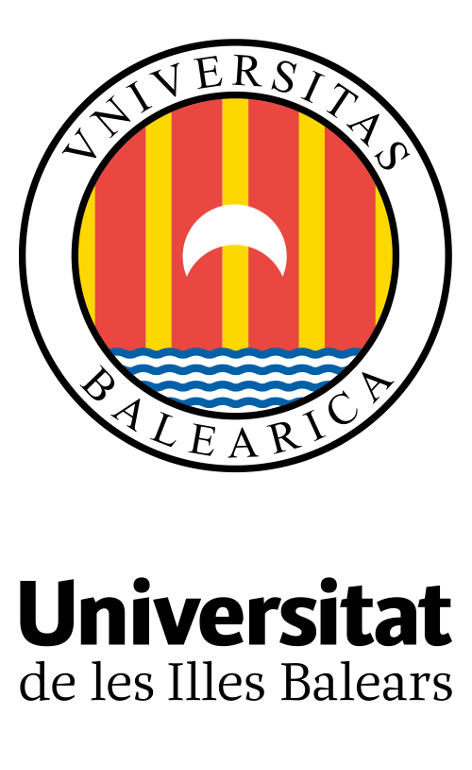}
        
            \vspace*{.05\textheight}

			{\Huge\bfseries DOCTORAL THESIS} \\
			{\Huge \bfseries 2024}

            \vspace*{.15\textheight}

			\begingroup
				{\Huge\bfseries AGING AND MEMORY EFFECTS IN SOCIAL AND ECONOMIC DYNAMICS \par}
			\endgroup

            \vfill

            {\Large \bfseries David Abella Bujalance}
        \end{center}
\end{titlepage}


\begin{titlepage}
    \large

    \hfill
    \begin{minipage}[c][0.2\textheight]{0.45\textwidth}
        \includegraphics[width=.75\textwidth]{Figs/uib.png}
	\end{minipage}
    \hfill
    \begin{minipage}[c][0.2\textheight]{0.45\textwidth}
        \includegraphics[width=1\textwidth]{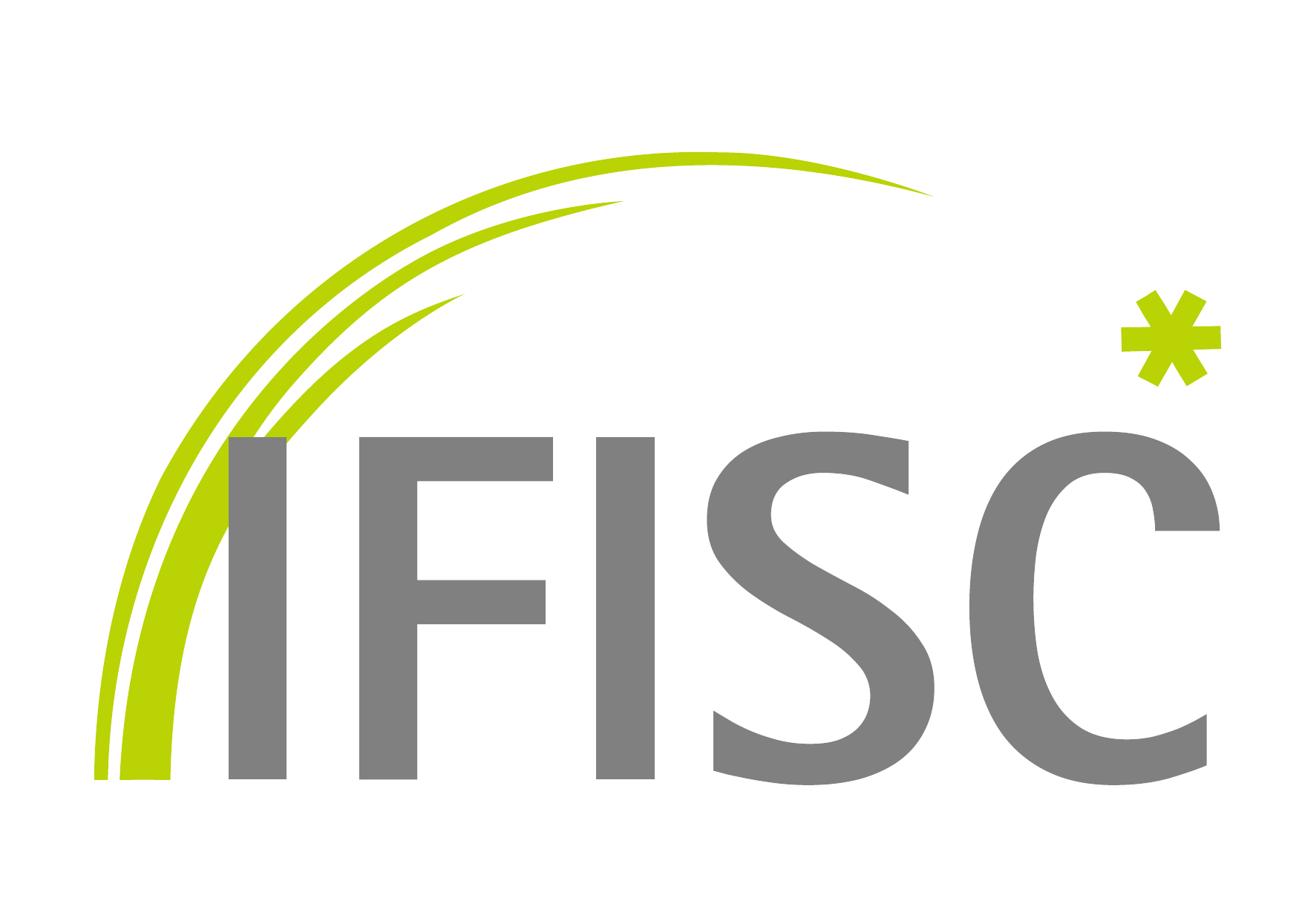}
    \end{minipage}
    \hfill
	{\centering\sffamily
    \vspace*{.1\textheight}
        
    {\Huge\bfseries DOCTORAL THESIS} \\
    {\Huge\bfseries 2024}
        
    \bigskip

    {\Huge\bfseries Doctoral programme in Physics} \\

    \vfill
        
    \begingroup
        {\Huge\bfseries AGING AND MEMORY EFFECTS IN SOCIAL AND ECONOMIC DYNAMICS \par}
    \endgroup

    \bigskip
            
    {\Large\bfseries David Abella Bujalance}
            
    \vfill
	}

    \begin{flushleft}
		\large\sffamily
        \textbf{Thesis Supervisor:} Jos\'e Javier Ramasco Sukia \\
        \textbf{Thesis Supervisor:} Maxi San Miguel \\
        \textbf{Thesis Tutor:} Crist\'obal L\'opez S\'anchez\\

        \vfill

        \textbf{ Doctor by the Universitat de les Illes Balears }\\
    \end{flushleft}
\end{titlepage}


\thispagestyle{empty} 

~\vfill 
\fontsize{13pt}{13pt}\selectfont\sffamily

\noindent \textbf{Collaborators:}

\noindent Jos\'e Javier Ramasco Sukia

\noindent Maxi San Miguel Ruibal\\

\noindent \textbf{Author:}

\noindent David Abella Bujalance\\

\noindent \textit{Aging and memory effects in social and economic dynamics.} \copyright

\noindent Palma de Mallorca, July 2024
\pagebreak

\vspace*{8 cm}
\thispagestyle{empty}

\begin{flushright}
    \sffamily\large
    \textit{
    A les meves germanes, Adriana i Ivet,\\
    per ser-hi sempre, per estimar-me, per fer-me riure, \\
    i per totes les vegades que heu intentat esbrinar què és un sistema complex.\\
    }
\end{flushright}


\newpage
\thispagestyle{plain} 
\mbox{}

\pagebreak
\frontmatter
\thispagestyle{empty}
\phantomsection
\addcontentsline{toc}{chapter}{Acknowledgements}
\textbf{ \huge Acknowledgements}

\vspace*{0.5 cm}

First, I would like to express my gratitude to my two directors and main collaborators through this thesis, Jos\'e J. Ramasco and Maxi San Miguel. I am very grateful for their educational and professional guidance through this path, which has been fundamental to reach this point. In particular, Jose has been a constant guide, and I am very grateful that his door always remained open for me, available for any question or doubt that could take 5 minutes or 5 hours to solve. Maxi has been a great supervisor, I want to thank him for his critical view, his constant push to improve my work, and his questions that always made me think in a different way. I must also thank them for their financial support and generosity, which allowed me to do this three-year PhD without worrying much about money. I would also like to thank Vincenzo Nicosia, with whom I had the pleasure to do a research stay during the past year. Collaborating with him allowed me to be part of the process of developing new ideas and projects, which has motivated me to face new challenges. In addition, throughout this period at IFISC, I thank having great conversations with scientists such as Juan Fern\'andez-Gracia, Konstantin Klemm and Damià Gomila, which have served as a source of inspiration and learning.

My firsts steps began in the Master in Physics of Complex Systems, where I had the opportunity to meet great people. I would like to thank my classmates, Alex, Miguel, Marius, Robert, Javi, Medi, Joje, Ana. I want to acknowledge the great environment we had during those years, which motivated me to follow the academic path. Alex has not only been my master's colleague, I consider him as a partner though this journey. I thank all the moments helping each other, being victims of FPI-CAIB's injustices, promoting Julia as the best programming language and doing ``rutillas chill'' to both Massanella and Sabotage. I am very grateful that we did this path together, and I hope one day we will work together at the Alex's Research Institute. Also, I want to thank Miguel for sharing good moments during the pandemic, the curiosities in Google Maps and the funny random videos.

I also want to thank the really nice environment between PhD students and PostDocs at IFISC. It may seem something small, but you just need to go to work once alone to realize how important it is. From all IFISC Basement, I need to highlight a special place for me, the ``Zulo'', which was an almost-empty temporary lab of Nonlinear Physics when I got there, and we turned into a cozy place to work and fight for justice via posters. I want to thank being part of this group and having worked in (probably) the only theoretical physics office with a giant extractor hood. 

Durante el doctorado, estoy agradecido de haber conocido a gente maravillosa como Manu. Agradezco las muchas (muchas muchas) conversaciones que hemos tenido a lo largo de estos años, ya sean a nivel profesional como personal, ayudándonos con nuestras epidemias raras. Recordaré las vueltas a casa quedándonos en la esquina charlando de todo, incluso sobre lo mucho que charlamos. Has sido de verdad mi Recursos HuManus y con mucho cariño te digo que parte de esta tesis también es tuya. También quiero agradecer haber conocido a Mar, por todo el apoyo, por todas las risas y por animarme a ser más fuerte. Sabes que yo siempre seré el primer MAG y tú mi MAH, ya que con nadie más tienes un póster de ensaimadas. También agradecer a Gorka, por las risas, por las conversaciones sobre porque ningún método de clustering va bien y por mantenerme al día sobre www.internet.com. Por supuesto, Fer ha sido un gran apoyo durante este periodo, agradezco las largas discusiones sobre mil temas, creo que no hay proyecto en el IFISC que no hayamos debatido y pensado como lo haríamos nosotros. Gracias por todas las referencias que me has dado y por demostrarme empíricamente que uno puede dormirse donde quiera. Cuando llegasteis al doctorado estaba un poco perdido y os agradezco haberme ayudado a encontrar mi camino.

També vull agrair a l'Adrià, per haver compartit moments graciosos, pel Philomeno Moreno, per l'ADSL i per les millors versions d'"Un Mundo Ideal". También estoy muy agradecido por haber compartido momentos con Bea (gracias por los memes de patitos), Pau, Maria (gracias por todas las respuestas), Mar F., Pablo, Irene, Lisa, Rodri, Jose. Estoy agradecido de haber podido hacer el doctorado con vosotros, pero sobre todo por todos los momentos que no eran doctorado. Y a los coleguillas del doctorado, muchas gracias porque me he sentido muy cómodo en este entorno, ha sido un placer compartir esta etapa con vosotros. Una mención especial a mi equipo de la "Grieta del Invocador": Jesus, Jun, Manu, Giovani. Ojalá saquemos un paper de esto eventualmente, pero si no lo hacemos, nos habremos llevado unas risas. Y por último, a mis compañeros del Barco, Alba y Manu P., que aunque estuvimos a punto de hundirnos varias veces, fue un placer navegar con vosotros. 

Vull també agrair als meus amics de tota la vida: Jaume, Vernet, Alberto i Marcel. Us agraeixo haver estat allà sempre per desfogar-me de la feina i per donar-me un punt de vista diferent. En aquesta feina, a vegades et perds una mica de la realitat i m'heu ajudat a tocar de peus a terra. També un agraïment molt fort i sincer a la Rosa i el Pau C., per tot el suport permetent-me parlar de qualsevol tema sense filtre i sentir-me escoltat i entès. Us agraeixo molt haver estat allà per mi i haver-me ajudat a superar moments difícils.

I per descomptat, a la meva família, al meu pare, a la meva mare, a les meves germanes Adriana i Ivet, al meu fillol Aleix, que és més jove que aquesta tesi, a la Jose, al Pau, a la Mimi. Gràcies per tot el suport i per la paciència quan us parlo d'alguna cosa que no enteneu. De debò que no treballo amb ocells però és un bon exemple de sistema complex. Gràcies també per tot el suport tant financer com emocional durant tots els anys que m'han dut a aquest punt. Òbviament, sense valtros no hi hauria tesi. També vull agrair tant als Abella com als Bujalance, per tot el suport mostrat i l'interés en aquestes coses que faig.

Finalment, vull agrair a una persona en especial que ha estat al meu costat des de l'inici d'aquesta tesi i amb la que hem construït junts un projecte de vida en comú. Riki, gràcies per ser la pedra angular que ha aguantat aquest procés, ho dic de debò quan dic que sense tu, aquesta tesi no existiria. Estic molt agraït per tots els missatges d'àudio llargs que t'ha tocat escoltar, per totes les converses sobre compaginar feina i vida, per donar-me suport en moments de frustració i entendre que a vegades necessitava estar sol, i per intentar trobar solució a problemes que ni tan sols jo sabia per on començar. Per tot això i per molt més, mereixes més que un simple agraïment. Aquesta tesi és tan teva com meva.

\thispagestyle{empty}

\vfill


\pagebreak
\thispagestyle{empty}
\phantomsection
\addcontentsline{toc}{chapter}{Preface}
\textbf{ \huge Preface}

\vspace*{0.5 cm}

This thesis is an original work that addresses the understanding of temporal interactions in social and economic systems through two distinct approaches: theoretical modeling in social systems and empirical analysis of spatial and temporal dynamics in the housing market. While these two parts are related to each other and approached from a similar perspective, their separation is due to an external factor. Unlike the continuous funding provided by a 4-year grant from the Spanish government, my funding came from various scientific projects. This required me to adapt my work to the research lines of these projects, resulting in a thesis that does not follow a continuous research line with a central problem to address but instead explores different topics and problems.

However, I do not see this diversity of topics as a drawback. On the contrary, it has allowed me to experience different topics and learn various tools throughout the process. Moreover, in my opinion, addressing both theoretical and practical issues from different points of view enriches its interdisciplinary character.

It is also worth noting that such theses are becoming more common. In my view, the doctorate has evolved from a purely formative period to one focused on developing the skills necessary to work as an academic and researcher. This raises whether it is more useful to compile a manuscript from already reviewed and published scientific articles or to reformulate the system to better suit contemporary needs.

Personally, I have enjoyed writing this thesis, organizing the work in a manner that seems appropriate for a reader and revisiting all the results from these 3 years of work. Nonetheless, writing a thesis requires significant time and effort, and the reward is just a title that allows you to continue doing the same work you did during the previous years - the work of a researcher.

\vfill
\newpage
\thispagestyle{plain} 
\mbox{}


\pagebreak
\thispagestyle{empty}
\phantomsection
\addcontentsline{toc}{chapter}{Resum}
\textbf{\huge Resum}

\vspace*{0.5 cm}

En aquesta tesi doctoral, investiguem la complexa interacció entre les dinàmiques temporals associades amb la memòria i el costum, i els seus efectes en els sistemes socials i econòmics. Per a això, combinem models teòrics, per a explorar les implicacions del costum en models de llindars (pressió social), i l'anàlisi empírica, per a abordar l'impacte dels patrons temporals i espacials en sistemes complexos reals, prenent com a cas d'estudi el mercat immobiliari.

La recerca en aquesta tesi s'estructura en dues parts principals. En la primera part, ens centrem en fer models teòrics per a entendre com l'acostumar-se a un estat (representatiu d'una opinió, comportament, etc.) influeix altres mecanismes de canvi social i quin implicacions té aquest mecanisme en el comportament emergent del sistema. Acostumar-se en aquest context s'entén com una resistència creixent a canviar l'estat actual, la qual cosa també pot entendre's com un record dels estats passats. En altres paraules, com a més temps porti un agent amb un estat, menys probable és que canviï est. En aquesta tesi, analitzem el costum en models de llindars, on el mecanisme de canvi social és la pressió social (modelada com un llindar). Aquests models són usats per a descriure 3 fenòmens socials diferents: segregació, difusió d'innovacions i l'arribada al consens. En el model de segregació de Sakoda-Schelling, els efectes d'acostumar-se s'entenen com una persistència a quedar-se en la residència actual com a més temps un agent hagi estat satisfet allí. Aquesta modificació és capaç de portar el sistema d'un estat mixt a un segregat, per tant, és capaç de trencar la transició de fase mixta-segregada present en el model original. A pesar que el costum promou la segregació, el creixement de dominis en la fase segregada és lent, sent capaç de trencant la invariància temporal. A continuació, introduïm un nou marc matemàtic, estenent l'equació mestra aproximada per a models binaris en xarxes complexes per a incloure els efectes del costum. Aquest marc ens permet escriure en termes d'un conjunt d'equacions diferencials la dinàmica del sistema i entendre que mecanisme rellevant causa el seu estat final. Testem els resultats d'aquestes equacions en el model de Granovetter-Watts, per a investigar com el costum modifica els processos de difusió d'innovacions. Ens trobem que el costum, entesa en aquest model com una resistència a adoptar la innovació, pot alterar significativament la dinàmica de contagi complex del model, on la cascada d'adopció exponencial és reemplaçada per un creixement o exponencial estiratge o en llei de potència, depenent de com modelitzem el costum. Per a aquest model, trobem una solució analítica per a la condició de cascada i els exponents, oferint una comprensió de com el costum i l'estructura de la xarxa influeixen en els processos de contagi complex. Finalment, estudiem un model de llindar simètric, un model on tots dos estats són simètrics i intenten arribar al consens. Els resultats revelen que acostumar-se afecta de manera important en la dinàmica del model, portant a noves fases no presents en la versió original, caracteritzades per un desordre inicial seguit d'un creixement lent de dominis. En aquesta fase, el mecanisme de costum és capaç de portar al consens a l'estat de la minoria inicial. El costum també introdueix un procés de creixement de dominis més lent amb estats transitoris de llarga durada, indicant que els efectes del costum, malgrat promoure l'ordre, poden retardar significativament la convergència del sistema a l'estat estacionari.

En la segona part, passem a una anàlisi empírica de dades reals d'una plataforma en línia de pisos en venda que ens permet analitzar les interaccions espacials i temporals del mercat immobiliari. Usem anuncis que han estat publicats en algun moment durant 2 anys en 3 províncies espanyoles, de manera que puguem oferir una visió objectiva de la dinàmica del mercat, incloent-hi el paper de les agències immobiliàries i la seva influència en els patrons espacials emergents. Comencem explorant la segmentació espacial dins del mercat immobiliari, causada per la presència i influència de les agències immobiliàries. Representem el mercat com una xarxa tripartida que connecta anuncis, agències i cel$\cdot$les espacials, de manera que ens permeti identificar la division del mercat mitjançant diferents algorismes de detecció de comunitats. La nostra anàlisi revela que la segmentació del mercat és consistent a través de diferents resolucions espacials i algorismes, i trobem patrons similars en dades d'Espanya i França (en tots dos països els mercats detectats estan connectats i són més grans que els municipis). A més, quant a la dinàmica temporal, analitzem les ràfegues d'anuncis, els patrons setmanals i la publicació dels anuncis per part de les diferents agències immobiliàries. Observem que la dinàmica dels anuncis exhibeix patrons temporals irregulars, influenciats per un efecte de memòria similar al d'acostumar-se en sistemes socials, però és la probabilitat que un anunci sigui eliminat la que disminueix amb el temps. Aquest efecte de memòria és consistent a través de diferents regions i tipus de propietats, suggerint que és una característica general del mercat immobiliari. També trobem que la publicació d'anuncis per part de les agències està influenciada per la grandària de la seva cartera d'anuncis (preferència per agències grans), el seu preu mitjà (similitud de preus agencia - nou casa) i la seva proximitat espacial (especialització).

\thispagestyle{empty}

En resum, a través de dos punts de vista complementaris, models teòrics i anàlisi empírica, aquesta tesi contribueix a la comprensió de com el costum i la memòria donen forma als sistemes socials i econòmics. Els nostres resultats subratllen el profund impacte de les dinàmiques temporals en els sistemes socioeconòmics, revelant com els efectes no Markovianos alteren els comportaments, portant a nous fenòmens en la dinàmica de la segregació, processos de contagi i problemes de consens. A més, l'anàlisi de dades reals del mercat immobiliari destaca la importància de les dinàmiques temporals (memòria) i espacials (especialització) de les agències immobiliàries en la formació de les estructures del mercat i en els processos de presa de decisions dels venedors. La fortalesa d'aquesta recerca radica en la combinació d'enfocaments teòrics i empírics, basats en l'ús de grans conjunts de dades, teoria de xarxes i models matemàtics simples. Aquest enfocament interdisciplinari crida a futurs desenvolupaments d'aquest tipus, que acaben de començar a revelar els secrets del comportament humà.

\vfill


\pagebreak
\thispagestyle{empty}
\phantomsection
\addcontentsline{toc}{chapter}{Resumen}
\textbf{ \huge Resumen}

\vspace*{0.5cm}

En esta tesis doctoral, investigamos la compleja interacción entre las dinámicas temporales asociadas con la memoria y la costumbre, y sus efectos en los sistemas sociales y económicos. Para ello, combinamos modelos teóricos, para explorar las implicaciones de la costumbre en modelos de umbrales (presión social), y el análisis empírico, para abordar el impacto de los patrones temporales y espaciales en sistemas complejos reales, tomando como caso de estudio el mercado inmobiliario.

La investigación en esta tesis se estructura en dos partes principales. En la primera parte, nos centramos en hacer modelos teóricos para entender como el acostumbrarse a un estado (representativo de una opinión, comportamiento, etc.) influye otros mecanismos de cambio social y qué implicaciones tiene este mecanismo en el comportamiento emergente del sistema. Acostumbrarse en este contexto se entiende como una resistencia creciente a cambiar el estado actual, lo que también puede entenderse como un recuerdo de los estados pasados. En otras palabras, como más tiempo lleve un agente con un estado, menos probable es que cambie este. En esta tesis, analizamos la costumbre en modelos de umbrales, donde el mecanismo de cambio social es la presión social (modelada como un umbral). Estos modelos son usados para describir 3 fenómenos sociales diferentes: segregación, difusión de innovaciones y la llegada al consenso. En el modelo de segregación de Sakoda-Schelling, los efectos de acostumbrarse se entienden como una persistencia a quedarse en la residencia actual como más tiempo un agente haya estado satisfecho allí. Esta modificación es capaz de llevar el sistema de un estado mixto a uno segregado, por lo tanto, es capaz de romper la transición de fase mixta-segregada presente en el modelo original. A pesar de que la costumbre promueve la segregación, el crecimiento de dominios en la fase segregada es lento, siendo capaz de rompiendo la invariancia temporal. A continuación, introducimos un nuevo marco matemático, extendiendo la ecuación maestra aproximada para modelos binarios en redes complejas para incluir los efectos de la costumbre. Este marco nos permite escribir en términos de un conjunto de ecuaciones diferenciales la dinámica del sistema y entender que mecanismo relevante causa su estado final. Testeamos los resultados de estas ecuaciones en el modelo de Granovetter-Watts, para investigar como la costumbre modifica los procesos de difusión de innovaciones. Nos encontramos que la costumbre, entendida en este modelo como una resistencia a adoptar la innovación, puede alterar significativamente la dinámica de contagio complejo del modelo, donde la cascada de adopción exponencial es reemplazada por un crecimiento o exponencial estirado o en ley de potencia, dependiendo de como modelizamos la costumbre. Para este modelo, encontramos una solución analítica para la condición de cascada y los exponentes, ofreciendo una comprensión de como la costumbre y la estructura de la red influyen en los procesos de contagio complejo. Finalmente, estudiamos un modelo de umbral simétrico, un modelo donde ambos estados son simétricos e intentan llegar al consenso. Los resultados revelan que acostumbrarse afecta de forma importante en la dinámica del modelo, llevando a nuevas fases no presentes en la versión original, caracterizadas por un desorden inicial seguido de un crecimiento lento de dominios. En esta fase, el mecanismo de costumbre es capaz de llevar al consenso al estado de la minoría inicial. La costumbre también introduce un proceso de crecimiento de dominios más lento con estados transitorios de larga duración, indicando que los efectos de la costumbre, a pesar de promover el orden, pueden retrasar significativamente la convergencia del sistema al estado estacionario.

En la segunda parte, pasamos a un análisis empírico de datos reales de una plataforma online de pisos en venta que nos permite analizar las interacciones espaciales y temporales del mercado inmobiliario. Usamos anuncios que han estado publicados en algún momento durante 2 años en 3 provincias españolas, de forma que podamos ofrecer una visión objetiva de la dinámica del mercado, incluyendo el papel de las agencias inmobiliarias y su influencia en los patrones espaciales emergentes. Empezamos explorando la segmentación espacial dentro del mercado inmobiliario, causada por la presencia e influencia de las agencias inmobiliarias. Representamos el mercado como una red tripartita que conecta anuncios, agencias y celdas espaciales, de forma que nos permita identificar la division del mercado mediante diferentes algoritmos de detección de comunidades. Nuestro análisis revela que la segmentación del mercado es consistente a través de diferentes resoluciones espaciales y algoritmos, y encontramos patrones similares en datos de España y Francia (en ambos países los mercados detectados están conectados y son más grandes que los municipios). Además, en cuanto a la dinámica temporal, analizamos las ráfagas de anuncios, los patrones semanales y la publicación de los anuncios por parte de las diferentes agencias inmobiliarias. Observamos que la dinámica de los anuncios exhibe patrones temporales irregulares, influenciados por un efecto de memoria similar al de acostumbrarse en sistemas sociales, pero es la probabilidad de que un anuncio sea eliminado la que disminuye con el tiempo. Este efecto de memoria es consistente a través de diferentes regiones y tipos de propiedades, sugiriendo que es una característica general del mercado inmobiliario. También encontramos que la publicación de anuncios por parte de las agencias está influenciada por el tamaño de su cartera de anuncios (preferencia por agencias grandes), su precio medio (similitud de precios agencia - nuevo casa) y su proximidad espacial (especialización).

\thispagestyle{empty}

En resumen, a través de dos puntos de vista complementarios, modelos teóricos y análisis empírico, esta tesis contribuye a la comprensión de como la costumbre y la memoria dan forma a los sistemas sociales y económicos. Nuestros resultados subrayan el profundo impacto de las dinámicas temporales en los sistemas socio-económicos, revelando como los efectos no Markovianos alteran los comportamientos, llevando a nuevos fenómenos en la dinámica de la segregación, procesos de contagio y problemas de consenso. Además, el análisis de datos reales del mercado inmobiliario destaca la importancia de las dinámicas temporales (memoria) y espaciales (especialización) de las agencias inmobiliarias en la formación de las estructuras del mercado y en los procesos de toma de decisiones de los vendedores. La fortaleza de esta investigación radica en la combinación de enfoques teóricos y empíricos, basados en el uso de grandes conjuntos de datos, teoría de redes y modelos matemáticos simples. Este enfoque interdisciplinario llama a futuros desarrollos de este tipo, que acaban de empezar a desvelar los secretos del comportamiento humano.

\vfill


\pagebreak
\thispagestyle{empty}
\phantomsection
\addcontentsline{toc}{chapter}{Abstract}
\textbf{ \huge Abstract}

\vspace*{0.5cm}

In this doctoral thesis, we investigate the complex interplay between temporal dynamics associated with aging and memory and their effects on social and economic systems. To do so, we combine theoretical modeling, to explore the aging implications in threshold (peer pressure) models, and empirical analysis, to address the impact of temporal and spatial patterns in real complex systems, taking housing market as a case study.

The research in this thesis is structured into two main parts. In the first part, we focus on theoretical models to elucidate how aging influence other social mechanisms and which are the implications of this mechanism to the emergent behavior. Aging in this context is understood as an increasing resistance to change the current state (representative of an opinion, behavior, etc.), which can also be understood as a memory to past states. In other words, the longer an agent has been in a state, the less probable is to change it. We analyze aging in threshold models, where the social change is driven by peer pressure (modeled as a threshold). These models are used to describe 3 different social phenomena: segregation, innovation diffusion, and consensus formation. Starting with the Sakoda-Schelling segregation model, we incorporate aging effects as an increasing persistence in a residential location the longer an agent has been satisfied there. This modification leads the system from a mixed state towards segregation, making disappear the mixed-segregated phase transition present in the non-aging version. Even though aging promotes order, the coarsening dynamics in the segregated phase are found to decay slowly, breaking the time translational invariance. We also introduce a novel mathematical framework, extending the approximate master equation for binary-state dynamics in networks to include aging. This framework allows us to write in terms of a set of differential equations the dynamics of the system and understand the relevant mechanism that drives it to the final state. We test the results of these equations for the Granovetter-Watts model, to investigate how aging modifies innovation diffusion processes. We find that aging, understood in this model as a resistance to adopt the innovation, can significantly alter the complex contagion dynamics of the model, where the exponential cascade of adoption is replaced by a stretched exponential or a power-law increase, depending on the aging mechanism. For this model, an analytical solution was derived for the cascade condition and exponents, offering a comprehensive understanding of how aging and the network structure influences complex contagion processes. Finally, we study a Symmetrical threshold model, a consensus model where both states are symmetric. The results reveal that aging dramatically impacts the model's dynamics, leading to new phases not present in the non-aging version, characterized by initial disordering followed by slow coarsening. In this phase, aging mechanism is able to lead to consensus in the state of the initial minority. Aging also introduces a slower coarsening process and long-lived transient states, indicating that aging effects, despite promoting order, can significantly delay the system's convergence to the steady state.

In the second part, we transition to an empirical analysis using real data from online platforms to analyze the spatial and temporal interactions in the housing market. We use a 2-years dataset, covering listings from 3 Spanish provinces, to offer comprehensive insights into market dynamics, including the roles of real estate agencies and emerging spatial patterns. We start by exploring spatial segmentation within the housing market, driven by the presence and influence of the real estate agencies. By a tripartite network representation of the market, connecting listings, agencies, and spatial units, we identify robust submarkets via consensus clustering of different community detection algorithms. Our analysis reveals that market segmentation is consistent across various spatial resolutions and algorithms, and we find similar patterns in datasets from both Spain and France (submarkets are connected and larger than municipalities). Moreover, regarding the temporal dynamics, we analyze the burstiness of listings, the weekly patterns and the attachment and detachment of listings to the real estate agencies. We observe that the listings' dynamics exhibit irregular temporal patterns, influenced by a memory effect akin to aging, where the probability of a listing being removed decreases over time. This memory effect is consistent across different regions and property types, suggesting it is a general feature of the housing market. We also find that the attachment of listings to agencies is influenced by the agencies' portfolio size (preferential attachment), their mean price (price similarity), and their spatial proximity (specialization).

\thispagestyle{empty}

Overall, this thesis via two complementary points of view, theoretical modeling and empirical analysis, contributes to the understanding of how aging and memory shape social and economic systems. Our findings underscore the profound impact of temporal dynamics on socio-economic systems, revealing how non-Markovian effects associated with aging alter behaviors, leading to new phenomenology to segregation dynamics, contagion processes, and consensus problems. Additionally, the analysis of real-world data from the housing market highlights the importance of temporal (memory) and spatial (specialization) dynamics of the real estate agencies in shaping market structures and decision-making processes. The strength of this research lies in the combination of theoretical and empirical approaches, relying on the use of large datasets, network theory and simple mathematical models. This interdisciplinary approach calls for further developments of this kind, which are just starting to unveil the secrets of human dynamics.

\vfill


\chapterimage{} 
\chapterspaceabove{6.75cm} 
\chapterspacebelow{7.25cm} 

\chapter*{List of publications}
\addcontentsline{toc}{chapter}{List of publications} 

The list of articles detailed below, in chronological order by date of publication, form the basis of the present thesis.
\vspace{0.5 cm}

\begin{enumerate}
	\item \fullcite{Abella-2022}
	\vspace{0.5 cm}
	\item \fullcite{Abella-2022-AME}
	\vspace{0.5 cm}
	\item \fullcite{Abella_2024}
	\vspace{0.5 cm}
	\item \fullcite{abella2024exploring}
	\vspace{0.5 cm}
	\item \fullcite{abella2024dynamics}
	\vspace{0.5 cm}
\end{enumerate}

\vfill
\pagebreak
Other publications published during the PhD period are also included in the following list.

\vspace{0.5 cm}
\begin{itemize}
	\item \fullcite{abella2023many}
	\vspace{0.5 cm}
	\item \fullcite{abella2023unraveling}
	\vspace{0.5 cm}
\end{itemize}


\pagestyle{empty} 

\tableofcontents 

\addcontentsline{toc}{chapter}{Contents} 

\listoffigures 

\addcontentsline{toc}{chapter}{List of figures} 


\pagestyle{fancy} 

\cleardoublepage 


\chapterimage{} 
\chapterspaceabove{6.75cm} 
\chapterspacebelow{7.25cm} 

\chapter*{List of acronyms}
\addcontentsline{toc}{chapter}{List of acronyms} 
\begin{abbreviations}
    \item[ABM] Agent-Based model
    \item[ER] Erd\H{o}s-R\'enyi
    \item[BA] Barab\'asi-Albert
    \item[RAU] Random Asynchronous Update
    \item[SAU] Sequential Asynchronous Update
    \item[MC] Monte Carlo
    \item[AME] Approximate Master Equation
    \item[PDF] Probability Density Function
    \item[RR] Random Regular
    \item[ODbL] Open Data Commons Open Database License
    \item[OSLOM] Order Statistics Local Optimization Method
    \item[CM] Center of Mass
    \item[HMF] Heterogeneous Mean Field
    \item[HMFA] Heterogeneous Mean Field with Aging
\end{abbreviations}

\mainmatter
\chapterimage{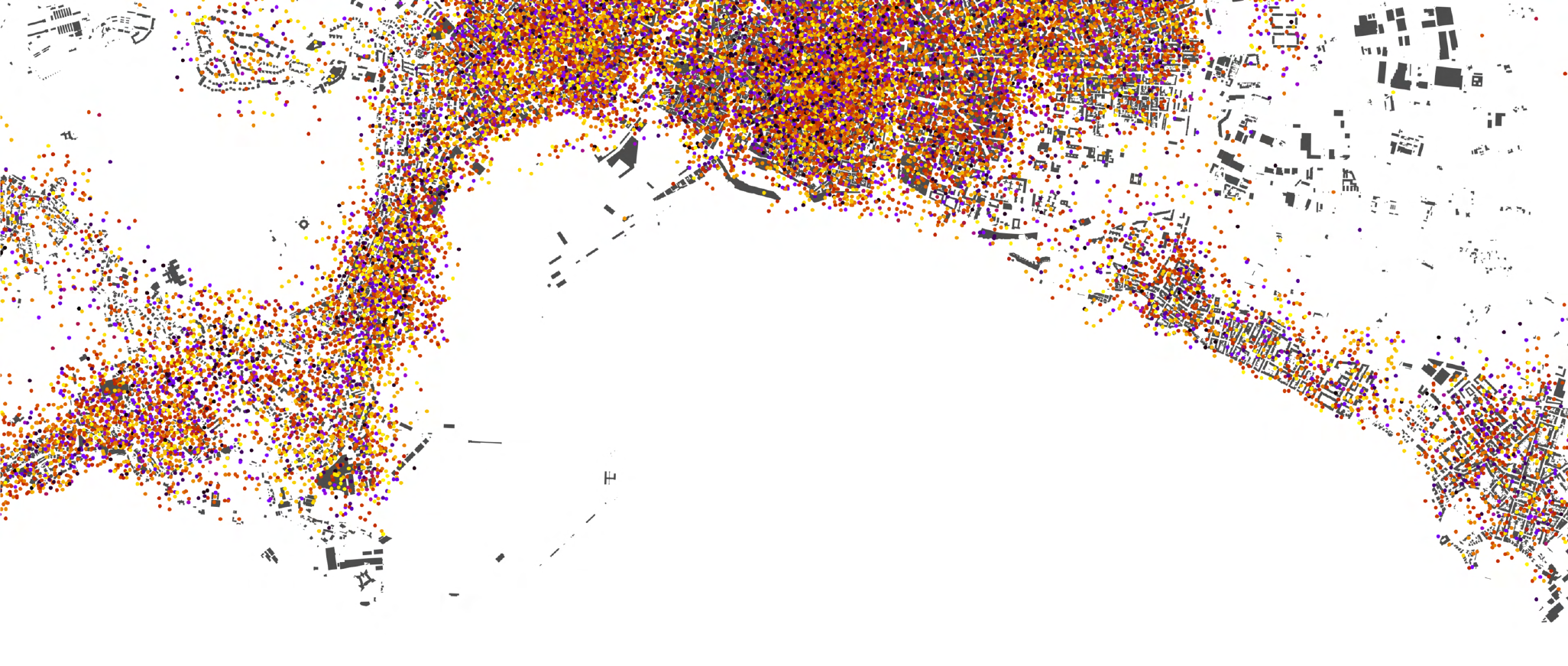}
\chapterspaceabove{6.75cm}
\chapterspacebelow{7.25cm}

\chapter{Introduction and positioning}
In this introductory chapter, we motivate the study of social and economic dynamics from the complex systems' perspective. We begin by reviewing the scientific landscape in which this thesis is developed, and we introduce the challenges that the study of human behavior faces. This chapter aims to serve as a solid ground in which the concepts and tools employed in this thesis are positioned and understood, trying to make this thesis as self-consistent as possible. We introduce some basic concepts of complex networks, and we describe some tools that are used to model social and economic systems. Finally, we introduce the datasets that we use to study the temporal interactions in complex economic systems.

\section{\label{sec:scie_lands} Scientific Landscape and outline of the thesis}

This thesis addresses the study of human behavior and social systems from a \textit{complex systems} perspective, in which the emergence of collective phenomena, that arise from the interactions of many individuals, cannot be understood by studying the behavior of individual agents in isolation (the so-called \textit{reductionist} approach)~\cite{anderson1972more}. The study of collective phenomena has a long history in the natural sciences, specially in the branch of statistical physics~\cite{stanley1971phase}. This branch traditionally studies the emergence of collective phenomena in physical systems, such as the phase transitions in magnetic materials via spin models~\cite{onsager-1944}, the turbulence in fluids~\cite{frisch1995turbulence}, the synchronization in oscillatory systems~\cite{pikovsky2001universal}, or percolation~\cite{stauffer-1985}. However, in recent years, the study of complex systems has evolved into the study of emergent phenomena beyond physical systems, such as biological~\cite{brown2000scaling,aderem-2005,alon2019introduction}, ecological~\cite{may-2001}, economic~\cite{limburg2002complex}, and social systems~\cite{castellano2009statistical}. Even though this branch of physics is relatively young, in 2021, the Nobel Prize in Physics was awarded to Syukuro Manabe, Klaus Hasselmann, and Giorgio Parisi for their contributions to the study of complex systems~\cite{nobel-2021}, giving a recognition to this field in the scientific community~\cite{bianconi2023complex}.

From the migration of birds~\cite{roche-1997} to the spreading of fake news through social media~\cite{yamir-2004,vosoughi-2018}, there are many examples of collective phenomena beyond the realm of traditional physics at which the concepts and tools from complex systems and statistical physics can be applied~\cite{perc2013evolutionary}. The high share effects of renewable energy in power grids~\cite{maria-2023}, the spread of a disease in a population~\cite{anderson1991infectious}, the consensus in political elections~\cite{anderson-2000}, the emergence of social norms~\cite{ellickson-1999} are some examples of social or economic collective behavior in which the global phenomena cannot be understood by looking at a single individual. These collective human behavior has been studied from a variety of perspectives (sociology, psychology, economics, political sciences...), which often relies on qualitative methods, such as interviews, discourse analysis, or ethnographic studies~\cite{bryman-2010}. However, the complex systems' perspective aims to provide a quantitative framework to understand the collective behavior based on methodologies from statistical mechanics and network theory~\cite{newman-book,barabasi-2013}. This approach relies on the use of large amount of detailed data to extract significant correlations, validate theories and develop models. Reliable sources of human activity data have been a limitation to the study of social systems historically. It is in fact surprising how other branches of physics, where the typical scale of the phenomena is very large, as astrophysics, or very small, as particle physics, do not suffer from a lack of data, while the study of social systems, where the typical scale is human, has been historically limited by the lack of data.

Thankfully, the digital revolution has changed this picture, allowing the storage of large amounts of data generated by human activities, such as social media, mobile phones, or online platforms. Nowadays, every two year, more human socio-economic data is produced than during all the preceding years of human history together~\cite{karsai2019computational}. This data, often referred to as \textit{Big data}, has opened a new era for the study of social systems from a computational perspective, together with a paradigm shift in the way we understand human behavior~\cite{manyika-2011}. Nevertheless, this new era comes with an awareness, as the use of big data for the study of human behavior raises important ethical and privacy concerns, which need to be addressed in order to ensure the responsible use of data for the study of social systems~\cite{boyd-2012}. Moreover, from the technical point of view, this huge amount of data needs computational and mathematical resources to be analyzed and modeled. Satisfying this demand, the field of \textit{Computational Social Science} has emerged, with the aim to develop new methods to study human behavior from real data significant correlations~\cite{Lazer2009CompSocSci}. This branch of the complex systems science was born as a combination of methodologies borrowed from social sciences, such as sociology, psychology, or economics, with computational methods from hard sciences (physics, mathematics, computer science)~\cite{watts-2007}, such as network theory, statistical mechanics, machine learning, data mining, etc. This interdisciplinary approach has allowed developing new methods for forecasting social phenomena and understanding the basic mechanism behind human interactions. 

One can differentiate two main approaches to build a modeling framework from the data source. The first focuses on the prediction and forecasting of a certain social phenomenon, such as the spread of a disease or the price of a stock. In this approach, the data is seen as a necessary input to our methodology to make quantitative predictions~\cite{Lazer2009CompSocSci}. However, in this approach, the mechanisms behind the phenomena are often hidden in the data, and the model is seen as a black box that provides accurate predictions~\cite{rudin-2019}. In this context, the use of machine learning~\cite{murphy-2012} and deep learning~\cite{goodfellow-2016} models are  very popular, as they are able to capture complex patterns in the data and reproduce it with high levels of accuracy. The second approach is to focus on the understanding of the mechanisms behind the phenomena. In the later approach, the data is seen as a problem to be understood, an observation from which we can extract qualitative behaviors and patterns~\cite{axelrod2006agent}. In this context, the aim is to develop very simple models that are able to reproduce the main features of the data, and to extract the basic mechanisms behind the phenomena.

Following the second approach, network science has a critical role in the study of socio-economic systems, as it provides a natural framework to study the interactions between individuals. A network, or graph, is a mathematical representation of a set of nodes connected by links or edges, which allows to study the structure of the interactions between the different elements. The study of networks has a long history in the natural sciences, from the neurons network in the brain~\cite{sporns-2004} to food webs in an ecosystem~\cite{ings-2008, elith-2009, bastolla-2009}. However, in recent years, new data sources lead to the discovery that complex properties and heterogeneity, present in most social systems, need for a topological description in terms of a complex network~\cite{newman-book, dorogovtsev2002evolution, boccaletti2006complex}. Social networks, such as a social media~\cite{dunbar-2015}, the collaboration network of scientists~\cite{newman-coll-2001,radicchi-2008}, or international conflicts~\cite{hafnerburton-2009, diaz2023network}, are found to exhibit non-trivial topological properties and often are referred as \textit{complex networks}, which will be explained later on this chapter. In particular, the study of information spreading as a dynamical system is an example of a system in which network theory has allowed to understand how information spreads and how consensus emerges (and if it does).

Spreading of information has been a topic of interest for many social scientists. Early theoretical frameworks, influenced by psychological and sociological theories, show how individuals in a crowd lose their sense of self and are more susceptible to the ideas and emotions of the crowd~\cite{le2023crowd}. Social imitation of behaviors and ideas was proposed as a mechanism for social change, facilitated by close contact and communication among individuals~\cite{kanter-1971}. Similarly, peer pressure was also proposed another possible mechanism, where individuals are influenced by their peers to adopt certain behaviors or ideas~\cite{granovetter-1978, brown-1986}. However, until now, these theoretical dissertations were not supported by quantitative results with real data analysis (besides controlled experiments). The spread of innovations~\cite{rogers2014}, the diffusion of information~\cite{valente-1996}, or the spread of diseases~\cite{anderson1991infectious} are some examples of social phenomena where the traditional theoretical frameworks can now be tested or modified to explain correlations observed from data sources.

Beyond the network structure and the contagion dynamics, there is a third ingredient that is critical for the study of human dynamics: the temporal patterns. Human interactions exhibit complex activity patterns that are difficult to understand and to model, since there are a lot of mechanisms that drive the human behavior. For example, the circadian rhythms~\cite{roenneberg-2013}, bursty interactions~\cite{Barabasi2005Bursts}, cascades~\cite{watts-2002}, periodic commuting behaviors~\cite{gonzalez2008understanding}, recurring patterns in online behavior~\cite{Lazer2009CompSocSci}, etc., are effects embedded to the socio-economic environment and the social network structure, so they cannot be ignored. In fact, through this thesis, we will show how the temporal patterns of human interactions affect processes such as information spreading ~\cite{Holme2012Temporal}, changing the nature of the dynamic process dramatically~\cite{karsai-2011}.

In this thesis, we use the complex systems approach to try to understand the consequences of the irregular human activity patterns (temporal interactions) in socio-economic models. In the first part of the thesis, we incorporate memory effects associated with aging to models where social change is driven by peer-pressure (\textit{threshold models}). Modelling in this part of the thesis is seen as a tool to understand the basic mechanisms behind the phenomena and to extract the main features of the data, not with the aim to make accurate predictions. Under this perspective, we will apply tools borrowed from computational social science, network theory and statistical mechanics to study some problems under this umbrella: the emergence of segregation, the spreading of information and the consensus problem. In the second part, data from a real economic system takes a relevant role. We use a dataset of listings from a real estate platform from which, via network theory, we extract the relevant information about the temporal interactions in the system, highlighting the memory effects. In particular, we develop a methodology to address the spatial segmentation and the decision-making mechanisms present in the housing market.

\subsection{\label{sec:Thesis outline} Thesis outline}

This thesis is divided in 3 main parts:

\begin{itemize}
    \item In the \textbf{Introductory part}, we explain the basic concepts necessary to understand the findings in this thesis. Chapter 1 introduces the field in which this thesis is developed. Network theory and agent-based modeling are also introduced in this chapter. Chapter 2 describes and reviews the so-called threshold models for this thesis, and introduce the notion of burstiness and aging in the socio-economic context. 

    \item In \textbf{Part I}, we investigate the aging implications in threshold models. We will see that aging has been proposed as a possible explanation to the heterogeneous human activity patterns. We explore how aging modifies the threshold model dynamics in 3 different threshold models designed for 3 different social systems. In Chapter 3, we study the aging implications in a popular segregation model. In Chapter 4, we introduce a new mathematical framework to include aging in binary state dynamics, which is applied in Chapters 5, to see the aging effects in a complex contagion model, and in Chapter 6, to explore aging in a consensus model. Chapter 6 is divided in two main ``subchapters'': The first one dedicated to understand the phase diagram of this consensus model and the second one to explore the aging implications in both the statics and dynamics of the model.
    
    \item In \textbf{Part II}, we study the temporal interactions in a real economic system. We will see an example of irregular temporal patterns in the housing market via the listings from an online platform.  From this data, Chapter 7 analyzes the statics of the system, exploring how the real estate agencies specialization segments the housing market. Following this approach, we develop a methodology to assess this segmentation, which is found to be robust in the different markets. In Chapter 8, we will analyze the dynamics of the listings, focusing on which mechanisms drive the decision-making process of house sellers when they choose a real estate agency to sell their property. 
    
    \item Finally, in the \textbf{Part III}, we summarize the main findings of this thesis and discuss the implications of the results in the field (Chapter 9). In Chapter 10, we discuss the limitations of the study and propose future possible research along the lines of this thesis.
\end{itemize}

\section{\label{sec:Challenges of Computational Social Science} Challenges of Computational Social Science}

Since the research in Computational Social Science is relatively new, the methodology is not yet well established, but follows the traditional approach from natural sciences, like physics, but with new relevant techniques adapted to the specific problem to address. In this section, we introduce some challenges the study of human behavior faces, and how these challenges are addressed in the context of our research.

Regarding the \textbf{data availability}, digital behavioral data provides a detailed look at human activities in their usual environments, capturing information about where, when, and how people behave~\cite{Eagle2006RealityMining}. This type of data is great because it avoids the common biases seen in older research methods, such as surveys, and gives us a clearer picture across a large number of people~\cite{Lazer2009CompSocSci,chen-2014}. Instead of relying on limited lab studies or broad census data, the data is now collected from real-world interactions, stored from digital devices and platforms that people use in their daily lives~\cite{Eckmann2004Entropy,blondel-2015,artime-2017}. However, this data is similar to a picture of a black hole, neither you can observe it in a controlled environment nor it is possible to analyze the response to a stimuli~\cite{lazer-2014}. Also, not everyone is equally represented in digital data, especially in underdeveloped countries, where the use of recent technology is less popularized~\cite{zook-2017}. Ethically, there are also big concerns about collecting and using personal data without people knowing it~\cite{boyd-2012, de-montjoye-2013}, which has pushed for stricter laws to protect privacy and ensure people have control over their own data.

Regarding  \textbf{data analysis}, few decades ago several statistical methodologies were designed to manage small data sets in social and economic research, giving rise to meaningful models and techniques~\cite{stevens-2012, gelman-2006}. These traditional techniques have evolved to accommodate the large data volumes seen in modern statistical analysis~\cite{hastie-2013}, where the challenge is no longer sample size but the data heterogeneity. Using tools such as data mining, data curation and statistical tests, we can convert large noisy data sets into intelligible knowledge from which extract significant patterns~\cite{witten-2005}. Furthermore, network analysis is helpful to analyze several datasets from a different perspective, highlighting the crossed interactions between the different elements~\cite{newman-book, clauset-2008}. The use of null models~\cite{perry2012null,gauvin-2022} is an example of a common practice in network analysis, in which the interactions are rewired (reshuffled), breaking the correlations, such that we can discriminate if the macroscopic properties of the network are meaningful or not.

Regarding to \textbf{understanding and modelling human behavior}, there are 3 main approaches one can follow: Statistical Models are used to identify correlations and make individualized predictions. For example, machine learning~\cite{murphy-2012}, Bayesian inference~\cite{gelman1995bayesian}, deep learning~\cite{goodfellow-2016}, etc. Mechanistic Models, inspired by physics, simulate emergent behaviors from simpler entity interactions, offering robust scenario testing but often simplifying individual differences~\cite{axelrod2006agent}. Agent-based modeling (ABM) emerged as a particularly influential mechanistic approach, enabling scientists to create and study systems of interacting agents (individuals or collective entities) and observe emergent behaviors from simple rules of interaction~\cite{epstein1999agent}.On the other hand, Data-driven Models blend real-world data with synthetic frameworks to focus on specific mechanisms, allowing for realistic simulations and in-depth analysis of the dynamics. The availability to address problems from these 3 perspectives is crucial in an interdisciplinary field as computational social science, and enhances understanding by forming, testing, and refining hypotheses~\cite{watts2004new}.

Finally, the scope of \textbf{applications} in Computational Social Science is very broad, including managing crowd dynamics, improving cooperative behaviors, optimizing traffic and public transport systems, facilitating product and service adoption, and forecasting global pandemics. This multifaceted field is becoming increasingly relevant in modern techno-social societies~\cite{vespignani2009predicting}, enhancing the quantitative nature of social sciences and leading the shift towards a digital, data-driven paradigm across science, technology, and various applied domains. One example of these applications, related to contagion dynamics, are the fake news observatories~\cite{Polis-observatory, EDMO-observatory, committed-observatory-2023}, which are platforms that monitor the spread of fake news in social media, and provide tools to detect and give feedback to the users.

\subsection{Thesis positioning in Computational Social Science\label{subsec:Thesis positioning in Computational Social Science}}

Through this thesis, the main challenge faced is the understanding and modeling of human dynamics. In the first part of the thesis, the data is used primarily as inspiration to develop reasonable models that explain the emergent properties of the data. We use tools borrowed from network theory and statistical mechanics to study the impact of aging on social processes driven by peer pressure, focusing on the emergence of segregation, the spread of information, and the consensus problem. The main challenge addressed in this part of the thesis is modeling the non-Markovian nature of human dynamics. The aging mechanism introduces a memory effect in the system, preventing us from using traditional tools to describe Markovian dynamics. Therefore, throughout the thesis, we extend mathematical frameworks to a formulation that allows us to describe the dynamics of non-Markovian binary-state models in complex networks.

In the second part of the thesis, we use a real dataset from an economic system. The main challenge addressed in this part is data curation and the analysis of temporal patterns and correlations. The dataset comprises listings from a real estate platform, containing information about interactions between sellers and real estate agencies during a period of time. We use network theory to extract relevant information from the data and develop a methodology to address spatial segmentation and decision-making mechanisms in the housing market. The main challenge in this part of the thesis is the development of new methodologies to extract relevant information from the data and the creation of new algorithms for clustering and studying the system's dynamics. However, there is also modelling involved, since this part is related with the first part of the thesis precisely because one can understand the temporal irregularities in the housing market as an aging/resistance mechanism.

\section{\label{sec:Terminology and general concepts} Terminology and general concepts}

Here, we briefly introduce general definitions and concepts, which will be used through the thesis. We introduce basic concepts of complex networks, and modeling techniques used in the study of social and economic systems.
\subsection{\label{subsec:Complex networks} Complex networks}
A network is a mathematical representation of a set of nodes connected by links. This simple definition allows to map the architecture of many complex systems by identifying the interactions between the different elements. This topology allows us to quantify how close or far two nodes are, how central a node is, how dense a system is, etc. As I understood it through these years,  what makes a network complex in my opinion is having a short path length, such that average distance between two nodes is relatively small, high levels of clustering, result from a triadic closure effect, and a scale free degree distribution, such that several nodes with low degree coexist with few nodes with a very high degree. These properties are often found in many social, biological and economic complex systems, such as social media, ecological networks, or financial systems.

\begin{figure}
    \centering
    \includegraphics[width=\textwidth]{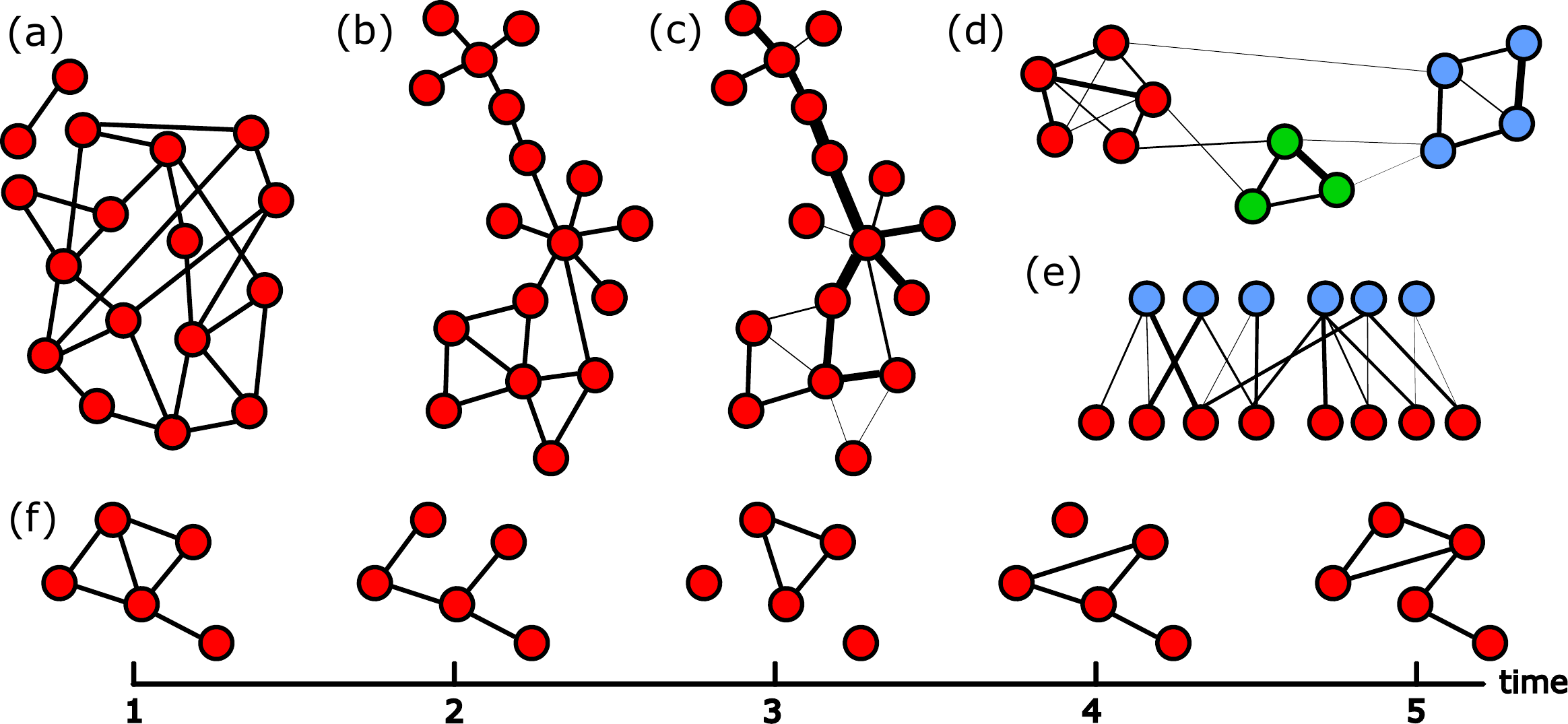}
    \caption[Different network types]{Representation of some different types of networks: \textbf{(a)} Simple Erd\H{o}s-R\'enyi network, \textbf{(b)} Simple Barab\'asi-Albert network, \textbf{(c)} Weighted Barab\'asi-Albert network, \textbf{(d)} Weighted network with community structure, \textbf{(e)} Weighted bipartite network and \textbf{(f)} Simple temporal network evolving during 5 time steps. The nodes are represented by colored circles, and the links are represented by lines, where the width of the line represents the weight of the link. In (d) the color of the nodes represents the community to which they belong and in (e) the type of the nodes to which they belong.}
    \label{fig:netwotk_types}
\end{figure}

To create synthetic networks with the same properties as real complex networks, we use stochastic models that, from few nodes, describe how we need to grow the network to reach our goal~\cite{posfai2016network}. Historically, the first attempt was the Erd\H{o}s-Renyi (ER) model~\cite{erdos1960evolution}, which grows the network by simply adding links between nodes with a certain probability $p$. This model is useful to study the emergence of the giant component in a network, which is the largest connected component in the network (see an ER network in Fig. \ref{fig:netwotk_types}a). However, as a result of this process of adding links according to $p$, the degree distribution is a binomial distribution, and the clustering, which decays with the network size $N$ as $p/N$, is too low to reproduce a real complex network. Another popular example is the Watts-Strogatz model~\cite{watts1998collective}, which starts from a regular network, typically a ring lattice, and rewires the links with a certain probability $r$. This model is useful to study the emergence of small-world networks, which have a short path length and a high clustering. However, rewiring does not change the degree distribution, so it is still binomial, which is a limitation to reproduce the fat tail distribution found in real complex networks~\cite{newman2003structure}. The Barab\'asi-Albert model~\cite{barabasi1999emergence} grows the network by adding nodes that attach to the present nodes with a probability proportional to the degree of the nodes. This process is called \textit{preferential attachment}, and it is known to be useful to reproduce the fat tail distributions~\cite{merton1968matthew} or, in other words, the emergence of hubs (nodes with a very high degree) in a network (see Fig. \ref{fig:netwotk_types}b). If instead of adding nodes according to a certain rule, we want to directly impose a certain observed degree distribution, we can use the configuration model~\cite{newman-book}, which is a generalization of the Erd\H{o}s-R\'enyi model in which we start from a set of nodes disconnected, each one with a certain number of stubs, and then we connect the stubs with a certain probability. It can be shown that the clustering in networks generated using the configuration model for any degree distribution is proportional to $1/N$, where $N$ is the number of nodes in the network. This means that for large networks, the clustering is very small, a feature that we will see may be useful to describe agent based models in networks.

Until now, we considered only simple networks, in which the links are binary (there is link or there is not). However, we can consider that links are weighted, which means that each link has a certain number (or another quality) associated to it~\cite{barrat2004architecture} (see Fig. \ref{fig:netwotk_types}c). This representation is useful to describe the strength of the interactions between nodes. For example, the railway network is a simple network that connects stations via the railway lines, but the weight of each line could be number of trains or the number of passengers that travel through it~\cite{latora-2002}. This representation is very useful when we have a system very densely connected, but there are some links that have a higher importance or are more saturated than others, depending on the context. Also, this representation allows us to map any correlation matrix as a weighted network, which might be useful to identify the more correlated and anticorrelated elements in a system~\cite{onnela-2003,tumminello-2005}.

In fact, a common trait that is found in complex systems is the presence of communities, which are groups of nodes that show a favoring bias to connect with nodes of the same group~\cite{girvan-2002} (see Fig. \ref{fig:netwotk_types}d). For example, communities can represent groups of friends, families, or work colleagues, which are more likely to share information and to interact between them~\cite{newman2003structure}. To identify communities in a network, we can use community detection algorithms, which are computational tools that group nodes in the network in a way that maximize the community structure, such as maximizing the number of links inside the groups and minimizes the number of links between the groups~\cite{lancichinetti-2008,fortunato2010community}. One popular example is the Louvain algorithm~\cite{blondel-2008}, which is a greedy algorithm that optimizes the modularity function, which is a measure of the quality of the partition of the network. Another popular example is the Infomap algorithm~\cite{rosvall-2008}, which is a flow-based algorithm that optimizes a map equation, which is a measure of the quality of the partition of the network. Another recent example more sophisticated is the OSLOM algorithm~\cite{OSLOM}, which is a stochastic algorithm that optimizes the likelihood of the partition of the network and is able to identify overlapping communities. These algorithms are useful to identify the structure of the network, and to understand the dynamics of the system.

On the other hand, there are networks in which by definition there are two natural types of nodes, and there are just links between the two nodes of different types (not between nodes of the same type). These networks are called bipartite networks~\cite{newman2003structure}, and they are useful to represent systems in which there are two types of elements, and there are only interactions between them~\cite{latapy-2008} (see example in Fig. \ref{fig:netwotk_types}e). For example, online platforms to see movies and TV shows, such as Netflix~\cite{netflix} and HBO~\cite{HBO}, can be considered as bipartite network in which there are users and movies, and there are links between users and movies if the user has watched the movie. This representation is useful to study the recommendation systems based on the user preference~\cite{ricci-2011}. Moreover, bipartite networks allow us to project into a weighted network of nodes of the same type, where the weight of the link is typically computed and normalized by the number of common neighbors (of the other type) between the two nodes~\cite{newman-2001-collaboration}. In the case of the movie recommendation system, we can project the bipartite network into a weighted network of users, where the weight of the link between two users is the number of movies that they have watched in common. The projection is useful to study the social interactions between users, and to identify communities in the network.

Finally, there are networks in which the interactions between nodes are not static, but they evolve in time (temporal networks)~\cite{Holme2012Temporal} (see evolution of a simple network in Fig. \ref{fig:netwotk_types}f). These networks are useful to represent systems in which the interactions between nodes change in time~\cite{Perra2012ActivityDriven}. For example, the phone calls in a group of friends is a temporal network, in which the links are the phone calls that take place at a certain time window. This representation is useful to study the dynamics of the system, and to understand the network activity~\cite{karsai-2011} or if there are periodic patterns~\cite{Jo2012Circadian}.

Through this thesis, the types of networks described above will be used to model the interactions between the agents in the system, and to understand the dynamics of the system. In the first part of this thesis, we just use artificial simple networks created via the models described above. However, at the second part, we use bipartite network (weighted and unweighted), projected networks, community detection algorithms, and temporal networks to study the dynamics of a real economic system. 

Besides these networks representations, there are plenty of other types of networks very popular to describe social and economic complex systems. One important example are multiplex networks~\cite{gomez-2013,kivela2014multilayer,de2013mathematical}, which are networks in which there are several layers of interactions between the nodes. This representation is useful to study the interplay between the different layers of interactions. Another example is the network of networks~\cite{gao2011robustness, d2014networks}, which are networks in which there are several networks that are interconnected between them. There are hypergraphs (or higher-order networks)~\cite{battiston-2021}, which are networks in which the links are not between two nodes, but between a set of nodes, and signed networks~\cite{leskovec2010signed}, which are networks in which the links can be positive or negative.

\subsection{\label{subsec:Agent-based models} Agent-based models}

Agent-based models are a class of computational models for simulating the actions and interactions of autonomous agents such that we can assess their effects on the system as a whole~\cite{duffy-1998, bonabeau-2002}. In the context of our thesis, agent based models are used to understand the collective behavior of the agents and the emergent properties of the system from a simple set of rules. Agents can be individuals, groups, or organizations, and if the network topology is necessary, the agents are placed in the nodes of a network~\cite{macal-2010, railsback-2011}. The agents are characterized by its state, which may be a set of  discrete or continuous variables, and its behavior or update rules. The rules, deterministic or stochastic, are the recipe that the agents need to follow to interact with its environment. For example, for the Voter model, an agent has a state that represents one of the two opinions and the update rules are just to imitate the state of a neighbor~\cite{castellano2009statistical}. Despite these simple rules, the Voter model shows a very rich phenomenology which nowadays is mainly studied via agent-based simulations.

An important ingredient when we perform agent-based simulations is the order in which nodes are updated. The order can be synchronous, in which all nodes are updated at the same time, or asynchronous, in which nodes are updated one by one~\cite{macal-2010, railsback-2011}. Nevertheless, the synchronous update may lead to problematic schemes, in which the system may get stuck in a periodic or chaotic solution. Performing simulations in asynchronous order, the order can also be random, in which nodes are updated one by one in a random order, or sequential, in which nodes are updated in a certain order. A common order is to just follow a random asynchronous update (RAU), which avoids biases and allows to explore the space of solutions in a more efficient way. Nevertheless, the sequential asynchronous update (SAU) is useful to reduce the computational cost, because going through the system sequentially may lead faster to the stationary state.

Usually, in agent based-models, these simulation steps are repeated until a stationary state is reached (if exists). To do so, one can perform Monte Carlo simulations~\cite{metropolis-1949,cronin2006monte}, which are a class of computational algorithms that rely on repeated random sampling for many time steps to reach the stationary state. In these simulations, a time step (Monte Carlo step) is defined as the time in which, on average, each node is updated once~\cite{newman-1999, landau-2014}. These are useful to explore the space of solutions, and to understand the dynamics of the system. A different approach is to use the Gillespie algorithm~\cite{gillespie-1977}, which is a stochastic simulation algorithm that generates a trajectory of the system by sampling the time of the next event and the event that will take place. This algorithm is useful because, since the time is not fixed, we avoid time steps in which nothing happens, and we can explore the dynamics of the system in a more efficient way~\cite{gibson-2000}.

Finally, the output of these simulations is a set of relevant statistics that describe the system's trajectory or the stationary state. These statistics can be the average opinion of the agents, the number of clusters, the number of links between clusters, the average interface density, etc. An ensemble of simulations is often performed to explore the space of possible trajectories and solutions, and to understand the robustness of the results. This ensemble can be performed by simply repeating the stochastic simulation, or by changing a trait of the system, such as the initial conditions of the system, or the realization of the network. This procedure is repeated for the relevant parameters of the model to explore the phase diagram, a methodology borrowed from statistical physics that refers to the parameter space in which, for each set of parameter values, we can identify the stationary solution~\cite{goldenfeld-1992}.

\begin{figure}[ht]
    \centering
    \includegraphics[width=0.80\textwidth]{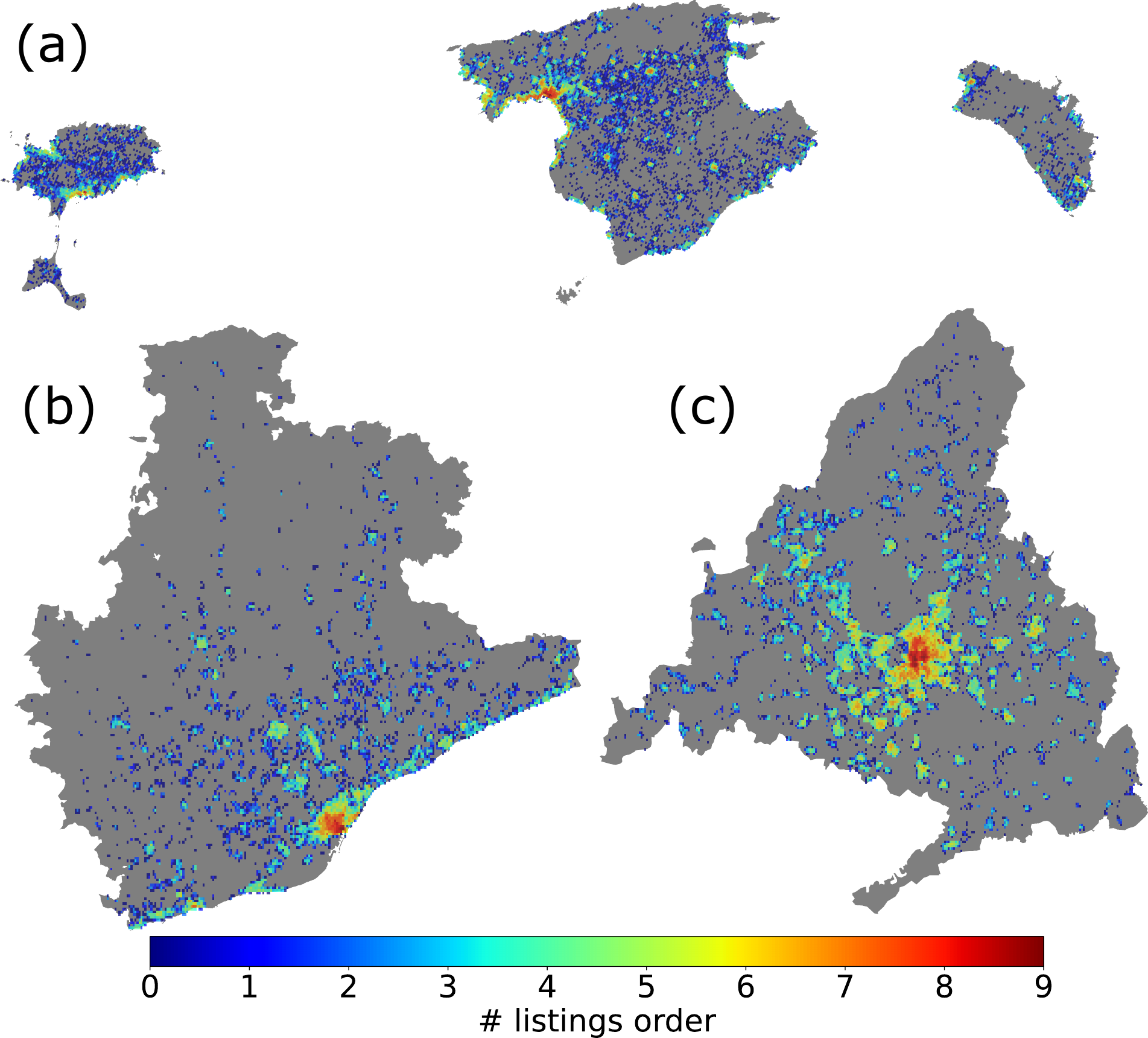}
    \caption[Listings distribution in space]{Distribution of the listings in the dataset in the Spanish provinces of Balearic Islands \textbf{(a)}, Barcelona \textbf{(b)}, and Madrid \textbf{(c)}. The space is divided into $1 \times 1$ km$^2$ cells. The color of each cell is the natural logarithm of the number of listings in the cell. The maps show that the listings are concentrated in the city centers. }
    \label{fig:maps_adds}
\end{figure}

\section{\label{sec:Datasets} Application to a real case scenario}

\subsection{\label{subsec:Idealista dataset} Idealista dataset}

A central part of research in computational social sciences is the use of data to validate theories and develop models to understand the dynamics. In the second part of the thesis, we used mainly one dataset to study the temporal interactions in a complex economic system. This is a dataset of listings published on the online platform \texttt{Idealista.com}~\cite{idealista}, which is a Spanish real estate platform that connects buyers and sellers of real estate properties. The dataset covers a 2-year time period, from January 2017 to December 2018, and it comprises a comprehensive collection of online listings georeferenced with their (lat, long) coordinates in the Spanish provinces of Balearic Islands, Barcelona, and Madrid. These listings were posted by more than $50,000$ real estate agencies, each identified with its unique ID. For each listing, we have information about the price of the property, the type of property (flat, house, etc.), the number of rooms, the surface of the property, the real estate agency that posted the listing, and the dates at which the listing was posted and removed. This dataset comprises over one million listings for sales, and over $800,000$ for rentals. We focus on houses and apartments, and do not consider farms or rural parcels and remove listing with missing information or with an irregular price. The dataset is a snapshot of the real estate market in the 3 Spanish provinces, and it allows us to study the dynamics of the market, the interactions between the real estate agencies, and the emergence of spatial correlations. See Fig. \ref{fig:maps_adds} to see the irregular distribution of the listings in the space for the Spanish provinces of Balearic Islands, Barcelona, and Madrid.

The exploration of this dataset is an interesting case study because it provides a detailed, comprehensive view of the real estate market in key Spanish provinces, which a large metropolis in each province, over a substantial period. The high accuracy georeferenced nature of the data enables sophisticated spatial analyzes, uncovering regional patterns and disparities. This dataset not only aids in validating economic theories and developing predictive models but also offers valuable insights for policymaking, business strategy formulation, and academic research in computational social sciences. On the other hand, it has several limitations. First, the dataset only covers listings from a single platform, potentially missing data from other real estate websites or offline sources, leading to a biased view of the market~\cite{gonzalez2014assessing}. Additionally, the dataset spans just two years, which may not capture long-term trends or the impact of economic cycles. It is worth also to mention that the dataset reflects only the listings, providing a picture of the market's offer side but not the actual transactions, limiting insights into the full market dynamics. These limitations must be considered when interpreting the results and making broader generalizations.

\subsection{\label{subsec:SeLoger dataset} SeLoger dataset}

Moreover, for test properties of the real estate market spatial segmentation, we used a French dataset of listings posted on the online platform \texttt{SeLoger.com}~\cite{SeLoger}. This dataset covers a 6-month time period, from July to December 2019, including over 2 million sale listings across 3 French urban areas: Paris, Marseilles, and Toulouse. These listings are geolocalized by ZIP codes (``\textit{code postal}''), municipalities (``\textit{communes}''), and census tracts (``\textit{IRIS}''), the finest and basic scale for sub-municipal information in France (the lat-long coordinates of the listings are not available). For each listing, we have information about the price of the property, the type of property (flat, house, etc.), and the real estate agency that posted the listing. We focus on houses and apartments, and do not consider farms or rural parcels and remove listing with missing information or with an irregular price, as we did for the Spanish dataset.

\chapterimage{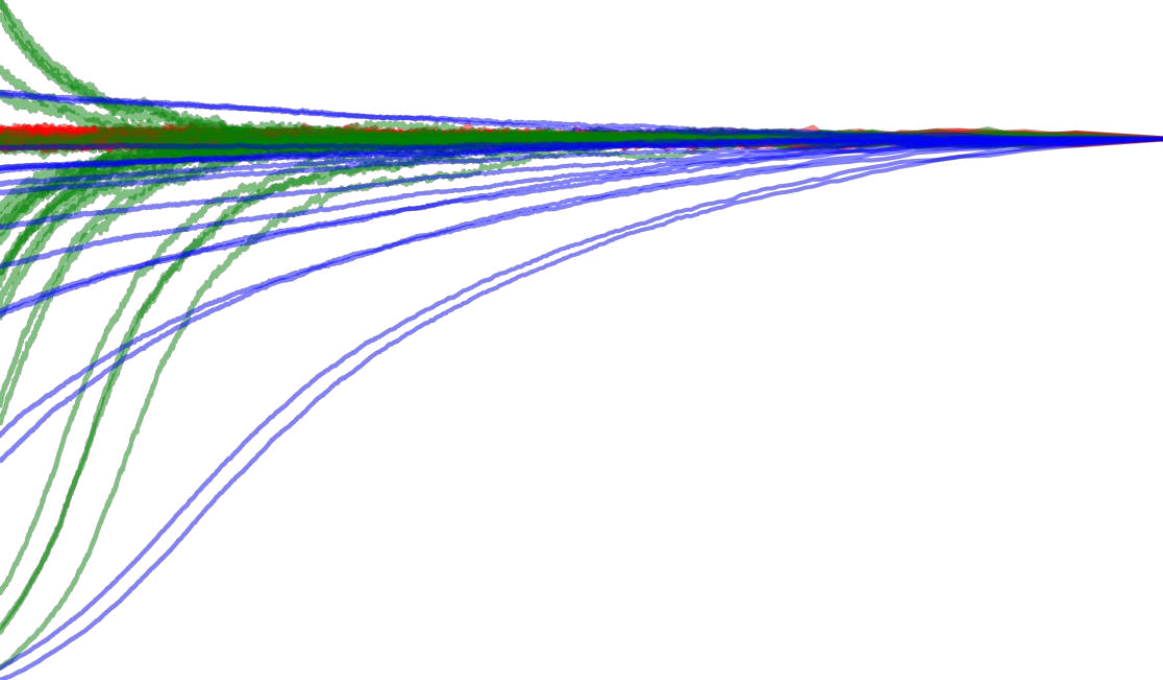}
\chapterspaceabove{6.75cm}
\chapterspacebelow{7.25cm}

\chapter{Threshold models, burstiness and aging}
In this chapter, we will explore in detail the, so called, threshold models and the concept of aging, which will be fundamental to understand the results in the following chapters of the thesis. We differentiate between simple and complex contagion models, two different information transmission mechanisms that have been widely studied in the literature. We introduce two well-known threshold models: the Granovetter-Watts threshold model, a fundamental model for understanding the dynamics of complex contagion in social networks, and the Sakoda-Schelling model, a segregation model that was a precursor of the nowadays agent-based simulations. Furthermore, we also introduce topics such as the bursty dynamics in human interactions and the role of aging, a non-Markovian mechanism that captures the tendency of individuals to stick to their previous beliefs or habits. It is shown how incorporating these factors to traditional Markovian models has important implications to the phase diagram and the dynamics. 

\section{\label{sec:Introduction} Introduction}

Information spreading is a process that has been studied for many years and is present in many social systems, ranging from small groups and communities to large networks and societies at a global scale. This process, often referred to as social contagion~\cite{christakis2013social}, involves the spread of ideas, behaviors, innovations, and emotions (spread of ``information'') among individuals and groups through various forms of social interaction. The metaphor of contagion highlights the similarities between the spread of infectious diseases and the transmission of information, where a single ``infected'' individual can influence multiple others, leading to widespread information.

In this context, binary-state models have emerged as a versatile tool to describe a variety of natural and social phenomena in systems formed by many interacting agents. Each agent is considered to be in one of two possible states: susceptible/infected, adopters/non-adopters, favor/against, etc., depending on the context of the model. In all cases, one of the states represent the presence/spreading of information and the other the absence of it. The interaction among agents is determined by the underlying network and the update rules of the model. Examples of binary-state models include processes of opinion formation and consensus~\cite{Voter-original,sood-2005,fernandez-gracia-2014,redner-2019}, disease or social contagion~\cite{granovetter-1978,pastor-satorras-2015}, among others. 

With the advent of network theory and the increasing availability of large-scale data from online platforms, researchers have been able to study the contagion of ideas with unprecedented precision and detail. Duncan Watts and Steven Strogatz's small-world model~\cite{watts1998collective} and Albert-László Barabási and Réka Albert's work on scale-free networks~\cite{barabasi2009scale} provided key insights into the complex structure of social networks and their role in facilitating or hindering the spread of information and ideas. These studies have shown that the structure of the network, including its degree distribution, clustering coefficient, and community structure, plays a crucial role in determining the dynamics of social contagion processes~\cite{newman2002spread, pastor-satorras-2015}.

On the other hand, the decision-making process in social systems is influenced by a variety of factors, including social media influence~\cite{online-platforms, jstor}, peer pressure~\cite{jensen-2015}, emotional engagement~\cite{ferrara-2015, steinert-2022} and individual preferences. Peer effects and social influence have been shown to play a significant role in the adoption of new technologies, with individuals more likely to adopt new products or services if they see others in their social network doing the same~\cite{rogers2014, valente-1996, bollinger-2012}.

In this chapter, we review the definitions of simple and complex contagion, two different mechanisms that describe how information spreads through social networks. Once we have defined these concepts, we move forward to introduce the Granovetter-Watts and the Sakoda-Schelling models, two fundamental models which update rules are based on a threshold mechanism, a particular case of complex interactions. Moreover, we review a popular theoretical framework useful to treat threshold models in complex networks called the Approximate Master Equation (AME), which will be used and extended in this thesis. Finally, we review the concepts of bursty human dynamics and aging mechanism, which show that the Markovian assumption is not always valid in the study of social dynamics.

\section{\label{sec:Simple and Complex Contagion} Simple and Complex Contagion}

In the study of social contagion, researchers distinguish between two main types of contagion processes: simple contagion and complex contagion~\cite{centola-2007}. Simple contagion refers to the spread of ideas, behaviors, or innovations primarily through single exposures or interactions, much like the transmission of infectious diseases. This process is characterized by the principle that an individual's likelihood of adopting a new idea or behavior increases with each additional exposure to that idea or behavior within their social network~\cite{christakis2007spread, fowler2009cooperative}. In contrast, complex contagion involves multiple exposures or reinforcements from different sources within the network, often requiring a critical mass of adopters before an individual is influenced to adopt the idea or behavior~\cite{granovetter-1978,centola-2007,centola-2010}.

\begin{figure}
    \centering
    \captionsetup{font=sf}
    \includegraphics[width=\textwidth]{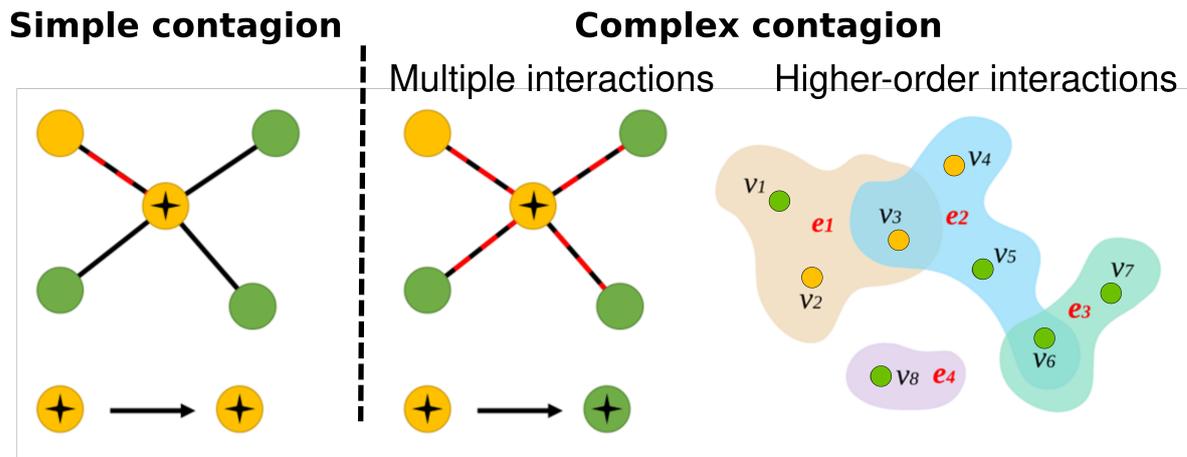}
    \caption[Simple and complex contagion processes]{Comparison between the different types of social interaction. {\bfseries Simple contagion}, where the agent considers just the pairwise interaction with one social contact (interaction highlighted with a dashed red line) and {\bfseries Complex contagion}, where the agent considers the interaction with multiple social contacts. There are two distinguishable types of Complex contagion: {\bfseries Multiple pairwise interactions}, where the agent considers the interaction with all social contacts (interactions highlighted with dashed red lines) and {\bfseries Higher-order interactions}, where the agent considers the interaction with a group of social contacts, all at once, in a single interaction (not pairwise). The green, yellow colors represent the state (idea, position, political party...). (The hypergraph representation is from Ref.~\cite{de-arruda-2020}).}
    \label{fig:SimpleComplexContagion}
\end{figure}

Simple contagion is often described as a process that involves only dyadic interactions, where the adoption of an idea or behavior is facilitated by direct contact between two individuals (see Fig. \ref{fig:SimpleComplexContagion}). This type of contagion is fundamental to understanding how information, beliefs, or diseases spread through populations via direct, pairwise connections~\cite{pastor2001epidemic, newman2002spread}. Features of simple contagion include the rapid dissemination of information, continuous phase transition between full adoption - no adoption states, and the efficient spread of both beneficial and detrimental behaviors across social ties~\cite{christakis2007spread, fowler2009cooperative}.

In contrast, Complex contagion takes place in scenarios where adoption is not merely a result of dyadic interactions but also involves group dynamics and/or the reinforcement from multiple sources within the network. This type of contagion often requires a critical mass or threshold of adopters at the individual's surroundings to trigger the adoption of information~\cite{centola-2007,centola-2010}. This condition that characterizes complex contagion can be understood in two ways: (i) as a reinforcement of the idea or behavior from multiple pairwise (dyadic) interactions~\cite{centola-2007,centola-2010}, or (ii) as a reinforcement from multiple sources in a single group interaction (higher-order interaction)~\cite{iacopini-2019,de-arruda-2020,battiston-2021}. In the first case, the peer pressure, characteristic of complex contagion processes, is included into the model via a threshold rule, which is designed to be used a simple network of dyadic social contacts. In the second case, the group interaction is included in a higher-order network or hypergraph~\cite{berge1984hypergraphs}, which is a more general representation of the social contacts where the interactions are not restricted to dyads. In this case, the complex contagion process takes place via a single group interaction. See Fig. \ref{fig:SimpleComplexContagion} for a graphical representation of the different examples of complex contagion. Features of complex contagion include global cascades of adoption~\cite{borge2013cascading,gleeson2017temporal}, discontinuous phase transitions, and the emergence of echo chambers and polarization in social networks~\cite{centola-2007,diaz-diaz-2022}.


Moreover, real-world processes are influenced not solely by either simple or complex contagion mechanisms but by a complex interaction between the two (Hybrid contagion)~\cite{min-2018,diaz-diaz-2022}. Such multifaceted interactions give rise to varied outcomes, including phenomena like discontinuous transitions, tricriticality, and echo chambers emergence, all of which profoundly affect how information is spread, how behaviors are adopted, and how collective actions are formed.

There have been attempts to extract the simple/complex nature of a process from real data. For example, by analyzing the correlation between the infection order of network nodes and their local topology, it is possible to infer the type of contagion process that is taking place~\cite{cencetti-2023}. Nevertheless, the classification of contagion processes remains a challenging task, as the dynamics of social contagion are influenced by a multitude of factors and high-quality data related to the infection process is often scarce.

\section{\label{sec:Granovetter-Watts threshold model} Granovetter-Watts threshold model}

In this thesis, we are interested in the dynamics of complex contagion driven by multiple interactions in a network of dyadic social contacts. In particular, we focus on a particular category of models called \textbf{threshold models}.

Threshold models represent a critical conceptual framework in understanding how individual behaviors aggregate to produce collective outcomes, especially in contexts where decisions are influenced by the actions of others~\cite{granovetter-1973,granovetter-1978}. By defining a "threshold" - the point at which an individual's perception of the collective behavior of others prompts them to act - these models offer insights into the pivotal role of social influence and network structure in driving large-scale changes from small initial actions~\cite{dodds-2004}. Rooted in the interdisciplinary nexus of sociology, economics, and network theory, these models illuminate the mechanics behind phenomena as diverse as social movements, technological adoption, market dynamics, and even cascading failures within infrastructures. All these phenomena share a common thread: the need for a critical mass of adopters to trigger a response, a threshold that must be crossed to initiate a cascade of adoption~\cite{centola-2007,centola-2010}.

When we talk about threshold models, the model that comes to our minds is the threshold model introduced by Mark Granovetter in 1978~\cite{granovetter-1978}, exploring how individual adoptions depend on the proportion of others adopting the behavior, highlighting the nonlinear nature of social influence and the importance of group interaction in complex contagion processes. In this model, each individual has a threshold that determines the number of neighbors they need to observe adopting a behavior before they themselves adopt it. This threshold can be interpreted as a measure of an individual's susceptibility to social influence, capturing the idea that some people are more likely to adopt a behavior if they see few other people doing the same, while others may require more convincing or reinforcement before they act. This concept of threshold was inspired from Thomas C. Schelling work on segregation~\cite{schelling-1969}, in which the threshold is understood as the maximum tolerance of different neighbors that an individual can withstand before changing their behavior. Duncan J. Watts, in 2002, built upon Granovetter's concept, applying mathematical analysis to explore the model within complex networks~\cite{watts-2002}. His work, particularly on how minor initial actions can lead to large cascades, further elucidated the cascade condition dependence on both individual thresholds and network structures. This model, named as the Granovetter-Watts threshold model, has since become a cornerstone of research on complex contagion and collective behavior, offering a powerful lens through which to study the cascade dynamics in complex networks.

\begin{figure}
    \centering
    \captionsetup{font=sf}
    \includegraphics[width=\textwidth]{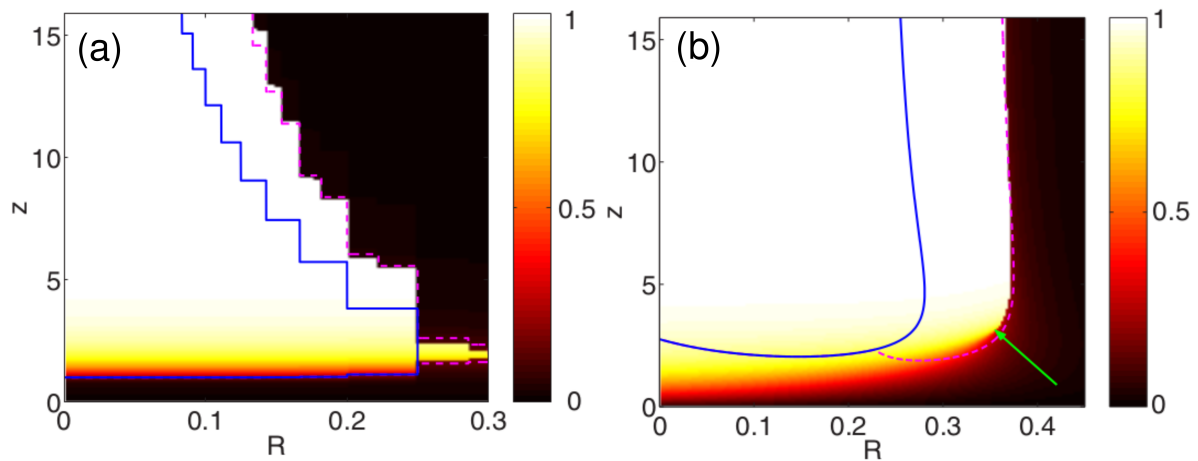}
    \caption[Cascade diagram of the Granovetter-Watts model]{Average density $n$ of nodes adopting as a heatmap for the Granovetter-Watts model. The simulations run in an ER graph of mean degree $z$ and uniform threshold value $R$  \textbf{(a)} and threshold distributed is Gaussian with mean $R$ and standard deviation $0.2$ \textbf{(b)}. Initial adopters seed is set $n_0 = 0.01$. Lines show a first (blue) and a second (purple) order approximations to the cascade condition (see equations in Ref.~\cite{gleeson-2007}). The phase transition is discontinuous. Image from Ref.~\cite{gleeson-2007}.}
    \label{fig:Cascade_gleeson}
\end{figure}

\begin{theorem}[Granovetter-Watts model]
    \sffamily\small
    An individual time step of the model is defined as follows:
    \begin{enumerate}
        \item Each node $i$ has a threshold $R_i$.
        \item At each time step, a node $i$ is selected at random.
        \item If the fraction of neighbors that have adopted of $i$ is greater than $R_i$, then $i$ adopts.
    \end{enumerate}
\end{theorem}

The Granovetter-Watts model exhibits a phase transition from a regime where the adoption is rare, where there are only small cascades of adoption and none of them is global, to a regime where the adoption is widespread, where there are large cascades that reach all the system. This phase transition is discontinuous~\cite{watts-2002,gleeson-2007}, and it is characterized by a critical threshold value $R_c$ that separates the two regimes (refer to Fig. \ref{fig:Cascade_gleeson}). The regime where the global cascades are rare, small and localized is a supercritical regime $R > R_c$ while the regime where cascades are fast and global is subcritical $R \leq R_c$. The  discontinuous transition between the two regimes is driven by the interplay between the individual thresholds and the network structure, and it is a result of the collective dynamics of the system (see dependence of $R_c$ with the average degree in Fig. \ref{fig:Cascade_gleeson}).

The exploration of this model has been widespread, encompassing studies on various types of networks including regular lattices and small-world networks~\cite{centola-2007}, as well as on random graphs~\cite{gleeson-2007}. It has also been examined within the contexts of networks with modular and community structures~\cite{gleeson-2008}, networks that exhibit clustering~\cite{hackett-2011,hackett-2013}, hypergraphs~\cite{de-arruda-2020}, and networks characterized by homophily~\cite{diaz-diaz-2022}, among others. In addition, the literature has expanded to cover the effects of varying the rules for adoption, such as incorporating social reinforcement across multiple layers~\cite{chen-2018}, examining the influence of opinion leaders and initial seed size on the process~\cite{liu-2018, singh-2013}, the introduction of on-off thresholds~\cite{dodds-2013}, and analyzing the dynamics when simple contagions compete with complex ones~\cite{czaplicka-2016, min-2018, diaz-diaz-2022}. Further, empirical data have been used to test the predictions of the Granovetter-Watts model, demonstrating its applicability across a wide range of real-world situations~\cite{centola-2010, karimi-2013, karsai-2014, rosenthal-2015, karsai-2016, mnsted-2017, unicomb-2018, guilbeault-2021}.

\section{\label{sec:The Sakoda-Schelling model} The Sakoda-Schelling model}

\begin{figure}
    \centering
    \captionsetup{font=sf}
    \includegraphics[width=0.7\textwidth]{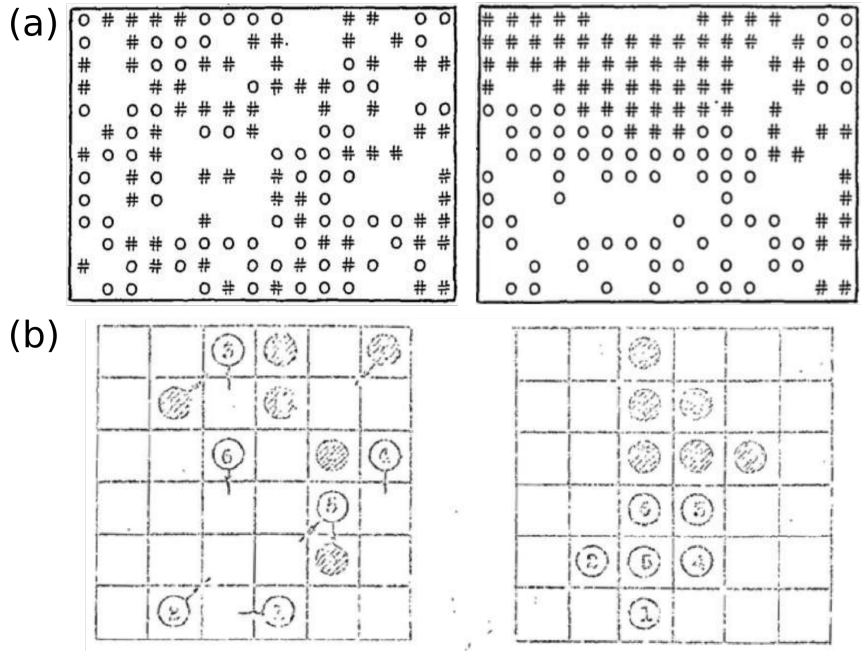}
    \caption[Schelling and Sakoda checkerboard examples]{\textbf{(a)} Examples of the dynamics in Schelling's Segregation Model (from Schelling's original work~\cite{Schelling}). \textbf{(b)} Examples of the dynamics in Sakoda’s Checkerboard Conceptual Model (from Sakoda's original work~\cite{sakoda1949minidoka}). In both cases, the author show at the left the initial configuration of the board and at the right the final segregated configuration after several iterations.}
    \label{fig:Schelling_fig}
\end{figure}

Thomas C. Schelling's segregation model~\cite{schelling-1969}, illustrates how individual preferences regarding neighbors can lead to significant segregation in urban areas, even when these preferences are relatively mild. The model utilizes a checkerboard setup where each agent (representing a household) prefers to live in a neighborhood where at least a certain percentage of neighbors are of the same type (see Fig. \ref{fig:Schelling_fig}). There are locations in the checkerboard that do not have an agent, these will be called vacancies. Agents move to a new vacancy location if their tolerance threshold is not met\footnote{\sffamily\small Note that this  tolerance threshold acts as a measure of a critical mass of different neighbors that the individual tolerates (large tolerance allows for more diverse neighborhood while small tolerance requires several similar neighbors to be satisfied). This threshold definition is complementary to the one in the Granovetter-Watts model, where it is a measure of the number of similar neighbors required to adopt (small threshold allows adopting with a more diverse neighborhood while large threshold requires several similar neighbors to adopt).}. This simple rule leads to complex patterns, showing that even a slight preference for similar neighbors can result in highly segregated communities, an insight that has profound implications for understanding social dynamics and urban planning. 

\begin{theorem}[Schelling's model]
    \sffamily\small
    An individual time step of the model is defined as follows:
    \begin{enumerate}
        \item Each node $i$ has a tolerance threshold $T_i$.
        \item At each time step, a node $i$ is selected at random.
        \item If the fraction of different kind neighbors of $i$ is greater than $T_i$, then $i$ moves to a neighboring location where the fraction of different kind neighbors is less than $T_i$.
            \begin{itemize}
                \item If there is no available location, then $i$ remains in the same location.
            \end{itemize}
    \end{enumerate}
\end{theorem}

On the other hand, James M. Sakoda's model, initially conceptualized in his 1949 dissertation and fully introduced in Ref.~\cite{sakoda1971checkerboard}, offers a more nuanced approach to modeling social interactions using a similar checkerboard framework. Unlike Schelling's, Sakoda's model incorporates a broader range of social interactions by allowing agents to exhibit positive, neutral, or negative attitudes towards their neighbors. These attitudes influence the agents' movements across the board, aiming to optimize their local environment according to specific utility functions that aggregate the effects of surrounding agents. Sakoda's model is capable of simulating a variety of social phenomena beyond segregation, such as the formation of stable social clusters and the dynamics of group interactions~\cite{hegselmann-2017}. 

\begin{theorem}[Sakoda's model]
    \sffamily\small
    An individual time step of the model is defined as follows:
    \begin{enumerate}
        \item Each node $i$ has an attitude matrix $A_i$, that defines the attitudes of $i$ towards all other agents on the board.
        \item At each time step, a node $i$ is selected at random.
        \item $i$ evaluates the total utility for each neighboring location based on the sum of influences (according to the attitudes) from all other agents on the board, weighted by distance.
        \item $i$ moves to the available location with the highest utility.
    \end{enumerate}
\end{theorem}

An important contribution associated to both models is the use of a checkerboard setup as the computational space where agents (or tokens) reside and interact according to predefined rules. The ``hand-made'' simulations performed by Sakoda and Schelling using this discrete spatial representation has since become a standard framework for studying agent-based models in social systems~\cite{hegselmann-2017}. This structure allows for the exploration of local interactions, emergence of global patterns and interface analysis, providing a powerful tool for understanding the dynamics of social systems.

The results of the Schelling's model demonstrated how even mild personal preferences can unexpectedly lead to significant spatial segregation. Its insights have been applied across economics, sociology, urban planning, and complexity science, profoundly influencing both academic research and practical policy discussions. Schelling's model became a foundational example in agent-based modeling, helping to educate countless researchers and practitioners about the impact of individual actions on broader social patterns. Nevertheless, when we check the update rules, we observe that the Schelling's model is a particular case of the previous Sakoda's model, where agents have a negative attitude towards different-kind agents and a fixed tolerance threshold $T_i$. To honor the original contributions of both authors, we refer to this model through the thesis as the Sakoda-Schelling model.

In particular, the Sakoda-Schelling model has been studied from a Statistical Physics point of view due to its close relation to different forms of Kinetic Ising-like models~\cite{stauffer-2007,stauffer-2013}, and also addressing general questions of clustering and domain growth phenomena, as well as for the existence of phase transitions from segregated to non-segregated phases. For example, the relation with phase separation in binary mixtures has been considered~\cite{Dall_Asta_2008,Vinkovic}, as well as the connection with the phase diagram of spin-1 Hamiltonians~\cite{BEG,BlumeCapel,Gauvin_2009,Gauvin_2010}. In this context, a useful classification of models is to distinguish between two possible types of dynamics~\cite{Dall_Asta_2008}: \textbf{``constrained''}, where unsatisfied agents just move to satisfying vacancies (if possible), and \textbf{``unconstrained''},  where each agent try to improve its satisfaction, but after a move they may remain unsatisfied. In addition, the motion can be short-range (only to neighboring sites, as in the original model) or long-range. Constrained motion has been named ``solid-like'' because it generally leads to frozen small clusters, while unconstrained motion has been considered ``liquid-like'' because it allows for large growing clusters~\cite{Vinkovic}. Including the motion of satisfied agents leads to a noisy effect playing the role of temperature in a statistical physics approach~\cite{Gauvin_2009,Gauvin_2010}.

Despite there has been attempts, the description of the phase diagram and the transitions in the Sakoda-Schelling model is a difficult task. In Ref.~\cite{lucquiaud2022modeliser}, the authors reduce the model to a simpler binary-state version that allows to obtain the segregated-mixed phase transition.

\section{\label{sec:Theoretical Framework} Theoretical Framework}

To explain the emergent properties exhibited by the agent based models and its computer simulations, we need to develop a theoretical framework that captures the essential features of the system. This framework should provide a mathematical description of the dynamics, allowing us to analyze the system's behavior and predict its evolution over time. In the context of social contagion and collective behavior, the theoretical framework typically involves a set of differential equations or master equations that describe the evolution of the system's state variables.

The theoretical framework for agent-based models running in complex networks can be broadly classified into two main categories: mean-field approaches and network-based approaches~\cite{barrat-2008}. The so-called mean-field\footnote{\sffamily\small Mean-field is a term borrowed from statistical physics, where it refers to the approximation of the interactions between particles by an average field. In the context of networks, mean-field is related to the approximations of ignoring the local effects and assuming all-to-all connections and infinite system size.} approaches treat the system as a homogeneous entity, where each agent interacts with the average behavior of the entire population. These approaches are well-suited for capturing the macroscopic dynamics of the system and are particularly useful for understanding the collective behavior that emerges from individual interactions. Network-based approaches, on the other hand, explicitly model the interactions between agents as a network structure, where nodes represent agents and edges represent interactions between them. These approaches are valuable for capturing the influence of the underlying network structure on the system's dynamics and for studying the impact of network properties on the spread of information and ideas.

\subsection{\label{sec:Approximate Master Equation} Approximate Master Equation}

A general framework for binary-state models in complex networks was developed by J. P. Gleeson~\cite{gleeson-2011,gleeson-2013}, which provides a general set of differential equations, the Approximate Master Equation (AME), to describe the dynamics of any Markovian binary-state model on a generic network. This framework has been widely used to study the dynamics of social contagion, opinion formation, consensus problems, and other collective behaviors in complex networks. The framework is particularly very useful in the context of thresholds models, which are not well suited for a mean-field approach~\cite{gleeson-2007}, allowing us to identify phase transitions, compute critical thresholds, and predict the final state of the system. The full AME description and derivation can be found in Ref.~\cite{gleeson-2013}, but we will provide a brief summary of the main concepts here.

Consider a system of $N$ nodes in a network, where each node can be in one of two states: $+1$ or $-1$. The state of each node evolves over time according to a set of rules that depend on the states of its neighbors. Let us consider a node $i$, with a degree $k$ (i.e., $k$ connections to other nodes) and $m$ neighbors of $i$ in state $-1$ (e.g., ``adopter''). If node $i$ is in state $+1$ (e.g., ``non-adopter''), the rate \( T^{+}_{k,m} \) defines the probability per unit time that $i$ will switch to state $-1$. Similarly, \( T^{-}_{k,m} \) defines the probability per unit time for $i$, in state $-1$, to switch to state $+1$. These rates are, in general, functions of both the degree $k$ and the number $m$ of neighbors in state $-1$, reflecting how the local network configuration influences state transitions.

Taking into account this framework, the AME can be written as:
\begin{flalign}
    \frac{d}{dt} x^{\pm}_{k,m} = & -T^{\pm}_{k,m} x^{\pm}_{k,m} + T^{\mp}_{k,m} x^{\mp}_{k,m} - (k-m) \beta^{\pm} x^{\pm}_{k,m} + (k-m+1) \beta^{\pm} x^{\pm}_{k,m-1} \nonumber\\
    & - m \gamma^{\pm} x^{\pm}_{k,m} + (m+1) \gamma^{\pm} x^{\pm}_{k,m+1}.
\end{flalign}

Here, $x^{+}_{k,m}$ and $x^{-}_{k,m}$ represent the fractions of nodes with degree $k$ and $m$ infected neighbors that are in state $+1$ and $-1$, respectively. $\beta^{\pm}$ and $\gamma^{\pm}$ are time-dependent rates that describe how the adoption process spreads and recedes across the edges of the network, encapsulating the network's dynamic connectivity and its influence on the spread of states.

A key advantage of the AME is its ability to capture the complex dynamics of networks by considering the interactions between neighboring nodes, making it more accurate than simpler models like the mean-field theory, which assumes independence between nodes. Moreover, from the AME, one can approximate the shape of the solutions $x^{\pm}_{k,m} (t)$ to reduce the number of differential equations, recovering the pair approximation and the heterogeneous mean field~\cite{gleeson-2011,gleeson-2013}.

On the other hand, the AME assumes a tree-like structure with negligible levels of clustering. This assumption implies that there are very few short loops in the network. This tree-like assumption simplifies the calculation and application of the AME by reducing the network's complexity, and becomes very useful for networks generated with the configuration model~\cite{newman-book}, with any given degree distribution, at the limit $N \to \infty$. Another limitation of the AME is based on the formulation itself, since framework is built assuming binary-state Markovian dynamics, which may not always accurately capture the real-world dynamics of social contagion. In fact, in next section, we will introduce the concept of bursty human dynamics, which is an empirical evidence of the presence of non-Markovian effects. These effects can significantly impact the dynamics of social contagion processes.

\section{\label{sec: Bursty Human Dynamics} Bursty Human Dynamics}

Bursty behavior refers to the irregular and sporadically temporal patterns of interactions that include natural phenomena, like earthquakes and neuron firing, as well as human activities, such as communication, mobility, and social dynamics. This section delves into the characteristics of bursty behavior, highlighting by empirical evidence, and discusses its significant implications for modeling human behavior.

\begin{figure}
    \centering
    \captionsetup{font=sf}
    \includegraphics[width=\textwidth]{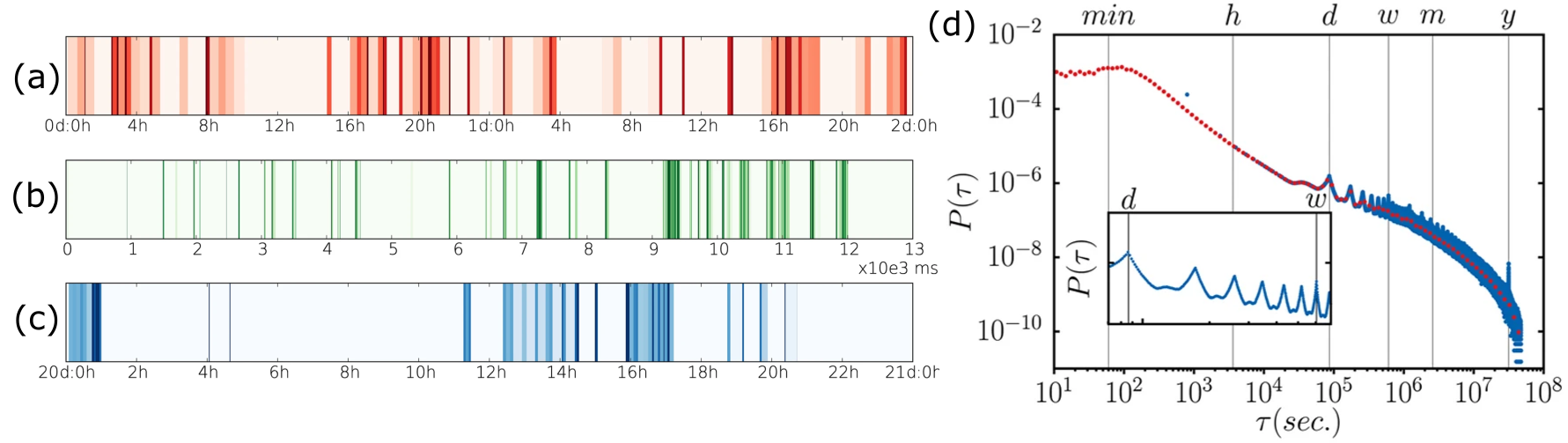}
    \caption[Bursty human dynamics: examples and distribution]{\textbf{(a)} Sequence of earthquakes with magnitude larger than two at a single location (South of Chishima Island, 8t-9th October 1994). \textbf{(b)} Firing sequence of a single neuron (from rat's hippocampus). \textbf{(c)} Outgoing mobile phone call sequence of an individual. Shorter the time between the consecutive events darker the color (a, b, c from Ref.~\cite{karsai2012universal}). \textbf{(d)} Distribution of inter-event times between tweets for several users. Blue and red dots represent the linear-binned and log-binned scales in the $\tau$ axis. The localized maxima in the tail of the distribution correspond to circadian rhythms, as shown in the bottom inset  (from Ref.~\cite{artime-2017}). The distribution is heavy-tailed, indicating bursty behavior.}
    \label{fig:bursty_human_dynamics}
\end{figure}

Human activities often exhibit complex temporal patterns characterized by bursts---short periods of high activity together with longer periods of inactivity (see examples Fig. \ref{fig:bursty_human_dynamics}(a-c)). This non-Poissonian behavior, referred to as burstiness, manifests across diverse human-driven processes and is extensively documented in communication dynamics, web browsing habits, and social interactions~\cite{Barabasi2005Bursts, Vazquez2006Bursts}. The seminal work, by A. L. Barabási~\cite{Barabasi2005Bursts}, shown from email communication's data that activity periods do not follow a regular pattern but are clustered in bursts. This phenomenon has since been observed universally across various platforms such as mobile phone calls, emails, tweets, text messaging and social media~\cite{karsai-2011, Miritello2013Capacity,artime-2017,rybski-2012,zignani-2016,kumar-2020,iribarren-2009}. Further research has analyzed burstiness, focusing on the persistence and periodicity of human interactions~\cite{Clauset2007Proximity} or the effects of circadian rhythms~\cite{Jo2012Circadian}, and has extended these analyses to web activity to predict behaviors across different online platforms~\cite{Radicchi2009WebActivity}.

An important evidence of bursty dynamics is the heavy-tailed distribution of inter-event times, indicating that the probability of short inter-event times is higher than expected from a Poisson process (see Fig. \ref{fig:bursty_human_dynamics}(c)). As a result of this bursty human behavior, there is an emergence of heterogeneous degree distributions~\cite{Muchnik2013PowerLaw}, which have been observed in many social systems~\cite{barabasi2009scale}. Further insights into the impact of burstiness on system dynamics come from studies linking it to memory and the structured nature of human dialogues, enhancing our understanding of how past interactions influence future activities~\cite{karsai2012universal, Goh2008Burstiness, Eckmann2004Entropy}.

Traditional models usually rely on numerical simulations to update agents following independent random Poisson processes, called Random Asynchronous Update (RAU), in which the characteristic time a node is updated is the Monte Carlo step, in which all agents have been updated once on average~\cite{fernandez-gracia-2011}. In this process, an exponential interevent time distribution is expected, which does not capture the bursty nature of human dynamics. To address this phenomenon, non-Poissonian models are necessary to provide a better fit for empirical observations~\cite{Vazquez2006Bursts}. We differentiate two main approaches to include bursty human dynamics in our models:

\begin{itemize}
    \item \textbf{Activity-driven models (nodes get activated):} These models incorporate the temporal aspects of human activity by assigning activity potentials to nodes within a network, dictating the likelihood of interactions based on observed human activity patterns~\cite{stark-2008,van-mieghem-2013,starnini-2017}.
    \item \textbf{Temporal networks (links get activated):} These models incorporate time-stamped interactions, such that at each time step our interaction network changes \cite{Holme2012Temporal, Perra2012ActivityDriven}.
\end{itemize}

While both approaches have been successful to include bursty human dynamics, they offer different perspectives on the underlying mechanisms driving these behaviors: activity-driven models emphasize the burstiness of individual attempts to interact with others, while temporal networks focus on the burstiness of the interactions themselves. The choice of model depends on the specific research question and the level of detail required to capture the dynamics of interest.

It has been shown that implications of bursty behavior are dramatic, influencing the dynamics of network processes such as the spread of epidemics and information diffusion~\cite{Rocha2013Bursts, Wang2009Viruses}. Understanding the mechanism behind this burstiness allows us to improve our predictions, aligning them more closely with natural human activity patterns.

\section{\label{sec:Aging mechanism} Aging mechanism}

`The concept of ``Aging'' is understood in many ways in the literature. Here, aging is one form of memory effect~\cite{jkedrzejewski2018impact} on which the rate of interactions depends on the persistence time of an agent in a state, modifying the transition to a different state~\cite{fernandez-gracia-2011,perez-2016,boguna-2014}. 

Aging was introduced in~\cite{stark-2008} taking as inspiration the non-equilibrium dynamics of spin glasses, where the effective temperature of the system changes with the time since a given perturbation was applied~\cite{cugliandolo1993analytical}. In a context of social systems, this resistance to change can be interpreted as conformism or laziness. In models of species competition~\cite{ravasz2004spreading}, this would imply that neighboring species are less likely to be displaced at a later stage of growth~\cite{stark-2008}. The motivation behind the aging mechanism here is to capture the tendency of individuals to stick to their previous beliefs or habits, a common feature in human behavior~\cite{granovetter-1973}. Moreover, the longer an individual holds a particular habit, the more he/she will accumulate experience, leading to a higher self-involvement and resistance to change~\cite{lejarraga2011let}. Furthermore, this emotional attachment balances the memory-less and purely rational considerations of traditional models~\cite{granovetter-1985}.

\begin{figure}[t]
    \centering
    \captionsetup{font=sf}
    \includegraphics[width=\textwidth]{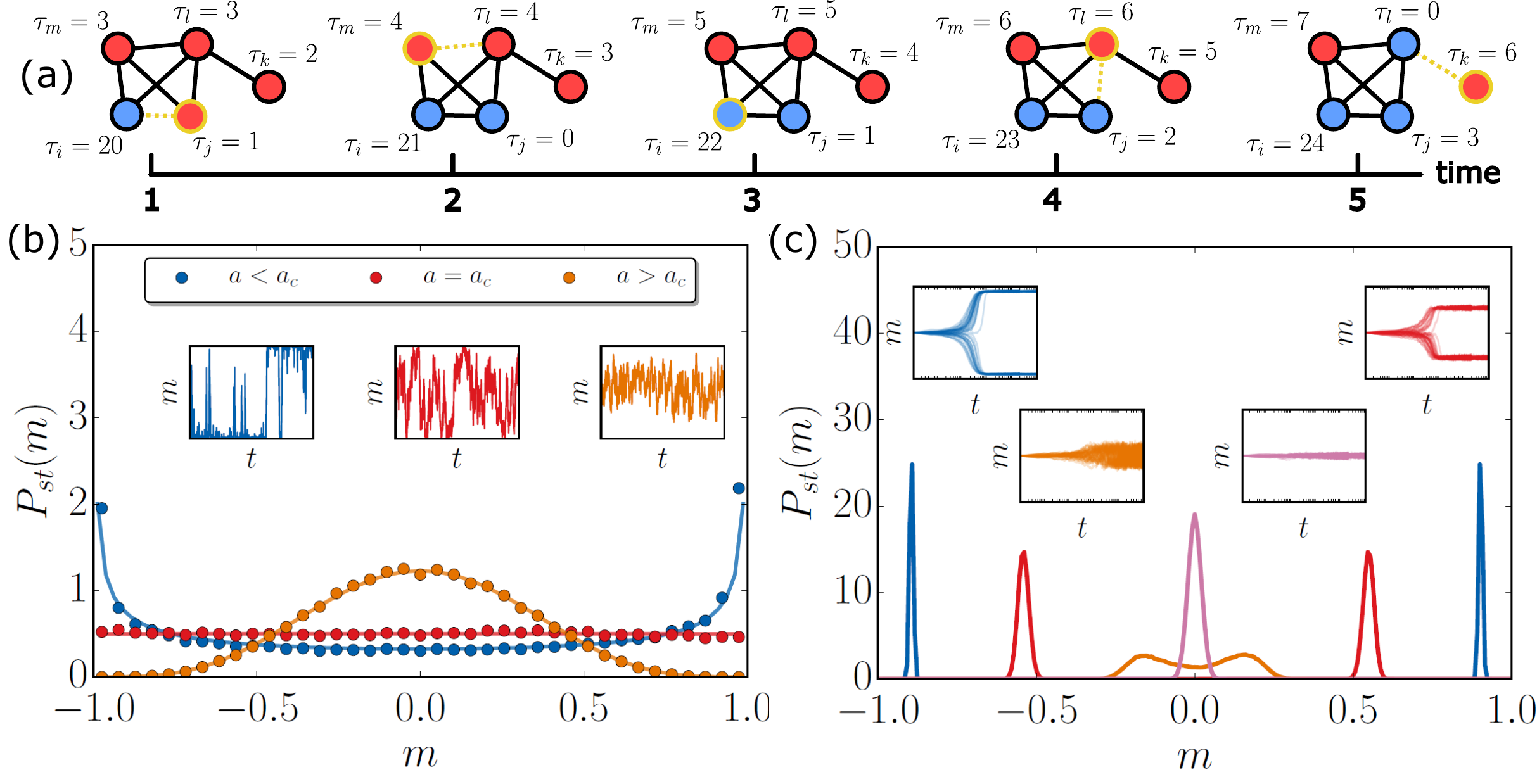}
    \caption[Aging in the Voter model]{\textbf{(a)} Schematic representation of the evolution of the Voter model with aging (given an initial configuration for both states and internal times): At 1, node $j$ activates, copies $i$ and resets age. At 2, node $m$ activates and copies $m$ (same state). At 3, $i$ does not activate due to the age. At 4, node $l$ copies $j$ and resets age. At 5, $k$ activates, copies $l$ and resets age. \textbf{(b)} Stationary probability density function (pdf) of the magnetization in the three different regimes. Points come from simulations, solid lines are the theoretical curves. The insets show one typical trajectory of the dynamics, in each of the regimes. \textbf{(c)} Stationary pdf for the noisy voter model with aging, in the different regimes. The insets show 50 trajectories of the magnetization. (b-c from Ref.~\cite{artime-2017}).}
    \label{fig:aging_pdf}
\end{figure}

The aging mechanism is a non-Markovian effect that can be included in a model via an activation function that modifies the transition rates between states (activity driven model). This function depends on the time since the last transition, allowing us to include bursty dynamics in the individuals' attempts to interact with others (given a proper choice of the activation function~\cite{fernandez-gracia-2011}). This activation function is build such that probability of an individual to interact with another individual decreases with the persistence time in a given state, even though there are studies that also account for anti-aging mechanisms (probability to interact increases)~\cite{peralta-2020C,chen-2020}.

\subsection{\label{sec:Aging in pairwise interactions} Aging in pairwise interactions}

Aging effects have been already shown to modify drastically the dynamics in the Voter model~\cite{Voter-original}, a popular framework for exploring consensus formation in statistical physics and social dynamic. The Voter model is a simple model of opinion dynamics, where agents update their state by copying the state of a randomly selected neighbor. These rules lead to dynamically active state that reaches consensus as a finite-size effect in a finite time, but consensus is not reached at the thermodynamic limit~\cite{Voter-original}. It is shown that, while aging can decelerate microdynamics by making state changes less frequent as agents' states age, it can accelerate macrodynamics, leading the dynamics to a well-defined coarsening process, allowing the system to reach consensus even in the thermodynamic limit~\cite{stark-2008,fernandez-gracia-2011, boguna-2014, perez-2016, peralta-2020C}. This effect is shown in Fig. \ref{fig:aging_pdf}(a), where the aging mechanism is able to modify the dynamics of the Voter model, leading to a faster consensus of the minority state.

In terms of stability, in Voter-like models, incorporating aging exhibits a higher tendency towards reaching consensus than their non-aging counterparts. The persistence of the majority state, reinforced by aging, contributes to this stabilization, making aging a significant factor in determining which is the consensus state~\cite{artime2019herding, peralta-2020C, baron2022analytical}. As an example, aging modifies the nature of the noise-driven phase transition in the noisy Voter model. Specifically, it transforms a finite-size discontinuous transition between ordered and disordered phases into a continuous transition that falls into Ising universality class~\cite{artime-2018} (see discontinuous transition for noisy Voter model in Fig. \ref{fig:aging_pdf}(b) and the continuous transition when aging is included in Fig. \ref{fig:aging_pdf}(c)).

Regarding models of multiple pairwise interactions or higher-order interactions, the aging implications are still an open question, and it is a topic of current research. Just for the specific case of the noisy majority vote model~\cite{chen-2020}, it is known that aging mechanism is able to modify the critical point of the noise-driven disordered-ordered phase transition\footnote{\sffamily\small The noisy majority voter model exhibits a continuous phase transition from an ordered phase, where the system reaches consensus according to the majority in the system, to a disordered phase, where consensus is not reached.}. Further research is needed to understand the joint effect of aging and multiple interactions.


\part{Aging in threshold models}

\chapterimage{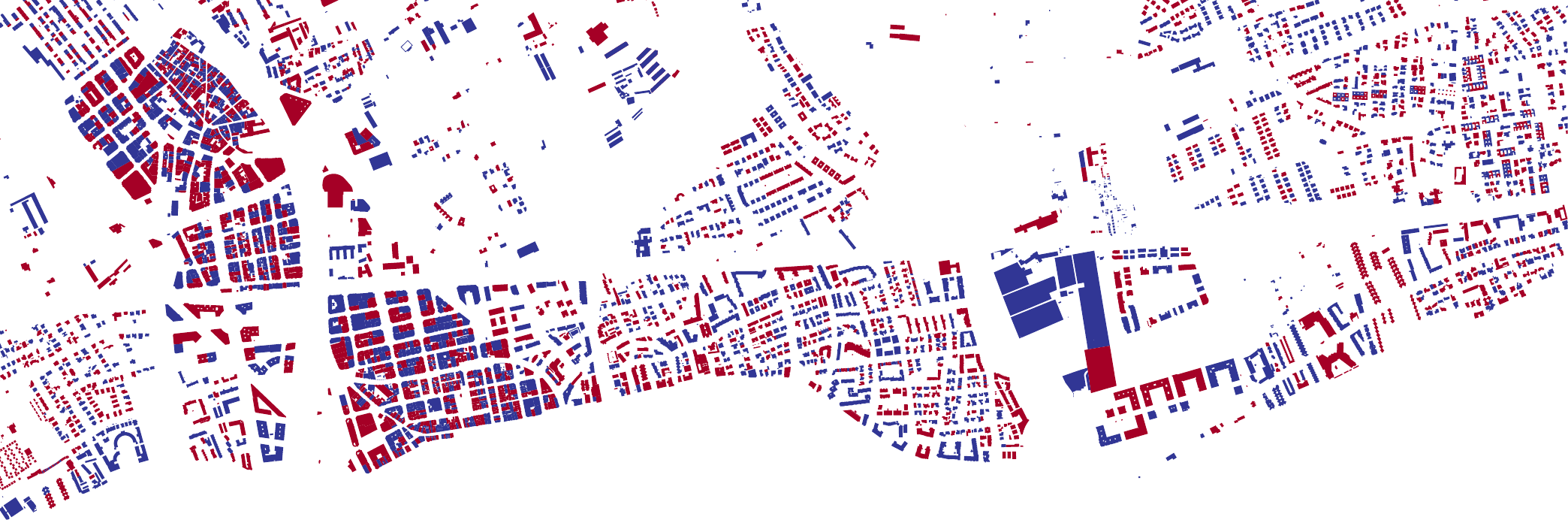}
\chapterspaceabove{6.75cm}
\chapterspacebelow{7.25cm}

\chapter{\label{ch:Aging effects in the Sakoda-Schelling segregation model} Aging effects in the Sakoda-Schelling segregation model}
\vspace{-1.2cm}
\small
\textbf{The results in this chapter are published as:}
\vspace{0.05 cm}

\fullcite{Abella-2022}
\fontsize{13pt}{13pt}\selectfont\sffamily
\vspace{0.5 cm}

We incorporate aging into the Sakoda-Schelling model by making the probability of agents to move inversely proportional to the time they have been satisfied in their present location. This mechanism simulates the development of an emotional attachment to a location where an agent has been satisfied for a while. The introduction of aging has several major impacts on the model statics and dynamics: the phase transition between a segregated and a mixed phase of the original model disappears, and we observe segregated states with a high level of agent satisfaction even for high values of tolerance. In addition, the new segregated phase is dynamically characterized by a slow power-law coarsening process similar to a glassy-like dynamics.

\section{Introduction}

As it was introduced in Section \ref{sec:The Sakoda-Schelling model}, a robust result of the Sakoda-Schelling model is that segregation occurs even when individuals have a very mild preference for neighbors of their own type, so collective behavior is not to be understood in terms of individual intentions. In addition, the model introduced the concept of behavioral threshold that inspired a number of other models of collective social behavior~\cite{granovetter}. But still currently, Schelling's model is at the basis of fundamental studies of the micro-macro paradigm  in Social Sciences~\cite{grauwin-2009}, while it continues to have important implications for social and economic policies addressing the urban segregation problem~\cite{clark-1991,Sassen,Clark,lamanna-2018}. A main limitation of the Sakoda-Schelling model is that it has no history or memory by which, for example, residents might prefer to maintain their present location~\cite{silver-2021}.

As a result of the notable implications of this model and the robustness of the emerging segregation, there exists a vast literature around Schelling's results. Many variants of the original Sakoda-Schelling model have been reported modifying the rules that govern the dynamics, the satisfaction condition, or including other mechanisms, network effects, or specific applications~\cite{Vinkovic,stauffer-2007,Dall_Asta_2008,gracia-lazaro-2009,gracia-2011,Gauvin_2009,Gauvin_2010,domic-2011,henry-2011,unified,Interfacial_roughening,stauffer-2013,lenormand-2015,barmpalias-2018,jensen-2018,holden-2019,sert-2020,agarwal-2020,vieira-2020,ortega-2021,ortega-2021.2}.

With the motivation of established relevant effects of aging in the previous chapter, our goal is to characterize how ``aging'' modifies the segregation dynamics of the Sakoda-Schelling model. In this context, aging must be understood as an emotional/economic attachment to a certain location linked to the persistence time in this location. This attachment balances the memory-less and purely rational considerations of the original model~\cite{granovetter-1985}. The aging-induced inertia, which results in resistance to movement, is a minimalist modeling of behavior with many possible causes. Besides the moving out cost due to the housing market fluctuations, aging accounts for the links established with the neighborhood's public goods, venues, schools, etc., which are known to be highly relevant in this context~\cite{wasserman-2001,chetty-2016,silver-2021}. These urban elements are also a major consideration when households locate~\cite{denton1995persistence,clark-2002,clark-2003,silver2016scenescapes} and aging also accounts for the memory of this decision.


In this chapter, aging is introduced in the Sakoda-Schelling model by considering that agents are less prone to change their location as they get older in a satisfying place. In other words, aging is introduced giving a smaller probability for  satisfied agents to ``move-out'' the longer they have remained in a satisfying neighborhood. We implement this aging mechanism in the long-range noisy constrained version of the Schelling Model, for which a detailed phase diagram was reported~\cite{Gauvin_2009}. We study how this phase diagram is modified by the aging mechanism, finding that aging inhibits a segregated-mixed phase transition. This implies that aging favors segregation, a counter-intuitive result. We also describe the coarsening dynamics in the segregated phase showing that aging gives rise to a slower coarsening that breaks the time-translational invariance.


\section{Aging in the Sakoda-Schelling model}

The model considered here is a variant of the noisy constrained Sakoda-Schelling model~\cite{Gauvin_2009} in which we explicitly include aging effects. For simplicity, we refer to this variant as the Sakoda-Schelling model during the rest of the paper to compare with the model presented here: the Sakoda-Schelling model with aging. For both, the system is established on an $L \times L$ Moore lattice with $8$ neighbors per site and periodic boundary conditions, where agents of two kinds (representing, for instance, wealth levels, race, language, etc) occupy the sites. There are also empty sites (vacancies), where agents can move to, depending on their state and on the vacancy neighborhood. The condition of each site $i$ of the lattice will be described with a variable  $\sigma_i$ that takes three possible values: $\sigma_i = \pm 1$ for the two kinds of agents and $\sigma_i = 0$ for vacancies. In addition, depending on the local environment, agents can be in two states: satisfied or unsatisfied. In our case, agents are satisfied if their neighborhood is constituted by a fraction of unlike agents lower than a fixed homogeneous threshold $T$. Otherwise, they are unsatisfied. Therefore, this control parameter $T$ is a measure of how tolerant the population of the system is. We also need a non-zero vacancy density, $n_0 > 0$, for agents to change their location. This $n_0$ is understood as an extra parameter of the model. The initial configuration is built by randomly distributing the agents ($N_{\rm agents} = L^2 \, (1 - n_0)$). We always consider initially one half of agents of each kind.

In the Sakoda-Schelling model considered in this study, an agent chosen by chance moves to a random satisfying vacancy (if any exists) independently of his/her initial state and of the distance. This process is repeated until the system reaches a stationary state. The movement of unsatisfied agents behaves as a driver for the system dynamics, while the motion of satisfied agents plays the role of noise. When tolerance $T$ becomes larger, more satisfying vacancies are present in the system and the noise consequently increases. 

The aging mechanism in our model is introduced by considering an activation probability of the agents inversely proportional to the time spent at a satisfied location, motivated by the definition for opinion dynamics~\cite{artime-2018}. This methodology was proposed to mimic the power-law like inter-event time distributions observed in real-world social systems~\cite{barabasi-2005,fernandez-gracia-2011}. If an agent $j$ is initially satisfied in her neighborhood, the internal time is set $\tau_j = 0$. Then, in every time step, a randomly chosen agent $j$ follows different rules depending on whether she is originally satisfied or not. If unsatisfied, $j$ moves to any random satisfying vacancy of the system. If satisfied, she moves to another satisfying vacancy with an activation probability $p_j = 1 / (\tau_j + 2)$. In both cases, if no vacancy has a satisfying neighborhood, the agent $j$ remains in the initial site. As before, these rules are iterated until the system reaches a stationary state (if possible). The time is counted in Monte-Carlo steps; after each Monte-Carlo step, that is after $N_{\rm agents}$ iterations, the internal time increases for all satisfied agents in one unit, $\tau_j \to \tau_j + 1$. Notice that, when an unsatisfied agent becomes satisfied due to the neighbor's motion, an internal time $\tau_j = 0$ is set for that agent. As for the Sakoda-Schelling model, there is a noise effect associated with the motion of satisfied agents. In this case, the intensity of this noise is related not only to the tolerance parameter $T$, but to the presence of aging as well. In fact, aging introduces more constraints to the movements and contributes to decreasing the noise.  

Given the number of neighbors available in the Moore lattice, numerical simulations are only performed for a finite set of meaningful tolerance values: $\{1/8,1/7, \cdots ,6/7,7/8 \}$. During all our analysis, we focus on the low vacancy density region of the phase diagram.

\section{Segregation coefficient}

Many metrics have been introduced in the literature to discern if the final state is segregated or not~\cite{Gauvin_2009,lenormand-2015,randomwalks,urban}. The number of clusters is known to be directly related to the segregation because a high presence of small clusters indicates a mixing between agents. As for the Sakoda-Schelling model\cite{Gauvin_2009}, we compute the following metric related to the second moment of the cluster size distribution:
\begin{equation}
s = \frac{2}{\left(L^{2} \, (1-n_0)\right)^{2}} \sum_{\{c\}} m_{c}^{2} ,
\end{equation}
where the index of the sum $c$ runs over all the clusters $\{c\}$ and $m_c$ is the number of agents in the cluster $c$. The average of $s$ over realizations after reaching a stationary state is defined as the segregation coefficient $\langle s \rangle$. This metric is bounded between 0 and 1: $\langle s \rangle \to 1$ if there are only 2 equally-sized clusters, and $\langle s \rangle \to 0$ if the number of clusters tends to the number of agents. The cluster detection is performed using the Hoshen-Kopelman algorithm~\cite{HoKo}.

Another metric of segregation is the interface density~\cite{Dall_Asta_2008}, defined as the fraction of links connecting agents of different kinds. The calculation is done in two steps: estimating the interface density for each agent $j$, $\rho_j$, and then the average over all the agents $\rho$:

\begin{equation}
    \rho_j = \frac{1}{2} \, \left( 1 - \frac{\sigma_j \,  \sum_{k \in \Omega_j} \sigma_k}{\sum_{k \in \Omega_j} \sigma_k^2 } \right) \quad \rm{and} \quad \rho = \frac{1}{N_{\rm agents}} \sum_{j = 1}^{N_{\rm agents}} \rho_j ,
\end{equation}

where the indices $k$ run over the neighborhood of agent $j$, $\Omega_j$. If an agent $j$ is surrounded only by vacant sites, we define by convention $\rho_j = 0$. Performing a realization average of $\rho$, we obtain the average interface density $\langle \rho \rangle$ in the stationary state is denoted as $\langle\rho_{\rm{st}} \rangle$. The time evolution of this metric, not present in literature, allows us to study the coarsening process.

\section{Results} 

\subsection{Phase diagram}

To discuss the phase diagram of our model, we focus on the region of parameters with a vacancy density $n_0 < 50 \%$ to avoid diluted states with a majority of vacancies.
For this region, the Sakoda-Schelling model presents 3 different phases~\cite{Gauvin_2009}: frozen, segregated and mixed. For low tolerance values, the system freezes in a disordered state, given that there are no satisfying vacancies for any kind of agent. With increasing tolerance, the system undergoes a transition toward a segregated state, which is characterized by a 2-clusters dynamical final state. Finally, for high values of $T$, after another transition, we find a dynamical disordered (mixed) state, in which a vast majority of vacancies are satisfying for both kinds of agents, and small clusters are continuously created and annihilated.

\begin{figure}
\centering \captionsetup{font=sf}
\includegraphics[width=0.95\linewidth]{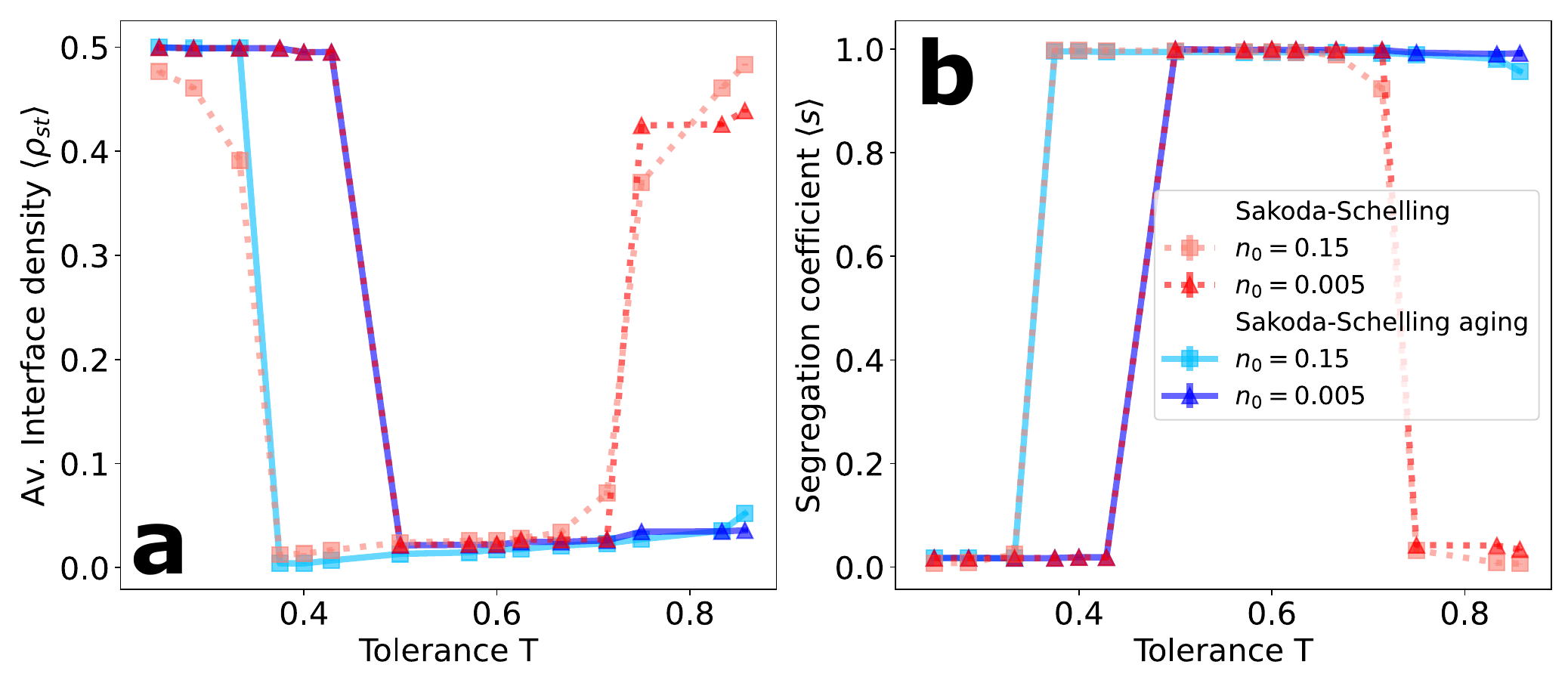} 
\caption[Average interface density and segregation coefficient]{Average interface density $\langle\rho_{\rm{st}} \rangle$ (\textbf{a}) and segregation coefficient $\langle s \rangle$ (\textbf{b}) at the stationary regime as a function of the tolerance parameter $T$ for two values of the vacancy density $n_0 = 0.5\%$ and $15\%$. Results are shown for both the Sakoda-Schelling model and the variant with aging introduced in this paper. Simulations are performed on an $80\times 80$ lattice and averaged over $5 \cdot 10^{4}$ realizations.}
\label{Fig1}
\end{figure}

These three phases are characterized by measuring the segregation coefficient $\langle s \rangle$ and the average interface density $\langle\rho_{\rm{st}} \rangle$ at the final state. The results for the original model are depicted as a function of the tolerance $T$ in Fig. \ref{Fig1}a for the interface density and in Fig. \ref{Fig1}b for the segregation coefficient. At low values of T, both indicators show a disordered state that falls in the frozen phase.  We also observe a dependence of the transition point with the vacancy density. On the other hand, for high $T$ values, the transition point between segregated and mixed states has no dependence on the parameter $n_0$. Notice that mixed and frozen states present a very similar value of $\langle s \rangle$ but can be differentiated by the stationary value of the average interface density $\langle\rho_{\rm{st}} \rangle$. These results are in agreement with the results reported for the Sakoda-Schelling model\cite{Gauvin_2009}, with the extra information provided by the average interface density.

The first quite dramatic effect of including aging in the system is the disappearance of the mixed state from the phase diagram (see blue lines in Fig. \ref{Fig1}). In both metrics, the difference between the models with and without aging is clearly manifested. For low $T$ values, the frozen-segregated transition behaves similarly to the original model since aging has no implications as the system gets quickly frozen. Nevertheless, for high values of the tolerance $T> 0.5$, the segregated-mixed transition disappears, and the segregated phase is always present. This is not an intuitive effect and one would think that aging, contributing to difficult agent's mobility, should prevent the system from forming fully developed segregated clusters. However, it is just the opposite, and it favors cluster prevalence.

\subsection{Segregated phase: final state}

\begin{figure}
    \centering \captionsetup{font=sf}
    \includegraphics[width=\linewidth]{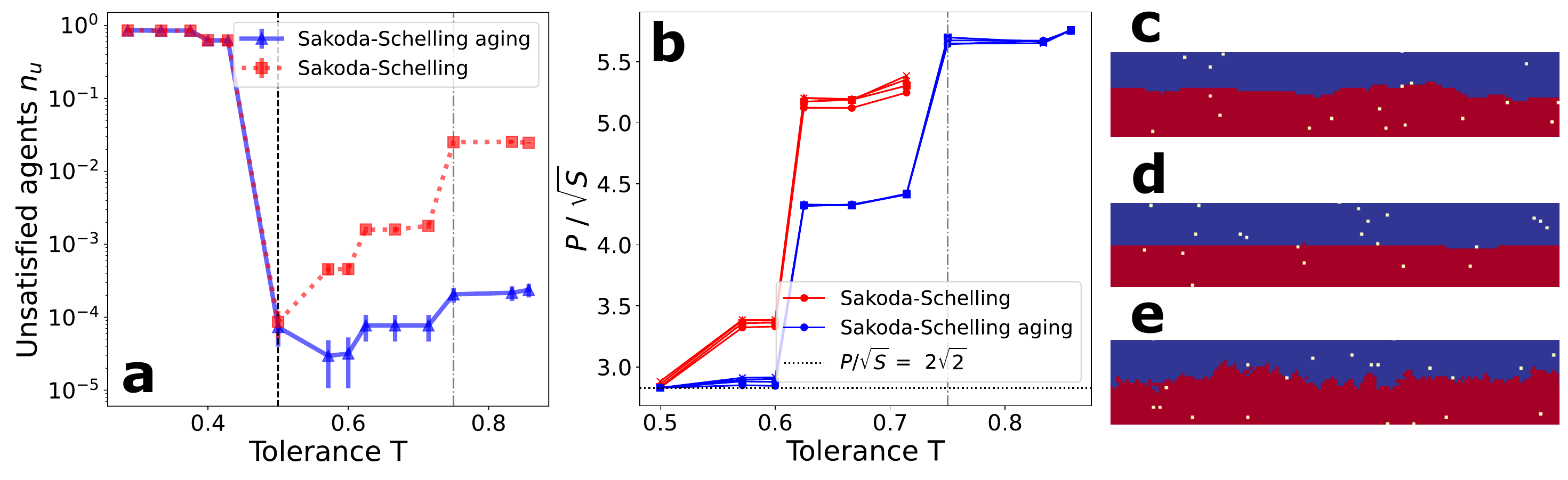} 
    \caption[Fraction of unsatisfied agents and roughness]{ (\textbf{a}) Fraction of unsatisfied agents $n_u$ at the stationary regime as a function of the tolerance parameter $T$. (\textbf{b}) Measure of the interface roughness between clusters of different kind of agents at the final stationary state $P/\sqrt{S}$ as a function of the tolerance parameter $T$. Different markers indicate different system sizes: $L = 40$ (circles), $60$ (squares), $80$ (triangles) and $100$ (crosses). Results are shown for both the Sakoda-Schelling model with and without aging. Numerical simulations are performed for $n_0 = 0.5\%$ and averaged over $5 \cdot 10^4$ realizations. The frozen-segregated transition (dashed black line) and the segregated-mixed transition (gray dot-dashed line) are highlighted to differentiate the phases that the Sakoda-Schelling model exhibits. There are no values of $P/\sqrt{S}$ for the Sakoda-Schelling model above $T = 3/4$ because the segregated-mixed transition occurs. (\textbf{c}) Final state interface zoom snapshot for $T = 0.57$ using the original model. (\textbf{d}) Final state interface zoom snapshot for $T = 0.57$ using the model with aging.  (\textbf{e}) Same as c for $T = 0.86$.}
    \label{Fig2}
    \end{figure}

To gain further insights into the differences in the system dynamics that lead to the extended segregated phase, we compute the fraction of unsatisfied agents at the stationary regime $n_u$ (see Fig. \ref{Fig2}a). This metric plays a role as a marker for the frozen-segregated transition, as shown for the 1D Sakoda-Schelling model~\cite{Dall_Asta_2008}. The frozen phase presents a big majority of unsatisfied agents for both models. After the transition, this parameter decays to very low values in the segregated phase, where a majority of agents are satisfied. In this phase, we observe a step-like increasing behavior of the unsatisfied agents with $T$. As the tolerance grows, the number of satisfying vacancies increases and the noisy movement of satisfied agents drives the system evolution, creating eventual unsatisfied agents in the sites that they abandon or target. However, in the Sakoda-Schelling model, the transition to a mixed state at $T = 0.75$ inhibits the creation of clear fronts between agents of different kinds, and it is also associated to a sharp increase of $n_u \simeq 0.05 $ (red squares in Fig. \ref{Fig2}a). The Sakoda-Schelling model with aging, on the other hand, shows a lower fraction of unsatisfied agents during all values of the tolerance above the frozen-segregated transition (blue triangles in Fig. \ref{Fig2}a). So much so, that many realizations reach $n_u = 0$ and this causes the large error bars in Fig. \ref{Fig2}a after the transition. In a counter-intuitive way, the introduction of aging causes a higher global satisfaction when compared with the original model in both the segregated and the mixed phases.

The creation of new unsatisfied agents at the final stationary state occurs at the interface, where different kind agents meet. This is why we study the interface roughness (perimeter) $P$ as a function of the tolerance parameter. To compute this measure, we compute the number of agents of one kind in contact with different kind agents. To perform this calculation, we smooth the interface by considering vacancies surrounded by a majority of agents of a certain kind as members of that kind. In our system of $L \times L$ with periodic boundary, the minimum interface size (perimeter) $P$ between clusters of agents of different kind is $P = 2 \, L$. To avoid the $L$ dependency, we calculate an adimensional magnitude $P/\sqrt{S}$, where $S$ is the number of agents of each kind $S = N_{\rm{agents}}/2 = L^2 \, (1 - n_0 )/2 $ (surface). This metric $P/\sqrt{S}$ is computed starting from a flat interface as an initial condition and evolving it for $t_{\rm{max}} = 10^4$ MC steps to reach well within the stationary state. With the metric $P/\sqrt{S}$, we are able to estimate how close is the final state interface of our system to the flat interface ($P/\sqrt{S} = 2 \, \sqrt{2}$). The results show an increasing dependence of roughness with the tolerance parameter $T$ (see Fig. \ref{Fig2}b). This growth can be explained as an increase in tolerance means that agents are satisfied with fewer ``same-kind'' neighbors. Therefore, the interface is able to be rougher, keeping the agents in a satisfied state. In addition, notice that all values with different $L$ collapse, so the dependence on the system size has been eliminated. 

Comparing both models, one observes a lower interface roughness for the Sakoda-Schelling model with aging, regardless of the value of $T$. The closest value to the flat interface occurs for the first values of $T$ after the frozen-segregated phase transition (shown in Fig. \ref{Fig2}d). In the original model, we observe higher values of $P/\sqrt{S}$ due to the noise produced by the satisfied agents' behavior (see Fig. \ref{Fig2}c). Moreover, aging allows us to obtain a segregated phase with even larger interface roughness than the maximum observed in the original model for large values of $T$ (see Fig. \ref{Fig2}e). We remark that, when aging is introduced, agents try to join those of their own kind but are less and less prone to change location as time passes. Thus, in the Sakoda-Schelling model with aging, agents in the bulk of the clusters mainly do not move and those moving more often are located at the interface between agent kinds. At medium and large scales, this phenomenon leads to ergodicity breaking in the final state dynamics.

\subsection{Segregated phase: coarsening dynamics}

\begin{figure}
\centering \captionsetup{font=sf}
\includegraphics[width=\linewidth]{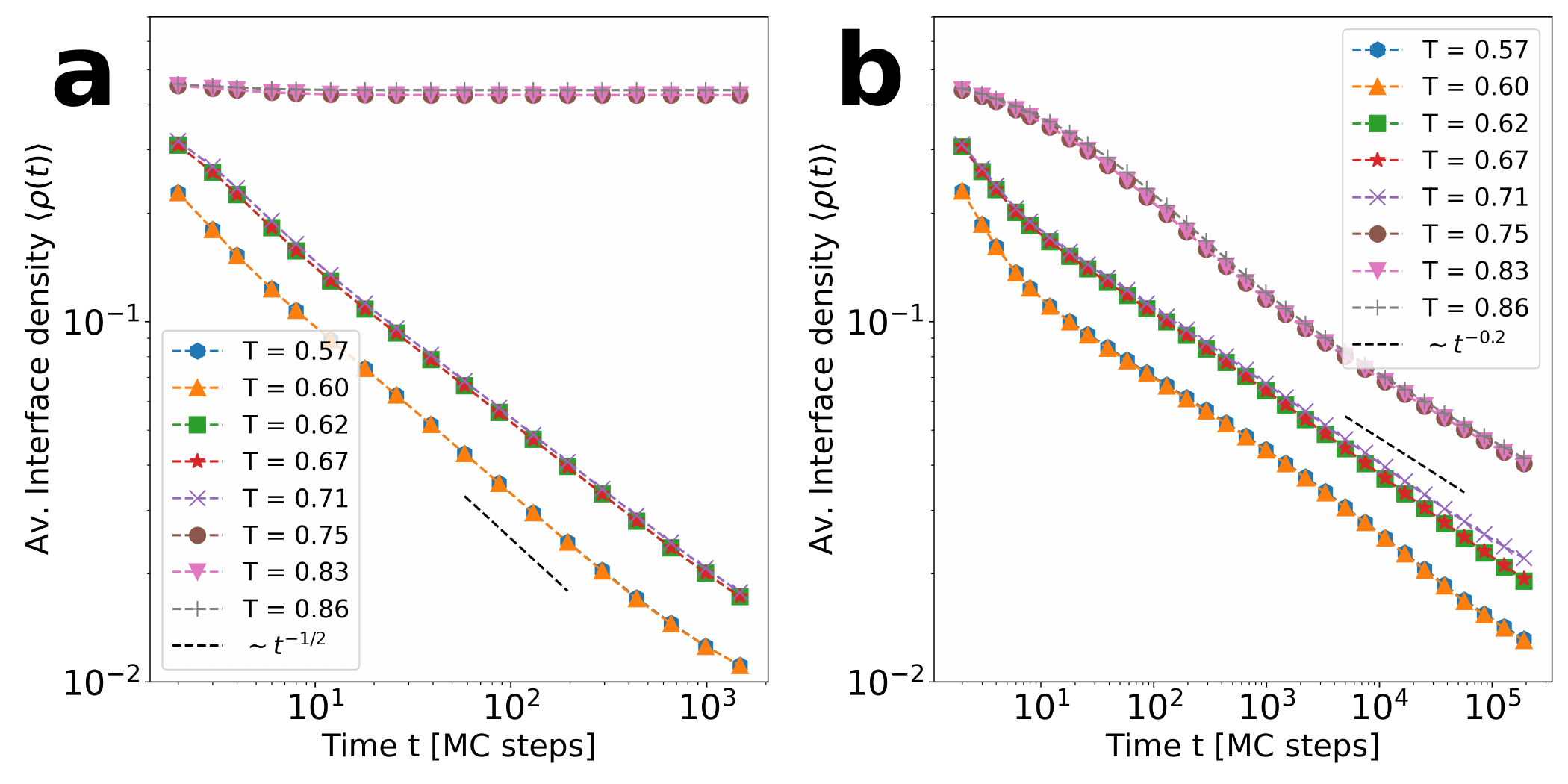} 
\caption[Average interface density evolution]{Average interface density $\langle \rho (t) \rangle$ as a function of time steps for different values of the tolerance parameter $T$ using the Sakoda-Schelling model (\textbf{a}) and the version with aging (\textbf{b}). Average performed over $5 \cdot 10^3$ realizations. Fitted power-law in a black dashed line highlighting the estimated exponent value. We set system size $L = 200$ and $n_0 = 0.005$.}
\label{Fig3}
\end{figure}

Diverse versions of the original Schelling Model exhibit different behaviors in terms of coarsening dynamics. Recent publications report a power-law like domain growth~\cite{Dall_Asta_2008,Interfacial_roughening}. We monitor here the evolution of the interface density $\langle \rho (t) \rangle$, which, in the segregated phase, decreases as $ \langle \rho (t) \rangle \sim t^{-\alpha}$, so the domains should grow in our model following a power-law with time. 

\begin{figure}[t!]
    \centering \captionsetup{font=sf}
    \includegraphics[width=0.9\linewidth]{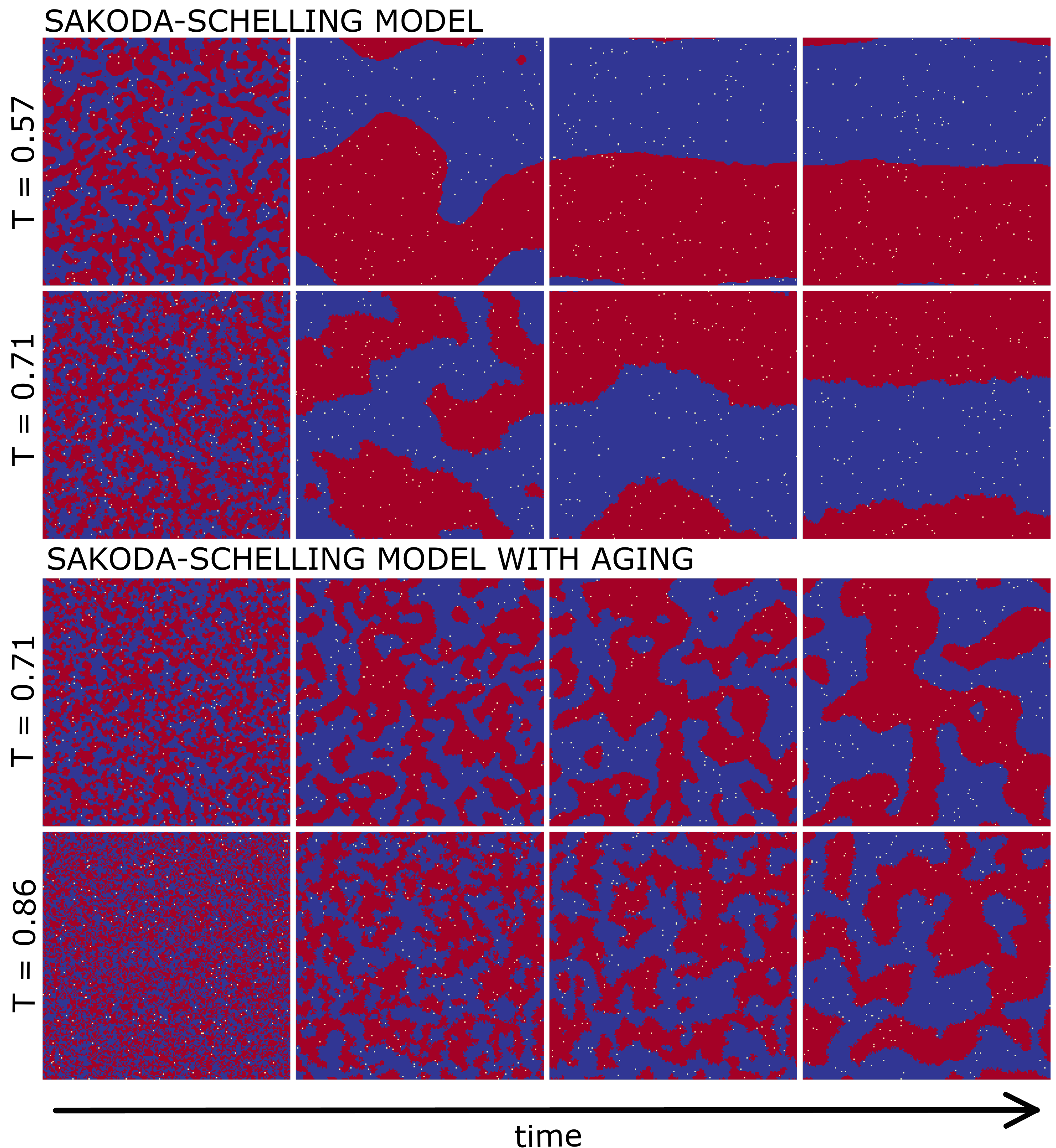}
    \caption[Coarsening towards the segregated state]{Coarsening towards the segregated state at two different values of $T$ for both models. Snapshots are taken for $5$, $500$, $5000$ and $50000$ time steps ordered from left to right. We set system size $L = 200$ and $n_0 = 0.005$.}
    \label{Fig4}
\end{figure}

The coarsening process of the Sakoda-Schelling model at the segregated phase ($0.5 \le T < 0.75$) is displayed in Fig. \ref{Fig3}a and Fig. \ref{Fig4}. We find that the average interface density follows a power-law decay with an exponent $\alpha \simeq 0.5$ for the limit of small vacancy density $n_0 \to 0$, in agreement with the value reported for close variants of the Sakoda-Schelling model~\cite{Dall_Asta_2008}. This exponent value is curious since the coarsening in the presence of a conserved quantity (but with local interactions) exhibits an exponent $\alpha = 1/3$~\cite{Maxi}. Nevertheless, the interactions in this model are not local, and the coarsening exponent is more similar to the one in systems with a non-conserved order-parameter ($\alpha = 1/2$). Fig. \ref{Fig3}a shows as well how coarsening changes with the tolerance parameter. Even though the exponent $\alpha$ does not depend on $T$, we observe a certain delay when increasing $T$ from $0.6$ to $0.62$. In the system evolution of Fig. \ref{Fig4}, one can see how the behavior of the satisfied agents for higher tolerance values is translated into rougher interfaces, causing such delay. For $T > 0.75$, the system exhibits a transition towards a mixed state where the interface density fluctuates around $\rho = 0.5$, indicating that the state is constantly disordered.

The Sakoda-Schelling model with aging shows very different behavior (Fig. \ref{Fig3}b). As expected, the average interface density exhibits a power-law decay with time for all values of the tolerance $T$ after the frozen-segregated transition. Still, the decay is slower than for the Sakoda-Schelling model, with $\langle \rho (t)\rangle \; \sim \; t^{-0.2}$. A mechanism that could be behind this behavior is that the model with aging counts more satisfied agents than the original model, and their probability to move becomes lower as time goes by. Moreover, satisfied agents inside a cluster will not move and the dynamics of the model take place at the interface. It is, therefore, more difficult for separated clusters to collide and merge, an effect that slows down the decay of the interface density. The persistence of small clusters becomes clear when the snapshots' evolution is compared for both models at the same tolerance value $T = 0.71$ (see Fig. \ref{Fig4}). Moreover, while for the original model the initial clustering for $t = 500$ steps does not determine the final state, in the case of aging the bigger clusters present at the beginning of the evolution are the ones that keep growing, determining the shape of the system configuration after $50000$ time steps. This is a dynamical effect, because the system in both cases tends to a final configuration with 2-clusters.

In the case of the Sakoda-Schelling model with aging, we observe an early cross-over in the dynamics (Fig. \ref{Fig3}b). For $T < 0.75$, the coarsening starts with an initial decay of $\langle \rho (t)\rangle$ faster than $t^{-0.2}$. This occurs because, due to the initial condition, the aging effects become relevant after a certain time from the initial state, and initially the system behaves as in the original model. Similarly, for $T \ge 0.75$, $\langle \rho (t)\rangle$ decays slowly for a moment before reaching the power-law behavior for large $t$ values.  Confirming this scenario, Fig. \ref{Fig4} shows that for $T = 0.86$, the system starts evolving similarly to a mixed state until some clusters are created. At this moment, aging prevents the small clusters' dissolution, leading the system very slowly to coarsening dynamics and, eventually, to a fully segregated state. 

Regarding the relaxation time to the final state, we see in Fig. \ref{Fig4} how for $T = 0.71$, the stationary state of the Sakoda-Schelling model is reached after approximately $t = 5000$ time steps. In contrast, the version with aging needs much more than $50000$ steps to attain it. This highlights the important temporal  difference between both models in terms of domain growth dynamics, which strongly increases the computational cost of the study of the stationary state of the model with aging. We have been thus able to study only medium and small system sizes in this final regime (see videos in Ref.~\cite{supplementary_Abella_2022}).

The dynamics studied thus far are performed considering the limit $n_0 \to 0$, but the analysis can be extended to higher vacancy densities. For the particular case of high $n_0$ and low $T$, aging leads to the formation of a vacancy cluster at the interface between domains (see details in Appendix \ref{app:Vacancy density effect on the Schelling model dynamics}).

\subsection{Aging breaks the asymptotic time-translational invariance}

Here, we explore further time translational invariance (TTI) of the model dynamics. For this, we start by defining the two-time autocorrelation function $C(\tau,t_{\rm{w}})$~\cite{spinglassbook} as

\begin{equation}
    C(\tau,t_{\rm{w}}) = \left\langle \frac{1}{M} \, \sum_{i = 1}^{N}  \sigma_i (t_{\rm{w}} + \tau) \,  \sigma_i(t_{\rm{w}}) \right\rangle ,
\end{equation}
where $N$ is the system size,  $\langle \cdot \rangle$ refers to averages over realizations, $t_{\rm{w}}$ is the waiting time to start the autocorrelation measurements, $\tau$ a time interval after $t_{\rm{w}}$ and $M$ is a normalization factor defined as
\begin{equation}
M =  \sum_{i = 1}^{N}  (\sigma_i (t_{\rm{w}} + \tau) \, \sigma_i(t_{\rm{w}}))^2 . 
\end{equation}
which is computed at each realization. 

\begin{figure}
\centering \captionsetup{font=sf}
\includegraphics[width=\linewidth]{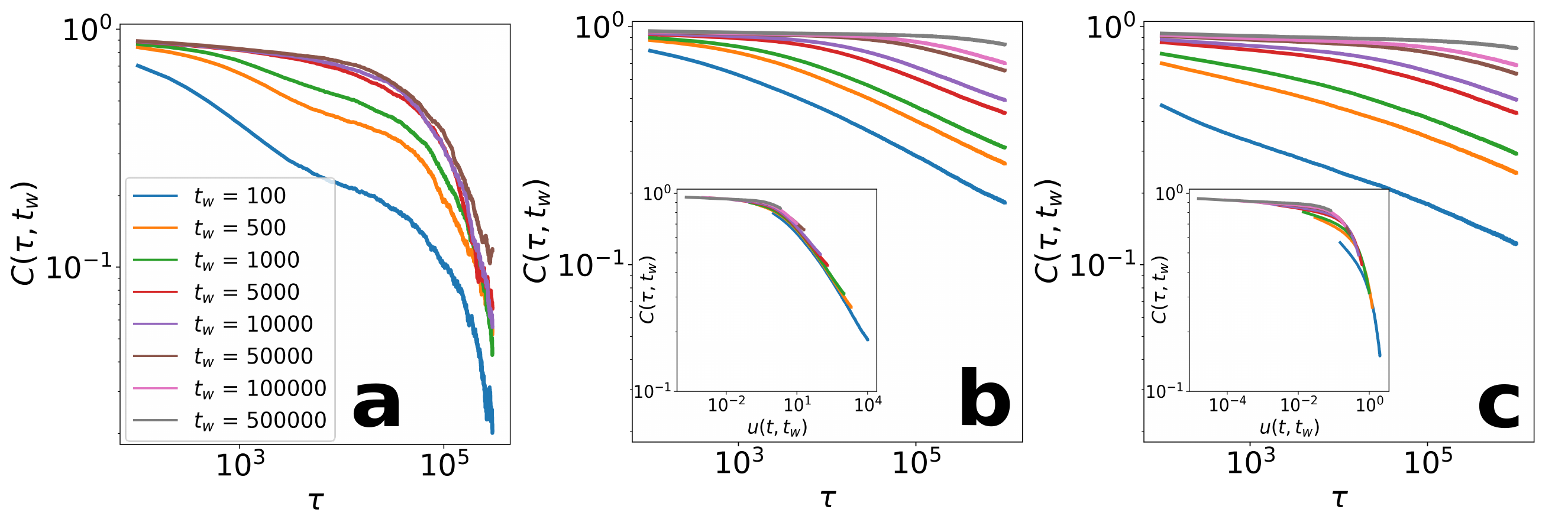} 
\caption[Two-times autocorrelation]{Two-times autocorrelation $C (\tau,t_{\rm{w}})$ as a function of the time period passed since the waiting time $t_{\rm{w}}$. First, the autocorrelation is shown for the Sakoda-Schelling model at $T = 0.71$ in \textbf{a}, and for the version with aging at $T = 0.71$ in \textbf{b} and $T = 0.86$ in \textbf{c}. The insets are the result of the collapse using $u(\tau,t_w) = \tau/t_w$ (\textbf{b}) and $u(\tau,t_w) = \log(\tau+t_w)/\log(t_w) - 1$ (\textbf{c}). The curves correspond to different values of the waiting time $t_{\rm{w}}$. Calculations performed on a $100 \times 100$ lattice averaged over $5 \cdot 10^{4}$ realizations.}
\label{Fig5}
\end{figure}

The autocorrelation function is displayed for the Sakoda-Schelling model with $T = 0.75$ in Fig. \ref{Fig5}a. We observe the curves decreasing with $\tau$ as expected, and that after a characteristic time period ($t_{\rm{w}}^* \approx 5000$ for a system size of $80\times 80$) they collapse into a single curve. This is the regime in which the dynamics becomes TTI, implying that the autocorrelation function does not depend any more on the waiting time, $C(\tau,t_{\rm{w}}) = C(\tau)$ for $ t_{\rm{w}} > t_{\rm{w}}^{*}$. 

For the Sakoda-Schelling model with aging, the dynamics show some different features (Figs. \ref{Fig5}b and \ref{Fig5}c). First, the autocorrelation functions decay slower with $\tau$ in all the cases, which is connected to the long-lived small clusters mentioned previously. We do not find in the simulations any value of $t_{\rm{w}}^*$ for the systems to fall into a TTI regime. Not only that, but a scaling relation including both $\tau$ and $t_{\rm{w}}$ can be applied to collapse the autocorrelation curves (see insets Figs. \ref{Fig5}b and \ref{Fig5}c). This behavior is similar to glassy systems~\cite{spinglassbook}, therefore it is useful to use the mathematical description for those systems in our case. In this type of dynamics, a final stationary state is not attainable in the thermodynamic limit, and it is possible to decompose the autocorrelation function into an equilibrium part and an ``aging'' part (aging in the sense of non-equilibrium dynamics in glassy systems) ~\cite{spinglassbook,Heisemberg}:

\begin{equation}
    C (\tau, t_{\rm{w}}) \simeq C_{\rm{eq}}(\tau) \; C_{\rm{aging}} u(\tau,t_w) = C_{\rm{eq}}(\tau) \; C_{\rm{aging}} \left( \frac{h (\tau) }{h(t_{\rm{w}})} \right),
\end{equation}

where $C_{\rm{eq}}$ describes the fast relaxation of the system components within each domain (TTI term), $C_{\rm{aging}}$ is a scaling function and $u(\tau,t_w)$ is a normalization factor which, in some cases, can be written as the quotient of an unknown function $h(t)$ at the two times $\tau$ and $t_{\rm{w}}$. This function $h(t)$ is known to be related to the dynamical correlation length ~\cite{Heisemberg,8Heisemberg}. In our case, we use $h(t) = t$ to scale the results in Fig. \ref{Fig5}b (see inset). This scaling is valid for values of $T \in [0.5,0.75)$. Nevertheless, higher values of $T$ do not hold a linear scaling, and we need to turn to other functional forms as the normalization factor $u(\tau,t_w) =  \log(\tau+t_w)/\log(t_w) - 1$ used in  Fig. \ref{Fig5}c. This indicates that for $T > 0.75$, the dynamical correlation length evolves in a different and slower way.

\section{\label{sec:Summary and Conclusions_Schelling} Summary and discussion}

We have studied the effect of aging on a 3-state threshold model (with two symmetrical states $\sigma_i = \pm 1$), which combines long-range mobility with local short-range interactions. Specifically, taking as basis the noisy constrained Sakoda-Schelling model, we assign to the agents an internal clock counting the time spent in the same satisfying location. The probability of changing state decreases then inversely proportional to this time. Therefore,  older satisfied agents are less prone to update resident locations. The original model displays a transition between a segregated phase and a mixed one as the tolerance control parameter $T$ increases. This transition disappears when aging is introduced into the system, the mixed phase is replaced by a segregated phase even for high values of the tolerance parameter $T$. As a result, the model with aging presents a higher global satisfaction than without this effect for all values of the tolerance.

On the dynamical perspective, the relaxation towards the segregated phase features a coarsening phenomena characterized by a power-law decay of the average interface density with time $\langle \rho \rangle \sim t^{-\alpha}$. For the original model in the limit of low vacancy density, the exponent is around $\alpha = 1/2$. This exponent is also reported in other variants of the Sakoda-Schelling model~\cite{Dall_Asta_2008,Interfacial_roughening}. Aging gives rise to long-lived small clusters and a slower coarsening, reducing the exponent to $\alpha \simeq 0.2$. We investigated the autocorrelation functions in the segregated phase and found that aging breaks the asymptotic time-translational invariance of the dynamics. This result, along with a nontrivial scaling of the autocorrelation functions, establish close similarities with low-coarsening systems, such as glassy systems, and our Sakoda-Schelling model with aging for high values of the tolerance parameter. Moreover, this work studies the case for equal size populations ignores effects arising from the competition between different population sizes. Further work would be to study a joint effect of minority population and aging.

As for the implications of our results from a social perspective, we must note that the fact that aging favors segregation, inhibiting the segregation-mixed phase transition, is rather counter-intuitive, but gives support to the argument that segregation is a stochastically stable state and may prevail in an all-integrationist world~\cite{Zhang}. Our model predicts the appearance of segregation even for tolerance  values close to one. Additionally, the model relaxation time multiplies manifold, which implies that if aging is present the natural state of this system seems to be generically out of equilibrium.   

\chapterimage{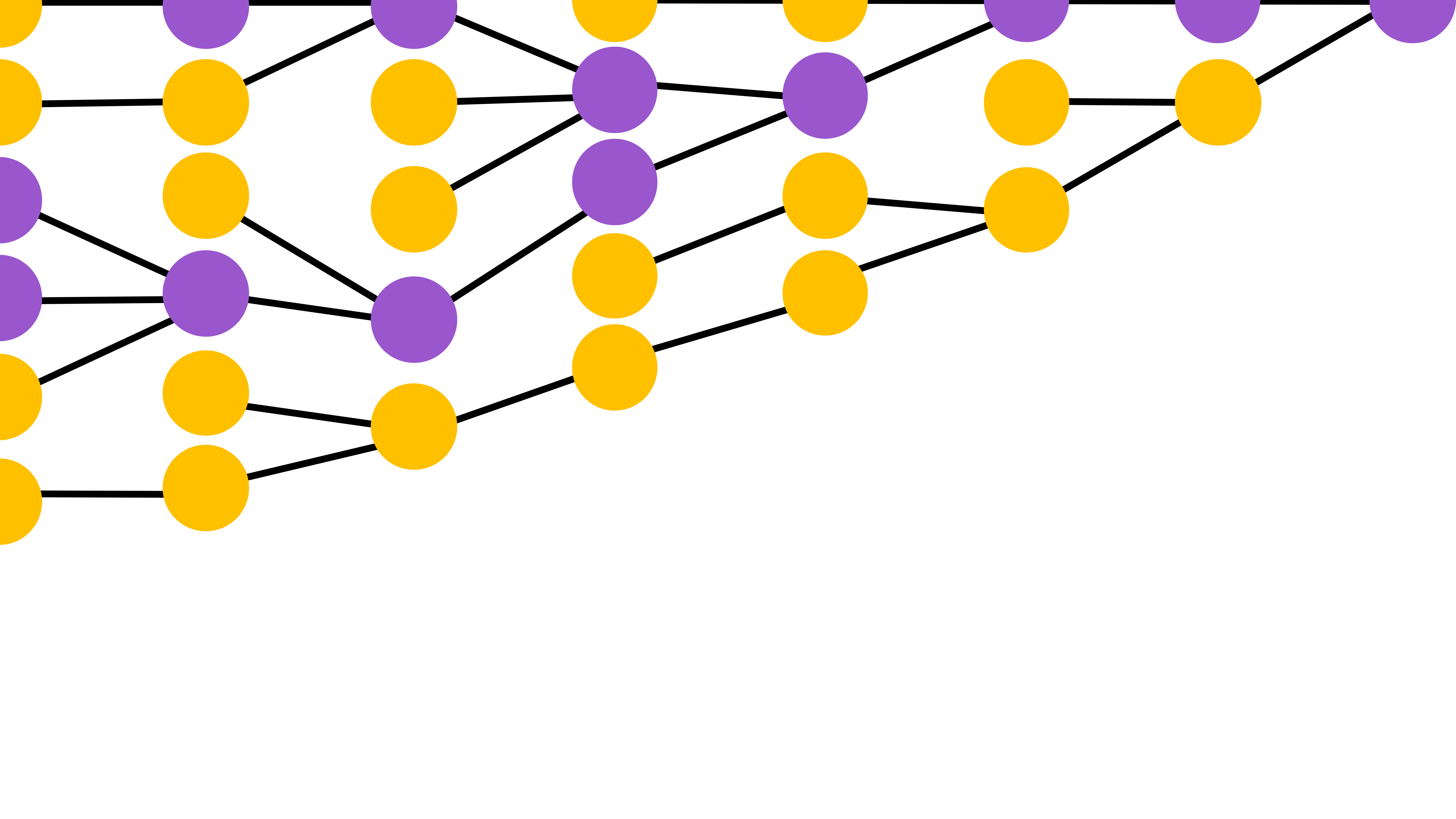}
\chapterspaceabove{6.75cm}
\chapterspacebelow{7.25cm}

\chapter{\label{ch:Aging in binary state dynamics} Aging in binary state dynamics: The Approximate Master Equation}
\vspace{-1.2cm}
\small
\textbf{The results in this chapter are published as:}
\vspace{0.05 cm}

\fullcite{Abella-2022-AME}
\normalsize
\vspace{0.5 cm}

The Approximate Master Equation (AME) is a general mathematical equation that allows to describe binary state models in complex networks. Here, we extend the traditional mathematical framework to include aging effects, which account for the influence of the persistence time of an agent in a given state on the transition rate to a different state. When aging is considered, the Markovian assumption is no longer valid, and the AME must be modified to include non-Markovian dynamics. We derive the AME for binary-state models with aging effects, including aging and resetting events, and show how it can be reduced to the original Markovian dynamics when the rates are not dependent on the internal time. We also demonstrate how the heterogeneous mean-field approximation can be derived from the master equation. The results presented in this chapter provide a comprehensive framework for studying aging effects in binary-state models, offering a more accurate description of the dynamics of complex networks. In the following chapters of this thesis, we apply this framework to describe the aging implications in two binary-state models: the Granovetter-Watts model and the Symmetrical Threshold model.

\section{\label{sec:Introduction_binary} Introduction}

Binary-state models are a versatile tool to describe a variety of natural and social phenomena in systems formed by many interacting agents. Each agent is considered to be in one of two possible states: susceptible/infected, adopters/non-adopters, democrat/republican, etc, depending on the context of the model. The interaction among agents is determined by the underlying network and the dynamical rules of the model. There are many examples of binary-state models, including processes of opinion formation~\cite{Voter-original,sood-2005,fernandez-gracia-2014,redner-2019}, disease or social contagion~\cite{granovetter-1978,pastor-satorras-2015}, etc. Extended and modified versions of these models can lead to very different dynamical behaviors than in the original model. As examples, the use of multi-layer ~\cite{diakonova-2014,diakonova-2016,amato-2017} or time-dependent networks~\cite{vazquez-2008}, higher-order interactions~\cite{de-arruda-2020, iacopini-2019, cencetti-2021}, non-linear collective phenomena~\cite{castellano-2009,peralta-2018}, noise~\cite{carro-2016} and non-Markovian~\cite{van-mieghem-2013,starnini-2017,peralta-2020A,chen-2020} effects induce significant changes to the dynamics.

Theoretical and computational studies of stochastic binary-state models usually rely on a Markovian assumption for its dynamics. This implies that events depend only on the present state, i.e., dynamical rules are memoryless. Markovian processes exhibit exponential distributions in the upcoming events times and the number of events in a given time interval follows a Poisson distribution. However, there is strong empirical evidence against this assumption in human interactions.  For example, bursty non-Markovian dynamics with heavy-tail inter-event time distributions, reflecting temporal activity patterns,  have been reported in many studies~\cite{iribarren-2009,karsai-2011,rybski-2012,zignani-2016,artime-2017,kumar-2020}. The understanding of these non-Markovian effects is in general a topic of current interest~\cite{van-mieghem-2013,starnini-2017,peralta-2020C,peralta-2020A}. In particular, for the threshold models, memory effects have been included as past exposures' memory~\cite{dodds-2004}, message-passing algorithms~\cite{shrestha-2014}, memory distributions for retweeting algorithms~\cite{gleeson-2016} and timers~\cite{oh-2018}.

Aging is an important non-Markovian effect that we address in this chapter for binary-state models. We here provide a general theoretical framework to discuss aging effects building upon a general Markovian approach for binary-state models~\cite{gleeson-2011,gleeson-2013}. We build an Approximate Master Equation (AME)\footnote{\sffamily\small We use here the term  ``master equation'' for consistency with  Refs.~\cite{gleeson-2011,gleeson-2013}, but the word ``master'' has a different meaning than the one used to describe an equation for the probability distribution~\cite{toral2014stochastic, peralta-2020B}} for any binary-state model with aging effects (including aging and resetting events). We show how the AME can be reduced to the original Markovian dynamics when the rates are not dependent on the age of the agents. Moreover, we also show how the heterogeneous mean-field approximation can be derived from the master equation. The results presented in this chapter provide a solid foundation for future studies on aging effects in binary-state models, offering a more accurate description of the dynamics of complex networks. 

\section{Derivation of the AME for binary-state models with aging \label{sec:Derivation of the Approximate Master Equation for binary-state models with aging}}
    
We consider  binary-state dynamics on static, undirected, connected networks assuming a locally tree-like structure and in the limit of $N \to \infty$, following closely the approach used in Ref.~\cite{gleeson-2013} for binary-state dynamics in complex networks. The new ingredient is to consider the nodes with different "age" or "internal time" as different sets, what allows us to treat as Markovian the memory effects introduced by aging~\cite{peralta-2020C,peralta-2020A}. We define $x^{\pm}_{k,m,j} (t)$ as the fraction of nodes that are in state $\pm 1$ and have degree $k$, $m$ infected neighbors and age $j$ at time $t$. The networks have degree distribution $p_k$ and have been generated by the configuration model~\cite{molloy-1995,newman-2001}. For the models considered in this thesis, the initial condition is set such that all agents have age $j = 0$ and there is a randomly chosen fraction $x^{-}_{0}$ of nodes in state $-1$:
\begin{flalign}
    \label{initial_condition} 
    \textrm{For } j > 0 & \quad \quad    x^{+}_{k,m,j} (0) = 0 \quad \quad \quad \quad \quad \quad \quad \; \; x^{-}_{k,m,j} (0) = 0, \nonumber\\
    \textrm{For } j = 0 & \quad \quad    x^{+}_{k,m,0} (0) = (1 -  x^{-}_{0})\, B_{k,m}[x^{-}_{0}] \quad  x^{-}_{k,m,0} (0) = x^{-}_{0}\, B_{k,m}[x^{-}_{0}],
\end{flalign}
where $B_{k,m}[x^{-}_{0}]$ is the binomial distribution with $k$ attempts, $m$ successes and $x^{-}_{0}$ is the initial fraction of agents in state $-1$ (as the probability of success of the binomial). Now, we examine how $x^{+}_{k,m,j}$ changes in a time step. We consider 3 possible events:

\begin{itemize}
    \item An agent changes state from $+1$ to $-1$ and resets the internal time to $j = 0$, with probability $T^{+} (k,m,j)$.
    \item An agent remains at its state and resets its internal time to $j = 0$, with probability $R^{+} (k,m,j)$.
    \item An agent remains at its state and ages, with probability $A^{+} (k,m,j)$.
\end{itemize}

The probability to change state and to age make sense in the context of aging. The reset probability is introduced to account for ``exogenous'' aging, in which an external influence forces the node to attempt a change of state, but the node remains in its current state. Moreover, notice that we assume all the probabilities to be a function of the degree $k$, the number of neighbors in state $-1$ $m$ and the time spent in the actual state (or since a reset) $j$.
\begin{figure}
    \centering
    \captionsetup{font=sf}
    \includegraphics[width=\columnwidth]{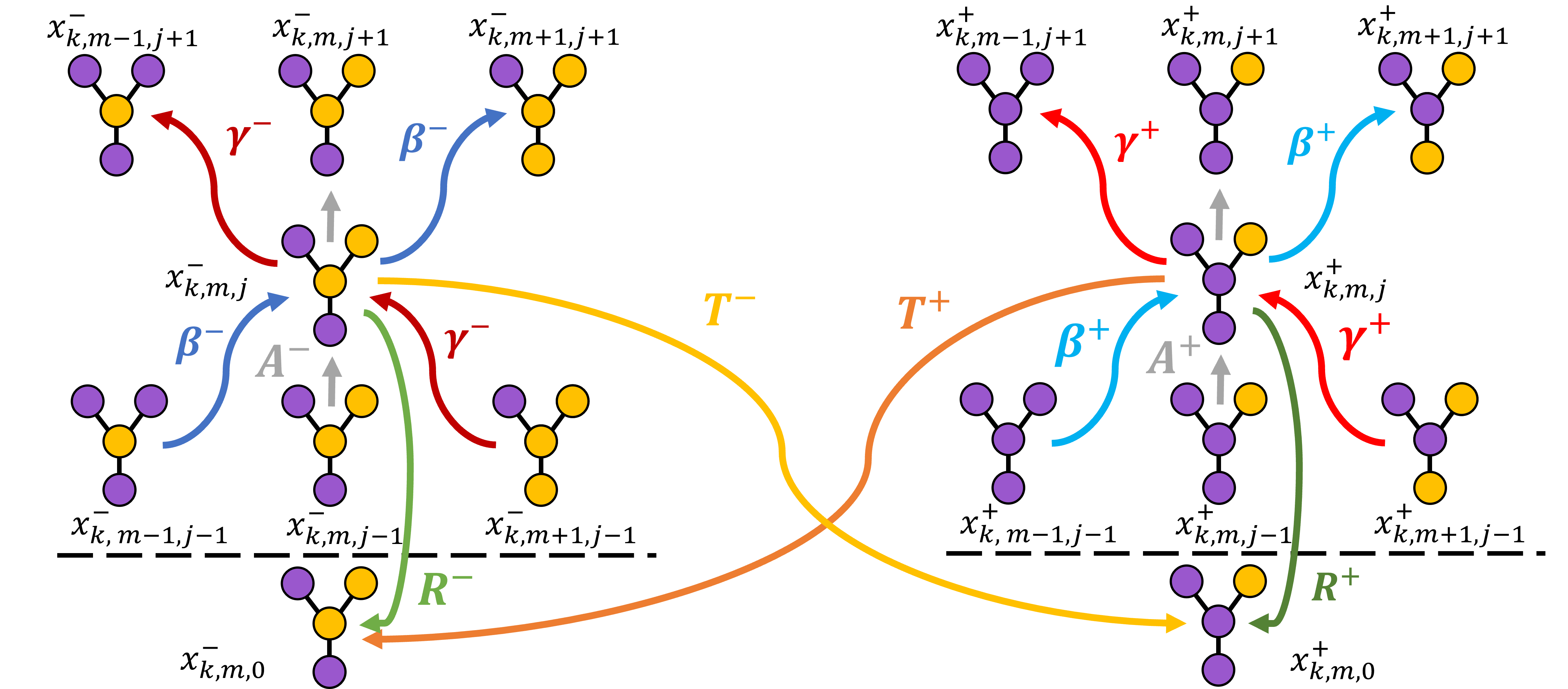}
    \caption[Schematic representation of the transitions to or from the set $x^{\pm}_{k,m,j}$]{\label{fig:ame_plot1} Schematic representation of the transitions to or from the set $x^{-}_{k,m,j}$ (left) and $x^{+}_{k,m,j}$ (right) ($j > 0$). We show the central node with some neighbors ($k = 3$) for different values $m$ (from left to right) and age $j$ (from bottom to top). Purple nodes are in state susceptible or non-adopters or $+1$, and yellow are in state infected or adopters or $-1$. The transitions are represented by colored arrows: The probability to swap and reset age $T^{\pm}$ (orange/gold), the probability to reset age $R^{\pm}$ (green), the probability to age $A^{\pm}$ (grey). The rates $\beta^{\pm}$ (blue) and $\gamma^{\pm}$ (red) are associated with the change of state of the neighbors.} 
\end{figure}
Now, taking into account this possible events, we write the possible transitions for the set $x^{+}_{k,m,j}$, for $j>0$ (see Fig. \ref{fig:ame_plot1} for a schematic representation of transitions involving $x^{+}_{k,m,j}$ and $x^{-}_{k,m,j}$):
\begin{flalign} \label{eq:pre_AME}
        x^{+}_{k,m,j} (t + dt) = & \, x^{+}_{k,m,j}(t) - T^{+} (k,m,j)\, x^{+}_{k,m,j}\, dt - R^{+} (k,m,j)\, x^{+}_{k,m,j} \, dt - A^{+} (k,m,j) \, x^{+}_{k,m,j} \, dt \nonumber \\
        & + A^{+} (k,m,j-1)\,  x^{+}_{k,m,j-1} \, dt - \omega (x^{+}_{k,m,j} \to x^{+}_{k,m+1,j+1}) \, x^{+}_{k,m,j}\, dt \\
        & - \omega (x^{+}_{k,m,j} \to x^{+}_{k,m-1,j+1})\,  x^{+}_{k,m,j} \, dt + \omega (x^{+}_{k,m+1,j-1} \to x^{+}_{k,m,j}) \, x^{+}_{k,m+1,j-1} \, dt \nonumber \\
        & + \omega (x^{+}_{k,m-1,j-1} \to x^{+}_{k,m-1,j-1}) \, x^{+}_{k,m-1,j-1}\,  dt. \nonumber
\end{flalign}
The case $j = 0$ needs to be treated differently from $j > 0$ because there is an injection of nodes into this set due to the resetting and changing state events. We write the possible transitions for the set $x^{+}_{k,m,0}$ as follows:
\begin{flalign}
        \label{eq:pre_AME_0}
        x^{+}_{k,m,0} (t + dt) = &\,  x^{+}_{k,m,0}(t) - T^{+} (k,m,0) \, x^{+}_{k,m,0} \, dt + \sum_{l = 0}^{\infty} T^{-} (k,m,l)\,  x^{-}_{k,m,l} \, dt + \sum_{l = 1}^{\infty} R^{+} (k,m,l)\,  x^{+}_{k,m,l}\,  dt   \nonumber\\
        & - T^{+} (k,m,0)\,  x^{+}_{k,m,0}\,  dt - \omega (x^{+}_{k,m,0} \to x^{+}_{k,m+1,1}) \, x^{+}_{k,m,0}\,  dt - \omega (x^{+}_{k,m,0} \to x^{+}_{k,m-1,1})\,  x^{+}_{k,m,0} \, dt.
\end{flalign}
Similar equations can be found considering transitions for $x^{-}_{k,m,j}$ and $x^{-}_{k,m,0}$. Notice that we have considered no transition increasing (or decreasing) the number of $-1$ neighbors $m$, keeping constant the age $j$. This is because the age $j$ is defined as the time spent in the current state (or since a reset). Therefore, if a node remains in its state and the number of neighbors in state $-1$ changes ($m \to m \pm 1$), the age of the node must increase ($j \to j + 1$). To determine the rate of these events, we use the same assumption as in Ref.~\cite{gleeson-2013}: we assume that the number of $++$ (edges between agents in state $+1$) edges change to $+-$ edges at a time-dependent rate $\beta^{+}$. Therefore, the transition rates are:
    \begin{align} \label{rate_beta_s}
    &  \omega (x^{+}_{k,m,j} \to s_{k,m+1,j+1}) = (k - m) \, \beta^{+}, \nonumber \\
    & \omega (s_{k,m-1,j-1} \to x^{+}_{k,m,j}) = (k - m + 1)\, \beta^{+} . 
    \end{align}
To determine the rate $\beta^{+}$, we count the change of $++$ edges that change to $+-$ in a time step. This change is produced by a neighbor of a node in state $+1$ changing state from $+1$ to $-1$. Thus, we can extract this information from the transition probability $T^{+}(k,m,j)$:
\begin{equation}
        \label{beta_s}
        \beta^{+} = \frac{\sum_{j=0}^{\infty} \sum_{k=0}^{\infty} p_k \sum_{m = 0}^{k} (k - m)\, T^{+} (k,m,j) \, x^{+}_{k,m,j}}{\sum_{j=0}^{\infty} \sum_{k=0}^{\infty} p_k \sum_{m = 0}^{k} (k - m) \, x^{+}_{k,m,j}}.
\end{equation}
A similar approximation is used to determine the transition rates at which $+-$ edges change to $++$ edges. We write:
\begin{align} \label{rate_gamma_s}
    &  \omega (x^{+}_{k,m,j} \to x^{+}_{k,m-1,j+1}) = m\, \gamma^{+}, \nonumber \\
    & \omega (x^{+}_{k,m+1,j-1} \to x^{+}_{k,m,j}) = (m + 1)\, \gamma^{+} ,
\end{align}
where the rate $\gamma^{+}$ is computed using the opposite transition probability $T^{-}(k,m,j)$:
\begin{equation}
        \label{gamma_s}
        \gamma^{+} = \frac{\sum_{j=0}^{\infty} \sum_{k=0}^{\infty} p_k \sum_{m = 0}^{k} (k - m)\, T^{-} (k,m,j) \, x^{-}_{k,m,j}}{\sum_{j=0}^{\infty} \sum_{k=0}^{\infty} p_k \sum_{m = 0}^{k} (k - m)\,  x^{-}_{k,m,j}}.
\end{equation}
    
Taking the limit $dt \to 0$ of Eqs. \eqref{eq:pre_AME}-\eqref{eq:pre_AME_0}, we obtain the approximate master equation (AME) for the evolution of the different sets $x^{\pm}_{k,m,j}$ and $x^{\pm}_{k,m,0}$:
\begin{align}
\label{eq:AME}
    \frac{d x^{\pm}_{k,m,j}}{dt} = & \, - \left( T^{\pm} (k,m,j) + A^{\pm} (k,m,j) + R^{\pm} (k,m,j) \right) x^{\pm}_{k,m,j} - (k - m)\, \beta^{\pm}\,  x^{\pm}_{k,m,j} - m \, \gamma^{\pm}\, x^{\pm}_{k,m,j} \nonumber\\
    & + (k-m+1)\, \beta^{\pm} \,   x^{\pm}_{k,m-1,j-1} + (m+1)\, \gamma^{\pm} \,  x^{\pm}_{k,m+1,j-1} + A^{\pm} (k,m,j-1)\,  x^{\pm}_{k,m,j-1},  \\
    \frac{d x^{\pm}_{k,m,0}}{dt}  = & \, - \left( T^{\pm} (k,m,0) + A^{\pm} (k,m,0) + R^{\pm} (k,m,0) \right) x^{\pm}_{k,m,0} - (k - m) \, \beta^{\pm}\,  x^{\pm}_{k,m,0} - m\, \gamma^{\pm} \,  x^{\pm}_{k,m,0}\nonumber\\
    & + \sum_{l = 0}^{\infty} T^{\mp} (k,m,l)\,  x^{\mp}_{k,m,l} + \sum_{l = 0}^{\infty} R^{\pm} (k,m,l)\, x^{\pm}_{k,m,l},\nonumber
\end{align}
where $\beta^{-}$ ($\gamma^{-}$) are time-dependent rates that account for the transitions at which $-+$ ($--$) edges change to $--$ ($-+$) edges:
\begin{align}
    \beta^{-} = \frac{\sum_{j=0}^{\infty} \sum_{k=0}^{\infty} p_k \sum_{m = 0}^{k} m \, T^{+} (k,m,j) \, x^{+}_{k,m,j}}{\sum_{j=0}^{\infty} \sum_{k=0}^{\infty} p_k \sum_{m = 0}^{k} m \,  x^{+}_{k,m,j}} \quad \quad \gamma^{-} = \frac{\sum_{j=0}^{\infty} \sum_{k=0}^{\infty} p_k \sum_{m = 0}^{k} m\, T^{-} (k,m,j) \, x^{-}_{k,m,j}}{\sum_{j=0}^{\infty} \sum_{k=0}^{\infty} p_k \sum_{m = 0}^{k} m\,  x^{-}_{k,m,j}}.
\end{align}
Equations \ref{eq:AME} define a closed set of deterministic differential equations that can be solved numerically using standard computational methods for any complex network and any model aging via the transition, reset and aging probabilities (a general script in Julia is available in a GitHub repository~\cite{link_git}).

\section{\label{sec:Reduction to Markovian dynamics}  Reduction to Markovian dynamics}

When there are neither resetting nor aging events ($R^{\pm} (k,m,j) = A^{\pm} (k,m,j) = 0$) and the transition probabilities do not depend on the internal time $j$, $T^{\pm} (k,m,j) = T^{\pm} (k,m)$, our dynamics are Markovian. In this case, if we are not interested in the solutions $x^{\pm}_{k,m,j} (t)$, Eq. \ref{eq:AME} can be reduced by summing variable $j$. We define $x^{\pm}_{k,m} = \sum_{j} x^{\pm}_{k,m,j}$ as the fraction of nodes that are in state $\pm 1$ and have degree $k$ and $m$ infected neighbors. The equations for the variables $x^{\pm}_{k,m}$ are:

\begin{align}
\label{eq:AME_noage}
    \frac{d x^{\pm}_{k,m}}{dt} = \frac{d x^{\pm}_{k,m,0}}{dt} + \sum _{j=1}^{\infty} \frac{d x^{\pm}_{k,m,j}}{dt}  = & \, - T^{\pm} (k,m) \, x^{\pm}_{k,m} + T^{\mp}(k,m) \, x^{\mp}_{k,m} - (k - m)\, \beta^{\pm} \,  x^{\pm}_{k,m} - m\, \gamma^{\pm} \,  x^{\pm}_{k,m} \nonumber\\
    & + (k-m+1)\, \beta^{\pm} \,  x^{\pm}_{k,m-1} \, + (m+1)\, \gamma^{\pm} \,  x^{\pm}_{k,m+1},
\end{align}
where the rates $\beta^{\pm}$ and $\gamma^{\pm}$ are redefined as follows:
\begin{align}
    & \beta^{+} = \frac{\sum_{k=0}^{\infty} p_k \sum_{m = 0}^{k} (k - m)\, T^{+} (k,m) \, x^{+}_{k,m}}{\sum_{k=0}^{\infty} p_k \sum_{m = 0}^{k} (k - m)\,  x^{+}_{k,m}}, \quad \quad \gamma^{+} = \frac{\sum_{k=0}^{\infty} p_k \sum_{m = 0}^{k} (k-m)\, T^{-} (k,m) \, x^{-}_{k,m}}{\sum_{k=0}^{\infty} p_k \sum_{m = 0}^{k} (k-m)\,  x^{-}_{k,m}}, \nonumber\\
    & \beta^{-} = \frac{\sum_{k=0}^{\infty} p_k \sum_{m = 0}^{k} m \, T^{+} (k,m) \, x^{+}_{k,m}}{\sum_{k=0}^{\infty} p_k \sum_{m = 0}^{k} (k - m)\,  x^{-}_{k,m}}, \quad \quad \quad \quad \; \gamma^{-} = \frac{\sum_{k=0}^{\infty} p_k \sum_{m = 0}^{k} m\, T^{-} (k,m) \, x^{-}_{k,m}}{\sum_{k=0}^{\infty} p_k \sum_{m = 0}^{k} m\,  x^{-}_{k,m}}.
\end{align}
Notice that this reduction is not an approximation and there is no loss of accuracy. The reduction to Markovian dynamics is a consequence of the chosen model. Eq. \ref{eq:AME_noage} correspond to the equations derived by J.P. Gleeson~\cite{gleeson-2013} for Markovian binary-state dynamics in complex networks. This is a set of $(k_{\rm max}+1)(k_{\rm max}+1)$ differential equations that can be solved numerically using standard computational methods for any complex network and any model via the transition probabilities $T^{\pm} (k,m)$.



\section{\label{sec:Heterogeneous mean-field approximation}  Heterogeneous mean-field approximation (HMF)}

Moreover, from Eqs. \eqref{eq:AME_noage}, we can perform a heterogeneous mean-field approximation (HMF) to reduce our system to $k_{\rm max}+1$ differential equations~\cite{gleeson-2013}. This approximation assumes a solution $x^{\pm}_{k,m} = x_{k}^{\pm} B_{k,m} [\omega]$, where $\omega = \langle k x^{-}_{k} \rangle / \langle k \rangle$ is the probability that one end of a randomly chosen edge is in state $-1$. Using this ansatz, the AME can be reduced to the following set of equations:

\begin{align}
    \frac{d}{d t} x^{-}_{k}= &- x^{-}_{k} \sum_{m=0}^{k} T^{-}_{k, m} B_{k, m}[\omega] +\left(1-x^{-}_{k}\right) \sum_{m=0}^{k} T^{+}_{k, m} B_{k, m}[\omega],
    \label{eq:HMF}
\end{align}
This system of $(k_{\rm max}+1)$ differential equations, coupled via $\omega$, cannot be solved analytically in general. However, it can be solved numerically using standard computational methods.


\section{\label{sec:Summary} Summary and discussion}

This chapter has expanded traditional binary-state models to include the effects of aging, providing a deeper insight into the dynamic behavior of complex networks. The mathematical framework developed here lays the groundwork for future studies to explore various aspects of aging in other dynamic systems, potentially leading to more accurate predictions and control strategies in both natural and engineered networks. The results presented in this chapter are a significant step forward in understanding the role of aging in binary-state dynamics, and we hope that they will inspire further research in this area.

Further work in this area could include the inclusion of other non-Markovian effects, such as memory kernels~\cite{saeedian2017memory} or extended to 3-state dynamics, which could be useful to understand the phase diagram of the Sakoda-Schelling model with aging, described in previous chapter. Moreover, the AME was improved in Ref.~\cite{peralta-2020C} to include finite-size and stochastic effects in binary-state models, which could be extended to include aging effects. Finally, the framework developed in this chapter could be useful to describe aging dynamics in threshold models, since these models need for a mathematical framework that needs to go beyond mean-field approximations to capture the dynamics of the system~\cite{gleeson-2007}.

\chapterimage{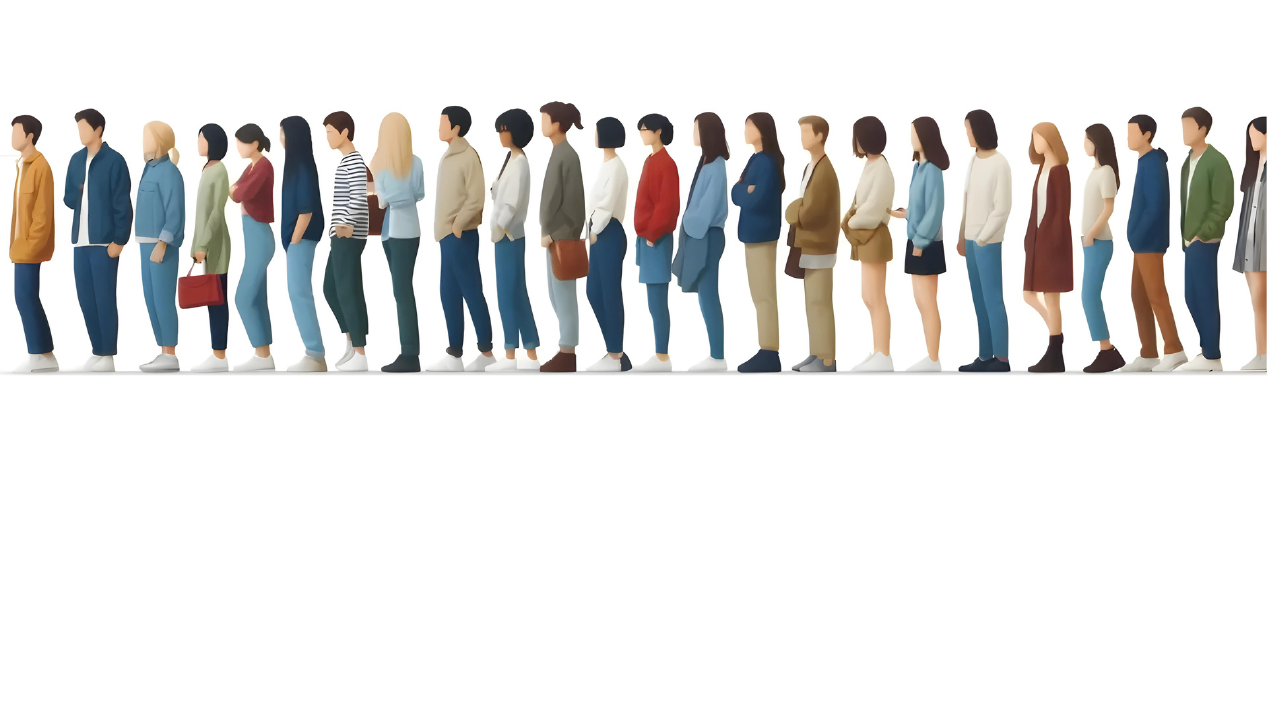}
\chapterspaceabove{6.75cm}
\chapterspacebelow{7.25cm}

\chapter{\label{ch:Aging in the Granovetter-Watts model} Impact of Aging in the Granovetter-Watts model}
\vspace{-1.2cm}
\small
\textbf{The results in this chapter are published as:}
\vspace{0.05 cm}

\fullcite{Abella-2022-AME}
\normalsize
\vspace{0.5 cm}

In this chapter, we analyze the aging implications in one of the simplest binary-state threshold models: the Granovetter-Watts model. Our analytical approximations give a good description of extensive Monte Carlo simulations in Erd\H{o}s-R\'enyi, random-regular and Barab\'asi-Albert networks. While aging does not modify the cascade condition, it slows down the cascade dynamics towards the full-adoption state: the exponential increase of adopters in time from the original model is replaced by a stretched exponential or power law, depending on the aging mechanism. Under several approximations, we give analytical expressions for the cascade condition and for the exponents of the adopters' density growth laws. Beyond random networks, we also describe by Monte Carlo simulations the effects of aging for the Granovetter-Watts model in a two-dimensional lattice.

\section{\label{sec:Introduction_Threshold} Introduction}

The Granovetter-Watts model~\cite{granovetter-1978,watts-2002}, is a well-known binary-state model for Complex contagion processes, such as rumor propagation,  riots, stock market herds, adoption of new technologies, political and environmental campaigns, etc. The discontinuous phase transition and the cascade condition exhibited by the Granovetter-Watts model were predicted with analytical tools in Ref.~\cite{watts-2002}. This model has been extensively studied in regular lattices and small-world networks~\cite{centola-2007}, random graphs~\cite{gleeson-2007},  modular and community structure~\cite{gleeson-2008}, clustered networks~\cite{hackett-2011,hackett-2013}, hypergraphs~\cite{de-arruda-2020}, homophilic networks~\cite{diaz-diaz-2022}, etc. Moreover, recent studies also include variants of the adoption rules including
the impact of opinion leaders~\cite{liu-2018} and seed-size~\cite{singh-2013}, on-off threshold~\cite{dodds-2013} and the competition between simple and complex contagion~\cite{czaplicka-2016,min-2018,min-2018-dual,diaz-diaz-2022,min2023threshold}. Additionally, the Granovetter-Watts model has been confronted with several sources of empirical data~\cite{centola-2010,karimi-2013,karsai-2014,rosenthal-2015,karsai-2016,mnsted-2017,unicomb-2018,guilbeault-2021}.

Previous studies of the Granovetter-Watts model usually rely on a Markovian assumption for the dynamics. This implies that events depend only on the present state, i.e., dynamical rules are memoryless. Markovian processes exhibit exponential distributions in the upcoming event times and the number of events in a given time interval follows a Poisson distribution. However, there is strong empirical evidence against this assumption in human interactions and thus, the understanding of these non-Markovian effects is in general a topic of current interest~\cite{van-mieghem-2013,starnini-2017,peralta-2020C,peralta-2020A}. In particular, for the threshold models, memory effects have been included as past exposures' memory~\cite{dodds-2004}, message-passing algorithms~\cite{shrestha-2014}, memory distributions for retweeting algorithms~\cite{gleeson-2016} and timers~\cite{oh-2018}.

Regarding the specific context of innovation adoption from the complex systems point of view~\cite{przybyla2014diffusion}, mechanisms of inertia or resistance to adopt the technology have been already introduced. In fact, the original approach of Rogers~\cite{rogers2014} considers a fraction of ``laggards'' that will resist innovating until a large majority of the population has already adopted it. Other works highlight the importance of timing interactions~\cite{bass1969} and the effect of ``contrarians'' (tendency to act against the majority), which has an important impact on the dynamics~\cite{galam-2008,goncalves-2012}. In Ref.~\cite{goncalves-2012}, it is discussed how different technologies may show different adoption cascades regarding the balance between advertisement and resistance to change.

In this chapter, we incorporate the aging mechanism into the Granovetter-Watts model, characterizing both the cascade condition and dynamics towards the fully adopted state. We propose two different aging mechanisms giving rise to heterogeneous activity patterns, characterized by flat-tail inter-event time distributions. To describe the results, we use the general master equation for any binary-state model with temporal activity patterns previously described in Chapter \ref{ch:Aging in binary state dynamics}. For the particular case of the Granovetter-Watts model, we are able to reduce the dimensionality of the full system without loss of accuracy. Theoretical predictions are matched with extensive Monte Carlo simulations in different networks. For completeness, the role of both aging mechanisms is also studied in a two-dimensional Moore lattice.

The chapter is organized as follows. In the next section, we describe the original Granovetter-Watts model and introduce exogenous and endogenous aging in the model. In Section \ref{sec:Complex networks}, numerical results are reported and contrasted with theoretical predictions for different complex networks. For completeness, in Section \ref{sec:Lattice} the case of a 2D-lattice is analyzed. The final section contains a summary and a discussion of the results.

\begin{figure}
    \centering \captionsetup{font=sf}
    \includegraphics[width=0.65\columnwidth]{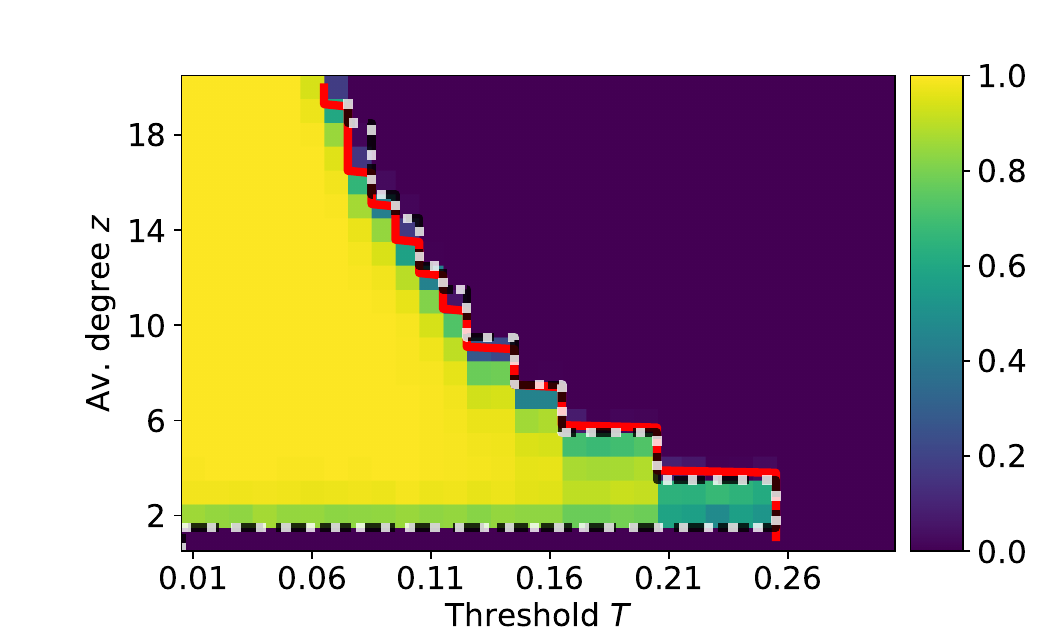}
    \caption[Average density $x^{-}$ of adopters for an Erd\H{o}s-R\'enyi graph]{\label{fig:umbral} Average density $x^{-}$ of adopters for an Erd\H{o}s-R\'enyi graph of mean degree $z$ using a model with threshold $T$. Color-coded values of $x^{-}$ are from Monte Carlo simulations of the model without aging in a graph with $N = 10,000$ agents.  Black dashed and white dotted lines correspond to $T_c$ value obtained numerically for the model with exogenous and endogenous aging, respectively.  Monte Carlo simulations are averaged over $M = 5 \times 10^4$ realizations. The red solid line is the analytical approximation of the cascade boundary, from Eq. \eqref{eq:linear}, which is the same with and without aging.}
\end{figure}

\section{\label{sec:Threshold model with aging} Aging in the Granovetter-Watts model}

As it was introduced before (see Section \ref{sec:Granovetter-Watts threshold model}), the standard Granovetter-Watts model~\cite{granovetter-1978,watts-2002} considers a network of $N$ interacting agents, where each node of the network represents an agent $i$ with a binary-state variable $\sigma_i = \left\{ 0,1 \right\}$  and a given threshold $T$ ($0<T<1$). The state indicates if the agent has adopted a technology (or joined a riot, spread a meme or fake news, etc.) or not. We use the wording of a technology adoption process for the rest of the chapter. If a node $i$ (with $k$ neighbors) has not adopted  ($\sigma_i = 0$) the technology, becomes adopter ($\sigma_i = 1$) if the fraction $m / k$ of neighbors adopters exceeds the threshold $T$. Adopter nodes cannot go back to the non-adopter state.

In the Granovetter-Watts model with aging, each agent has an internal time $j = 0,1,2,...$  (in Monte-Carlo units) as in Refs.~\cite{fernandez-gracia-2013,artime-2018,peralta-2020C,peralta-2020A,chen-2020,fernandez-gracia-2011,perez-2016,stark-2008}.  As initial condition, we set $j = 0$ for all nodes. In Monte Carlo simulations, we follow a Random Asynchronous Update in which agents are activated in discrete time steps with probability $p_{A} (j) = 1/(j+2)$. When a non-adopter agent is activated, she changes state according to the threshold condition $m/k > T$. We will consider two different aging mechanisms, endogenous and exogenous aging~\cite{fernandez-gracia-2011}, which account for the power law inter-event time distributions empirically observed in human interactions~\cite{artime-2017}. For endogenous aging,  the internal time measures the time spent in the current state: If an agent in an updating attempt is not activated or does not adopt, the internal time increases by one unit. Therefore, the longer an agent has remained without adopting the technology, the more difficult it is for her to adopt it. 

For exogenous aging, the internal time accounts for the time since the last attempt to change state: In each updating attempt in which the agent is activated, the internal clock resets to $j = 0$ even if there is no adoption. In this case, aging is understood as a resistance to adopt the technology the longer the agent has not been induced to consider adoption by some external influence.  

\section{\label{sec:Complex networks} Results  on Complex networks}

In this section we discuss the Granovetter-Watts model with endogenous and exogenous aging in three different complex networks: random-regular (RR)~\cite{wormald_1999}, Erd\H{o}s-R\'enyi (ER)~\cite{erdos1960evolution} and Barab\'asi-Albert (BA)~\cite{barabasi2009scale}.

\begin{figure}
    \centering \captionsetup{font=sf}
    \includegraphics[width=0.55\columnwidth]{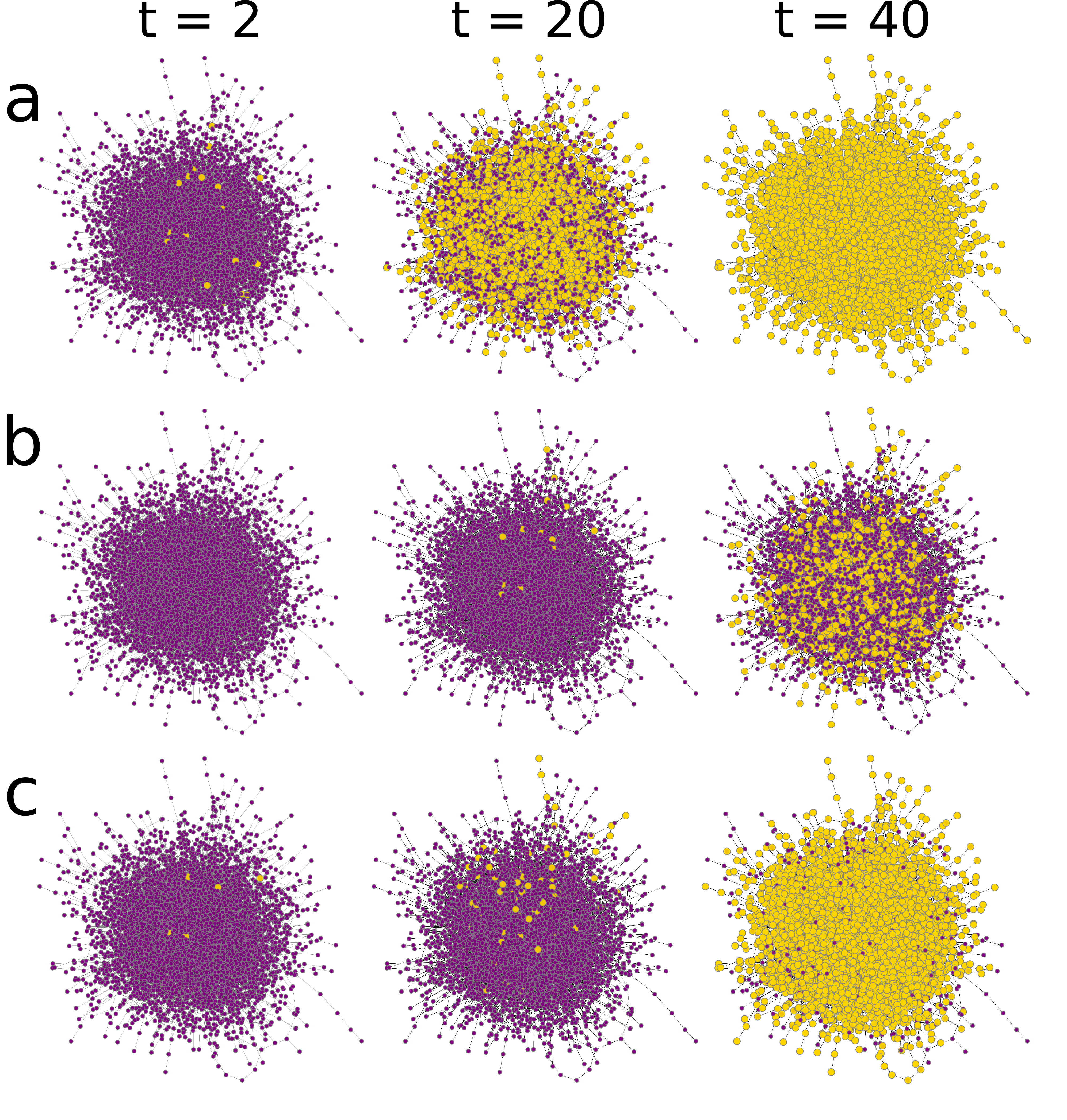}
    \caption[Cascade spreading for the Granovetter-Watts model]{\label{fig:graph_plot} Cascade spreading for the original Granovetter-Watts model \textbf{(a)}, and the versions with endogenous (reset if adopts) \textbf{(b)} and exogenous (reset if activates) \textbf{(c)} aging. Yellow nodes are adopters and purple nodes are non-adopters. Time increases from left to right. Monte Carlo simulations are performed in an Erd\H{o}s-R\'enyi network with mean degree $z = 3$ and $T = 0.22$. System size is $N = 8,000$.}
\end{figure}

\subsection{\label{subsec:Numerical results} Numerical results}

For the networks considered, the Granovetter-Watts model undergoes a discontinuous phase transition at a certain critical value $T_{c}$ (cascade condition)~\cite{watts-2002}. For $T<T_c$, a small initial seed of adopters triggers a global cascade where, on average, a significant proportion of agents in the system adopt the technology (change from $\sigma_i = 0 \mbox{ to } \; 1$). In our analysis, the initial condition is set to favor cascades: one agent $i$ with degree $k_i = z$ is selected randomly and all her neighbors are initially adopters, as in Ref.~\cite{centola-2007,singh-2013}. For $T>T_c$, there are few cascade occurrences and none of them is global. The cascade condition dependence with the average degree $z$ of the underlying network has been studied in Refs.~\cite{watts-2002, gleeson-2007}. For the two aging mechanisms considered, Monte Carlo simulations in random graphs show that the $T_c$ dependence on $z$ is very similar to the one for the model without aging (see Fig. \ref{fig:umbral}). Therefore, for random networks, tends to the same cascade condition derived for the original model (which for ER graphs is $T_{c} = 1 / z$~\cite{watts-2002}). Similar results were found for RR and BA graphs. This result is not obvious a priory because aging has been shown to modify the final state in several models~\cite{fernandez-gracia-2013,artime-2018,peralta-2020C,peralta-2020A,chen-2020,fernandez-gracia-2011,perez-2016,stark-2008}. 

\begin{figure}
\centering \captionsetup{font=sf}
\includegraphics[width=0.8\columnwidth]{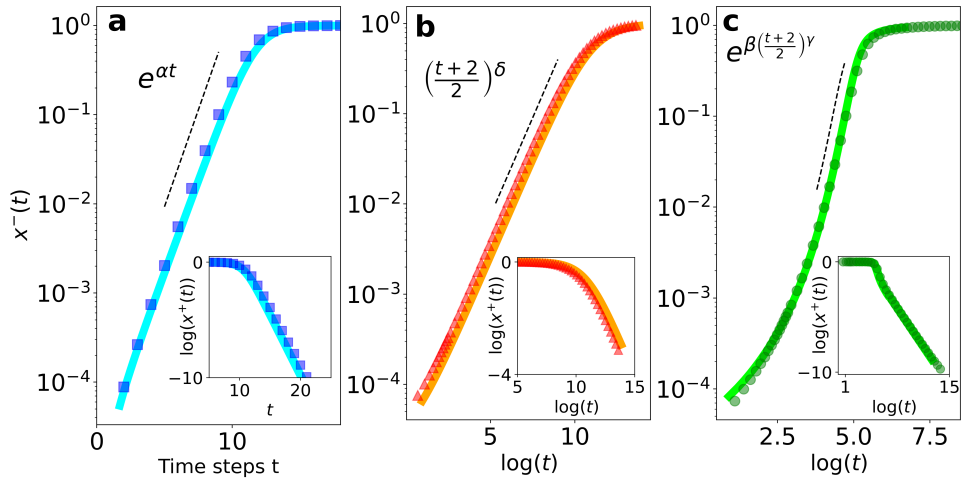}
\caption[Cascade dynamics and fall to the full-adopt state ($x^{-} \sim 1$)]{\label{fig:models} Cascade dynamics and fall to the full-adopt state ($x^{-} \sim 1$) of the Granovetter-Watts model without aging \textbf{(a)} and the versions with endogenous \textbf{(b)} and exogenous \textbf{(c)} aging effects. At (b-c), the evolution is plotted as a function of the logarithm of time $\log{(t)}$ in Monte Carlo steps, as in the insets. The underlying network is a 3-regular random graph and the threshold is $T = 0.2$. The exponent values are $\alpha \simeq 1.0$, $\beta \simeq 1.14$, $\gamma \simeq 0.38$ and $\delta \simeq 1.0$. Numerically integrated solutions of Eq. \eqref{eq:AME_Threshold} (solid lines) describe accurately the numerical results. Monte Carlo simulations are averaged over $M = 5 \times 10^4$ realizations in a network of $N = 1.6 \times 10^5$ nodes.}
\end{figure}

Even though aging in the Granovetter-Watts model does not modify the cascade condition, it has a large impact in the complex contagion cascade dynamics (Fig.\ref{fig:graph_plot}). 
From Monte Carlo simulations in a random regular graph we find that, without aging,  the average fraction of adopters, denoted by $x^{-}$, follows an initial exponential increase with time (see Fig. \ref{fig:graph_plot}a and \ref{fig:models}a), 
\begin{equation}
x^{-}(t) \sim x^{-}_{0} \, e^{\alpha \, t},
\label{eq:exponential}
\end{equation}
where $x^{-}_{0}$ is the initial fraction of adopters (seed). This behavior is universal for all values of the control parameters $z$ and $T$ below the cascade condition. In addition, we investigated the approach to the full-adopt state ($x^{-} = 1$) and we found that the fraction of non-adopters, denoted by $x^{+}$, follows an exponential decay for all values of the control parameters (see inset in Fig.\ref{fig:models}a).

When aging is introduced, the cascade dynamics are much slower than an exponential law (see Fig. \ref{fig:graph_plot}b). For endogenous aging, all non-adopters agents have the same activation probability $p_A(j)$, which decreases at each time step. This gives rise to cascade dynamics well-fitted by a power law increase (see Fig. \ref{fig:models}b),
\begin{equation}
x^{-}(t) \sim x^{-}_0 \, \left( \frac{t + 2}{2}\right)^\delta .
\label{eq:power law}
\end{equation}
For exogenous aging, we observe a slow adoption spread at the beginning followed by a cascade where almost all agents adopt the technology (Fig. \ref{fig:graph_plot}c). This behavior is well-fitted with a stretched exponential increase of the number of adopters (see Fig. \ref{fig:models}c),
\begin{equation}
x^{-}(t) \sim  x^{-}_0\,  e^{\beta \, ((t + 2) / 2)^{\gamma}} .
\label{eq:streched_exp}
\end{equation}
For both aging mechanisms, in the last stages of evolution, a few ``stubborn'' non-adopters remain, although the environment favors the adoption. Due to the chosen activation probability, the number of non-adopters decay with a power law $x^{+}(t) \sim 1/(t+2)$ in both cases (see insets at Fig. \ref{fig:models}(b-c)).

\begin{figure}
\centering \captionsetup{font=sf}
\includegraphics[width=0.6\columnwidth]{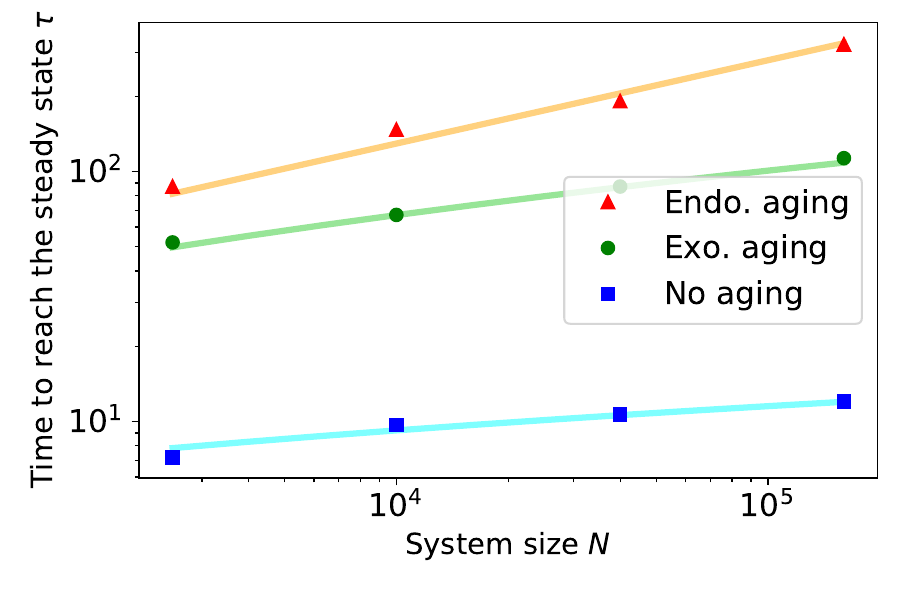}
\caption[Average time to reach the steady state $\tau$]{\label{fig:time_steady} Average time to reach the steady state ($x^{-} > 0.9$) $\tau$ as a function of the system size $N$ for the original Granovetter-Watts model and the versions with endogenous and exogenous aging. The underlying network is a 5-regular random graph and the threshold is $T = 0.12$. Monte Carlo simulations are averaged over $M = 5 \times 10^4$ realizations. Solid lines are the system size-dependent timescale: For the original model, $\tau_{\rm{NO AG.}} = (1/\alpha)\log(N)$, for the endogenous $(\tau_{\rm{ENDO}} = 2N^{1/\delta} - 2)$ and for the exogenous aging ($\tau_{\rm{EXO}} = 2(\log(N)/\beta)^{1/\gamma} - 2$), which follows from the dynamics from Eq. \eqref{eq:exponential}, \eqref{eq:power law} and \eqref{eq:streched_exp}. The exponents $\alpha$, $\beta$, $\gamma$ and $\delta$ are fitted exponents from numerical simulations.}
\end{figure}

Comparing the evolution of the original model with one of the versions with aging, we observe an important separation of time scales. While for the original model, the time to reach the steady state follows a logarithmic increase with the system size, the versions with endogenous and exogenous aging show a power law and a power-logarithmic dependence, respectively (see Fig.\ref{fig:time_steady}). Therefore, the time scale separation between the original model and the versions with aging increases as we increase the system size, and thus, the aging effects are more relevant for large systems.

The power law and the stretched exponential dynamics for endogenous and exogenous aging, respectively, are observed for $z$ and $T$ below the cascade condition ($T < T_c$) and for many different system sizes. This is shown in Fig. \ref{fig:exo_endo_evo} for a random regular, Erd\H{o}s-R\'enyi and  Barab\'asi-Albert networks. In particular, we show that the time-dependent behavior for different system sizes collapses to a single curve when time is scaled with the system size-dependent timescale (previously analyzed in Fig. \ref{fig:time_steady}) that follows from either the power law dynamics $(\tau_{\rm{ENDO}} = 2N^{1/\delta} - 2)$  or the stretched exponential law  $(\tau_{\rm{EXO}} = 2( \log(N)/\beta )^{1/\gamma} - 2)$. Notice that the scaling of the y-axis is necessary for Fig.\ref{fig:exo_endo_evo}(d-f) to show a linear dependence due to the stretched exponential increase.


A different question is the dependence of the exponents of the power law and stretched exponential with the parameters $z$ and $T$. Numerical results from fitted Monte Carlo simulations for $\alpha(z,T)$, $\delta(z,T)$ and $\gamma(z,T)$ are shown in Figs. \ref{fig:endo_exp} and \ref{fig:exo_exp}. For a random-regular graph, as apparent from Fig. \ref{fig:exo_endo_evo}, the exponents do not depend on the parameter $T$ up to $T_c$ (so the exponents are dependent only on $z$, $\alpha(z)$, $\gamma(z)$ and $\delta(z)$), while for Erd\H{o}s-R\'enyi and Barab\'asi-Albert networks the value of the exponents decrease with $T$ when approaching $T_c$, indicating a slowing down of the dynamics. Also, for these two latter networks, the exponents present a maximum value at a certain value of $z$. This maximum value at a certain $z$ for a fixed $T$ can be understood as being between the two critical lines of Fig. \ref{fig:umbral}.

\subsection{\label{subsec:Approximate master equation and solutions} General mathematical description}

\begin{figure}[t]
    \centering \captionsetup{font=sf}
    \includegraphics[width=\linewidth]{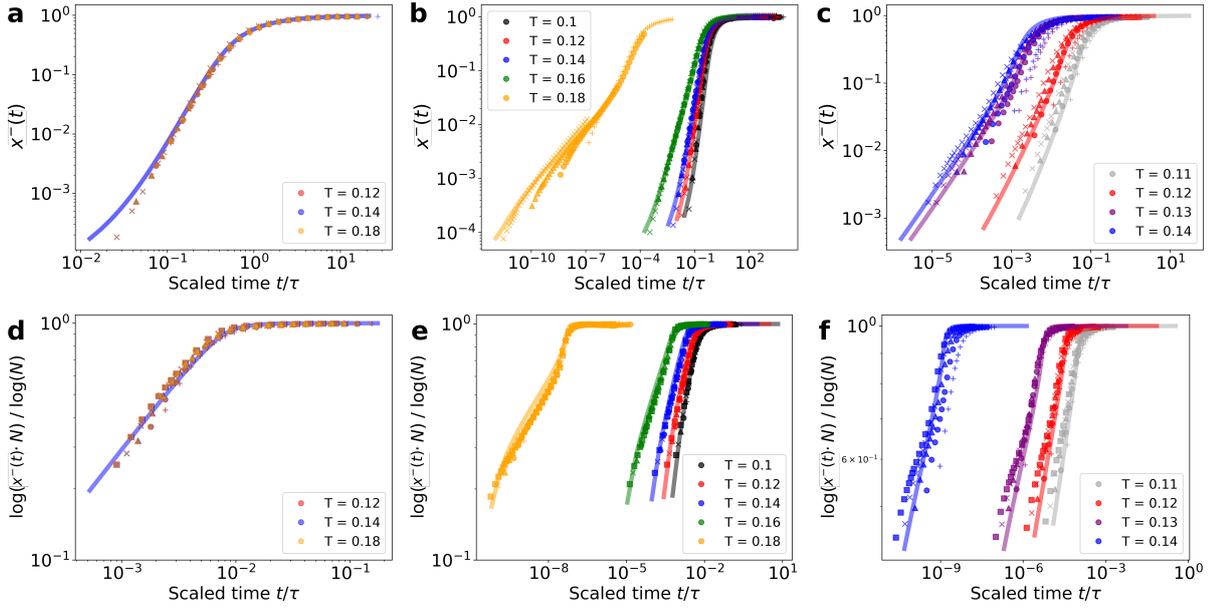}
    \caption[Cascade dynamics of the Granovetter-Watts model in graphs]{\label{fig:exo_endo_evo} Cascade dynamics of the Granovetter-Watts model with endogenous (a - c) and exogenous (d - f) aging. From the left column to the right: a random regular graph with degree $z=5$ (a and d), an Erd\H{o}s-R\'enyi graph with average degree $z = 5$ (b and e) and a Barab\'asi-Albert graph with average degree $z = 8$ (c and f). Different colors indicate different values of $T$ and markers correspond to different system sizes: $N = 2,500$ (plus), $10,000$ (circles), $40,000$ (triangles), $160,000$ (crosses) and $640,000$ (squares). Time is scaled according to the system size for each model: $\tau_{\rm{EXO}} = 2(\log(N)/\beta)^{1/\gamma} - 2$, $\tau_{\rm{ENDO}} = 2N^{1/\delta} - 2$, where $\beta$,$\gamma$ and $\delta$ are the fitted exponents from the behavior according to Eq. \eqref{eq:power law} and \eqref{eq:streched_exp}. Solid lines are obtained from the solutions of Eq. \eqref{eq:PA}. Monte Carlo simulations are averaged over $M = 5 \times 10^4$ realizations.}
\end{figure}

To account for the non-Markovian dynamics introduced by the aging mechanism, we need to go beyond the standard mathematical descriptions of the Granovetter-Watts model~\cite{gleeson-2007,gleeson-2008,gleeson-2013}. We do so using a Markovian description by enlarging the number of variables~\cite{peralta-2020C,peralta-2020A}. Namely, we classify the agents with degree $k$, number of adopter neighbors $m$ and age $j$ as different sets in a compartmental model in a general framework for binary-state dynamics in complex networks, as described in Chapter \ref{ch:Aging in binary state dynamics}. To write down the AME for the Granovetter-Watts model with aging, we need to consider the following possible transitions:
\begin{itemize}
    \item A node $i$, in state $\sigma_i = \pm 1$, changes state and resets internal age with probability $T^{\pm} (k,m,j)$;
    \item A node $i$, in state $\sigma_i = \pm 1$, remains in the same state and resets internal age to zero ($j \to 0$) with probability $R^{\pm} (k,m,j)$;
    \item A node $i$, in state $\sigma_i = \pm 1$, remains in the same state and ages ($j \to j+1$) with probability $A^{\pm} (k,m,j)$.
\end{itemize}
For the specific case of the Granovetter-Watts model, dynamics are monotonic and $T^{-} (k,m,j) = 0$ (no adopter becomes a non-adopter). Moreover, when an agent becomes an adopter, there are neither resetting nor aging events $R^{-} (k,m,j) = A^{-} (k,m,j) = 0$. This means as well that equations for the non-adopters $x^{+}_{k,m,j}$ and adopters $x^{-}_{k,m,j}$ nodes are independent. Thus, we can write the following rate equations for the evolution of the fraction $x^{+}_{k,m,j} (t)$ of $k$-degree non-adopters nodes with $m$ infected neighbors and age $j$:
\begin{align}
\label{eq:AME_Threshold}
\frac{d x^{+}_{k,m,j}}{dt} = & \,  - x^{+}_{k,m,j} - (k-m)\, \beta^s \, x^{+}_{k,m,j} + (k-m+1) \, \beta^s \, x^{+}_{k,m-1,j-1} + A^{+} (k,m,j-1)\, x^{+}_{k,m,j-1},  \\
\frac{d x^{+}_{k,m,0}}{dt}  = & \,   - x^{+}_{k,m,0} - (k - m)\, \beta^s   \,x^{+}_{k,m,0} + \sum_{l = 0} R^{+} (k,m,l)\, x^{+}_{k,m,l}, \nonumber 
\end{align}
where $\beta^s$ is a non-linear function of $x^{+}_{k',m',j'}$ for all values of $k'$,$m'$ and $j'$  (see Eq. \eqref{beta_s}). The remaining step is to define explicitly the transition probabilities for our aging mechanisms. For both exogenous and endogenous aging, the adoption probability is the probability that an agent is activated and has a fraction of adopters that exceeds the threshold $T$, which means that 
\begin{equation}
T^{+}(k,m,j) = p_A(j) \, \theta(m/k - T),
\end{equation} 
where $\theta(\cdot)$ is the Heaviside step function. 

The reset and aging probabilities for endogenous and exogenous aging mechanisms are different. The simplest case is endogenous aging where there is no reset $R^{\pm} (k,m,j) = 0$ and agents increase by one the age with probability 
\begin{equation}
A^{+} (k,m,j) = \,  1 - T^{+}(k,m,j) = \, 1 - p_{A}(j)\, \theta \left( m/k - T \right).
\end{equation}
When aging is exogenous, the reset probability is the probability to activate and not adopt 
\begin{equation}
R^{+} (k,m,j) = p_A (j)\, \left(1 - \theta \left(m/k - T\right)\right). 
\end{equation}
Thus, agents that age are just the ones that do not activate, $A^{+} (k,m,j) = 1 - p_A(j)$.

Using these definitions, we have integrated numerically Eq. \eqref{eq:AME_Threshold} for the Granovetter-Watts model with both endogenous and exogenous aging. Numerical solutions give  good agreement with Monte Carlo simulations (see Fig. \ref{fig:models}). However, in a general network, considering a cutoff for the degree $k = 0,\dots,k_{\rm{max}}$ and age $j = 0,\dots,j_{\rm{max}}$, the number of differential equations to solve is $(k_{\rm{max}} + 1)\, (k_{\rm{max}} + 1)\, (j_{\rm{max}} + 1)$ according to the three subindexes of the variable $x^{+}_{k,m,j}$. This number grows with the largest degree square and largest age considered and thus, some further approximations are needed to obtain a convenient reduced system of differential equations. 

\begin{figure}
    \centering \captionsetup{font=sf}
    \includegraphics[width=0.7\columnwidth]{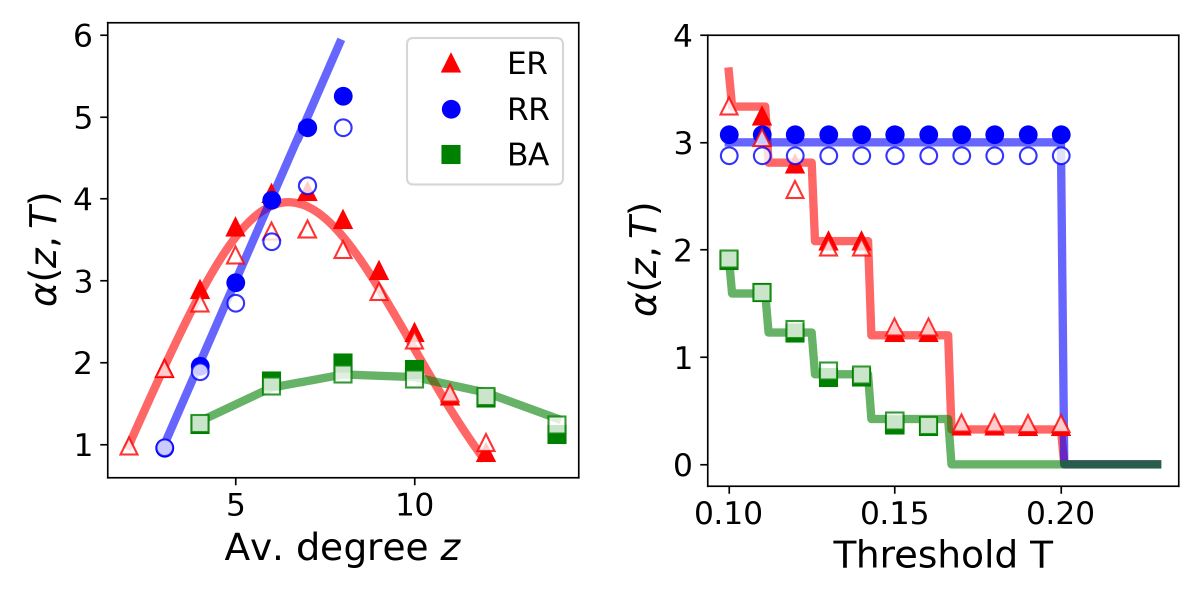}
    \caption[Exponent for the Granovetter-Watts model]{\label{fig:endo_exp} Exponent $\alpha$ for the original Granovetter-Watts model (empty markers) and $\delta$ for the version with endogenous aging (filled markers) for different values of the average degree $z$ (and $T = 0.1$) \textbf{(left)} and as a function of $T$ for fixed $z$ \textbf{(right)}. Different markers indicate results from Monte Carlo simulations with different topologies: red triangles indicate an Erd\H{o}s-R\'enyi (ER) graph, blue circles indicate a random regular (RR) graph and green squares indicate a Barab\'asi-Albert (BA) graph. In the right panel, the average degree is fixed $z = 5$ for ER and RR, and $z = 8$ for the BA. Predicted values by Eq. \eqref{eq:alpha} (solid lines) fit the results for each topology. System size is fixed at $N = 4 \times 10^6$ for the original model and $N = 3.2 \times 10^5$ for the version with aging.}
\end{figure}

As an ansatz, we assume that timing interactions can be effectively decoupled from the adoption process so the solution of Eq. \eqref{eq:AME_Threshold} can be written as
\begin{equation}
    \label{eq:assumption1}
    x^{+}_{k,m,j}(t) = x^{+}_{k,m}(t) \, G_{j} (t),
\end{equation}
where $x^{+}_{k,m}$ is the fraction of non-adopters with degree $k$ and $m$ infected neighbors $x^{+}_{k,m} = \sum_{j} x^{+}_{k,m,j}$ and there is an age distribution $G_{j} (t)$, independent of the adoption process.

If we sum over the variable age $j$ in Eq. \eqref{eq:AME_Threshold}, we can rewrite the following rate equations for the variables $x^{+}_{k,m}$
\begin{equation}
    \label{eq:threshold_AME_red}
    \frac{d x^{+}_{k,m}}{dt}  = \,  - \langle p_A \rangle \, \theta(m - kT)\, x^{+}_{k,m} - (k - m) \, \beta^s \,  x^{+}_{k,m} + (k - m + 1)\, \beta^s \,  x^{+}_{k,m-1},
\end{equation}
where aging effects are  just included in $\langle p_A \rangle(t)$: 
\begin{equation}
    \langle p_A \rangle(t) = \sum_{j = 0}^{\infty} p_A(j) \, G_j (t).
\end{equation}

Using the definition of the fraction of $k$-degree adopters $x^{-}_{k} (t)$,
\begin{equation}
    x^{-}_{k}(t) = 1 - \sum_{j=0}^{\infty} \sum_{m = 0}^k x^{+}_{k,m,j},
\end{equation}

and along the lines of Ref.~\cite{gleeson-2013}, we use the following ansatz

\begin{equation}
    x^{+}_{k,m} = (1 - x^{-}_{k} (0)) \, B_{k,m}[\phi],
\end{equation}

\begin{figure}
    \centering \captionsetup{font=sf}
    \includegraphics[width=0.7\columnwidth]{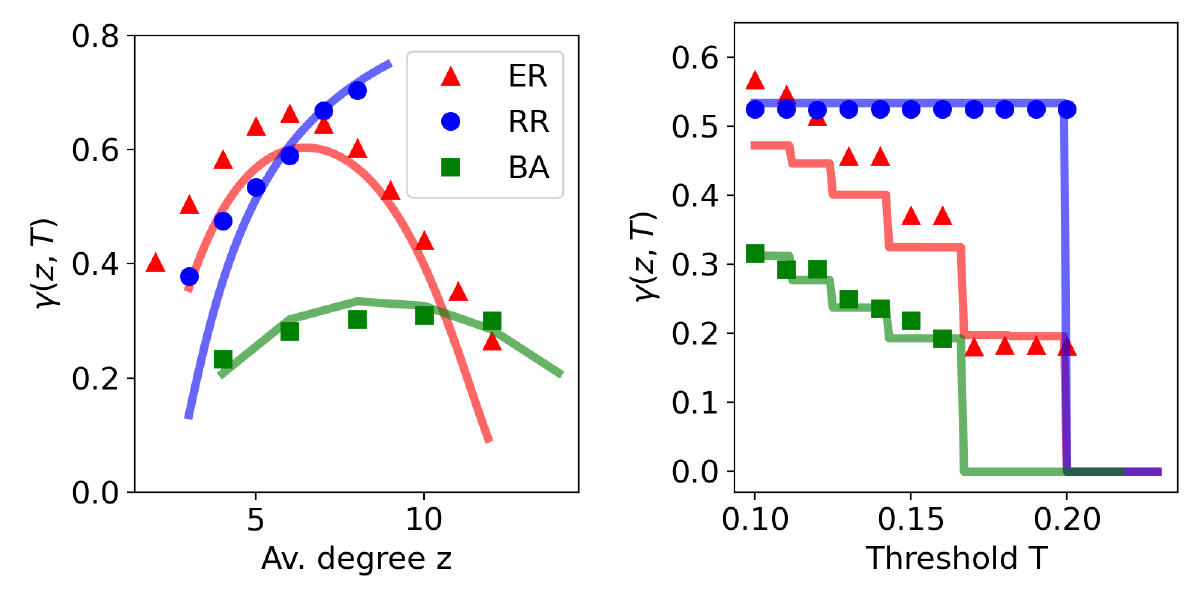}
    \caption[Exponent $\gamma$ for the Granovetter-Watts model with exogenous aging]{\label{fig:exo_exp} Exponent $\gamma$ for the Granovetter-Watts model with exogenous aging for different values of the average degree $z$ ($T = 0.1$) \textbf{(left)} and as a function of  $T$ for fixed $z$ \textbf{(right)}. Different markers indicate results from Monte Carlo simulations with different topology: red triangles indicate an Erd\H{o}s-R\'enyi (ER) graph, blue circles indicate a random regular (RR) graph and green squares indicate a Barab\'asi-Albert (BA) graph. In the right panel, the average degree is fixed $z = 5$ for ER and RR, and $z = 8$ for the BA. Predicted values by numerical integration of Eqs. \eqref{eq:PA} (solid lines) fit approximately the results for each topology. System size is fixed at $N = 3.2 \times 10^5$.}
    \end{figure}

where $B_{k,m}[\phi]$ is the binomial distribution with $k$ attempts, $m$ successes and with success probability $\phi$. From this point, we derive from Eq. \eqref{eq:threshold_AME_red} a reduced system of two coupled differential equations for the fraction of adopters $x^{-}(t) = \sum_k p_k x^{-}_{k} (t)$ and an auxiliary variable $\phi (t)$ (see details in Ref.~\cite{gleeson-2013}):
\begin{equation}
    \label{eq:PA}
        \frac{d x^{-}}{dt} = \langle p_A \rangle [ h(\phi) - x^{-} ], \quad \quad \frac{d \phi}{dt} = \langle p_A \rangle [ g(\phi) - \phi ],
\end{equation}
where $\phi(t)$ can be understood as the probability that a randomly chosen neighbor of a non-adopter node is an adopter at time $t$. The functions $h(\phi)$ and $g(\phi)$ are nonlinear functions of this variable $\phi$
\begin{align}
    h (\phi)  = & \,  \sum_{k=0}^{\infty} p_k\,  \left( x^{-}_{k} (0) + (1 - x^{-}_{k} (0))\,  \sum_{m = kT}^{k} B_{k,m}[\phi]\right),\nonumber\\
    \\
    g (\phi)  = & \, \sum_{k=0}^{\infty} \frac{k}{z}\,  p_k \,  \left( x^{-}_{k} (0) + (1 - x^{-}_{k} (0)) \, \sum_{m = kT}^{k} B_{k-1,m}[\phi]\right). \nonumber
\end{align}
 When $\langle p_A \rangle$ is replaced by a constant, Eqs. \eqref{eq:PA} reduce to previous results for the original model~\cite{gleeson-2008}.
 
Determining the distribution $G_j (t)$ a priori is not easy. For endogenous aging, all non-adopters have the same age at each time step and $G_j (t) = \delta(j-t)$ (where $\delta(\cdot)$ is the Dirac delta function). Therefore, $\langle p_A \rangle = 1/(t+2)$. The numerical solution of Eq. \eqref{eq:PA} gives a good agreement with Monte Carlo simulations (see Fig. \ref{fig:exo_endo_evo}(a-c)). For the case of exogenous aging, the reset of the internal clock makes more difficult a choice for $G_j (t)$.  Inspired on the stretched exponential behavior of $x^{-}(t)$ observed from Monte Carlo simulations, we propose $\langle p_A \rangle = 1/(t+2)^\mu$. For $\mu = 0.75$, the numerical solutions of Eq. \eqref{eq:PA} gives a very good agreement with our Monte Carlo simulations (see Fig. \ref{fig:exo_endo_evo} (d-f)).

\subsection{\label{subsec Analytical results} Analytical results}

To obtain an analytical result for the cascade condition and for the exponents of the predicted exponential, stretched-exponential and power law cascade dynamics that we fitted from Monte Carlo simulations, we need to go a step beyond the numerical solution of our approximated differential equations (Eqs. \eqref{eq:AME_Threshold} and \eqref{eq:PA}). 

For a global cascade to occur, it is needed that the variable $\phi(t)$ grows with time. If we assume a small initial seed ($x^{-}_{k} (0) \; \to \; 0$), Eq. \eqref{eq:PA} can be rewritten as in Ref.~\cite{gleeson-2007}
\begin{equation}
    \label{eq:pre_lin}
    \frac{d \phi}{dt}  = \langle p_A \rangle \, \left( -\phi + \sum_{k=1}^{\infty} \frac{k}{z} \, p_k \, \sum_{m = k\, T}^{k} B_{k-1,m} [\phi] \right).
\end{equation}
Rewriting the sum term as $\sum_{l=0}^{\infty} C_l \, \phi^l$, with coefficients 
\begin{equation}
    \label{eq:coef_phi}
    C_l = \sum_{k=l}^{\infty} \sum_{m=0}^{l} { k-1 \choose l} \, {l \choose m} \, (-1)^{l+m} \, \frac{k}{z} \, p_k \, \theta\left(m/k - T \right),
\end{equation}
we linearize Eq. \eqref{eq:pre_lin} around $\phi = 0$:
\begin{equation}
    \label{eq:linear}
    \frac{d \phi}{dt} \approx  \langle p_A \rangle \, ( C_1 -1) \, \phi.
\end{equation}
The solution for Eq. \eqref{eq:linear} is then
\begin{equation}
    \label{eq:phi_general}
    \phi(t) = x^{-}_{0}\,  e^{(C_1 - 1) \, \int_0^t \langle p_A \rangle (s) \, ds},
\end{equation}
given that $ \phi(0) = x^{-}_{0}$.

Linearization is useful to determine the time dependence of the cascade process.  Assuming a small initial seed and rewriting the term $h(\phi)$ as  $ \sum_{l=0}^{\infty} K_l\,  \phi^l $, the linearized equation for the fraction of adopters $x^{-}(t)$ becomes
\begin{equation}
    \label{eq:linear_r}
    \frac{d x^{-}}{dt} \approx  \langle p_A \rangle\,  ( K_1 -1)\, \phi,
\end{equation}
where the coefficients $K_l$ are
\begin{equation}
    \label{eq:coef_rho}
    K_l = \sum_{k=l}^{\infty} \sum_{m=0}^{l} { k \choose l} \, {l \choose m} \, (-1)^{l+m}\,  p_k \, \theta\left( m/k - T \right).
\end{equation}

A solution for the fraction of adopters $x^{-}(t)$ can be obtained from  Eqs. \eqref{eq:phi_general} and \eqref{eq:linear_r}.  For the case of the Granovetter-Watts model without aging, setting $\langle p_A \rangle = 1$,  the solution is an exponential cascade dynamics
\begin{equation}
    x^{-}(t) = x^{-}_{0} \, e^{(C_1 - 1)\, t}.
\end{equation}
Therefore, the number of adopters $x^{-} (t)$ follows an exponential increase with exponent $\alpha(z,T)$:
\begin{equation}
    \label{eq:alpha}
    \alpha(z,T) = C_1 - 1 = \sum_{k=0}^{\lfloor 1/T \rfloor} \frac{k \, (k - 1)}{z}\, p_k - 1,
\end{equation}
where $C_1$ is computed from Eq. \eqref{eq:coef_phi}. 

For endogenous aging, the same derivation is valid to determine the exponents $\delta(z,T)$. Using $\langle p_A \rangle = 1/(t+2)$, the fraction of adopters follows a power law dependence,
\begin{equation}
    \label{rho_endo}
    x^{-}(t) = x^{-}_{0} \, \left( \frac{t+2}{2} \right)^{(C_1 - 1)}.
\end{equation}
The exponent reported for the power law cascade dynamics $\delta(z,T)$ turns out to be, therefore, the same exponent as the one for the exponential behavior where there is no aging:  $\delta(z,T)= \alpha(z,T)= C_{1} - 1$. Fig. \ref{fig:endo_exp} compares the prediction of Eq. \eqref{eq:alpha} with the results computed from Monte Carlo simulations. There is a good agreement for both Barab\'asi-Albert and Erd\H{o}s-R\'enyi networks for all values of $T$ and $z$. For a random-regular graph, the predicted dependence, $\alpha(z) = z - 2$, is not a good approximation for large $z$. This is because the presence of small cycles increases importantly in a random-regular graph as the average degree $z$ grows~\cite{wormald_1999} and the locally-tree assumption made for the derivation of the rate equations (Eq. \eqref{eq:AME_Threshold}) is not valid anymore. A different approach is necessary for clustered networks (as in Ref.\cite{Leah2022} for the Granovetter-Watts model). 

\begin{figure}[b!]
    \centering \captionsetup{font=sf}
    \includegraphics[width=0.6\columnwidth]{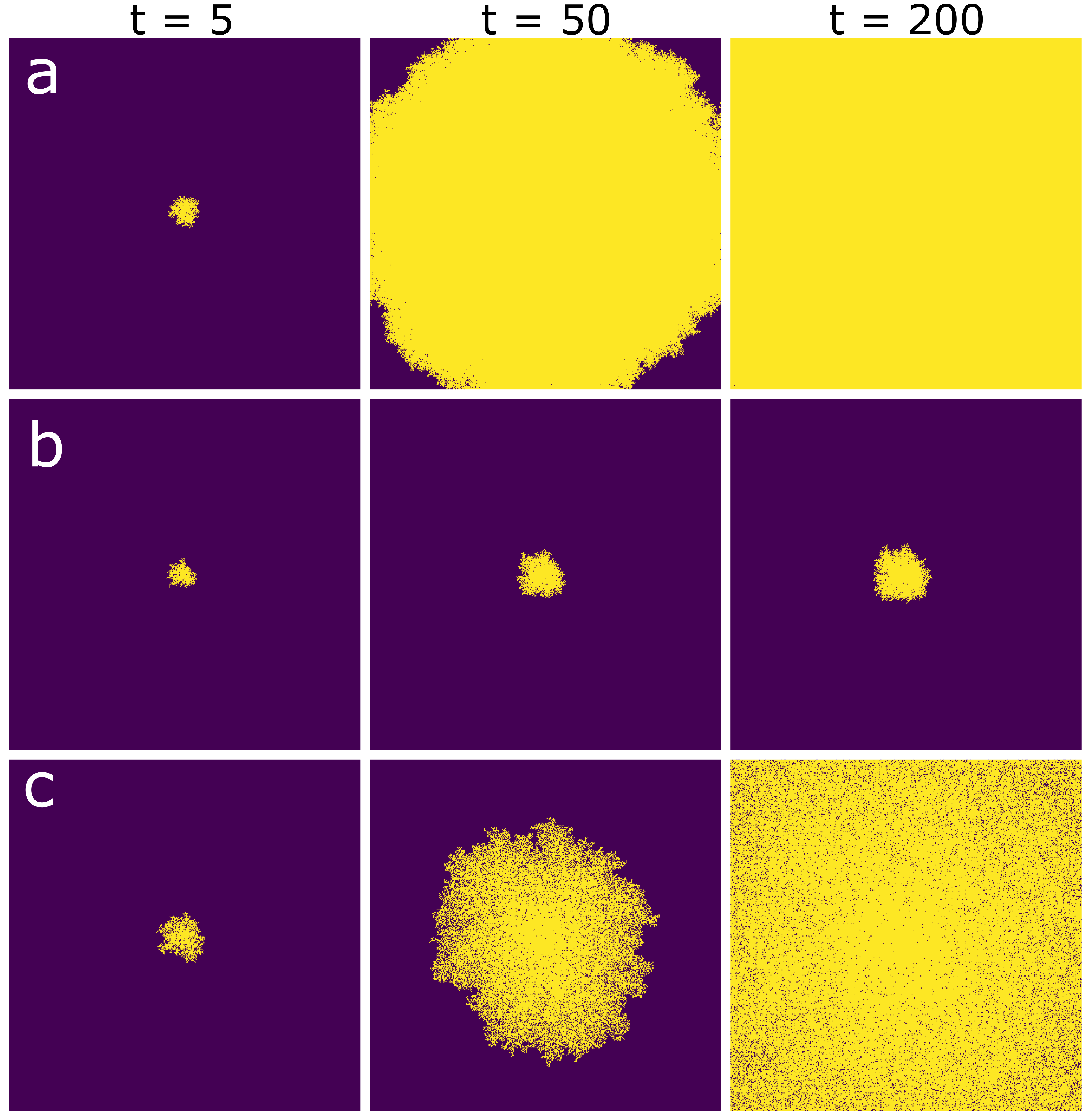}
    \caption[Cascade spreading of the Granovetter-Watts model in a lattice]{\label{fig:evo_lat} Cascade spreading of the original Granovetter-Watts model \textbf{(a)} and the versions with endogenous (reset if adopts) \textbf{(b)} and exogenous (reset if activates) \textbf{(c)} aging on a Moore neighborhood lattice with size $N = L \times L$, $L = 405$. Yellow and purple nodes are adopters and non-adopters, respectively. Time increases from left to right. Initial seeds are selected favoring cascades: one agent and all him/her neighbors are set as adopters at the center of the system.}
\end{figure}

Moreover, from Eq. \eqref{eq:linear}, we can extract the cascade condition for the Granovetter-Watts model in general. Since $\langle p_A \rangle(t)$ is always positive, global cascades occur when $(C_1 - 1) > 0 $, so the cascade condition is:
\begin{equation}
    \label{eq:umbral}
    \sum_{k=0}^{\lfloor 1/T_c \rfloor} \frac{k \, (k - 1)}{z}\, p_k = 1
\end{equation}
This cascade condition does not depend on the aging term $\langle p_A \rangle(t)$ and thus, it is the same as for the Granovetter-Watts model without aging. In Fig. \ref{fig:umbral}, the red solid line is the result of this analytical calculation, and it is in good agreement with the numerical results. 

For exogenous aging, an analytical expression for the exponent $\gamma(z,T)$ is not obtained following this methodology. Still, we can fit the exponent from the numerical solutions in Fig. \ref{fig:exo_endo_evo} (d-f). Fig.\ref{fig:exo_exp} shows a good comparison between the exponent calculated from the numerical solutions and the one calculated from  Monte Carlo simulations. The dependence of $\gamma(z,T)$ with the parameters $z$ and $T$ is qualitatively similar to the dependence of  $\alpha(z,T)$.

\section{\label{sec:Lattice} Results on a Moore lattice}

The Granovetter-Watts model in a two-dimensional regular lattice with a Moore neighborhood (nearest and next nearest neighbors) has a critical threshold (cascade condition) $T_c = 3/8$~\cite{centola-2007}. Below this value, cascade dynamics follows a power law increase in the density of adopters $x^{-}(t)$, which does not depend on the threshold value $T$~\cite{centola-2007}. In Fig. \ref{fig:evo_lat}a, we show a typical realization of this model: From an initial seed, the adoption radius increases linearly with time until all agents adopt the technology.

\begin{figure}[b!]
    \centering \captionsetup{font=sf}
    \includegraphics[width=\columnwidth]{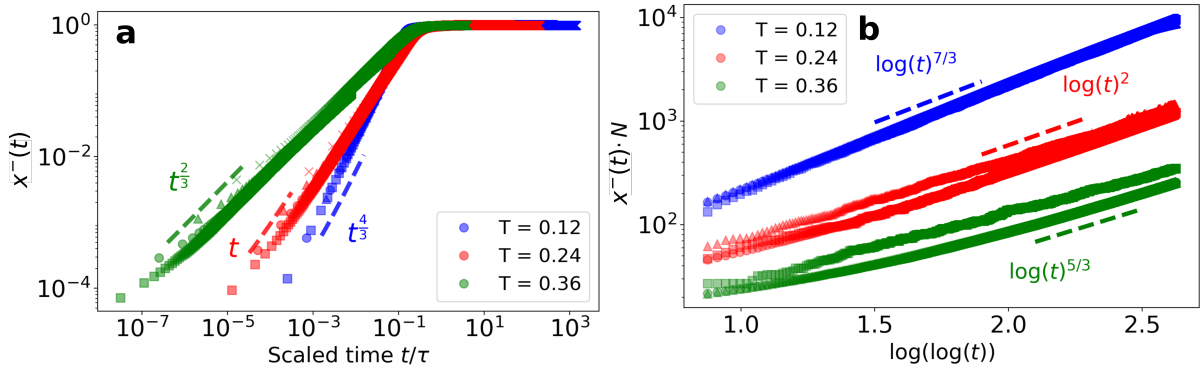}
    \caption[Cascade dynamics snapshots in a lattice]{\label{fig:lattice} Cascade dynamics of the Granovetter-Watts model with exogenous \textbf{(a)} and endogenous \textbf{(b)} aging on a Moore neighborhood lattice. Different colors indicate different values of the threshold $T$. Different markers indicate the results of Monte Carlo simulations with different system size $N = L \times L$:  $L = 50$ (crosses), $100$ (triangles), $200$ (circles) and $400$ (squares). In (a), time is scaled according to size $\tau = L^{2 / \zeta}$. Discontinuous solid lines indicate a power law behavior with exponent $ \zeta = 4/3$ (blue), $1$ (red) and $2/3$ (green). In (b), the system sizes are not scaled due to the slow dynamics. Discontinuous solid lines indicate a power-logarithmic behavior, $x^{-}(t) \, N \sim \log(t)^{\nu} $, with exponent $ \nu = 7/3$ (blue), $2$ (red) and $5/3$ (green).}
\end{figure}

When aging is considered, cascade dynamics become much slower and a dependence on $T$ appears. When the aging mechanism is exogenous, Monte Carlo simulations indicate cascade dynamics following a power law $x^{-}(t) \approx t^{\zeta(T)}$. Qualitatively, we observe that while in the case without aging there was a soft interface between adopter and non-adopters, aging causes a strong roughening in the interface and the presence of non-adopters inside the bulk (see Fig. \ref{fig:evo_lat}c). In addition, the exponent values fitted from Monte Carlo simulations allow us to collapse curves for different system sizes (see Fig. \ref{fig:lattice}a). Due to finite size effects, the interface between adopters and non-adopters eventually reaches the borders of the system and the remaining non-adopters, in the bulk, will slowly adopt with the density of adopters following the functional shape $x^{-}(t) = 1- 1/(t+2)$.

Fig.\ref{fig:evo_lat}b shows the dynamics towards global adoption for endogenous aging. In comparison with the case of exogenous aging, we do not observe strong interface roughening between adopters and non-adopters, because non-adopters are not present in the bulk. Monte Carlo simulations indicate a very slow increase of the density of adopters $x^{-}$, similar to a power-logarithmic growth  $x^{-}(t) \approx (\log(t))^{\nu}$, with a threshold dependent exponent $\nu(T)$  (Fig. \ref{fig:lattice}b). Our general approximation used for complex networks assumes a tree-like network, and it is not appropriate for the Moore lattice.

\section{\label{sec:Summary and Conclusions} Summary and discussion}

We have addressed in this chapter the role of aging in general models with binary-state agents interacting in a complex network. Temporal activity patterns are incorporated by means of a variable that represents the internal time of each agent. We have developed an approximate Master Equation for this general situation. In this framework, we have explicitly studied the effect of aging in the Granovetter-Watts model as a paradigmatic example of Complex Contagion processes. Aging implies a lower probability to change state when the internal time increases. We considered  two aging mechanisms: endogenous aging, in which the internal time measures the persistence time in the current state, and exogenous aging, in which the internal time measures the time since the last update attempt.


Our theoretical framework with some approximations to attain analytical results provide a good description of the results from Monte Carlo simulations for Erd\H{o}s-R\'enyi, random-regular and Barab\'asi-Albert networks. For these three types of complex networks, we found that the cascade condition $T_c$ (critical value of the threshold parameter $T$ as a function of mean degree $z$ of the network) for the full spreading from an initial seed is not changed by the aging mechanisms. However, aging modifies in non-trivial ways cascade dynamics of the process. The exponential growth with exponent $\alpha(z,T)$ of the density of adopters in the absence of aging becomes a power law with exponent $\delta(z,T)$ for endogenous aging, and a stretched exponential characterized by an exponent $\gamma(z,T)$ for exogenous aging. We have analyzed the exponents' dependence with the order parameters $\alpha(z,T)$, $\delta(z,T)$, $\gamma(z,T)$ and shown that $\delta(z,T)=\alpha(z,T)$, for which an analytical expression is obtained.

Our general theoretical framework, based on the assumption of a tree-like network, is not appropriate for a regular lattice. In this case, we have been only able to run Monte Carlo simulations. Our results indicate that  exogenous aging gives rise to adoption dynamics characterized by an increase in the interface roughness, by the presence of non-adopters in the bulk, and by a power law growth of  the density of adopters with exponent $\zeta (T)$, while in the absence of aging $\zeta = 2$ independently of $T$. Endogenous aging, on the other hand, produces very slow (logarithmic-like) dynamics, with a threshold-dependent exponent $\nu(T)$.

This study highlights the importance of non-Markovian dynamics in general  binary-state dynamics and, specifically, in the Granovetter-Watts model. For the problem of innovation adoption that this model addresses, we show how persistence times have an important impact on the adoption cascade. Further work in this direction would be to categorize technologies according to the adoption curve, to show if the system has important resistance to the previous technology (endogenous aging) or a balance between memory and external influence or advertisement (exogenous aging). Furthermore, the theoretical framework presented here gives a basis for further investigations of the memory effects and non-Markovian dynamics in networks, and in particular for  binary-state models with aging. Still, a number of theoretical developments remain open for future work, such as the consideration of stochastic finite size effects~\cite{peralta-2020B}, or extending this framework to tri-state models. Also, proper approximations need to be developed to account for some of our numerical results for random-regular networks with high degree, as well as for high clustering, degree-degree correlations networks and for regular lattices, including continuous field equations for this latter case.

\renewcommand{\thechapter}{6A} 
\chapterimage{Images/STM2.pdf}
\chapterspaceabove{6.75cm}
\chapterspacebelow{7.25cm}

\chapter{\label{ch:Ordering dynamics in the Symmetrical Threshold model} Symmetrical Threshold model: Ordering dynamics}
\vspace{-1.2cm}
\small
\textbf{The results in this chapter are published as:}
\vspace{0.05 cm}

\fullcite{Abella_2024}
\normalsize
\vspace{0.5 cm}

In the previous chapters, we have studied the aging effects in two different models: the Sakoda-Schelling segregation model, a 3-state threshold model with 2 symmetric states, and the Granovetter-Watts model, a binary-state threshold asymmetric model. Despite both models being threshold models, the aging implications are different in both models. In this chapter and the next, we investigate a symmetric version of the threshold model and the aging implications in this model. In this first chapter, we explore the Symmetrical Threshold model in different network topologies and for different initial conditions. We find that the model exhibits three different phases: a mixed one (dynamically active disordered state), an ordered one, and a heterogeneous frozen phase. For random interaction networks, we develop a theoretical description based on an AME that describes with good accuracy the results of numerical simulations for the model.

\section{\label{sec:Introduction_Schelling} Introduction}

In recent decades, various techniques of probability and statistical physics have been employed to measure and explain social phenomena~\cite{castellano2009statistical,jusup2022social,bianconi2023complex}. A variety of social collective phenomena can be well understood through models of interacting agents. For example, the consensus problem consists of determining under which circumstances the agents end up sharing the same state or when the coexistence of both states prevails. This is characterized by a phase diagram that provides the boundaries separating domains of different behaviors in the control parameter space.

As we have seen in this thesis, an important binary-state model is the Granovetter-Watts model~\cite{granovetter-1978, watts-2002}, In this model, multiple exposures, or group interaction, are necessary to update the current state, a characteristic of complex contagion models~\cite{centola-2007,unknown-author-2018}. A main difference between the Granovetter-Watts model and other binary-state models, such as the Voter~\cite{Voter-original}, majority vote (MV)~\cite{de1992isotropic,pereira2005majority,campos2003small}, and nonlinear Voter model~\cite{castellano-2009,mobilia2015nonlinear,mellor2016characterization,Min-2017,jewski-2017,peralta-2018}, is the lack of symmetry between the two states. In the Granovetter-Watts model, changing state is only possible in one direction, representing the adoption forever of a new state that initially starts in a small minority of agents. A symmetric version of the Granovetter-Watts threshold model, with possible changes of states in both directions, shows hysteresis when the noise is introduced into the model~\cite{nowak2019homogeneous,nowak2020symmetrical}. However, a complete characterization of the Symmetrical Threshold model and its ordering dynamics have not been addressed so far.

In this chapter, we present a comprehensive analysis of the Symmetrical Threshold model, including its full phase diagram. The model is examined in various network topologies, such as the complete graph, Erd\H{o}s-Rényi (ER) ~\cite{erdos1960evolution}, random regular (RR)~\cite{wormald_1999}, and a two-dimensional Moore lattice. The possible phases of the system are defined by the final stationary state as well as by the ordering/disordering dynamics characterized by the time-dependent magnetization, interface density, persistence, and mean internal time. The results of Monte Carlo numerical simulations are compared with results from the theoretical framework provided by the Approximate Master Equation (AME) (see details in Chapter \ref{ch:Aging in binary state dynamics}), which is general for any random network. We also derive a mean-field analysis to describe the outcomes in a complete graph.

\section{\label{Symmetrical Threshold model} Symmetrical Threshold model}

The system consists of a set of $N$ agents located at the nodes of a network. The variable describing the state of each agent $i$ takes one of the two possible values: $s_i = \pm 1$. Every agent has assigned a fixed threshold $0 \leq T \leq 1$, which determines the fraction of different neighbors required to change state. Even though this value might be agent-dependent, we will consider here homogeneous $T$ for all the agents. In each update attempt, an agent $i$ (called active agent) is randomly selected, and if the fraction of neighbors with a different state is larger than the threshold $T$, the active agent changes state $s_i \to -s_i$. In other words, if $m$ is the number of neighbors in state $-1$ out of the total number of neighbors $k$, the condition to change is $\theta(m/k - T)$, for a node in state $+1$, and $\theta((k-m)/k - T)$, for a node in state $-1$, where $\theta(x)$ is the Heaviside step function. Notice that this update rule is equivalent to ``shifted'' Glauber dynamics~\cite{glauber1963time}, with swapping probability $1/(1+\exp[\beta(\Delta E + C)])$ (where $\beta$ is the inverse temperature, $\Delta E$ the energy loss to swap the state of a node according to Ising Hamiltonian and $C$ a shifting constant), at the limit of zero temperature ($\beta \to \infty$). Numerical simulations of the model run until the system reaches a frozen configuration (absorbing state) or until the average magnetization, $m = (1/N) \sum_i s_i$, fluctuates around a constant value.   Simulation time is measured in Monte Carlo (MC) steps, i.e., $N$ update attempts. 

\section{\label{sec:Results on Complex networks} Results on Random networks}

\begin{figure}
	\centering \captionsetup{font=sf}
	\includegraphics[width=0.9\textwidth]{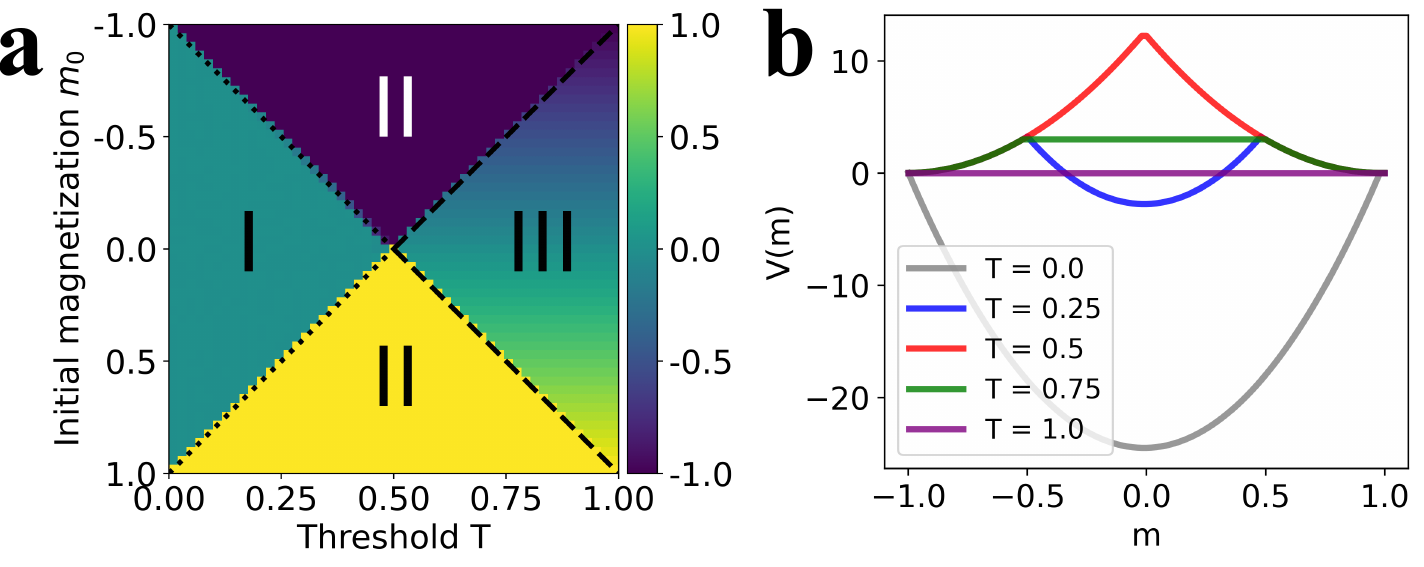}
	\caption[Phases of the Symmetrical Threshold model]{\textbf{(a)} Phase diagram of the Symmetrical Threshold model in a Complete graph of $N = 2500$ nodes. Dotted and dashed lines correspond to $T = (1-|m_0|)/2$ and $T = (1+|m_0|)/2$, respectively. The color map indicates the value of the average final magnetization $m_f$.	Average performed over 5000 realizations. \textbf{(b)} Potential representation from Eq. (\ref{eq:pot}) for a set of values of the threshold $T$, shown in different colors.}
	\label{COM_LAT_PD}
\end{figure}

\subsection{Mean-field}

We first consider the mean-field case of the complete graph (all-to-all connections). We take an initial random configuration with magnetization $m_0$ and run numerical simulations for various values of $T$ to construct the phase diagram (shown in Fig. \ref{COM_LAT_PD}a). We find three different phases based on the final state:

\begin{itemize}
	\item \textbf{Phase ${\rm {\bf I}}$ or Mixed}: The system reaches an active disordered state (final magnetization $m_f = 0$) where the agents change their state continuously;
	\item \textbf{Phase ${\rm {\bf II}}$ or Ordered}: The system reaches the ordered absorbing states ($m_f = \pm 1$) according to the initial magnetization $m_0$;
	\item \textbf{Phase ${\rm {\bf III}}$ or Frozen}: The system freezes at the initial random state $m_f = m_0$.
\end{itemize}

For a given initial magnetization $m_0 \neq 0$ and increasing $T$, the system undergoes a mixed-ordered transition at a critical threshold $T_{c} = (1-|m_0|)/2$, and an ordered-frozen transition at a critical threshold $T_{c}^{*} = (1 + |m_0|)/2 > T_{c}$ (indicated by dotted and dashed black lines in Fig. \ref{COM_LAT_PD}a, respectively). In this mean-field scheme, if the fraction of nodes in state $+1$ is denoted by $x$, the condition for a node in state $-1$ to change its state is given by $\theta(x - T)$, where  $\theta$ is the Heaviside step function. Thus, in the thermodynamic limit ($N\to \infty$), the variable $x$ evolves over time according to the following mean-field equation:
\begin{equation}
	\frac{dx}{dt} = (1 - x) \; \theta(x - T) - x \; \theta(1 - x - T) = - \frac{\partial V(x)}{\partial x}.
\end{equation}
Here, $V(x)$ is the potential function. The stationary value of $x$, $x_{\rm st}$, is the solution of the implicit equation resulting from setting the time derivative equal to $0$. The stationary solutions are $x_{\rm st} = 1/2$ ($m =0$), the absorbing states $x_{\rm st} = 0,1$ ($m = \pm 1$) or a degenerate continuum of solutions. The stability of these solutions can be understood in terms of the potential $V(x)$:
\begin{flalign}
	V(x) &=-\int (1 - x) \; \theta(x - T) - x \; \theta(1 - x - T) \; dx \nonumber\\
	&=\frac{x^2}{2} + \frac{1}{2} \left( T^2 - 2T - x^2 + 1\right) \; \theta(T+x-1) - \frac{1}{2} \left( T^2 - 2T - x(x-2)\right) \; \theta(x - T)
	\label{eq:pot}
\end{flalign}
The minimum and maximum values of $V(x)$ correspond to stable and unstable solutions, respectively. Figure \ref{COM_LAT_PD}b shows the potential's dependence on the magnetization, obtained after a variable change $m = 2x-1$ in Eq. (\ref{eq:pot}). For $T < 0.5$, $m = 0$ is a stable solution, but increasing the threshold reduces the range of values of the initial magnetization from which this solution is reached, enclosing Phase ${\rm I}$ between the unstable solutions $m = 1-2\, T$ and $2\, T-1$. In fact, if $m_0 > 1-2\, T$, the system reaches the absorbing solution $m=+1$, while if $m_0 < -1+2\, T$, it reaches $m=-1$ (Phase ${\rm II}$). For $T = 0.5$, there is just one unstable solution at $m=0$, and all the initial magnetization values reach the absorbing states $m=\pm 1$. For $T > 0.5$, the potential is equal to a constant value for a range of $m_0$, which means that an initial condition will remain in this state forever (Phase ${\rm III}$). The range of values of the initial condition from which this phase is reached grows linearly with $T$ until $T=1$, where all initial conditions fulfill $\frac{dm}{dt}=0$.

\begin{figure}
	\centering \captionsetup{font=sf}
	\includegraphics[width=\linewidth]{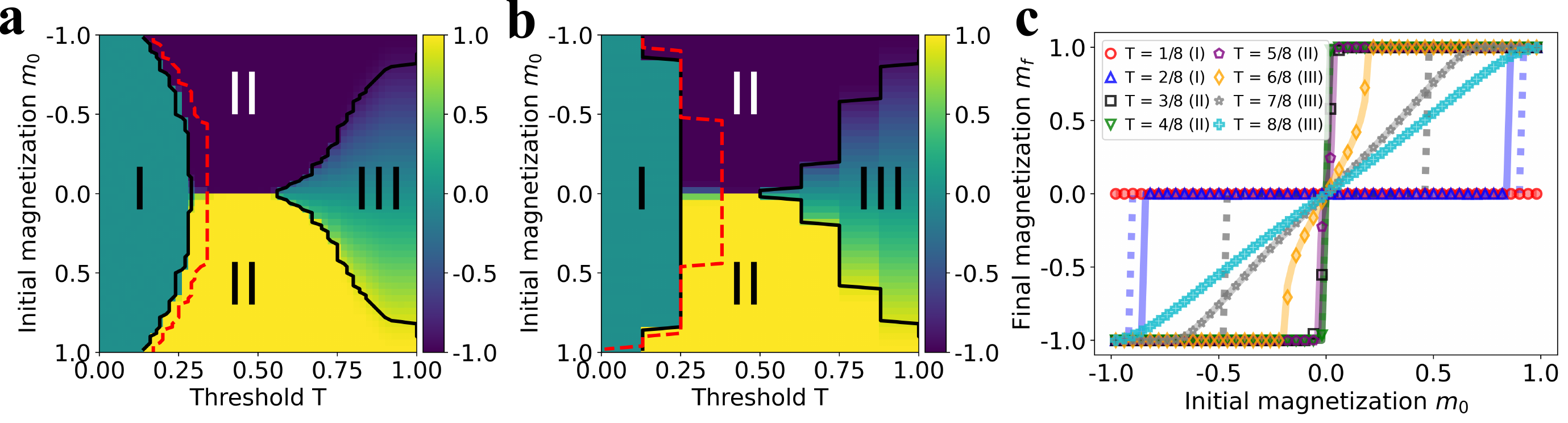}
	\caption[Phase diagram in random networks]{\label{ER_REG_PD} Phase diagram of the Symmetrical Threshold model in an ER \textbf{(a)} and a RR \textbf{(b)} graph, both of $N=4\cdot10^4$ nodes and mean degree $\langle k \rangle=8$. The color map indicates the value of the average final magnetization $m_f$. The red dashed line is the HMF prediction of the mixed-ordered critical line. The black solid lines correspond to the AME prediction of the borders of Phase ${\rm II}$. \textbf{(c)} Average final magnetization $m_f$ as a function of the initial magnetization $m_0$ for different $T$ values (indicated with different colors and markers) in the RR graph. The average is performed over 5000 realizations. The dotted and solid lines are the HMF (for $T=1/8 - 4/8$) and AME predictions (for all $T$), respectively.}
\end{figure}

Note that the mean-field Symmetrical Threshold model for $T=1$ shows the same potential profile as the mean-field Voter model~\cite{Suchecki-2005, Voter-original,castellano2009statistical}. The important difference is that for the Voter model, any initial magnetization is marginally stable, while in our model any initial magnetization is an absorbing state in Phase ${\rm III}$. In the Voter model finite size fluctuations will take the system to the absorbing states $m=\pm 1$. 


\subsection{Random networks}

We analyze the phase diagram of the Symmetrical Threshold model in two random networks: Erd\H{o}s-Rényi (ER)~\cite{erdos1960evolution} and random regular (RR)~\cite{wormald_1999} graphs with mean degree $\langle k \rangle = 8$. Figures \ref{ER_REG_PD}a and \ref{ER_REG_PD}b show the phase diagram for both networks, where it is shown that the existence of the three phases previously described is robust to changes in network structure. The main difference from the all-to-all scenario is that Phase ${\rm III}$ does not freeze exactly at the same initial magnetization. Instead, the system reaches an absorbing state with a higher magnetization $m_f > m_0$. In this phase, the value of $m_f$ depends on the threshold such that increasing $T$, increases the disorder in the system, until $T = 1$, where $m_f = m_0$ (see Fig. \ref{ER_REG_PD}c). On the other hand, phases ${\rm I}$ and ${\rm II}$ reach the same stationary state as in the mean-field case. Furthermore, the critical thresholds $T_{c}$ and $T_{c}^{*}$ show a different dependence on $m_0$ depending on the network structure.
\begin{figure}[ht]
	\centering \captionsetup{font=sf}
	\includegraphics[width=\textwidth]{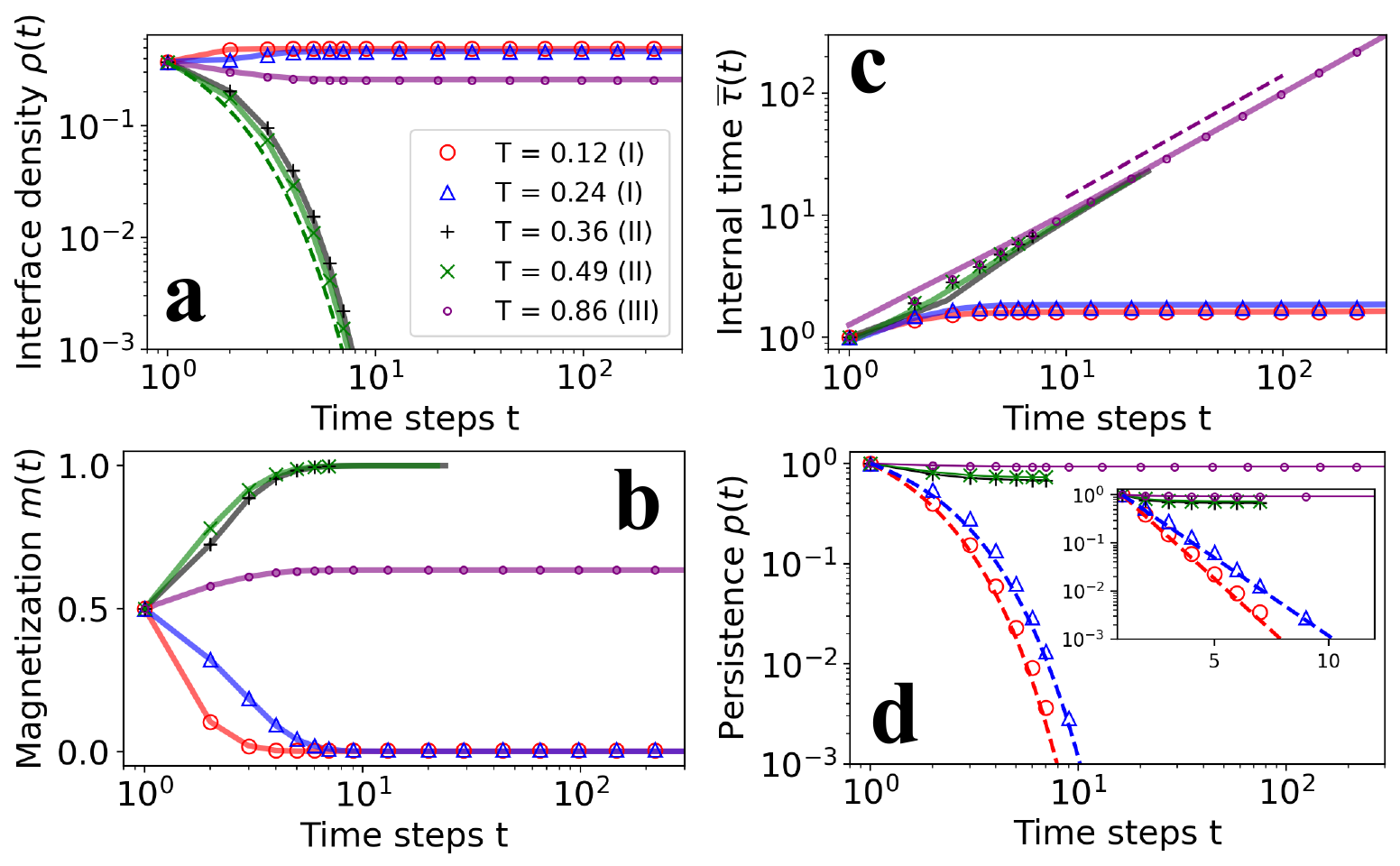}
	\caption[Symmetrical Threshold model dynamics in random networks]{\label{fig:evolution_random} Evolution of the average interface density $\rho(t)$ \textbf{(a)}, the average magnetization $m(t)$ \textbf{(b)}, the mean internal time $\bar{\tau}(t)$ \textbf{(c)}, and the persistence $p(t)$ \textbf{(d)} for the Symmetrical Threshold model. The average is computed over $5000$ surviving trajectories (simulations stop when the system reaches the absorbing ordered states). Results for different values of $T$ are plotted with diverse markers and colors: red ($T = 0.12$) and blue ($T = 0.24$) belong to Phase ${\rm I}$, green ($T = 0.36$) and gray ($T = 0.49$) belong to Phase ${\rm II}$ and purple ($T = 0.86$) belongs to Phase ${\rm III}$. Solid colored lines are the AME integrated solutions, using Eqs. (\ref{eq:interface})-(\ref{eq:time}). The initial magnetization is $m_0 = 0.5$. The system is on an ER graph with $N = 4 \cdot 10^4$ and mean degree $\langle k \rangle = 8$. The dashed green line in (a) shows $\rho(t) \sim \rho_0 \, e^{-t}$, the dashed purple line in (c) shows $\bar{\tau}(t) = t$ and the dashed lines in (d) show $p(t) \sim e^{-\alpha t}$, where $\alpha = 1$ (red) and $\alpha = 3/4$ (blue). 
	}
\end{figure}
To explain the transitions exhibited by the model, we use the AME, described in detail in Chapter \ref{ch:Aging in binary state dynamics}, which considers agents in both states $\pm 1$ with degree $k$, $m$ neighbors in state $-1$ that have been $j$ time steps in the current state (called ``internal time'' or ``age'') as different sets in a compartmental model. For the Symmetrical Threshold model, the dynamics are Markovian, since the rates do not depend on the internal time. Nevertheless, we keep this formalism to study the evolution of the mean internal time $\bar{\tau}(t)$, and to compare with the version with aging in the next chapter. According to the update rules of the model, the rates are defined as follows:
\begin{flalign}
	&T^{+}_{k,m,j} = \theta(m/k - T) \quad \quad T^{-}_{k,m,j} = \theta((k-m)/k - T)\\
	&A^{\pm}_{k,m,j} = 1 - T^{\pm}_{k,m,j} \quad \quad \quad R^{\pm}_{k,m,j} = 0.\nonumber
\end{flalign}
Thus, the AME for the Symmetrical Threshold model is:
\begin{flalign}
	\frac{d}{d t} x^{\pm}_{k, m, 0}(t)=&- x^{\pm}_{k, m, 0}(t) + \sum_l T^{\mp}_{k, m,l} \, x^{\mp}_{k, m, l}(t) - (k-m) \,\beta^{\pm} \, x^{\pm}_{k, m, 0}(t) - m \,\gamma^{\pm} \, x^{\pm}_{k, m, 0}(t), 
	\nonumber\\
	\frac{d}{d t} x^{\pm}_{k, m, j}(t)=&- x^{\pm}_{k, m, j}(t)+ A^{\pm}_{k, m,j} \, x^{\pm}_{k, m, j-1}(t) - (k-m) \,\beta^{\pm} \, x^{\pm}_{k, m, j}(t)\label{eq:AME_age}\\
	&+ (k-m+1) \,\beta^{\pm} \, x^{\pm}_{k, m-1, j-1}(t)+ (m+1) \,\gamma^{\pm} \, x^{\pm}_{k,m+1,j-1}(t) - m \,\gamma^{\pm} \, x^{\pm}_{k, m, j}(t), \nonumber
\end{flalign}
where variables $x^{+}_{k,m,j}(t)$ and $x^{-}_{k,m,j}(t)$ are the fractions of $k$-degree nodes that are in state $+1$ (respectively, $-1$), have $m$ neighbors in state $-1$, and have age $j$. The configuration-dependent rates $\beta^{\pm}$ account for the change of state of neighbors ($\pm$) of a node in state $+1$. The rates $\gamma^{\pm}$ are equivalent but for nodes in state $-1$. If we were not concerned with the internal time dynamics, we can simplify our AME to the one reduced Markovian binary-state models (see the reduction in the section \ref{sec:Reduction to Markovian dynamics}).

The mixed-ordered and ordered-frozen transitions predicted (solid black lines in Figs. \ref{ER_REG_PD}a and \ref{ER_REG_PD}b, respectively) are in agreement with the numerical simulations. The predicted lines represent the initial and final values of $T$ at which the AME reaches the ordered absorbing states $m_f = \pm 1$. In Fig. \ref{ER_REG_PD}c, we also observe a good agreement between numerically integrated solutions (solid colored lines) and numerical simulations (markers).

An alternative simpler approximation is to consider a heterogeneous mean-field approximation (HMF) (refer to section \ref{sec:Heterogeneous mean-field approximation}). This approximation is very useful when we work with networks with high clustering, close to the complete graph scenario ($\langle k \rangle /N \to 1$), a regime where the AME does not work properly because the clustering is not negligible. For our networks, HMF captures the qualitative behavior, but the numerically integrated solutions do not agree with numerical simulations (see red dashed lines in Figs. \ref{ER_REG_PD}a and \ref{ER_REG_PD}b, and the colored dotted lines in Fig. \ref{ER_REG_PD}c), and the frozen phase is not predicted by this framework. These findings demonstrate that threshold models (in networks far from $\langle k \rangle/N = 1$) need approximations beyond mean-field to achieve accuracy~\cite{gleeson-2007,gleeson-2013}.

Beyond the stationary states, the previous phases can be characterized by their ordering dynamical regimes. To describe the coarsening process, we use the time-dependent average interface density $\rho(t)$ (fraction of links between nodes in different states), the average magnetization $m(t)$, the mean internal time $\bar{\tau}(t)$ (mean time spent in the current state over all the nodes) and the persistence $p(t)$ (fraction of nodes that remain in their initial state at time $t$)~\cite{ben-naim-1996}. Fig. \ref{fig:evolution_random} shows the average results obtained from the numerical simulations, starting from an initial magnetization $m_0 = 0.5$. There are 3 regimes with different dynamical properties:

\begin{itemize}
	\item \textbf{Mixed regime (Phase ${\rm {\bf I}}$ in the static diagram):} It is characterized by fast disordering dynamics, which is reflected by an exponential decay of the persistence. The interface density, the magnetization, and the mean internal time exhibit fast dynamics towards their asymptotic values in the dynamically active stationary state (see $T = 0.12, 0.24$ in Fig. \ref{fig:evolution_random});
	\item \textbf{Ordered regime (Phase ${\rm {\bf II}}$ in the static diagram):} It is characterized by an exponential decay of the interface density. The magnetization tends to the ordered absorbing state based on the initial majority, and the mean internal time scales as $\bar{\tau}(t) \sim t$. Persistence in this phase decays until a plateau that corresponds to the initial majority that reaches consensus (since this fraction of nodes does not change state from the initial condition). When consensus is reached, the surviving trajectory is stopped (see $T = 0.36, 0.49$ in Fig. \ref{fig:evolution_random});
	\item \textbf{Frozen regime (Phase ${\rm {\bf III}}$ in the static diagram):} It is characterized by an initial ordering process followed by the stop of the dynamics, with constant values of the metrics. The only exceptions are the mean internal time that grows as $\bar{\tau}(t) \sim t$ (see $T = 0.86$ in Fig. \ref{fig:evolution_random}) and the persistence.
\end{itemize}

Using the numerically integrated solutions of AME ($x^{\pm}_{k,m,j}(t)$) from Eq. \ref{eq:AME_age}, we can compute the magnetization $m(t)$, the interface density $\rho(t)$, and the mean internal time $\bar{\tau}$:
\begin{flalign}
	\rho(t) &=  \frac{\sum_j \sum_k p_k \sum_m  m x^{+}_{k,m,j}}{\frac{1}{2} \sum_j \sum_k p_k \sum_m  k (x^{+}_{k,m,j} + x^{-}_{k,m,j})},\label{eq:interface}\\
	\nonumber\\
	m(t) &=  2 \sum_j \sum_k p_k \sum_m x^{+}_{k,m,j} - 1 = - 2 \sum_j \sum_k p_k \sum_m x^{-}_{k,m,j} + 1,\label{eq:magne}\\
	\nonumber\\
	\bar{\tau} (t) &=  \sum_j \sum_k p_k \sum_m j \left(x^{+}_{k,m,j} + x^{-}_{k,m,j}\right),\label{eq:time}
\end{flalign}
where $p_k$ is the degree distribution of the network. All metrics exhibit a strong agreement between the numerical simulations and the integrated solutions (see solid lines in Fig. \ref{fig:evolution_random}). However, the persistence cannot be directly calculated from the integrated solutions. This is because the fraction of persistent nodes at time $t$ corresponds to the fraction of nodes with internal time $j = t$, which is at an extreme of the age distribution at each time step, since $x^{\pm}_{k,m,j}(t) = 0$ for $j > t$. Therefore, the computation of this measure requires a more sophisticated analysis using extreme value theory~\cite{haan2006extreme}.

We note that the dynamical characterization discussed above holds for all possible $m_0$ except for the symmetric initial condition $m_0 = 0$. In this case, an order-disorder transition arises at a critical mean degree $k_c$, whose value depends on the size of the system $N$~\cite{Pournaki-2022}.

\section{\label{sec: Dynamics on a Moore Lattice}  Results on a Moore Lattice}

\begin{figure}
		\centering \captionsetup{font=sf}
		\includegraphics[width=\textwidth]{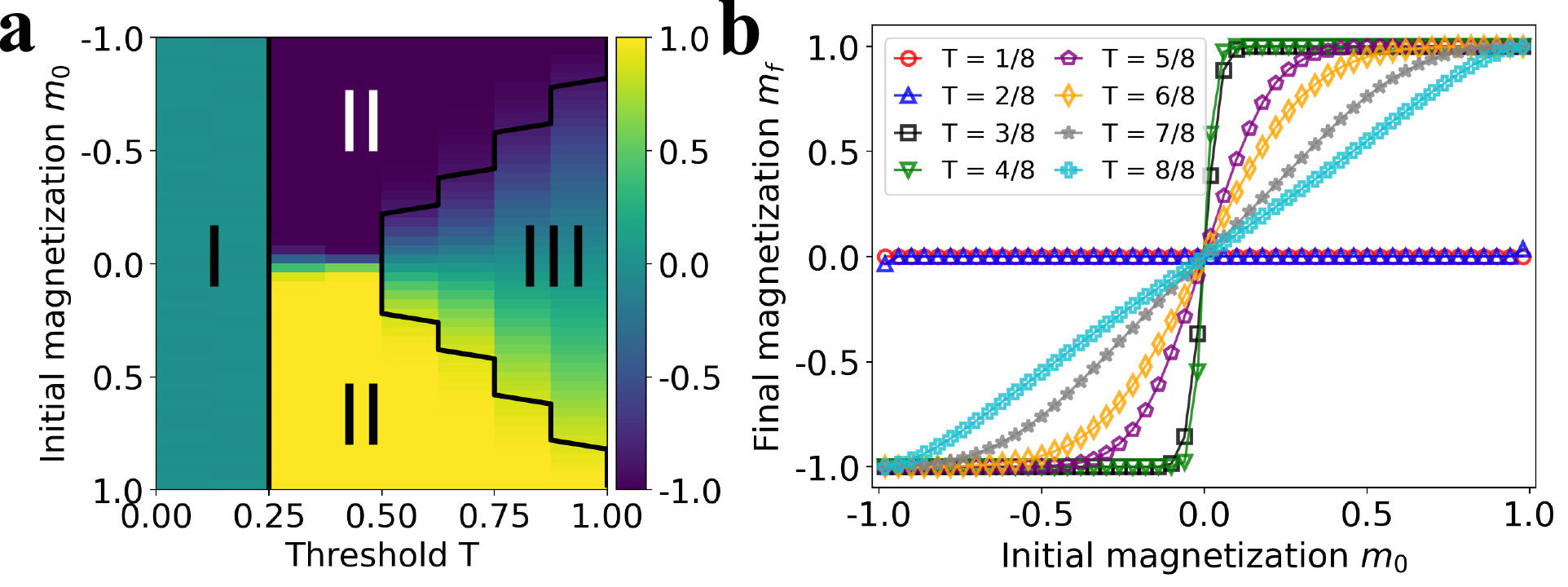}
		\caption[Symmetrical Threshold model in a Moore lattice]{\label{LAT_PD} \textbf{(a)} Phase diagram of the Symmetrical Threshold model in a Moore lattice of size $N = L \times L$, with $L = 100$. The color map indicates the value of the average final magnetization $m_f$. Solid black lines are the borders of Phase ${\rm II}$ (first and last value of $T$ where the system reaches the absorbing ordered state for each $m_0$), computed from the numerical simulations. \textbf{(b)} Average final magnetization $m_f$ as a function of the initial magnetization $m_0$ for the discrete values of the threshold $T$ (indicated with different colors and markers) in a Moore lattice of the same size. Average performed over 5000 realizations.}
\end{figure}

In this section, we consider the Symmetrical Threshold model in a Moore lattice, which is a regular 2-dimensional lattice with interactions among nearest and next-nearest neighbors ($k=8$).  From numerical simulations, we obtain a phase diagram (Fig. \ref{LAT_PD}a) that is consistent with our previous results in random networks. The system undergoes a mixed-ordered transition at a threshold value $T_{c} = 2/8$  which is independent of the value of the initial magnetization $m_0$. When $T > 4/8$, the system undergoes an ordered-frozen transition at a critical threshold $T_{c}^{*}$, which depends on $m_0$ (similarly to what happens in random networks). The final magnetization $m_f(m_0)$ (Fig. \ref{LAT_PD}b) also shows a dependence on $m_0$ similar to the one found in RR networks (Fig. \ref{ER_REG_PD}c).

\begin{figure}[t!]
	\centering \captionsetup{font=sf}
	\includegraphics[width=\linewidth]{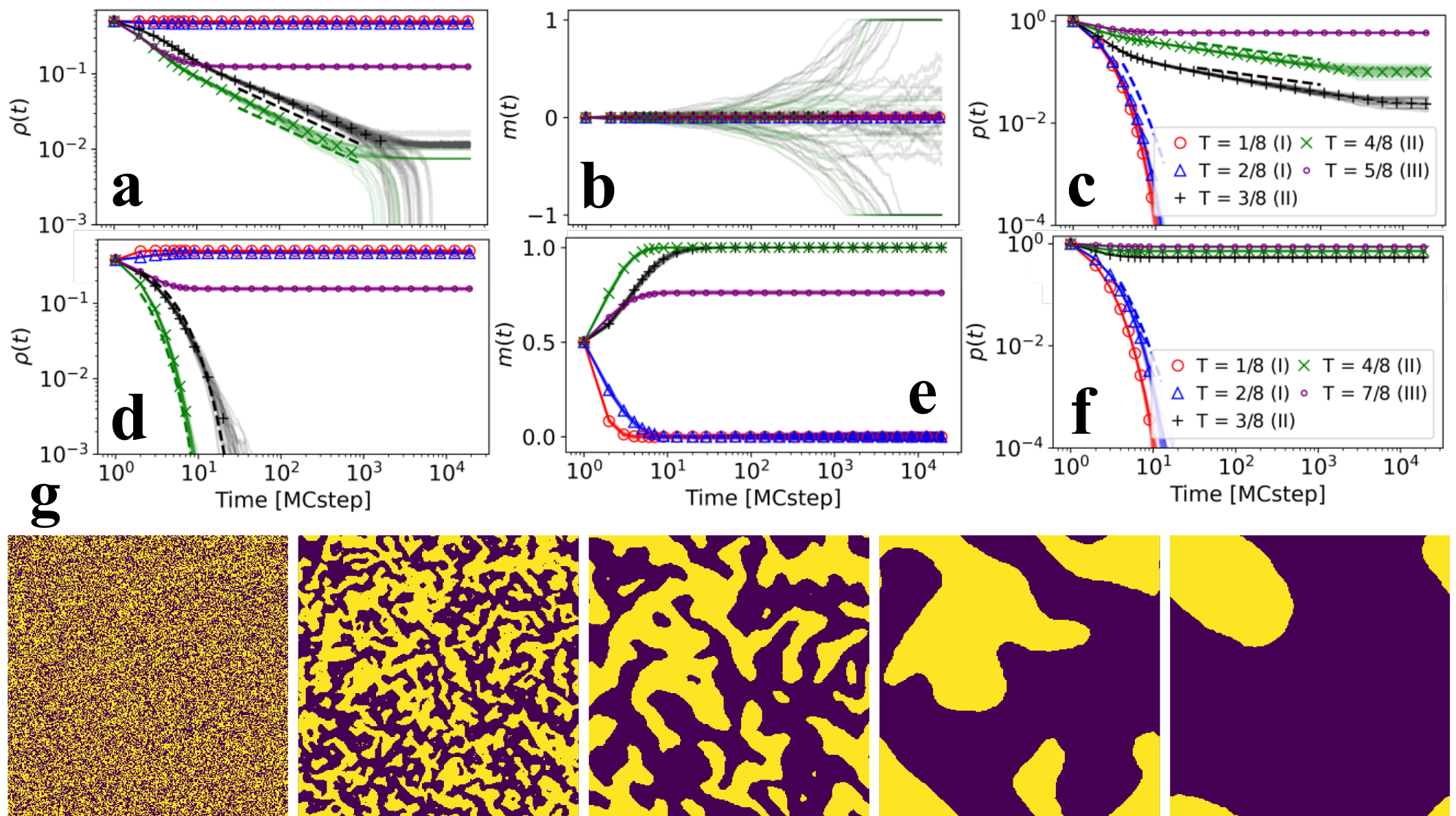}
	\caption[Dynamical regimes in a Moore lattice.]{\label{fig:evolution_lattice} Evolution of the average interface density $\rho(t)$ \textbf{(a-d)}, the average magnetization $m(t)$ \textbf{(b-e)}, and the persistence $p(t)$ \textbf{(c-f)} for the Symmetrical model in a Moore lattice starting from a random configuration with $m_0 = 0$ (a-b-c) and $m_0 = 0.5$ (d-e-f). We plot 50 different trajectories in solid lines and the average of $5000$ surviving trajectories (simulations stop when the system reaches the absorbing ordered states) in different markers. Different colors and markers indicate different threshold values: red ($T = 1/8$) and blue ($T = 2/8$) belong to Phase ${\rm I}$, green ($T = 3/8$) and black ($T=4/8$) belong to Phase ${\rm II}$, and purple ($T = 5/8, 7/8$) belong to Phase ${\rm III}$. The average magnetization $m(t)$ is computed according to the two symmetric absorbing states. System size is fixed at $N = L \times L$, $L = 200$. The dashed lines in (a) are $\rho \sim \exp(-\alpha \cdot t)$ with $\alpha = 0.5$ (black) and $\alpha = 0.8$ (green), in (d) are $\rho(t) \sim at^{-1/2}$ with $a = 0.36$ (black) and $a = 0.2$ (green), in (c-f) are $p(t) \sim \exp(- \ln(t)^2)$ (blue) and $p(t) \sim c*t^{-0.22}$, with $b = 0.12$ (black) and $b = 0.56$ (green). \textbf{(g)} Evolution of a single realization for $T = 0.5$ and $m_0 = 0$ using the Symmetrical Threshold model. Snapshots are taken after $1,10,60,440$ and $3300$ time steps increasing from left to right. System size is fixed to $N = L \times L$, $L = 256$.}
\end{figure}
Fig. \ref{fig:evolution_lattice} shows the results from numerical simulations (for $m_0 = 0$ and $0.5$) for the average interface density $\rho(t)$, the magnetization $m(t)$, and the persistence $p(t)$ (the internal time shows the same results as in random graphs). Dynamical properties change significantly for different values of the threshold and initial magnetization $m_0$. {Similarly to the case of random networks, we find three different regimes corresponding to the three phases, but with some properties different from the results on  random networks:}
\begin{itemize}
	\item \textbf{Mixed regime (Phase ${\rm {\bf I}}$):} It is characterized by fast disordering dynamics with a persistence decay $p(t) \sim \exp(- \ln(t)^2)$, consistent with the results of the Voter model~\cite{ben-naim-1996}. The interface density and the magnetization exhibit fast dynamics towards their asymptotic values in the dynamically active stationary state (see $T = 1/8,2/8$ in Fig. \ref{fig:evolution_lattice});
	\item \textbf{Ordered regime (Phase ${\rm {\bf II}}$):} It is characterized by an exponential or power-law decay of the interface density, depending on the initial condition (see details below). The magnetization tends to the absorbing ordered state (see $T = 3/8,4/8$ in Fig. \ref{fig:evolution_lattice});
	\item \textbf{Frozen regime (Phase ${\rm {\bf III}}$):} It is characterized by an initial ordering process, but the system freezes fast (see $T = 5/8$ in Fig. \ref{fig:evolution_lattice}).
\end{itemize}

In particular, in Phase ${\rm II}$ for $m_0 = 0$ the persistence and interface density decay are found to decay as a power law, $p(t) \sim t^{-0.22}$ and $\rho(t) \sim t^{-1/2}$, respectively (consistent with the results of the Ising model~\cite{stauffer-1994,derrida-1995A,derrida-1995B,derrida-1997}). For a biased initial condition ($m_0 = 0.5$), $p(t)$ decays to the initial majority fraction (which corresponds to the state reaching consensus), and $\rho(t)$ follows an exponential-like decay. Note that, for $m_0 = 0$, not all trajectories reach the ordered absorbing states ($m_f=\pm 1$). There exist other absorbing configurations as, for example,  a flat interface configuration for $T = 4/8$, no agent will be able to change, and the system remains trapped in this state. This result is not observed for $m_0 > 0$. Contrary, phases ${\rm I}$ and ${\rm III}$ show similar dynamics for balanced ($m_0 = 0$) and unbalanced ($m_0 = 0.5$) initial conditions. In Phase ${\rm I}$, the system shows disordering dynamics with a persistence decay similar to the one exhibited for the Voter model in a lattice~\cite{ben-naim-1996} while in Phase ${\rm III}$, the system exhibited freezing dynamics with an initial tendency towards the majority consensus. Due to the lattice structure and high clustering, the mathematical tools employed in the previous sections for random networks are inapplicable to regular lattices. Consequently, we limit ourselves to the results of numerical simulations.


\section{\label{sec:Summary and Conclusions_STM} Summary and discussion}

In this chapter, we have studied with Monte Carlo numerical simulations and analytical calculations the phase diagram of the Symmetrical Threshold Model. In this model, the agents, nodes of a contact  network, can be in one of the two symmetric states $\pm 1$.  System dynamics follows a complex contagion process in which a node changes state when the fraction of neighboring nodes in the opposite state is above a given threshold $T$. For $T=1/2$, the model reduces to a majority rule or the zero temperature Spin Flip Kinetic Ising Model. When the change of state is only possible in one direction, say from $1$ to $-1$, it reduces to the Threshold model~\cite{granovetter-1978,watts-2002}. We have considered the cases of a fully connected network, Erd\H{o}s-Rényi, and random regular networks, as well as a regular two-dimensional Moore lattice. 

We have found that, in the parameter space of threshold $T$ and initial magnetization $m_0$, the model exhibits three distinct phases, namely Phase ${\rm I}$ or mixed, Phase ${\rm II}$ or ordered, and Phase ${\rm III}$ or frozen. The existence of these three phases is robust for different network structures.
These phases are well characterized by the final state ($m_f$), and by dynamical properties such as the interface density $\rho(t)$, time-dependent average magnetization $m(t)$, persistence $p(t)$, and mean internal time $\bar{\tau}(t)$. These phases can be obtained analytically in the mean-field case of a fully connected network. For the random networks considered, we derive an approximate master equation (AME)~\cite{gleeson-2013} considering agents in each state according to their degree $k$,  neighbors in state $-1$, $m$, and age $j$. From this AME, we have also derived a heterogeneous mean-field (HMF) approximation. While the AME reproduces with great accuracy the results of Monte Carlo numerical simulations of the model (both static and dynamic), the HMF shows an important lack of agreement, highlighting the importance of high-accuracy methods necessary for threshold models.

The model exhibits a rich dynamical behavior, with different regimes for the interface density, magnetization, and persistence. In the mixed phase, the system shows fast disordering dynamics, with an exponential decay of the persistence. In the ordered phase, the system exhibits a decay of the interface density, which can be exponential or power-law, depending on the initial condition and the topology. The magnetization tends to the ordered absorbing state, and the persistence decays to a plateau that corresponds to the initial majority that reaches consensus. In the frozen phase, the system shows an initial ordering process followed by the stop of the dynamics, with constant values of the metrics, except for the mean internal time that grows linearly with time.

Further research with the general AME used in this study would involve to incorporate finite size effects~\cite{peralta-2020B}, which are relevant when $m_0$ is close to zero for ER graphs, and would provide a mathematical framework for further analysis of the results in Ref.~\cite{Pournaki-2022}. Regarding the model, this chapter reports the main features of the Symmetrical Threshold model dynamics. However, there are several areas for future research along these lines, such as investigating the impact of strongly heterogeneous~\cite{barabasi2009scale} or coevolving networks~\cite{Zimmermann,vazquez-2008}.

\renewcommand{\thechapter}{6B} 
\chapterimage{Images/STMA.pdf}
\chapterspaceabove{6.75cm}
\chapterspacebelow{7.25cm}

\chapter{\label{ch:Aging implications in the Symmetrical Threshold model} Symmetrical Threshold model: Aging implications}
\vspace{-1.2cm}
\small
\textbf{The results in this chapter are published as:}
\vspace{0.05 cm}

\fullcite{Abella_2024}
\normalsize
\vspace{0.5 cm}

In this second part of chapter 6, we explore the effects of introducing aging in the Symmetrical Threshold model, where agents become increasingly resistant to change their state the longer they remain in it. When aging is present, the mixed phase is replaced, for sparse networks, by a new phase with different dynamical properties. This new phase is characterized by an initial disordering stage followed by a slow ordering process towards a fully ordered absorbing state. In the ordered phase, aging modifies the dynamical properties. For random contact networks, we use a theoretical description based on the Approximate Master Equation that describes with good accuracy the results of numerical simulations for the model with and without aging. The aging implications in this model show similarities with the results in both the Granovetter-Watts model and the Sakoda-Schelling model.

\section{\label{sec:Introduction_Chapter6} Introduction}

As it was introduced in previous chapter, the Symmetrical Threshold model is a binary-state model where agents change their state if the fraction of neighbors in a different state exceeds a threshold $T$. The phase diagram of this model shows 3 different dynamical regimes: Phase ${\rm I}$ or mixed, Phase ${\rm II}$ or ordered and phase ${\rm III}$ or frozen. In the mixed phase, the system shows fast disordering dynamics, with an exponential decay of the persistence. In the ordered phase, the system exhibits an exponential decay of the interface density. The magnetization tends to the ordered absorbing state, and the persistence decays to a plateau that corresponds to the initial majority that reaches consensus. In the frozen phase, the system shows an initial ordering process followed by the stop of the dynamics, since the system gets trapped in a configuration where no threshold is exceeded.

In this chapter, we investigate the effects of aging in the Symmetrical threshold model. The model is examined in the same network topologies as in the previous chapter, complete graph, Erd\H{o}s-Rényi (ER) ~\cite{erdos1960evolution}, random regular (RR)~\cite{wormald_1999}, and a two-dimensional Moore lattice, such that we can compare the results with the version without aging. The possible phases of the system are defined by the final stationary state as well as by the ordering/disordering dynamics characterized by the time-dependent magnetization, interface density, persistence, and mean internal time, as in previous chapter for the original model. The results of Monte Carlo numerical simulations are compared with results from the Approximate Master Equation (AME). We also derive a heterogeneous mean-field framework to account for the effects of aging in a complete graph, where the AME cannot be used (does not fulfil the tree-like approximation).
	
\section{\label{Symmetrical Threshold model with aging} Symmetrical Threshold model with aging}
	
In contrast to the original Symmetrical Threshold model, which assumes that agents update their state at a constant rate, the Symmetrical Threshold model with aging introduces an activation function $p_A (j)$ that is inversely proportional to the agent's internal time $j$. At each time step, the following two steps are performed:

\begin{enumerate}
    \item A node $i$ with age $j$ is selected at random and activated with probability $p_A(j)$;
    \item If the fraction of neighbors in a different state is greater than the threshold $T$, the activated node changes its state from $s_i$ to $-s_i$ and resets its internal time to $j=0$.
\end{enumerate}

As in previous chapters in this thesis, we make the choice of $p_A(j) = 1/(j+2)$ for the aging probability. This particular choice is motivated by the fact that it allows reproducing inter-event time distributions observed empirically~\cite{rybski-2009,artime-2017}. 

\section{\label{sec:Dynamics on networks} Results on Random networks}

\subsection{Mean-field}
\begin{figure}
	\centering \captionsetup{font=sf}
	\includegraphics[width=\textwidth]{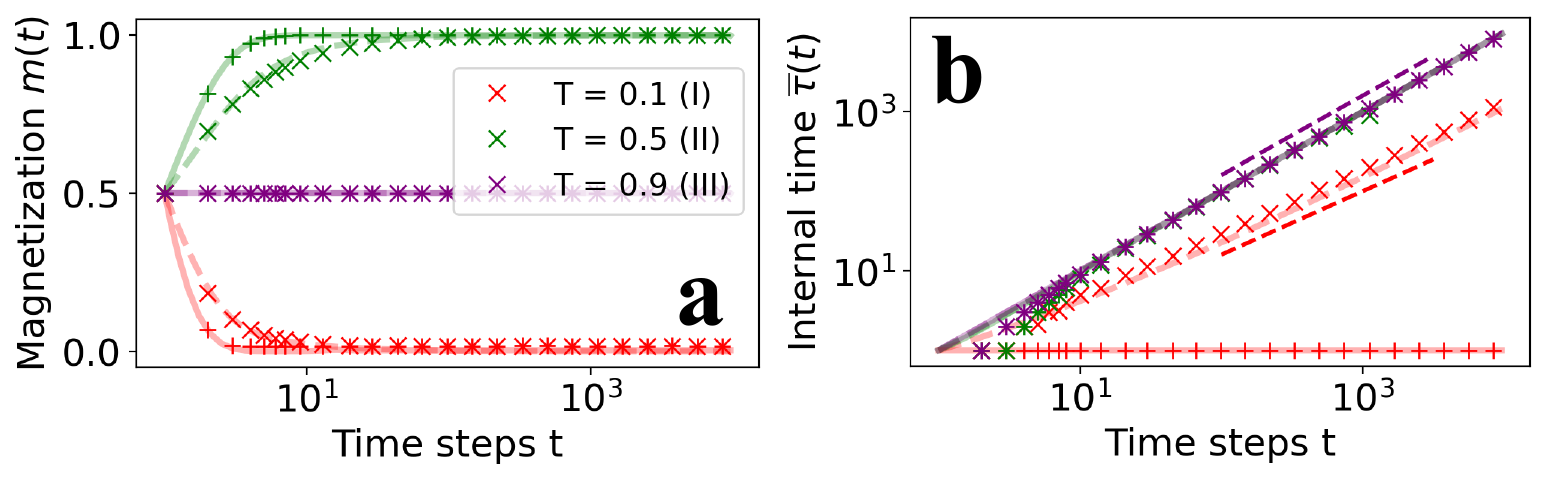}
	\caption[Aging effects in the complete graph]{\label{fig:COM_AGING} Evolution of the average magnetization $m(t)$ \textbf{(a)} and the mean internal time $\bar{\tau}(t)$ \textbf{(b)} in a complete graph of $N=2500$ nodes. Results are shown for the Symmetrical Threshold Model (pluses) and the version with aging (crosses) obtained from simulations. Different colors correspond to different values of the threshold $T$: red ($T = 0.1$) belongs to Phase ${\rm I}$, green ($T = 0.5$) belongs to Phase ${\rm II}$, and purple ($T = 0.9$) to Phase ${\rm III}$. The initial magnetization is fixed at $m_0 = 0.5$. The solid and dashed lines correspond to the numerically integrated solutions from Eq. \ref{eq:HMFaging2} for the original model ($p_A(j) = 1$) and the version with aging ($p_A(j) = 1/(t+2)$), respectively. The dashed lines in (b) show $\bar{\tau}(t) = t$ (purple) and the solution from the recursive relation in Eq. (\ref{eq:RR}) (red).}
\end{figure}

Figure \ref{fig:COM_AGING} compares the evolution of the average magnetization and mean internal time on a complete graph of the original Symmetrical Threshold model and the version with aging in phases ${\rm I}$, ${\rm II}$, and ${\rm III}$. We observe that, for all considered threshold values, aging introduces a delay. However, the final stationary magnetization coincides with the one observed for the original model. To explain these dynamics, we use a heterogeneous mean-field approach that considers the effects of aging (HMFA)~\cite{chen-2020}. Notice that, for a complete graph, the AME cannot be used, as it does not fulfill the tree-like approximation $\langle k \rangle = N$. We derive here this formalism for heterogeneous degree networks, such that we can compare the results with the AME in the next section: for a general network with degree distribution $p_k$, we define the fraction of agents in state $\pm 1$ with $k$ neighbors and age $j$ at time $t$ as $x^{\pm}_{k,j} (t)$. The probability of finding a neighbor in state $\pm 1$ is $\tilde{x}^{\pm}$, which can be written as 
\begin{equation}
    \tilde{x}^{\pm} = \sum_k p_k \frac{k}{\langle k \rangle} \,  \sum_{j=0}^{\infty} x^{\pm}_{k,j},
\end{equation}
where $\langle k \rangle$ is the mean degree of the network. The transition rate $\omega_{k,j}^{\pm}$ for a node with state $\pm 1$, degree $k$ and age $j$ to change state is given by 
\begin{equation}
    \omega_{k,j}^{\pm} = p_{A} (j) \,  \sum_{m=0}^{k} \theta\left(\frac{m}{k} - T\right) \,  B_{k,m}[\tilde{x}^{\mp}],
\end{equation}
where $B_{k,m}[x]$ is the binomial distribution with $k$ attempts, $m$ successes, and with the probability of success $x$. In our model, there are two possible events for a node with degree $k$ and age $j$:
\begin{itemize}
    \item It changes state and the age is reset to $j = 0$;
    \item It remains at its state and the age increases by one time step $j = j + 1$.
\end{itemize}
According to these possible events, we can write the rate equations for the variables  $x^{\pm}_{k,j}$ and $x^{\pm}_{k,0}$ as
\begin{flalign}
    \frac{dx^{\pm}_{k,0}}{dt} & = \sum_{j=0}^{\infty} x^{\mp}_{k,j} \,  \omega_{k,j}^{\mp} - x^{\pm}_{k,0},\nonumber\\
    \frac{dx^{\pm}_{k,j}}{dt} & =  x^{\pm}_{k,j-1} \,  ( 1 - \omega_{k,j-1}^{\pm}) - x^{\pm}_{k,j} \qquad j > 0. \label{eq:HMFaging2}
\end{flalign}
It can be shown from Eq. (\ref{eq:HMFaging2}) that the stationary solution  for the fraction of agents in state $+1$, $x_f$, obeys the following implicit equation for a complete graph (see Appendix \ref{appendix_HMFA} for a detailed explanation):
\begin{equation}
    x_f = \frac{F(1 - x_f)}{F(x_f) + F(1-x_f)},
    \label{eq:x_f}
\end{equation}
where,
\begin{equation}
    F(x) = 1 + \sum_{j=1}^{\infty} \prod_{a=0}^{j-1} \left( 1 - p_A(a) \, \sum_{m = (N-1)T}^{N-1} B_{N-1,m}[x] \right).
    \label{eq:F(A)}
\end{equation}

\begin{figure}
        \centering \captionsetup{font=sf}
        \includegraphics[width=\textwidth]{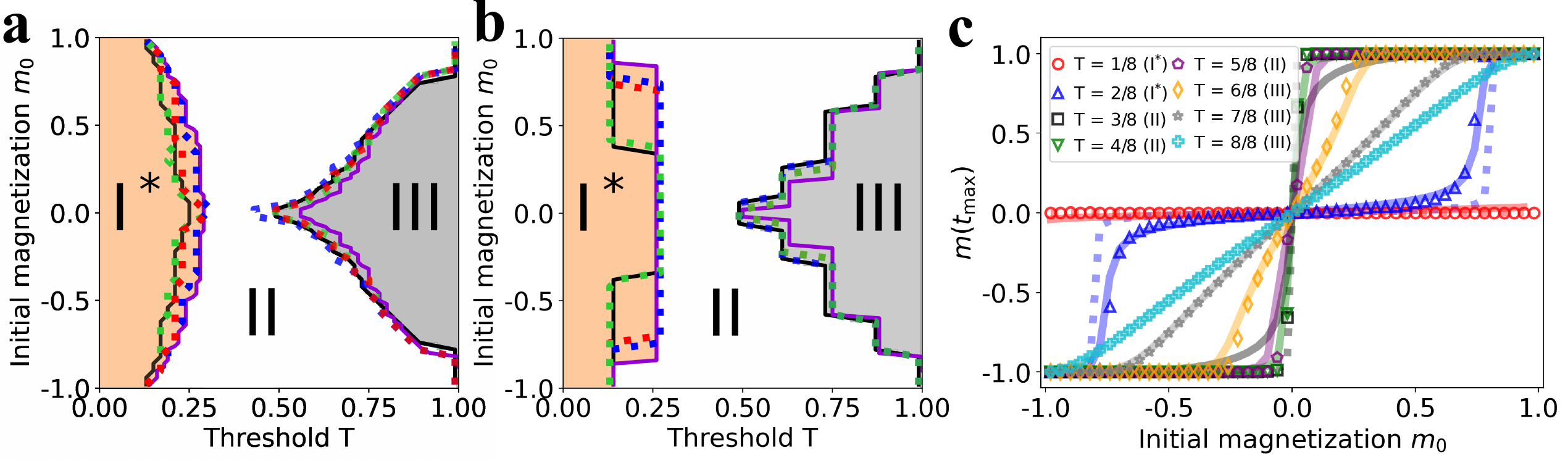}
        \caption[Phase diagram modified by aging.]{\label{ER_REG_PDAGING} Phase diagram of the Symmetrical Threshold with aging model in an ER \textbf{(a)} and RR \textbf{(b)} graph of $N = 4 \cdot 10^4$ nodes and $\langle k \rangle = 8$. The blue, red, and green dotted lines show the borders of Phase ${\rm II}$ (first and last value of $T$ where the system reaches the absorbing ordered state for each $m_0$) computed from numerical simulations evolving until $t_{\rm max} = 10^3$, $10^4$ and $10^5$ time steps, respectively. Black solid lines show AME solution integrated $10^5$ time steps. Phase ${\rm I}^{*}$, ${\rm II}$ and ${\rm III}$ correspond with the orange, white and gray areas, respectively. The solid purple lines are the mixed-ordered and ordered-frozen critical lines for the non-aging version of the model. \textbf{(c)} Average magnetization at time $t_{\rm max}$ ($m(t_{\rm max})$) as a function of the initial magnetization $m_0$ for different values of the threshold $T$ (indicated with different colors and markers) in an 8-regular graph of $N = 4 \cdot 10^4$. Average performed over 5000 realizations evolved until $t_{\rm max} = 10^4$ time steps. Dotted and solid lines are the HMFA (for $T = 1/8 - 4/8$) and AME (for all $T$) solutions integrated numerically $10^4$ time steps.}
\end{figure}

A solution of Eq. (\ref{eq:x_f}) can be obtained numerically using standard methods. The final magnetization is calculated as $m_f = 2 \,x_f - 1$. With this method, we obtain that the phase diagram for the model with aging is the same as for the model without aging (refer to Fig. \ref{COM_LAT_PD}a). As a technical point, we note that a truncation of the summation of the variable $j$ in Eq. (\ref{eq:F(A)}) is required for the numerical resolution of the implicit equation. The higher the maximum age considered $j_{\rm max}$, the higher the accuracy. With $j_{\rm max} = 5 \cdot 10^4$, the transition lines predicted by this mean-field approach show great accuracy. Moreover, by numerically integrating Eqs. (\ref{eq:HMFaging2}), the dynamical evolution of the magnetization and mean internal time can be obtained. Fig. \ref{fig:COM_AGING} shows the agreement between integrated solutions and Monte Carlo simulations of the system both for the aging and non-aging versions. It should be noted that, while aging introduces only a dynamical delay for the magnetization $m(t)$, the mean internal time $\bar{\tau}(t)$ in Phase ${\rm I}$ shows a different dynamical behavior with aging than in the original model (where $\bar{\tau}(t)$ fluctuates around a stationary value). In this phase, due to the low value of $T$, the agents selected randomly will change their state (as they fulfill the threshold condition) and reset their internal time. Consequently, while the internal time fluctuates around a stationary value for the original model, when aging is incorporated, due to the activation probability $p_A(j)$ chosen, the mean internal time increases following a recursive relation (Eq. (\ref{eq:RR})). We refer to Appendix \ref{appendix_RR} for a derivation of this result.

\subsection{\label{sec:Complex networks aging} Random networks}

In contrast to the results obtained in a complete graph, aging effects have a significant impact on the phase diagram of the model on random networks. In Fig. \ref{ER_REG_PDAGING}, we show the borders of Phase ${\rm II}$ (first and last value of $T$ where the system reaches the absorbing ordered state for each $m_0$) obtained from Monte Carlo simulations running up to a maximum time $t_{\rm max}$ (dotted colored lines). Reaching the stationary state in this model requires many steps (with a corresponding high computational cost). The two borders of Phase ${\rm II}$ exhibit different behavior as we increase the time cutoff $t_{\rm max}$: while the ordered-frozen border does not change with different $t_{\rm max}$, the mixed-ordered border is shifted to lower values of $T$ as we increase the time cutoff $t_{\rm max}$. Our results suggest that Phase ${\rm I}$ is actually replaced in a good part of the phase diagram by an ordered phase in which the absorbing state $m_f = \pm 1$ is reached after numerous time steps. These results occur for both ER (Fig. \ref{ER_REG_PDAGING}a) and RR (Fig. \ref{ER_REG_PDAGING}b) graphs. The ordered-frozen border is now slightly shifted to lower values of the threshold $T$ due to aging. Figure \ref{ER_REG_PDAGING}c shows the average magnetization on RR graphs with simulations running up to a time $t_{\rm max} = 10^4$. Upon comparison with Figure \ref{ER_REG_PD}c, the dependence on $m_0$ is quite similar, indicating the persistence of a transient mixed phase. The dependence of the results with $t_{\rm max}$ calls for a characterization of different phases in terms of dynamical properties rather than by the asymptotic value of the magnetization.



\begin{figure}[b!]
    \centering \captionsetup{font=sf}
    \includegraphics[width=\textwidth]{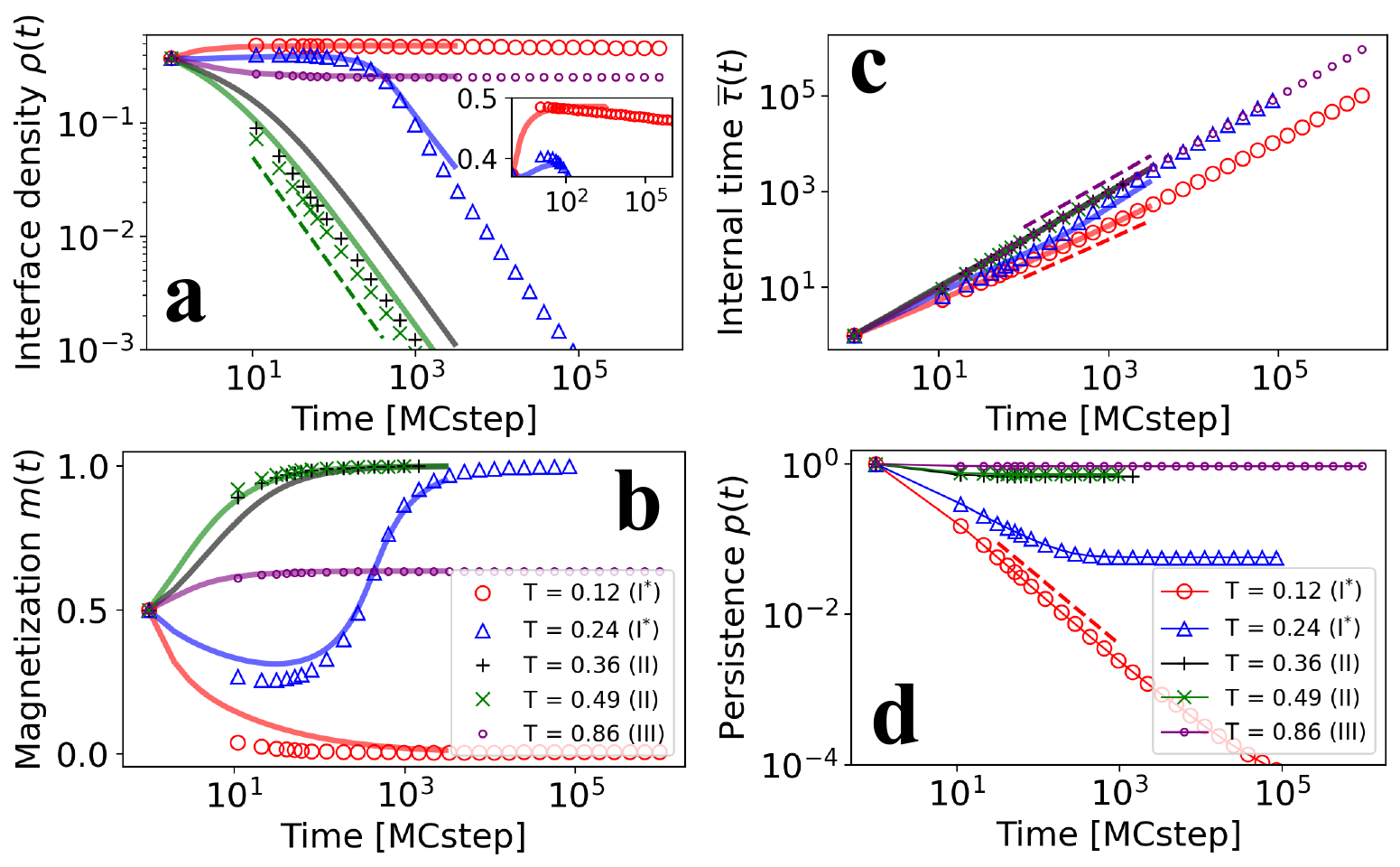}
    \caption[Symmetrical threshold model with aging dynamics in random networks]{\label{fig:evolution_random_aging} Evolution of the average interface density $\rho(t)$ \textbf{(a)}, the average magnetization $m(t)$ \textbf{(b)}, the mean internal time $\bar{\tau}(t)$ \textbf{(c)} and the persistence $p(t)$ \textbf{(d)} for the Symmetrical Threshold model with aging. The average is computed over $5000$ surviving trajectories (simulations stop when the system reaches the absorbing ordered states) for different values of $T$, shown by different markers and colors: red ($T = 0.12$) and blue ($T = 0.24$) belong to Phase ${\rm I}^{*}$, green ($T = 0.36$) and gray ($T = 0.49$) belong to Phase ${\rm II}$ and purple ($T = 0.86$) belong to Phase ${\rm III}$. The inset in (a) shows a close look to the evolution for $T = 0.12$, in linear-log scale. Solid colored lines are the AME integrated solutions for $10^4$ time steps, using Eqs. \ref{eq:interface} - \ref{eq:magne}. The initial magnetization is $m_0 = 0.5$. The system is on an ER graph with $N = 4 \cdot 10^4$ and mean degree $\langle k \rangle = 8$. The dashed green line in (a) shows $\rho(t) \sim \rho_0 \, t^{-1}$. The dashed lines in (c) $\bar{\tau}(t) = t$ (purple) and the solution from the recursive relation in Eq. (\ref{eq:RR}) (red). The dashed red line in (d) shows $p(t) = t^{-1}$.
    }
\end{figure}

\begin{figure}[t!]
    \centering \captionsetup{font=sf}
    \includegraphics[width=0.9\textwidth]{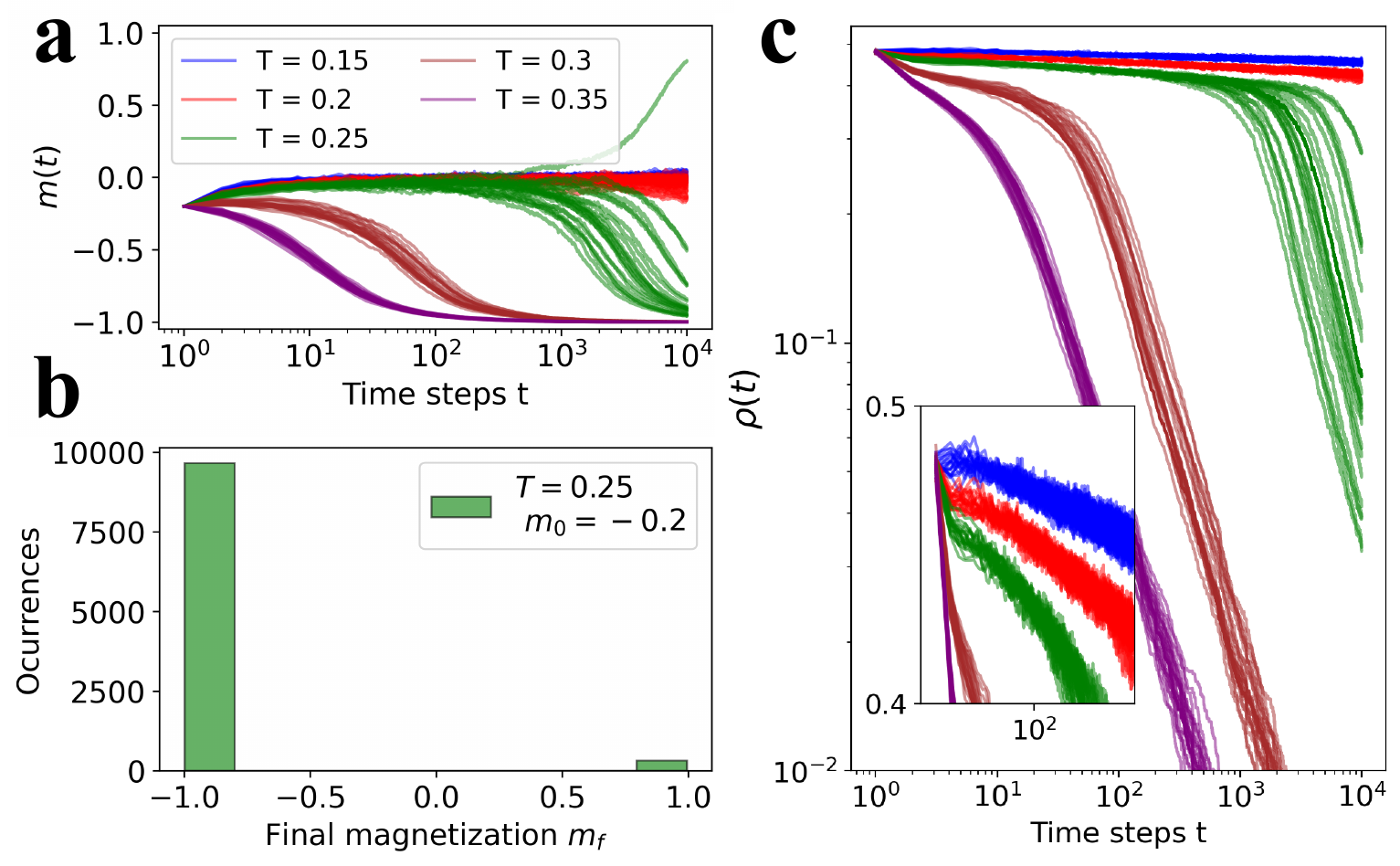}
    \caption[Phase ${\rm I}^{*}$ slow decay and minority consensus]{\label{fig:mixed_phase} Magnetization $m(t)$ \textbf{(a)} and interface density $\rho(t)$ \textbf{(c)} trajectories for different values of the threshold $T$ ($m_0 = -0.2$) using the Symmetrical Threshold model with aging. \textbf{(b)} Final magnetization histogram of $1000$ trajectories for the same system at $T=0.25$. Different colors indicate different values of $T$. The inset at (b) shows a close look at the logarithmic-like decay, shown in linear-log scale. The system is an ER graph with $N = 4 \cdot 10^4$ and mean degree $\langle k \rangle = 8$.}
\end{figure}

Figure \ref{fig:evolution_random_aging} shows the time evolution of the average interface density $\rho(t)$, the average magnetization $m(t)$, the mean internal time $\bar{\tau}(t)$, and the persistence $p(t)$ for the Symmetrical Threshold model with aging in an ER graph. The dynamical properties are largely affected by the aging mechanism. In terms of the evolution, we find the following regimes:
\begin{itemize}
    \item \textbf{Initial mixing regime (Phase ${\rm {\bf I}}^{*}$):} It is characterized by two dynamical transient regimes: a fast initial disordering dynamics followed by a slow ordering process. After the initial fast disordering stage, the average interface density exhibits a very slow (logarithmic-like) decay. When the system is dominated by a large majority of agents in the same state, the interface regime changes. The average interface density follows a power law decay with time, where $\rho(t)$ scales as $t^{-1}$. This phase exists for the same domain of parameters ($m_0$, $T$) as Phase ${\rm I}$ (orange region in Fig. \ref{ER_REG_PDAGING}) in the model without aging (see $T = 0.12, 0.24$ in Fig. \ref{fig:evolution_random_aging});
    \item \textbf{Ordered regime (Phase ${\rm {\bf II}}$):} According to the initial majority, the magnetization tends to the ordered absorbing state. This regime is characterized by a power-law interface decay, where $\rho(t)$ scales as $t^{-1}$. (see $T = 0.36, 0.49$ in Fig. \ref{fig:evolution_random_aging});
    \item \textbf{Frozen regime (Phase ${\rm {\bf III}}$):} Each individual realization is characterized by an initial tendency towards the majority consensus, but very fast reaches an absorbing frozen configuration (see $T = 0.86$ in Fig. \ref{fig:evolution_random_aging}).
\end{itemize}
The main effect of aging is that the mixed states of Phase ${\rm I}$ are no longer present, at least not for the type of networks that we are analyzing here. We will show later that Phase ${\rm I}$ reemerges in denser graphs (consistently with the results in a complete graph). Instead, for sparse graphs, we observe a new Phase ${\rm I}^{*}$ in which the system initially disorders and later orders until reaching the absorbing states $m_f = \pm 1$. This behavior is shown in Fig. \ref{fig:evolution_random_aging} for $T = 0.12$ and $0.24$. For $T = 0.12$, the system initially disorders, and then the interface density follows a logarithmic-like decay (see inset in Fig. \ref{fig:evolution_random_aging}a). Due to the slow decay, the system stays in this transient regime even after $10^{6}$ time steps, and the fall to the absorbing states is not observed in this figure. Similarly, for $T = 0.24$ the disordering process stops, and then the system gradually evolves towards a fully ordered state. For this value of $T$, the logarithmic-like decay is not appreciated, and we just observe the power-law decay due to the finite size of the system. The difference between $T = 0.12$ and $T = 0.24$ comes from the fact that in this Phase ${\rm I}^{*}$, the interface decay becomes faster as we increase the threshold $T$ (see Fig. \ref{fig:mixed_phase}(c)). Notice the different interface decay in Fig. \ref{fig:mixed_phase}c (inset) between values of $T < 0.3$ (Phase ${\rm I}^{*}$), where all trajectories show a logarithmic-like decay of $\rho(t)$ in a transient regime, and $T \geq 0.3$ (Phase ${\rm II}$), where trajectories from the initial condition exhibit fast ordering dynamics towards the majority consensus. Moreover, we observe that in Phase ${\rm I}^{*}$, the initial magnetization $m_0$ introduces a bias to the stochastic process, implying that the larger $m_0$ in absolute value, the larger the number of realizations that reach the absorbing state with the same sign of $m_0$. However, the system can still reach the absorbing state of the opposite sign of $m_0$ (initial minority), as shown in the trajectories with $T = 0.25$ in Fig. \ref{fig:mixed_phase}a. Due to the characteristic logarithmic decay of Phase ${\rm I}^{*}$, a statistical analysis of the inversion process incurs a significant computational cost. In Fig. \ref{fig:mixed_phase}b, we present the final magnetization histogram for $T=0.25$, a value proximal to the ${\rm I}^{*} - {\rm II}$ boundary where this analysis is computationally feasible. As depicted in this figure, the proportion of realizations in which consensus is reached in the initial minority state is approximately $3.3\%$. Fig. \ref{fig:evolution_random_aging}(c-d) shows the evolution of the temporal dynamics via the mean internal time and the persistence. The persistence in Phase ${\rm I}^{*}$ shows a power-law decay, where $p(t)$ scales as $t^{-1}$, and the internal time shows an increase following the recursive relation given in Appendix \ref{appendix_RR}, as it occurred for the mean-field scenario (Fig. \ref{fig:COM_AGING}).

In Phase ${\rm II}$, the system asymptotically orders for any initial condition as in the original model, but the dynamical properties are modified due to the presence of aging: the exponential decay of the interface density is replaced by a slow power-law decay, where the exponents of the exponential and the power-law are found to be similar. Contrary, the dynamical properties of Phase ${\rm III}$ are not affected by the presence of aging. 

As it occurred for the non-aging version of the model, the dynamical characterization discussed above holds for all possible
$m_0$ except for the symmetric initial condition $m_0 = 0$. The implications of the order-disorder transition (that occurs at a critical mean degree $k_c (N)$)~\cite{Pournaki-2022} are still present in the model with aging. Moreover, as it occurred for the Symmetrical Threshold model, the persistence cannot be predicted by this framework.

To account for the results of our Monte Carlo simulations, we use the same mathematical framework as described in Equation (\ref{eq:AME_age}). According to the update rules of the Symmetrical Threshold Model with aging, the transition probabilities now depend on the age $j$, as given by the activation probability  $p_A (j)$:
\begin{equation}
    T^{+}_{k,m,j} = p_A(j) \, \theta(m/k - T) \quad \quad T^{-}_{k,m,j} = p_A(j) \, \theta((k-m)/k - T) \quad \quad A^{\pm}_{k,m,j} = 1 - T^{\pm}_{k,m,j}.
\end{equation}
We show in Figure \ref{ER_REG_PDAGING} the mixed-ordered and ordered-frozen transition lines predicted by the integration of the AME equations until a time cutoff $t_{\rm max}$. We find good agreement between the theoretical predictions and the simulations both for ER and RR networks. Regarding dynamical properties, the AME integrated solutions exhibit a remarkable concordance with the evolution of all the metrics as shown in Figure \ref{fig:evolution_random_aging}. Minor discrepancies between the numerical simulations and the integrated solutions are attributed to the different assumptions, discussed previously, on which the AME is based.  
\begin{figure}
        \centering \captionsetup{font=sf}
        \includegraphics[width=0.95\textwidth]{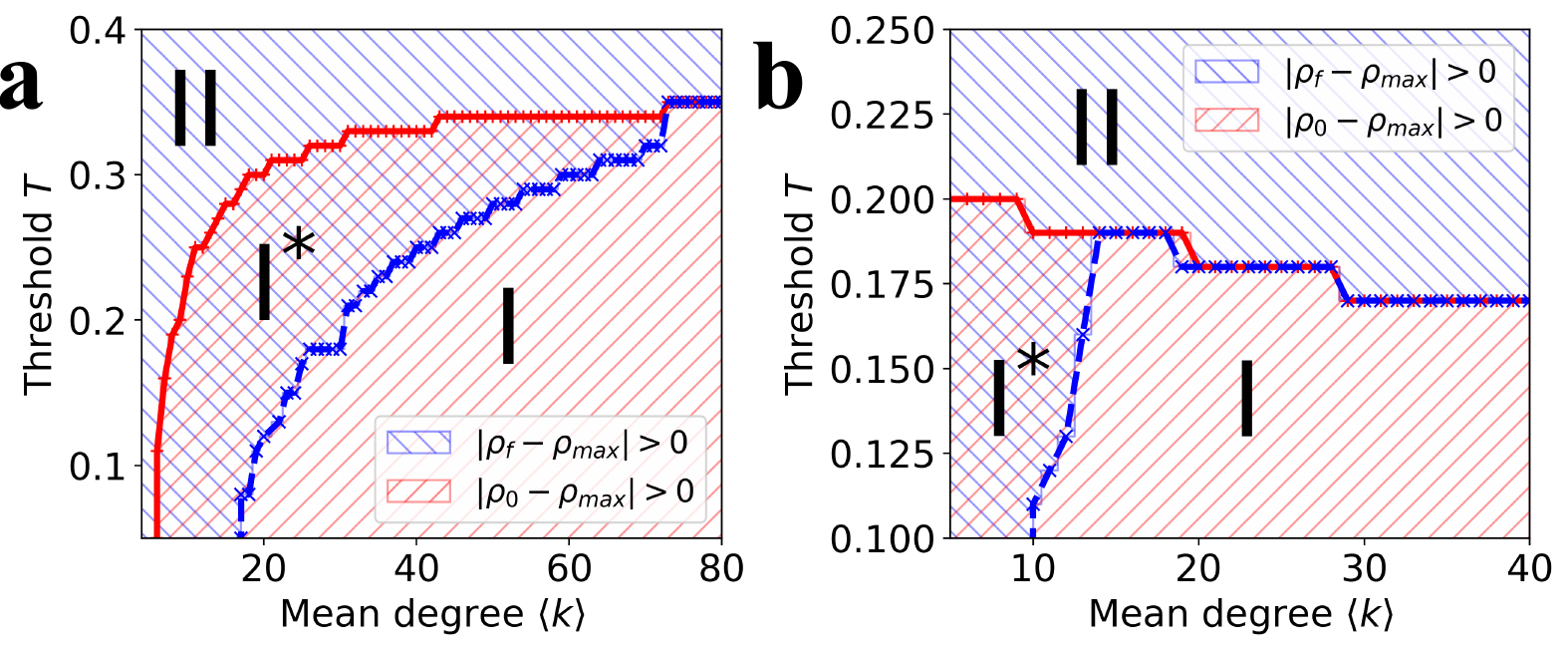}
        \caption[Phase ${\rm I}^{*}$ dependence with the network mean degree]{\label{fig:PD_Z} Critical threshold $T_c$ dependence with the mean degree $\langle k \rangle$ for the Symmetrical Threshold model with aging for an ER graph with $N = 4 \times 10^4$ nodes for an initial magnetization of $m_0 = 0.25$ \textbf{(a)} and $m_0 = 0.75$ \textbf{(b)}. The blue and red markers indicate the borders of phases ${\rm I}$ and ${\rm II}$, which coincide for a sufficiently large value of the mean degree. The hatched area corresponds to the fulfillment of the inequality in the legend.}
\end{figure}

The numerical results discussed so far are for random networks with average degree $\langle k \rangle = 8$. According to them and to the analytical insights, one can conclude that aging significantly changes the phase diagram for sparse networks. However, we know that the model with aging shows the same phase diagram as the model without aging for a fully connected network. This implies that, for ER graphs, as the mean degree $\langle k \rangle$ approaches $N$, Phase ${\rm I}^{*}$ must disappear. Therefore, the combined effects of increasing the mean degree and introducing aging need to be investigated in more detail. Phase ${\rm II}$ is distinguishable from phases ${\rm I}$ and ${\rm I}^{*}$ because the system initially orders, i.e., $|\rho_0 - \rho_{\rm max}| = 0$, where $\rho_{\rm max}$ is the maximum value attained by the interface density during the dynamical evolution. In contrast, Phase ${\rm I}$ is distinguished from Phases ${\rm I}^{*}$ and ${\rm II}$ because the system remains disordered, i.e., $|\rho_{\rm max} - \rho(t_{\rm max})| \approx 0$. Thus, Phase ${\rm I}^{*}$ is the only phase among these three where $|\rho_0 - \rho_{\rm max}| > 0$ and $|\rho_{\rm max} - \rho(t_{\rm max})| > 0$. Using this criterion, we studied the dependence of the critical threshold  $T_c$ on the mean network degree defining the transition lines between phases I, ${\rm I}^{*}$, and ${\rm II}$ (see Fig. \ref{fig:PD_Z}). In the absence of aging, the red line in Fig. \ref{fig:PD_Z} gives the value of the mixed-ordered transition line $T_c$. When aging is included, at low degree values, Phase ${\rm I}$ is replaced by ${\rm I}^{*}$, as expected. However, as the mean degree increases, Phase ${\rm I}$ emerges despite the presence of aging, leading to the range of mean degree values where the model exhibit 4 different phases: ${\rm I}$, ${\rm I}^{*}$, ${\rm II} $ and ${\rm III}$. As the mean degree is further increased, a critical value is reached where Phase ${\rm I}^{*}$ is no longer present, and the discontinuous transition I-${\rm II}$ occurs at the same value as in the model without aging. Importantly, this critical mean degree at which Phase ${\rm I}^{*}$ disappears, depends significantly on the initial magnetization $m_0$.

\section{\label{sec: Results on a Moore Lattice}  Results on a Moore Lattice}

\begin{figure}[t!]
        \centering \captionsetup{font=sf}
        \includegraphics[width=\textwidth]{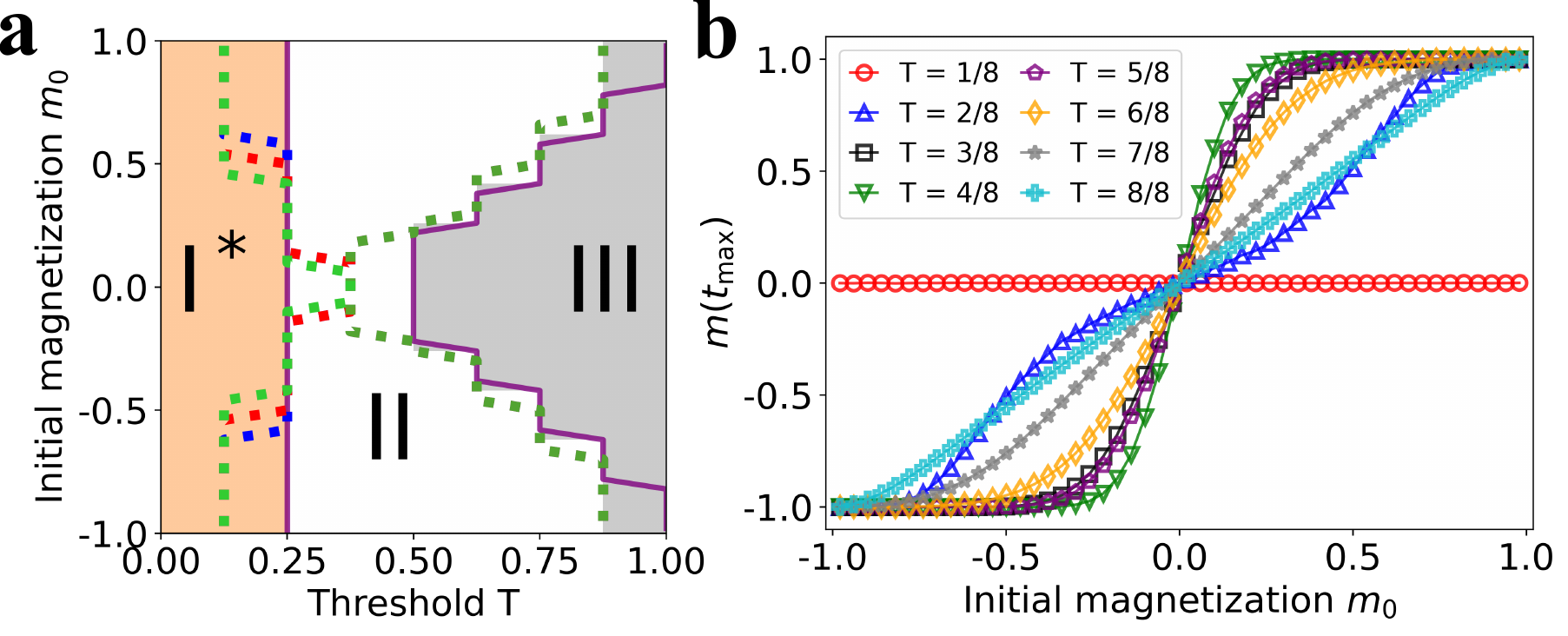}
        \caption[Symmetrical Threshold model with aging in a Moore lattice]{\label{LAT_PDAGING} \textbf{(a)} Phase diagram of the Symmetrical Threshold model with aging in a Moore lattice of $N = L \times L$, with $L = 100$. The blue, red and green dotted lines show the borders of Phase ${\rm II}$ (first and last value of $T$ where the system reaches the absorbing ordered state for each $m_0$) from numerical simulations evolving until $t_{\rm max} = 10^3$, $10^4$ and $10^5$ time steps, respectively. Phase ${\rm I}^{*}$, ${\rm II}$ and ${\rm III}$ correspond with the orange, white and gray areas, respectively. The solid purple lines are the mixed-ordered and ordered-frozen transition lines for the Symmetrical threshold model without aging (from Fig. \ref{LAT_PD}). \textbf{(b)} Average magnetization at time $t_{\rm max}$ ($m_f(t_{\rm max})$) as a function of the initial magnetization $m_0$ for different values of the threshold $T$ (indicated with different colors and markers) in a Moore lattice of $N = L \times L$, with $L = 100$. The numerical simulations are obtained until $t_{\rm max} = 10^4$ time steps. Average performed over 5000 realizations.}
\end{figure}

We show in Figure \ref{LAT_PDAGING}a the borders of Phase ${\rm II}$ obtained from numerical simulations of the Symmetrical threshold model with aging running up to a time $t_{\rm max}$ (dotted colored lines) in a Moore lattice. Similarly to the behavior observed in random networks, the mixed-ordered border is shifted to lower values of $T$ as we increase the simulation time cutoff $t_{\rm max}$. Thus, Phase ${\rm I}$ is replaced by an ordered phase due to the aging mechanism. Examining the dependence of the final value of the magnetization on its initial condition  $m_f(m_0)$  (Figure \ref{LAT_PDAGING}b), one can conclude that the mixed phase is still present, at least transiently, as in the initial disordering phase described in the previous section (Phase ${\rm I}^{*}$). Phase ${\rm II}$ is again characterized by an asymptotically ordered state where the initial majority reaches consensus. However, for this specific structure, near $m_0 = 0$ and $T = 1/2$, the ordered state is not reached for any threshold value. Furthermore, comparing with Fig. \ref{LAT_PDAGING}b with the results from the model without aging (Fig. \ref{LAT_PD}b), the discontinuous jump at $m_0 = 0$ for $T = 3/8, 4/8$ is replaced by a continuous transition, where a range of states with $0 < |m_f| < 1$ are present around $m_0 = 0$. To determine whether these states belong to Phase ${\rm I}^{*}$, ${\rm II}$ or ${\rm III}$, we need again a characterization of phases in terms of dynamical properties. According to the results in Figure \ref{fig:evolution_lattice_aging}, we find here the same regimes identified for random networks:
\begin{figure}[t!]
    \centering \captionsetup{font=sf}[t!]
    \includegraphics[width=\linewidth]{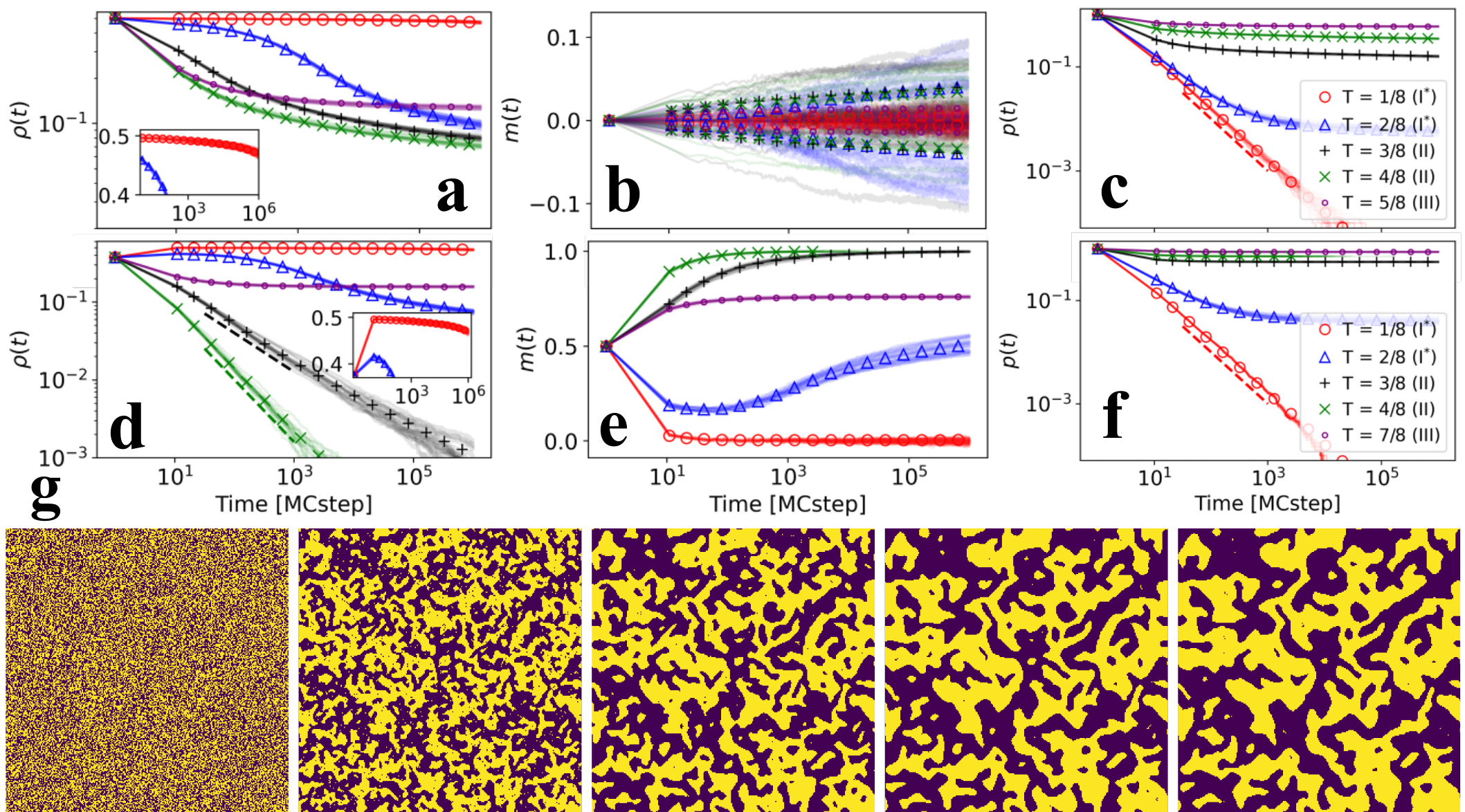}
    \caption[Modified dynamical regimes by aging in a Moore lattice]{\label{fig:evolution_lattice_aging} Evolution of the average interface density $\rho(t)$ \textbf{(a-d)}, the average magnetization $m(t)$ \textbf{(b-e)}, and the persistence $p(t)$ \textbf{(c-f)} for the Symmetrical model with aging in a Moore lattice starting from a random configuration with $m_0 = 0$ (a-b-c) and $m_0 = 0.5$ (d-e-f). We plot 30 different trajectories in solid lines and the average of over $5000$ surviving trajectories in symbols. Colors and symbols indicate different threshold values: red ($T = 1/8$) and blue ($T = 2/8$) belong to Phase ${\rm I}^{*}$, green ($T = 3/8$), and black ($T=4/8$) belong to Phase ${\rm II}$, and purple ($T = 5/8, 7/8$) belong to Phase ${\rm III}$. The average magnetization is computed according to the two symmetric absorbing states. The insets in (a-d) show a close look at the evolution for $T = 0.12$, in linear-log scale. System size is fixed at $N = L \times L$, $L = 200$. The dashed lines in (d) are $\rho \sim t^{-\alpha}$ with $\alpha = 0.5$ (black) and $\alpha = 0.8$ (green), and in (c-f) are $p(t) \sim t^{-1}$ (red). 
    Simulations stop when the system reaches the absorbing ordered states. \textbf{(g)} Evolution of a single realization for $T = 0.5$ and $m_0 = 0$ using the Symmetrical threshold model with aging. Snapshots are taken after $1,60,3300,2 \cdot 10^5$ and $5 \cdot 10^6$ time steps, increasing from left to right. System size is fixed to $N = L \times L$, $L = 256$.}
\end{figure}
\begin{itemize}
    \item \textbf{Initial mixing regime (Phase ${\rm {\bf I}}^{*}$):}  After the initial disordering stage, the average interface density shows a very slow decay reflecting the slow growth of spatial domains in each binary state. The persistence in this phase shows a power-law decay $p(t) \sim t^{-1}$ (see $T = 1/8,2/8$ in Fig. \ref{fig:evolution_lattice_aging});
    \item \textbf{Ordered regime (Phase ${\rm {\bf II}}$):} It is characterized by coarsening dynamics that end in the absorbing states $m_f = \pm 1$. The form of the decay of the interface density depends on the value of $m_0$ (see $T = 3/8,4/8$ in Fig. \ref{fig:evolution_lattice_aging});
    \item \textbf{Frozen regime (Phase ${\rm {\bf III}}$):} It is characterized by an initial tendency to order, but the system very fast reaches an absorbing frozen configuration (see $T = 5/8,7/8$ in Fig. \ref{fig:evolution_lattice_aging}).
\end{itemize}

The implications of aging become explicit by comparing the dynamical properties of the cases with aging (Figure \ref{fig:evolution_lattice_aging}) and without aging (Figure \ref{fig:evolution_lattice}). When the threshold is $T<3/8$, Phase ${\rm I}$ is replaced by Phase ${\rm I}^{*}$ in which there is an initial disordering process very fast followed by a slow coarsening process that accelerates when we increase the threshold. Although the aging implications in this phase are similar to those observed in the ER graph, the coarsening process is much slower in a Moore lattice (see insets in Fig. \ref{fig:evolution_lattice_aging}a-d).

In Phase ${\rm II}$ ($T=3/8, 4/8$) and when $m_0=0.5$, the system exhibits coarsening towards the ordered state $m_f=\pm 1$. In this case, the interface decay $\rho \sim \exp(-\alpha \, t)$, observed in the absence of aging is replaced, due to aging, by a power law decay $\rho \sim t^{-\alpha}$, as it ocurred in Chapter \ref{ch:Aging in the Granovetter-Watts model} for the endogenous aging. We find $\alpha=0.5$ and $0.8$ for $T=3/8$ and $4/8$, respectively. For $m_0=0$, the power law decay of the interface density vanishes with aging, and the system exhibits coarsening dynamics much slower than for $m_0 \neq 0$. In this region of the phase diagram, spatial clusters start to grow from the initial condition, but once formed, it takes a long time for the system to reach the absorbing state $m_f = \pm 1$. 
We note that for these parameter values, the system is not able to reach $|m|$ over $0.1$ even after $10^6$ time steps, but since there is coarsening from the initial condition, the expected stationary state as $t \to \infty$ is $m_f=\pm1$. There is neither initial disordering nor freezing, these values correspond to the defined Phase ${\rm II}$, even though the system exhibits ``long-lived segregation'' long transient dynamics (compare the coarsening process in Fig.\ref{fig:evolution_lattice}g with Fig.\ref{fig:evolution_lattice_aging}g). In Fig. \ref{LAT_PDAGING}a, we differentiate Phase ${\rm II}$ from Phase ${\rm III}$ by analyzing the activity in the system: If agents are changing, even though the interface decay is slow, the system is in Phase ${\rm II}$. If agents are frozen, it lies in Phase ${\rm III}$. 

Finally, it should be noted that in Phase ${\rm I}^{*}$, the initial disordering dynamics drive the system towards $m=0$. Therefore, the subsequent coarsening dynamics follow the slow interface decay observed in Phase ${\rm II}$ for $m_0 \sim 0$. Thus, the presence of aging implies that the system asymptotically orders for any initial condition, but due to the initial disordering, the coarsening dynamics fall into the ``long-lived segregation'' regime independently of the initial condition.

\section{\label{sec:Summary and Conclusions_STMA} Summary and discussion}

In this chapter, aging is incorporated in the model as a decreasing probability to modify the state as the time already spent by the agent in that state increases. The key finding is that the mixed phase (Phase ${\rm I}$), characterized by an asymptotically disordered dynamically active state, does not always exist: the aging mechanism can drive the system to an asymptotic absorbing ordered state, regardless of how low the threshold $T$ is set. A similar effect of aging was already described for the Sakoda-Schelling model in Chapter \ref{ch:Aging effects in the Sakoda-Schelling segregation model}. When the dynamics are examined in detail, a new Phase ${\rm I}^{*}$, defined in terms of dynamical properties, emerges in the domain of parameters where the model without aging displays Phase ${\rm I}$. This phase is characterized by an initial disordering regime ($m \to 0$) followed by a slow ordering dynamics, driving the system toward the ordered absorbing states (including the one with spins opposite to the majoritarian initial option). This result is counter-intuitive since aging incorporates memory into the system, yet in this phase, the system ``forgets'' its initial state. The network structure plays an important role in the emergence of Phase ${\rm I}^{*}$ since it does not exist for complete graphs. A detailed analysis reveals that Phase ${\rm I}^{*}$ replaces Phase ${\rm I}$ only for sparse networks, including the case of the Moore lattice. For ER networks we find that, as the mean degree increases, Phase ${\rm I}$ reappears and there is a range of values of the mean degree for which both phases, ${\rm I}$ and ${\rm I}^{*}$, are present in the same phase diagram for different values of $(m_0,T)$. Beyond a critical value of the mean degree, Phase ${\rm I}$ extends over the entire domain of parameters where Phase ${\rm I}^{*}$ was observed.

While aging favors reaching an asymptotic absorbing ordered state for low values of $T$ (Phase ${\rm I}$), in Phase ${\rm II}$ the ordering dynamics are slowed down by aging, changing, both in random networks and in the Moore lattice, the exponential decay of the interface density by a power law decay with the same exponent. The aging mechanism is found not to be important in the frozen Phase ${\rm III}$. All these effects of aging in the three phases are well reproduced for random networks by the AME derived in this work, which is general for any chosen activation probability $p_A (j)$.

For the Moore lattice, we have also considered in detail the special case of the initial condition $m_0=0$. In this case, Phase ${\rm I}^{*}$ emerges, and Phase ${\rm III}$ is robust against aging effects. However, in Phase ${\rm II}$ aging destroys the characteristic power law decay of the interface density, $\rho(t) \sim at^{-1/2}$, associated with curvature reduction of domain walls. This would be a main effect of aging in the dynamics of the phase transition for the zero temperature spin flip Kinetic Ising model~\cite{Maxi}.

As a final remark on the general effects of aging in different models of collective behavior, we note that the replacement of a dynamically active disordered stationary phase by a dynamically ordering phase is generic. In this chapter, we find the replacement of Phase ${\rm I}$ by Phase ${\rm I}^{*}$. Likewise in the Voter model, aging destroys long-lived dynamically active states characterized by a constant value of the average interface density, and it gives rise to coarsening dynamics with a power law decay of the average interface density~\cite{fernandez-gracia-2011}. In the same way, in the Sakoda-Schelling segregation model, a dynamically active mixed phase is replaced, due to the aging effect, by an ordering phase with segregation in two main clusters (refer to chapter \ref{ch:Aging effects in the Sakoda-Schelling segregation model}). 
Another aging effect that seems generic, in phases in which the system orders when there is no aging, is the replacement of dynamical exponential laws by power laws. This is what happens here in  Phase ${\rm II}$ for the decay of the average interface density but, likewise, exponential cascades in the Granovetter-Watts threshold model are replaced due to aging by a power-law growth with the same exponent (refer to chapter \ref{ch:Aging in the Granovetter-Watts model}).


\part{Assessing the housing market dynamics}

\renewcommand{\thechapter}{7} 
\chapterimage{Images/influence_bal.pdf}
\chapterspaceabove{6.75cm}
\chapterspacebelow{7.25cm}

\chapter{Exploring the housing market spatial segmentation \label{ch:Exploring the housing market spatial segmentation}}
\vspace{-1.2cm}
\small
\textbf{The results in this chapter are published as:}
\vspace{0.05 cm}

\fullcite{abella2024exploring}
\normalsize
\vspace{0.5 cm}

Real estate market has an inherent connection to space. In this chapter, we develop a methodology to detect the spatial segmentation emerging from data on housing online listings via multipartite networks. Considering the spatial information of the listings, we build a bipartite network that connects agencies and spatial units. This bipartite network is projected into a network of spatial units, whose connections account for similarities in the agency ecosystem. We then apply clustering methods to this network to segment  markets into spatially-coherent regions, which are found to be robust across different clustering detection algorithms, discretization of space and spatial scales, and across countries with case studies in France and Spain. This methodology addresses the long-standing issue of housing market segmentation, relevant in disciplines such as urban studies and spatial economics, and with implications for policymaking. 

\section{Introduction}
\label{sec:introduction}

The spatial dimension of housing markets is a crucial aspect for urban studies and planning. Understanding the spatial segmentation of the housing market into submarkets~\cite{morawakage2022housing,bourassa2003housing} has important implications for real estate valuation and investment decisions, which together affect urban development and social equity~\cite{bourassa2003housing}. Spatial segmentation is the product of many factors such as residential location and the proximity to amenities~\cite{bourassa2003housing}, differences in housing stock~\cite{keskin2017defining}, price levels~\cite{goodman1998housing}, and consumer preferences~\cite{leishman2013predictive}. 

Online real estate listings provide a rich source of data that can be used to study the temporal and spatial patterns of property transactions, the behavior of real estate agencies, and the preferences of buyers and sellers. With property portals being nowadays the dominant way to create and access market information, online listings constitute a new type of data to study housing markets~\cite{sawyer1999ict, boeing2017new, boulay2021moving}. Scholars studied the spatio-temporal distribution of housing prices~\cite{yao2018mapping,adolfsen_segmentation_2022}, revealed the persistence of spatial inequalities in the housing information landscape~\cite{boeing2020online}, predicted the social profile of neighborhoods~\cite{delmelle2021language}, or detected the segmentation of the market from online search patterns~\cite{rae2015online}.
Aside from price, pictures or textual descriptions, a listing includes a critical piece of information: the identity of the marketing agency that has posted the listing on the portal. As such, listings constitute digital traces~\cite{salganikbit2017} of the work performed by real estate agencies when acquiring, selling or marketing on property portals. It is therefore possible to reconstruct, for each agency, its own portfolio of listings, whose volume and location patterns result from and reflect the heterogeneous practices and market shares of real estate agencies. By informing on \textit{who sells where}, listings offer new ways to examine how real estate agencies unevenly operate and specialize across space~\cite{palm1976RealEstate}.

\begin{figure}[t]
    \centering
    \includegraphics[width =0.98\textwidth]{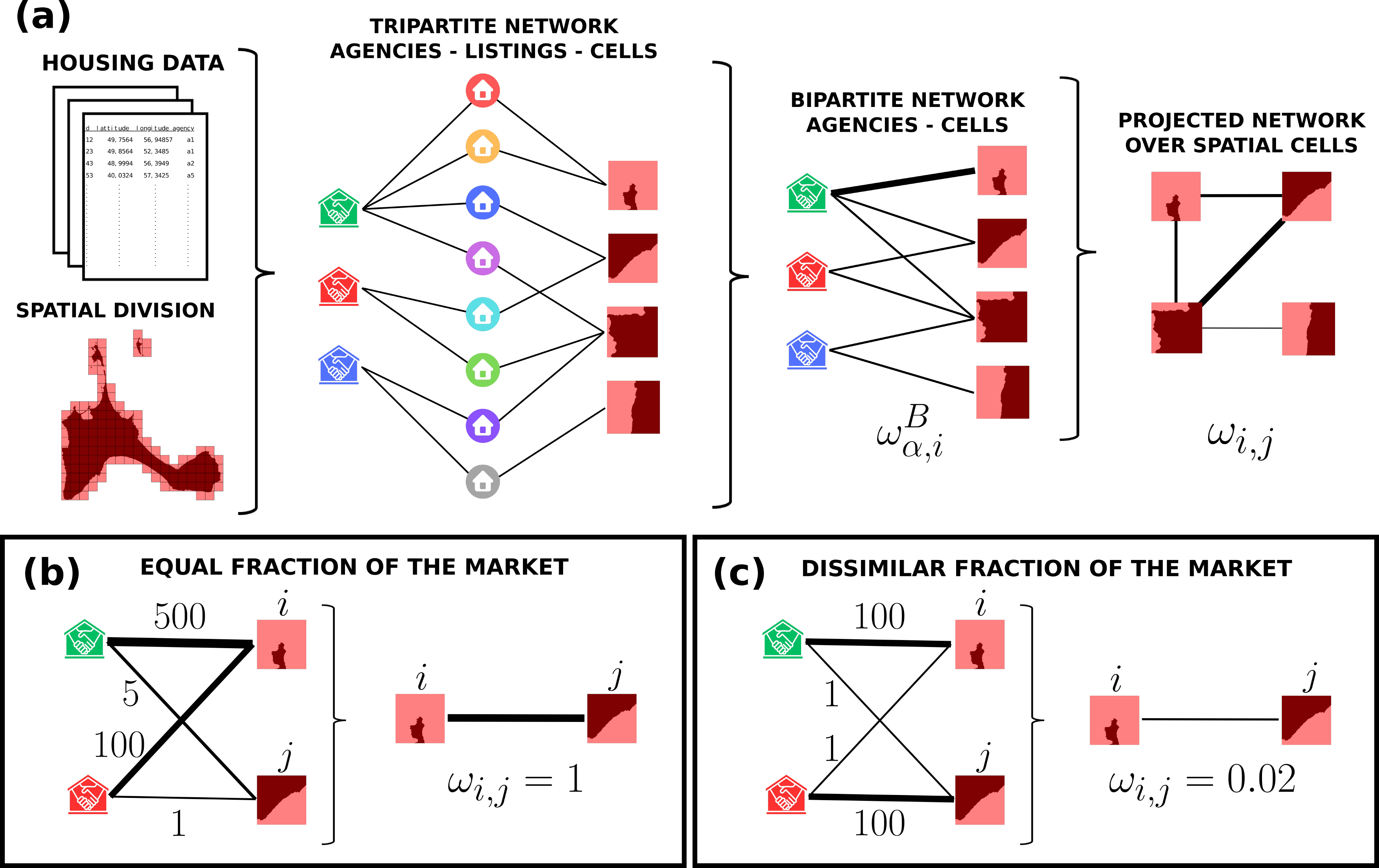}
	\caption[Bipartite network construction and projection.]{ \textbf{(a)} A tripartite network is constructed between real estate agencies, listings, and spatial units obtained from geolocalized housing data and the division of space in regular grid cells. In this network, each listing is connected to its real estate agency and the spatial cell where it is located. This simple tripartite network is contracted into a bipartite network linking agencies and cells, where the link weight $\omega^{B}_{\alpha,i}$ corresponds to the number of listings the agency $\alpha$ has in the spatial cell $i$. Finally, the network is then projected over the cells to form a weighed network of spatial units, where the weight $\omega_{i,j}$ of the link between cells $i$ and $j$ quantifies how much they are similar in the market -- $\omega_{i,j}$ is properly defined by Equation \eqref{eq:weight_eq}. Two simple examples of the projection process are shown below: with \textbf{(b)} equal and \textbf{(c)} complementary listings distributions for the agencies in the cells. \label{fig:network_construction}}
\end{figure}

The spatial division of the real estate market has been studied from different perspectives and with different methods in the literature. Some studies have examined the spatial segmentation of the urban housing market focusing on neighborhood correlations of housing prices~\cite{palm1978spatial}, the spatial effects of urban public policies on housing values~\cite{baumont2009spatial}, the neighborhood quality and accessibility effects on housing prices~\cite{dubin1992spatial}, while others have determined if a specific property market is spatially segmented into submarkets, and whether accounting for the existence of submarkets improves the accuracy of price modeling~\cite{keskin2017defining,usman2021priori}. This is especially important for hedonic pricing models that seek to incorporate spatial autocorrelation and heterogeneity to predict price evolution~\cite{usman2021priori,paez2009recent,bitter2007incorporating, case2004modeling}. Ref.~\cite{hu2022NovelApproach} distinguishes two main approaches for spatial segmentation: using pre-defined geographical boundaries based on \textit{a priori} knowledge, such as local administrative boundaries or expert areas used by market stakeholders, or relying on clustering methods to infer patterns from pricing similarities. For the latter, popular statistical approaches to divide space into submarkets are principal component analysis and hierarchical clustering~\cite{goodman1998housing,bourassa1999defining,bourassa2003housing}.

There is ample evidence underlining how real estate agencies influence market segmentation by determining housing prices, sorting homebuyers into different market channels, and specializing in certain types of neighborhoods and market segments~\cite{palm1976RealEstate,palm1978spatial,keskin2017defining,bonneval2017agents,besbris2017investigating}. Furthermore, it has been shown that the definition of submarkets based on agencies is far superior to other segmentation techniques~\cite{leishman2013predictive}.

This chapter introduces a new method to identify the housing market segmentation using geospatial data, complex network analysis techniques, and taking as a basis the local ecosystem of real estate agencies. We build a network structure based on two factors:  the \textit{presence} of an agency within a particular area, and the relative \textit{influence} of an agency in this area, determined by the agency's proportional share of all listings located in the area. Our methodology is applied to the residential property market in 3 Spanish provinces and 3 French urban areas, for which we have a rich, high resolution dataset sourced from property portals. We find that the market in those regions is divided into a hierarchy of subregions. We test the robustness of our results against different community detection algorithms, scales, and administrative boundaries in different countries.

\section{Data}

For Spain, we analyze listings published on the portal \texttt{Idealista.com}~\cite{idealista}, as in the previous chapter (see details in Section \ref{subsec:Idealista dataset})

French listings were obtained from the portal \texttt{SeLoger.com}~\cite{SeLoger} (see details in \ref{subsec:SeLoger dataset}). Geographical information is only available at the administrative and census levels, such as ZIP codes (``\textit{code postal}''), municipalities (``\textit{communes}''), and census tracts (``\textit{IRIS}''), the finest and basic scale for sub-municipal information in France. We focus on three major urban areas: Paris, Marseilles and Toulouse. 

For both datasets, we focus on houses and apartments, and do not consider farms or rural parcels.

\section{Building a network \label{sec:materials_and_methods}}

\begin{figure}
    \centering
    \includegraphics[width = 0.98\textwidth]{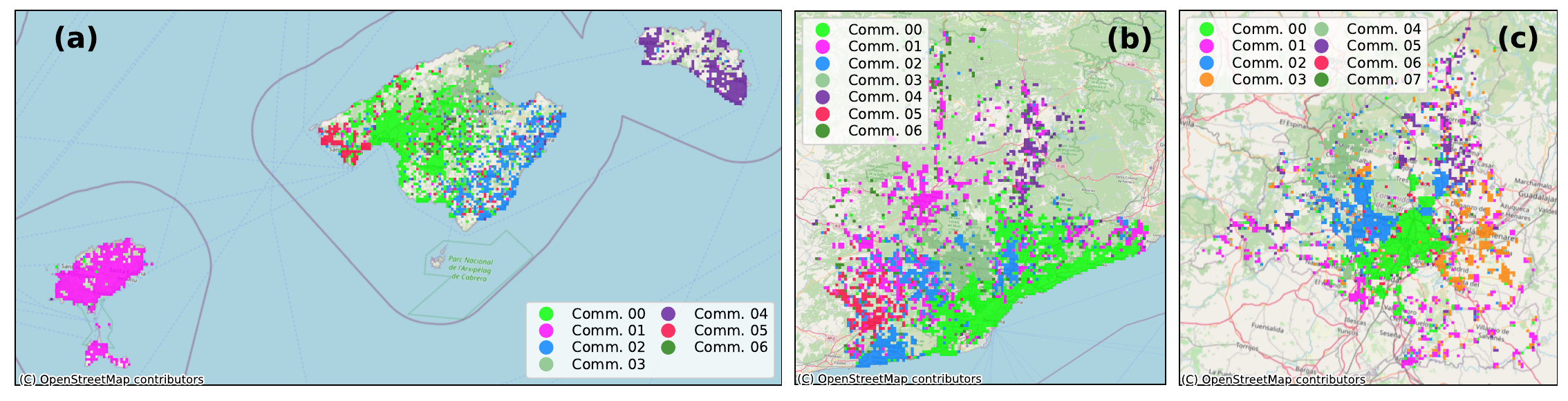}
    \caption[Market segmentation for $1 \; \textrm{km}$ square cells.]{ Communities from the projected network for the three Spanish provinces studied: Balearic Islands \textbf{(a)}, Barcelona \textbf{(b)}, and Madrid \textbf{(c)}. The spatial cells are $1 \; \textrm{km}$ square cells. The communities shown are detected using the Louvain algorithm with a consensus clustering of $1000$ realizations. The underground map data is rendered from OpenStreetMap under ODbL.\label{fig:cell_1000}}
\end{figure}

We begin by discretizing the space into spatial units (square grid cells, municipalities, districts, postal codes, census-tracts, etc.).
This allows us to label each listing according to the spatial unit it falls into, along with the agency that posted this listing. By doing so, we build a tripartite relation between agencies, listings, and spatial units. Based on this structure, we can build a weighted bipartite network that connects agencies and spatial units, where the link weight $\omega^{B}_{\alpha,i}$ accounts for the number of listings posted by agency $\alpha$ that are located in the spatial unit $i$.

Bipartite networks can be projected to create networks with a single type of nodes~\cite{newman2003structure,newman2001scientific,zhou2007bipartite}. In our case, we project it to build a new weighted network connecting spatial units (see Fig. \ref{fig:network_construction}(a) for schematic representation, taking as an example the discretization of space with square grid cells). Let us assume that we have $N$ spatial units and $N^a$ real estate agencies. The set of all agencies operating in the entire area is $\{ \alpha \}$, while the subset operating in the spatial unit $i$ is denoted by $\{ \alpha \}_i$. The fraction of listings in $i$ that belong to a certain agency $\alpha$ is
\begin{equation}
    f_{\alpha, i} = \frac{\omega^{B}_{\alpha, i}}{\sum_{\gamma \in \{ \alpha \}_i} \omega^{B}_{\gamma, i}},
\end{equation}
where the index $\gamma$ runs over all the agencies operating in $i$. In the projected network, we define the influence weight between two spatial units $i$ and $j$ as
\begin{equation}
\omega_{i,j} = \frac{\sum_{\gamma \in \left\{ \alpha \right\}_{ij}} f_{\gamma,i} \, f_{\gamma,j}}{ \frac{1}{2} \left[ \sum_{\gamma \in \left\{ \alpha \right\}_{i}} f^2_{\gamma, i} + \sum_{\beta \in \left\{ \alpha \right\}_{j}} f^2_{\beta, j} \right]},
\label{eq:weight_eq}
\end{equation}
where $\{ \alpha \}_{ij} \equiv \{\alpha\}_i \cap \{\alpha\}_j$ is the subset of agencies operating in $i$ and $j$. The weight $\omega_{i,j} =1 $ if the agencies operating in $i$ and $j$ are the same, and cover an equal fraction of the market in both spatial units. If the market distribution is similar, but not equal, the weight will deviate from $1$. Reciprocally, if no common agency is found across the two spatial units, the weight is zero and there is no link between them. Fig. \ref{fig:network_construction}(b) and \ref{fig:network_construction}(c) show examples of the influence weights between two spatial units with equal distribution of the listings in (b), for which $\omega_{i,j} =1 $, and a complementary distribution in (c) with a value of $\omega_{i,j} = 0.02$. Note that our influence weight is related to the participation ratio introduced by Derrida \textit{et al.} in~\cite{Derrida_1987}. 
\begin{figure}[t!]
    \centering
    \includegraphics[width = 0.98\textwidth]{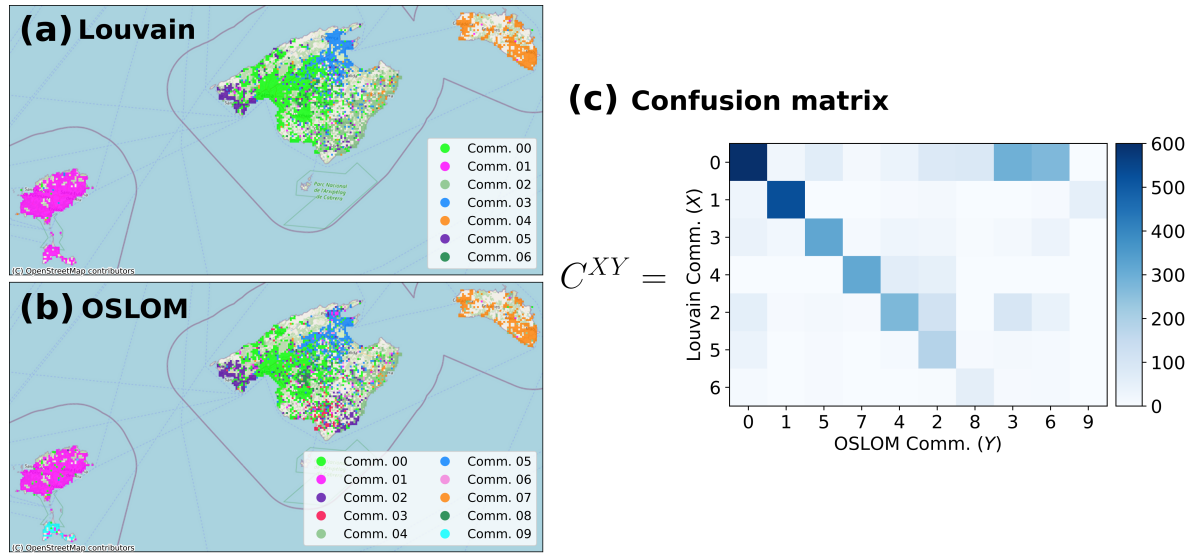}
	\caption[Agreement between different partitions.]{ Partition result of the community detection methods at the Balearic Islands using Louvain algorithm \textbf{(a)} and OSLOM method \textbf{(b)} $1 \textrm{km}$ square cells. The confusion matrix $C^{XY}$ of the two partitions \textbf{(c)}, is ordered according to the maximum overlap. The underground map data is rendered from OpenStreetMap, under ODbL. \label{fig:agreement_method}}
\end{figure}

The projected network is thus built with the spatial units as nodes, which are connected with links weighted according to Equation \ref{eq:weight_eq}. A group of spatial units strongly connected between them implies that they share a common ecosystem of agencies, that operate with a similar market share in these units.
Searching for clusters in this weighted spatial network should therefore inform us on the spatial segmentation of the housing market, the clusters corresponding to submarkets. In the network literature, such clusters are commonly referred to as communities, with numerous methods proposed to detect them~\cite{fortunato2010community}. We use several classic community detection algorithms~\cite{newman2004finding,rosvall-2008,blondel-2008,Louvain-Leiden,OSLOM} that account for network weights, including Louvain~\cite{blondel-2008}, Infomap~\cite{rosvall-2008}, and OSLOM~\cite{OSLOM}. These algorithms enable us to classify the spatial units into communities. Since these algorithms are stochastic, we perform several realizations of each method, and perform consensus clustering~\cite{lancichinetti2012consensus} for higher stability.
\begin{figure}[t!]
    \centering
    \includegraphics[width = 0.8\textwidth]{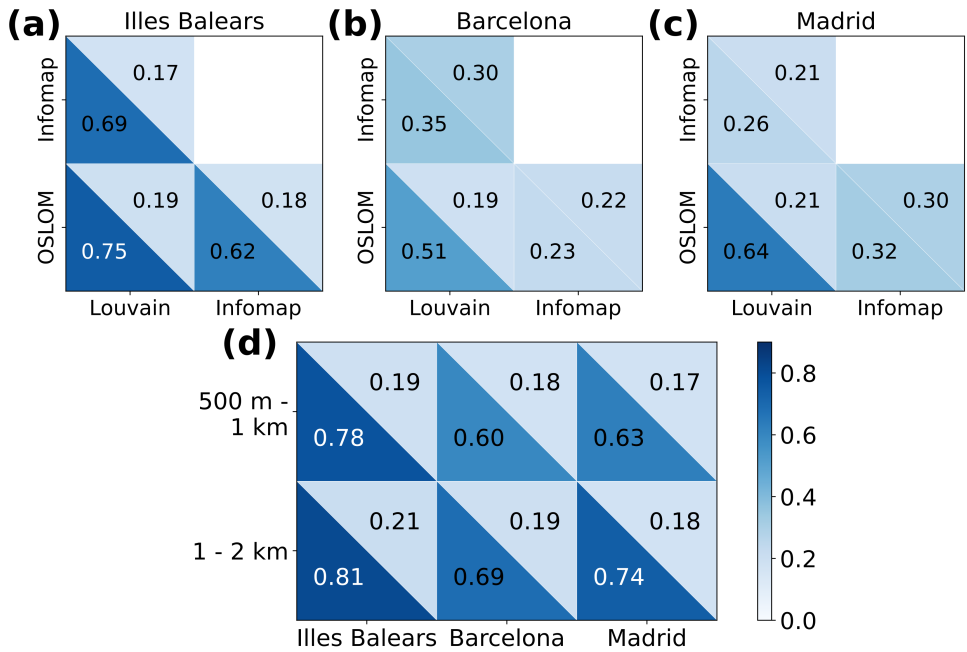}
	\caption[Agreement across three methods and cell sizes.]{ Agreement across the different community detection methods for the network in Balearic Islands \textbf{(a)}, in the province of Barcelona \textbf{(b)} and of Madrid \textbf{(c)}. The metric used to compute the agreement between method partitions is $H(X,Y)$, shown in the lower triangles for each pair of methods, denoted by $X$ and $Y$. The upper triangles display the value $H(X_r,Y_r)$, being $X_r$ and $Y_r$ the partitions randomized (preserving the communities size). In \textbf{(d)}, comparison of partitions obtained with the Louvain method for networks generated with different cell sizes: 500 m-sided vs  1 km-sided cells (top row), and 1 km-sided vs 2 km-sided cells (bottom row). \label{fig:agreement_scale_method}}
\end{figure}

\section{Results}
\label{sec:results}

\subsection{Segmenting the market according to agencies' operations}\label{sec:segmentation_cells}

We start by analyzing the spatial segmentation that arises from the data geolocated in the Balearic Islands, Barcelona, and Madrid using $1$ km-sided square cells. Fig. \ref{fig:cell_1000} presents the communities listed according to their size, from larger to smaller. Even though our methodology does not consider spatial proximity, we observe spatial segmentation in adjacent regions with few exceptions. For the Balearic Islands, we observe that spatial constraints, such as insular nature of the environment, affect the segmentation of the housing market: while the same submarket covers Minorca or Ibiza-Formentera, Majorca is divided into four different ones. It is noteworthy that the submarkets that emerge in all these three provinces are slightly larger than municipalities.

\begin{figure}
    \centering
    \includegraphics[width = 0.8\textwidth]{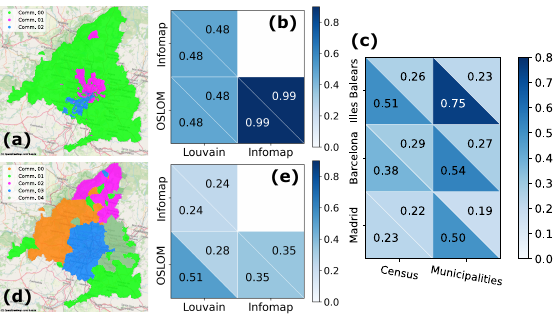}
	\caption[Community detection from networks using administrative spatial units.]{ Communities detected using census areas \textbf{(a)} and municipalities \textbf{(d)} as spatial units to build the network in Madrid. The clustering method employed is the Louvain algorithm. The agreement across the different methods for the census \textbf{(b)} and municipalities \textbf{(e)}. \textbf{(f)} shows the communities' agreement between $1$ km cells and administrative boundaries networks for all Spanish provinces. The agreement in (b)-(c)-(e) is computed using $H(X,Y)$ (lower triangles) compared with the value randomizing the communities (upper triangles).  The underground map data is rendered by OpenStreetMap, under ODbL. \label{fig:political_scales} \vspace{-1 cm}}
\end{figure}

To study the robustness of identified submarkets in each of the three provinces, we run several community detection algorithms, and compare the communities obtained across realizations of different algorithms. We define as a network partition the classification of the cells in communities, $X = \{x_0, x_1, \cdots , x_{|X|-1} \}$, where each community $x_i$ is a set of cells. The partition $X$ has $|X|$ communities in this notation. Every cell must be in at least one community, but in some clustering methods a cell may belong to several.
In order to compare two partitions $X$ and $Y$, we compute a confusion matrix $C^{XY}$ in which each element is defined as
\begin{equation}
    C^{XY}_{ij} = | x_i \cap y_j | ,
\end{equation}
where $x_i$ and $y_j$ are communities in the partitions $X$ and $Y$, respectively, and $| . |$ stands for the cardinal (number of elements) of a set. An element $C^{XY}_{ij}$ can be zero if there is no overlap between the communities, and it can be large if the two communities coincide across the partitions. We reorder then the elements of the matrix  $C^{XY}$ to have the largest values in the pseudo-diagonal. Note that $C^{XY}$ is not necessarily a squared matrix because the number of communities in each partition may differ. This process is essentially the identification of the communities in one partition that correspond to the communities in the other. This is a statistical match, given that the cells of a community in $X$ may be distributed in several communities in $Y$. As shown in Fig. \ref{fig:agreement_method}, if the partitions between the two methods are similar, we must observe a strong pseudo-diagonal in the confusion matrix. The sum of the elements of this pseudo-diagonal is the number of cells clustered in the same way in the two partitions. To compute a measure of the agreement between two partitions, we use the fraction of cells clustered in the same way in the two partitions $H(X,Y)$~\cite{newman2004finding,hric2014community} defined as 
\begin{equation}
H(X,Y) = \sum_{i = 0}^{\textrm{min}(|X|,|Y|)-1} \frac{C^{XY}_{ii}}{N} , 
\end{equation}
where the matrix $C^{XY}$ is ordered to maximize the pseudo-diagonal, and $N$ is the total number of cells. The fraction $H(X,Y)$ is a metric commonly used in the literature to compute the accuracy between community detection algorithms~\cite{danon2005comparing,duch2005community,li2008quantitative,darst2014improving,chen2015deep,saoud2016community,wang2017mitigation,fortunato2016community}, its value is bounded in the interval $(0,1]$, but it has the downside that $H(X,Y)$ depends on the size of the communities. To determine if the value of $H(X,Y)$ is significant, it is necessary to compare it with a randomized version of the partitions, $H(X_r,Y_r)$, in which the cells are reshuffled at random across the communities of each partition respecting the community sizes.

Figure \ref{fig:agreement_scale_method}(a-c) compares the three community detection algorithms (Louvain, OSLOM, and Infomap) used for different provinces. In all cases, the agreement between the communities detected from the real partition is higher than that of the randomized communities. The OSLOM-Louvain comparison exhibits the highest agreement, which is significant in all provinces. In the Balearic Islands, a robust and statistically significant agreement is evident among all methods. However, when examining Barcelona and Madrid, Infomap detects a large community probably due to the high density of the network, and this does not compare well with the other methods which detect more communities. In fact, the value of $H(X,Y)$ approaches the one of the randomized model. This issue is absent in the Balearic Islands, where the network has a stronger intrinsic spatial division into different islands.

\begin{figure}
    \centering
    \includegraphics[width = 0.7\textwidth]{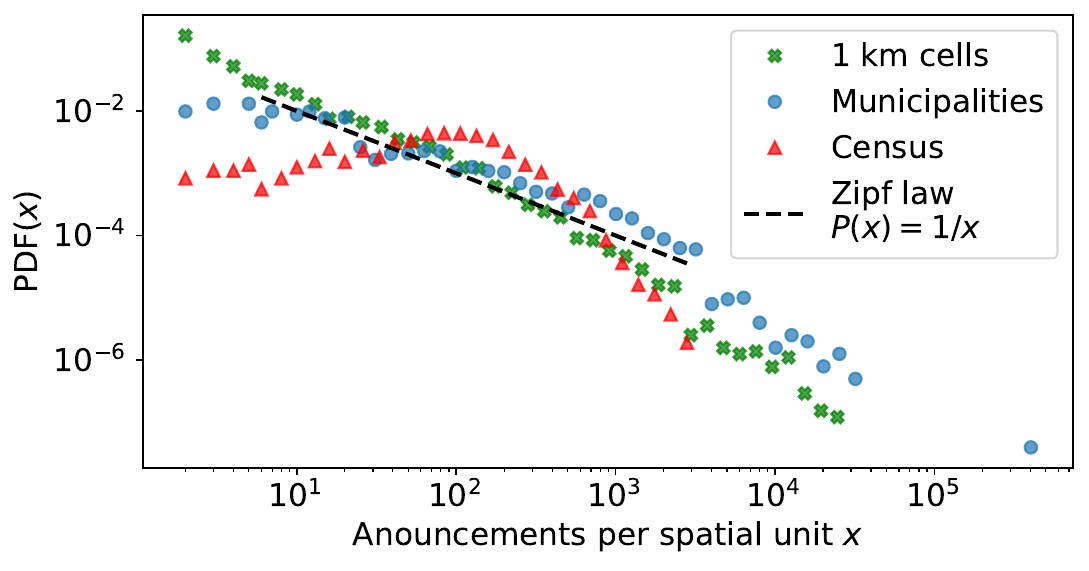}
	\caption[Distribution of listings for different spatial units.]{ Each spatial unit is shown by a different color and marker: green crosses (1 km-sided cells), blue circles (municipalities), and red triangles (census). The dashed black line shows the slope of a Zipf law distribution. \label{fig:distribution_announcements}}
\end{figure}

So far, we have focused on the results for the networks built with $1 \; \textrm{km}$-sided square cells. It is, nevertheless, important to check whether the results may vary depending on the scale of the unit cells. We thus recalculate the networks taking as basis square cells of side $500 \; \textrm{m}$ and $2 \; \textrm{km}$ and compute the communities using the Louvain method with consensus clustering. 
The cells of the different scales have been delimited to keep spatial coherence: four $500 \;\rm{m}$ cells form one of the $1  \;\rm{km}$ cells used in the previous figures, and four $1 \; \rm{km}$ cells aggregate to form a $2 \; \rm{km}$ cell. 
This hierarchical structure allows us to compare communities at various levels because we can identify the cells across scales. For example, if a $2 \;\rm{ km}$ cell belongs to a community, then the four $1 \;\rm{ km}$ cells composing it share the same community label.
In parallel, we also run the community detection algorithm in the network composed of $1 \; \rm{ km}$ cells, and then we can use the confusion matrix and $H(X,Y)$ to compare the partitions at these two scales using $1 \;\rm{ km}$ cells. Note that the calculation of $H(X,Y)$ requires the same number of basic units in the two partitions. Figure \ref{fig:agreement_scale_method}(d) shows the results of this analysis, where we use $1$ km-sided cells as a reference for comparison with the other scales. In all cases, we notice a consistently high and statistically significant level of agreement. This demonstrates that our methodology generates communities that remain robust across the three spatial scales.

\subsection{Comparison with networks obtained from administrative boundaries}\label{sec:political spatial units}

In this section, we examine how incorporating administrative spatial boundaries to build networks impacts the detection of communities. In many cases, the geographical information for listings is only available at the level of existing administrative boundaries and statistical units, which are by design more heterogeneous than square cells.

\begin{figure}
    \centering
    \includegraphics[width = 0.98\textwidth]{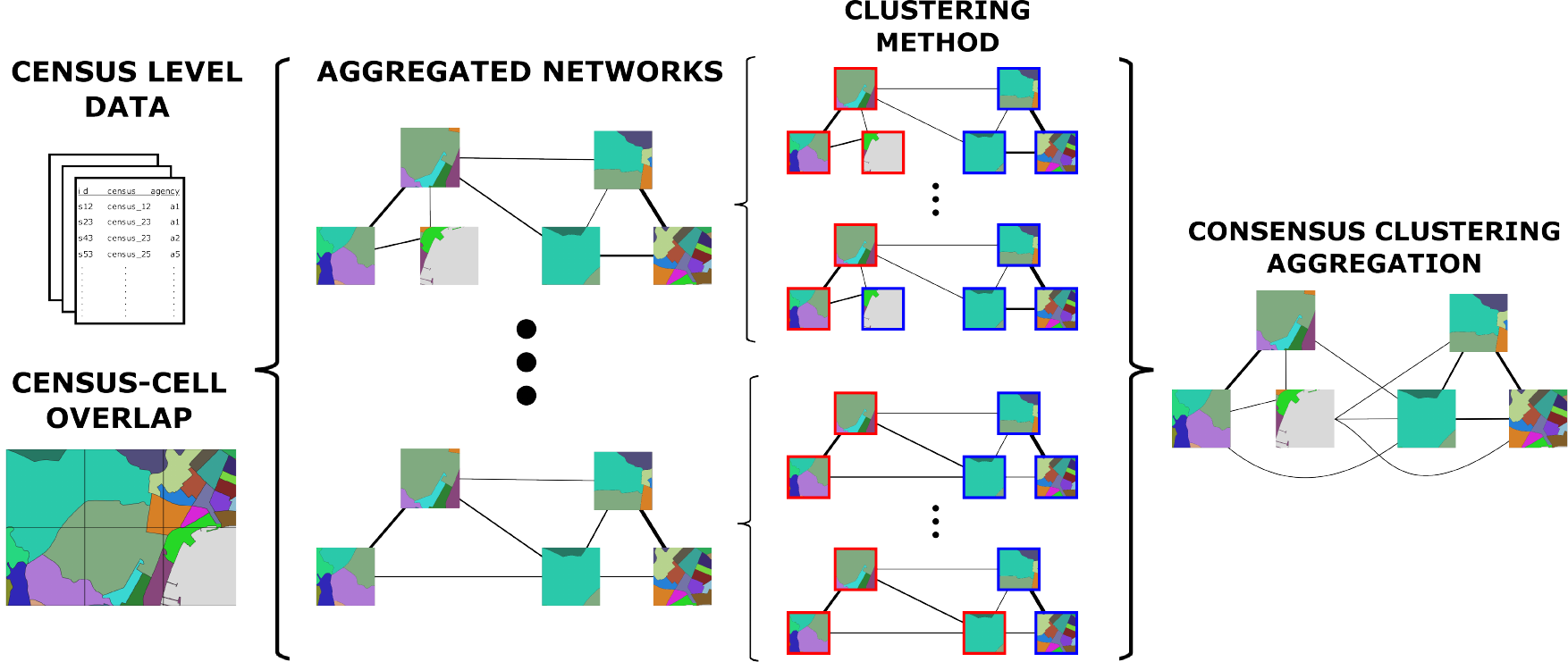}
	\caption[Stochastic aggregative method using census level data.]{ From a census-level listing and the spatial division of the census in square cells, we generate an ensemble of networks. In this ensemble, each listing within a census tract is associated to a cell with a probability based on the overlapping area between the census and the cell. For each of these cell networks, we run a community detection algorithm multiple times. The next step involves combining the results from these partitioned networks through consensus clustering, resulting in an aggregated network. \label{fig:stochastic_construction}}
\end{figure}

We aggregate listings into administrative and statistical spatial units to determine if the emergent submarkets are stable and consistent when comparing with the ones observed with the networks built with square cells. In this case, we consider municipalities and census tracts as they are the most common administrative divisions applied to spatial statistics. 

Fig. \ref{fig:political_scales} shows the communities found in the province of Madrid. We observe clear differences between the results obtained using census tracts (Fig. \ref{fig:political_scales}(a)-(b)) and using municipalities (Fig. \ref{fig:political_scales}(d)-(e)). The results for census tracts are characterized by a large community that covers almost all the territory and the agreement between methods is not significant. In contrast, the results using municipalities have a good and significant OSLOM-Louvain agreement. Keeping Louvain as the reference method, we compare the partitions of the networks originated from $1  \, \textrm{km}$, census tracts, and municipalities in Fig. \ref{fig:political_scales}(c). The communities in the networks using cells and municipalities show significant agreement, while those based on census tracts show non-significant values in Barcelona and Madrid.

While the distribution of listings per spatial unit in the other cases follows a heterogeneous distribution, well-described by a Zipf law, the one for census tracts follows a more homogeneous distribution (see Fig. \ref{fig:distribution_announcements}). This effect is a consequence of how the census tracts are built, forcing the population in each unit to be similar by a heterogeneous selection of the space included in each unit. This distribution is directly translated into the network weights and thus impacts the spatial segmentation method. 

\begin{figure}
    \centering
    \includegraphics[width = 0.7\textwidth]{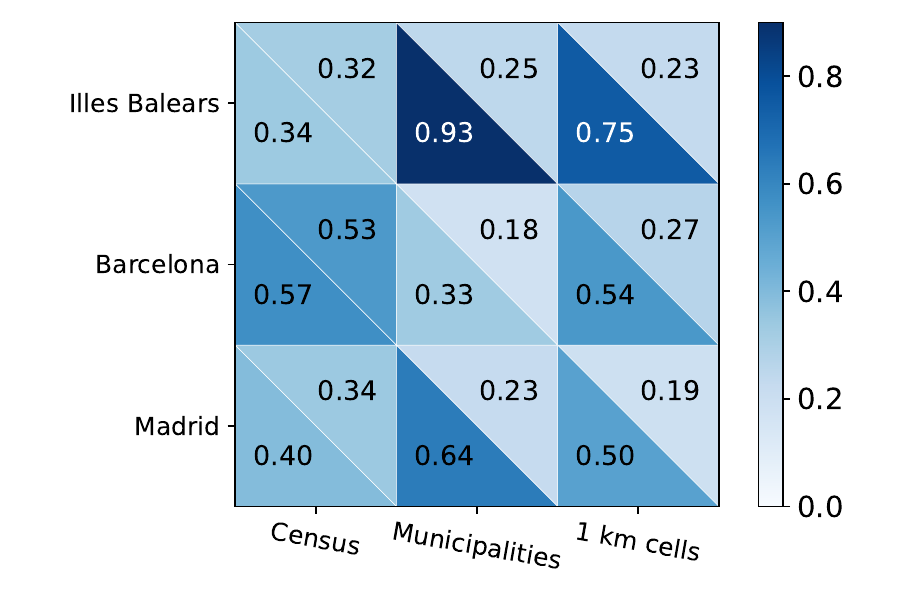}
	\caption[Comparison between the communities from aggregated cells network and other spatial units.]{ Each column shows the agreement between the communities of the $1 \; \textrm{km}$ aggregated cells networks (from census data) and the networks obtained from the other spatial units: Census, Municipalities, and $1 \; \textrm{km}$ cells from the original latitude longitude coordinate data. Each row shows the results for each province: Balearic Islands, Barcelona and Madrid. The agreement is computed via the fraction of correctly detected cells $H(X,Y)$ (lower triangles) compared with the value randomizing the communities (upper triangles). \label{fig:network_sector_distribution}}
\end{figure}

\subsection{Recovering the submarkets from census level data}\label{sec:data_agreggation}

\begin{figure}
    \centering
    \includegraphics[width = 0.98\textwidth]{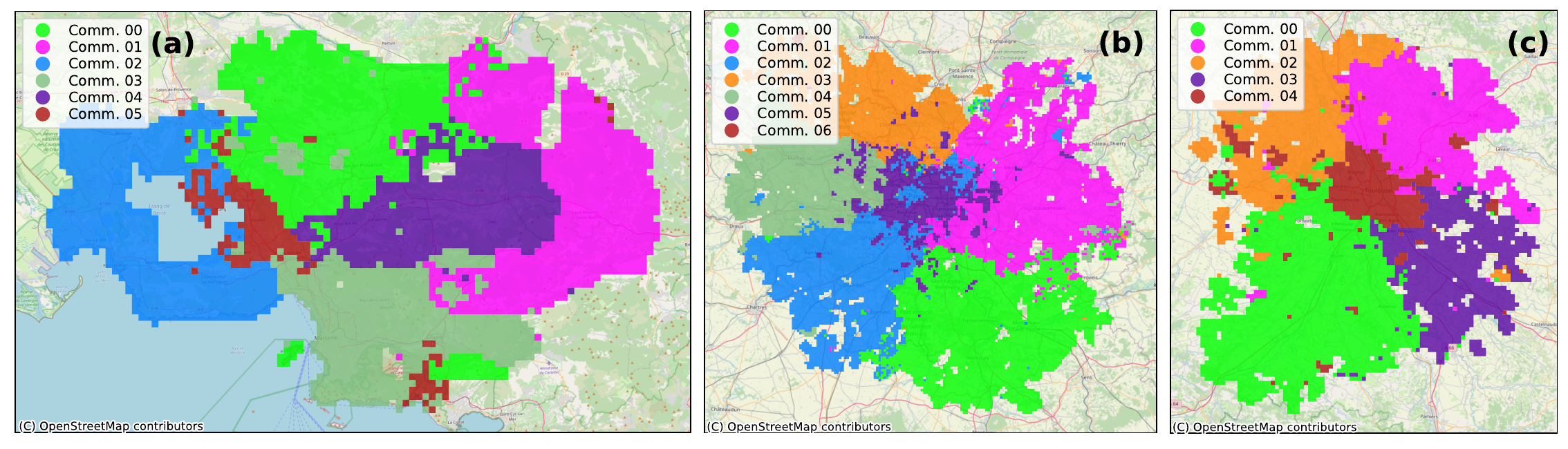}
	\caption[Spatial segmentation for $1 \; \textrm{km}$ aggregated cells constructed from IRIS level data for France.]{ Communities detected at the stochastic projected network for the 3 French FUA studied: Marseilles-Aix en Provence \textbf{(a)}, Paris \textbf{(b)}, and Toulouse \textbf{(c)}. The communities shown are detected using the Louvain algorithm with a consensus clustering of $200$ clustering method realizations for each of the $100$ stochastic networks generated in the IRIS to cell aggregative process. The underground map data is rendered by OpenStreetMap, under ODbL. \label{fig:france_comm} \vspace{-0.5 cm}}
\end{figure}

Multiple datasets, such as our French data, are available at census level. To maintain the broad applicability of our spatial segmentation methodology, we have devised a data aggregative method to recover the results obtained at the cell and municipality levels. This technique enables us to restore the Zipf law pattern using data gathered at the census level and to find similar segmentation results regardless of the basic spatial units.

We start with listings at a census scale, such that each listing is associated to an agency and a census tract. The first step is to divide the space into square cells, as we did in Section \ref{sec:materials_and_methods}. The cells intersect with the census tracts. We then associate each listing to a cell with a probability proportional to the overlapping area between the listing census tract and the cell. This process is repeated for all the listings to reconstruct a tripartite network of agencies-listings-cells, from which we can follow the methodology explained to reach a cell-cell network and a segmentation in submarkets (communities). 
We observe that in the final networks the Zipf law distribution of listings per cell is recovered.
Since the assignation of listings to cells is stochastic, the projected network is different each time the process is repeated. To avoid uncertainty, we construct an ensemble of these networks. For each network, we run the community detection algorithm multiple times. Once our cells are labeled with a community, we perform consensus clustering to aggregate all partitions from all aggregated networks of our ensemble into a single consensus aggregated network. We represented this process in detail on Fig. \ref{fig:stochastic_construction}.

To verify the results of the aggregative method, we perform a comparison of the submarkets obtained out of different networks. Starting with our Spanish data, where the listings are geolocated using exact coordinates, we build networks at the level of $1 \, \rm{km}$ cells, census tracts and municipalities. We then apply the method to aggregate the census tracts to the cells. This gives us a fourth family of networks, which we call aggregated cells network. We then run community detection methods and compare them across the networks, taking as a basis the partition obtained from the network of aggregated cells (see Fig. \ref{fig:network_sector_distribution}). For all cases, the agreement exhibited by partitions of the aggregated cells network and the original cells or the municipalities is very high (and significant compared to the randomized communities). Therefore, by reconstructing the network with the aggregative method, we recover the original communities at the cell and municipality levels and avoid the issues caused by the natural spatial heterogeneity of census tracts.

\subsection{Comparison across countries}\label{sec:France}

\begin{figure}
    \centering
    \includegraphics[width = 0.5\textwidth]{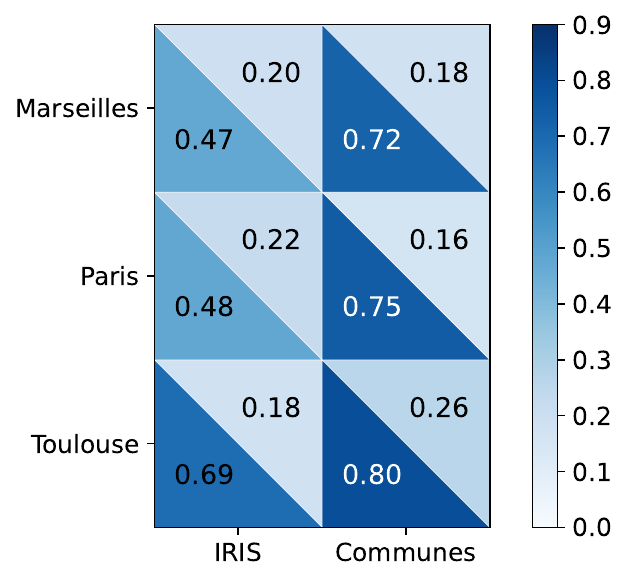}
	\caption[Comparison between the $1 \; \textrm{km}$ aggregated cells communities and the political units communities in France.]{ Each column shows the agreement between the $1 \; \textrm{km}$ aggregated cells and the French political spatial units: IRIS and Communes. Each row shows the results for each FUA: Marseilles-Aix en Provence, Paris, and Toulouse. The agreement is computed via the fraction of correctly detected cells (lower triangles) compared with the value randomizing the communities (upper triangles). \label{fig:Compare_france}}
\end{figure}

In this section, we investigate whether the emergent spatial segmentation revealed by our method is a unique feature of the Spanish market, or can be understood as a more general phenomenon across geographical contexts. To this end, we use listing data for three major French urban areas, namely, Marseilles, Paris, and Toulouse. Since we do not have exact coordinates for the listings, which are only located at a census tract level, we have to employ the stochastic aggregative technique described in the previous section to obtain the networks at the cell level or to aggregate the data at the municipality (\textit{commune}) level (since the census tracts can be grouped within each commune). 

Communities emerge in these French urban areas at aggregated cell level as well (see Fig. \ref{fig:france_comm}). The communities are contiguous in space, similar to the ones observed in Spain, suggesting that listings (as a source of information on listed properties and agencies) allow us to study the spatial segmentation of the housing market through a data-driven, bottom-up method that foregrounds the practices of key market intermediaries.

We repeat the exercise of comparing networks built from different spatial divisions. If France exhibits the same structures found in the Spanish dataset, we would expect the communities found from the aggregated cells and municipality networks to coincide, being the ones from the network of IRIS level very different. Fig. \ref{fig:Compare_france} displays the agreement between the communities using aggregated cells and administrative divisions (IRIS and communes). All values of the agreement are significant when compared with the randomized communities, but the largest agreement is found between aggregated cells and communes in all places, echoing results with the Spanish data. This indicates that our aggregative method is a general tool to compute a robust spatial segmentation of the housing market.

\section{Conclusions}
\label{sec:conclusions}

In this study, we present a new method for analyzing the spatial segmentation of housing markets through the activity of real estate agencies, using online listings to extract information on the location of both the property and the marketing agency. We apply this method to analyze comprehensive datasets of geolocated listings in two different countries: Spain and France. 

Our methodology is based on dividing space into spatial units, to construct a tripartite network between listings, real estate agencies, and spatial units. We project the network, taking into account the presence and influence of real estate agencies. These projected networks at the spatial resolution of $1$ km cell, census-tract, and municipality are available for further research at Zenodo~\cite{zenodo-2024} and GitHub~\cite{Abella-github-2024}. These links also include the code and additional code to perform the stochastic aggregative method from generic census data. Moreover, to divide our projected networks, we use different classic community detection algorithms that account for network weights, such as Louvain, Infomap, and OSLOM. Our methodology generates a spatial segmentation into regions that happen to be spatially connected and larger than municipalities. This segmentation into submarkets remains robust across different community detection algorithms, scales, and administrative boundaries across different countries.

We discovered a limitation of our method when the spatial units exhibit a highly heterogeneous area distribution, and the Zipf law of the distribution of listings per spatial unit is not fulfilled, as in the case of census tracts. To overcome this limitation and extend our methodology to heterogeneous-level data, we developed a method to create an aggregated network via stochastic reconstruction and consensus clustering aggregation. This methodology exhibits good accuracy when compared with the communities from the original high-precision data.

To summarize, we have developed a new methodology that uses listings data to evaluate the spatial segmentation of housing markets into spatially-coherent submarkets. This methodology is generally applicable to different datasets of geolocated listings to infer the submarkets that emerge from the uneven presence and influence of real estate agencies across space. The market-based supra-municipal communities that emerge from the data are found to be robust. Future research should investigate how identifying the submarkets created by market intermediaries can inform policymaking and improve price modeling.

\renewcommand{\thechapter}{8} 

\chapterimage{Cover_Agencies_Barcelona.pdf}
\chapterspaceabove{6.75cm}
\chapterspacebelow{7.25cm}

\chapter{Housing market dynamics revealed through online listings \label{ch:Housing market dynamics revealed through online listings}}
\vspace{-1.2cm}
\small
\textbf{The results in this chapter will be published as:}
\vspace{0.05 cm}

\fullcite{abella2024dynamics}
\normalsize
\vspace{0.5 cm}

In previous chapter, we have seen how the real estate market shows an inherent connection to space. Real estate agencies unevenly operate and specialize across space, price and type of properties, thereby segmenting the market into submarkets. This chapter examines the dynamics of real estate housing market from real data from an online listings platforms. By analyzing temporal patterns, we find that market dynamics are non-stationary, with weekly patterns and a steady increase in listings. The data reveals a preferential attachment mechanism, where the likelihood of new listings connecting to an agency depends on the agency's portfolio. Additionally, we observe fluctuation scaling in listing prices and distances, highlighting market heterogeneity. The study shows that new listings are more likely to be posted by certain agencies based on a combination of popularity, price similarity, and spatial specialization. These findings provide insights into the complex interplay of factors driving the housing market, offering a foundation for predicting market trends and improving efficiency.

\section{Introduction}

The use of Internet is widespread in our society and has transformed the way we interact, communicate, and consume information~\cite{berners-lee-2006,dorogovtsev2002evolution,pastor-satorras-2004,watts-2007}. New digital platforms have emerged, providing a wide range of services and products, from social media and e-commerce to online banking and food delivery~\cite{unknown-author-2013}. The data generated by the human interaction via these digital platforms has become a valuable source of information for researchers, policymakers, and businesses. However, despite the increasing reliance on digital platforms, the extract of useful knowledge from the data generated by these platforms is still a challenge. The data generated by the human activity in online platforms is often noisy, incomplete, and biased, making it difficult to analyze and interpret. On the other hand, a better understanding of the way in which users interact in these platforms has important economic consequences, as it can help both the development and monetization of the platforms and the understanding of the interactions between society and these new digital environments~\cite{choudary-2016}.

Previous studies on human dynamics using web analytics have demonstrated that, even though human activity in online listings is expected to be unpredictable at an individual level, there are common patterns that emerge at the aggregate level~\cite{Lazer2009CompSocSci}. The preferential attachment~\cite{barabasi1999emergence,goncalves-2008}, teams self-assembly~\cite{guimera-2005}, or the limited social capacity~\cite{goncalves-2011,dunbar-2012} are examples of emergent social mechanisms extracted from online platforms data. In this approach, the individual entity is ignored, focussing on the number of interactions, the time between them, or the entities involved in the interactions.

The advent of digital platforms has revolutionized various industries, and the real estate sector is no exception. The increasing reliance on online listings to buy, sell, and rent properties offers a unique opportunity to analyze the underlying dynamics of the housing market through a computational quantitative approach.

\begin{figure}
    \vspace{0.2 cm}
    \centering
    \includegraphics[width =\textwidth]{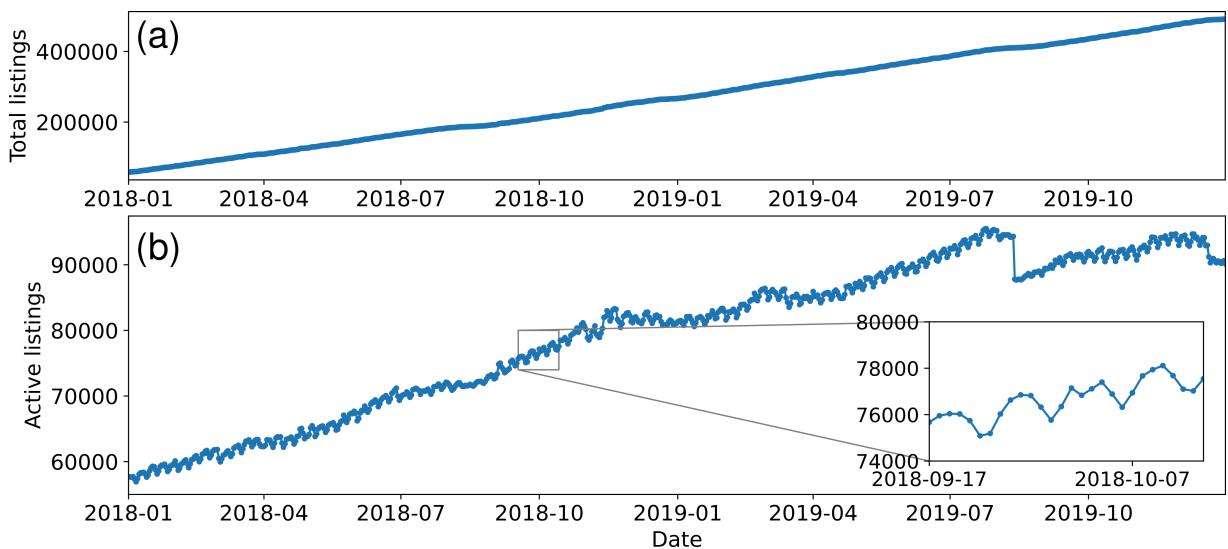}
	\caption[Active listings' evolution.]{\label{fig:active_adds} Daily number of total \textbf{(a)} and active \textbf{(b)} listings during the period 2018 - 2020 for the Barcelona province. The inset in (b) shows a zoomed view to a 1-month period (09-17 to 10-16), where the weekly patterns are more evident.}
\end{figure}

In this chapter, we aim to investigate the dynamics of real estate listings and agencies. Specifically, we seek to answer the following questions: What are the temporal patterns of the listings on the platform? Which is the decision-making process of a house seller when hiring a real estate agencies? How the real estate agency prestige, price and location affect this process? To give an answer to these questions, we examine the real estate agencies dynamics via a temporal bipartite network, where listings are connected to the agency that posted them in a certain temporal window. This methodology is applied to the Spanish market at 3 provinces: Madrid, Barcelona and Balearic Islands, using a comprehensive dataset of listings posted on the online platform \texttt{Idealista.com}. For all 3 regions, the dynamics of the market is non-stationary and show non-trivial correlations between the listings' metadata in which we are interested: the price, the location and the agency that posted the listing. The results presented in this chapter serve as a basis to understand quantitatively emergent processes such as spatial specialization o price segmentation of the agencies, and could be used to develop models of the housing market dynamics.

\section{Data \label{sec:Data}}

\begin{figure}[t!]
    \centering
    \includegraphics[width =\textwidth]{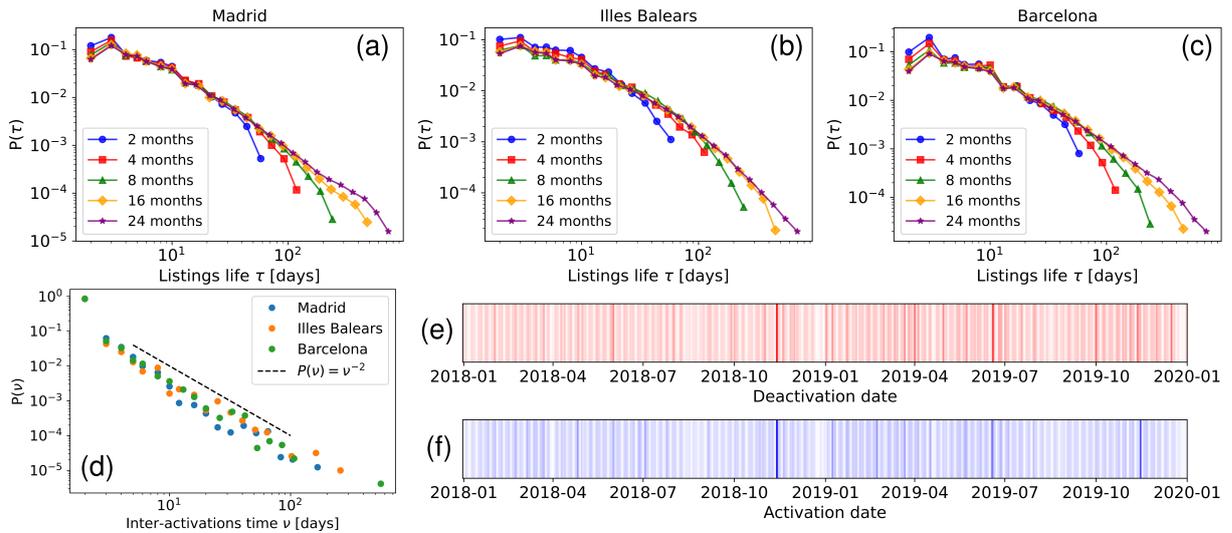}
	\caption[Temporal statistics of the listing dynamics.]{ \label{fig:panel_time} Temporal statistics of the listing dynamics. Listings life (time posted in the platform) distribution for different time windows at Madrid \textbf{(a)}, Barcelona \textbf{(b)} and Balearic Islands \textbf{(c)}. Different colors indicate different time windows. \textbf{(d)} Inter-activations time distribution for the 3 regions. Different colors indicate different region or province. The dashed black line shows a $\nu^{-2}$ power-law distribution. \textbf{(e)} and \textbf{(f)} show the bar sequence of the adds deactivations (date of a listing removal) and activations (date of a listing publication), respectively, for the Balearic Islands market. The color intensity indicates the number of adds.}
\end{figure}

We use the dataset, mentioned in Section \ref{sec:Datasets}, of listings published on the online portal \texttt{Idealista.com}~\cite{idealista}. The dataset covers a 2-year time period, from January 2017 to December 2018, and it comprises a collection of online geolocated listings in the Spanish provinces of Balearic Islands, Barcelona, and Madrid. These listings were posted by real estate agencies for renting or selling residential properties. Posted listings by private individuals are not included into the dataset.

Each listing contains a set of attributes such as the price (in $\textup{\euro}$), the surface area (in $m^2$), the real estate agency that posted it, its location and the dates of both the publication and the removal of the listing. The dataset also contains information about the type of property (e.g. flat, house, etc.), what allowed us to filter rural parcels and commercial properties. We also removed listings with missing or inconsistent information, such as irregular prices or surface areas. The temporal resolution of the dataset is daily. Through this chapter, the main results are shown for the selling market, but the same analysis stands for the renting market as well.

\section{Listings dynamics}

We begin by analyzing the temporal dynamics of the listings, ignoring the additional data about real estate agency, price and location. To do so, we consider as total listings, the number of listings that have been posted in the online platform before a given time $t$, and as active listings the number of those that are available in the platform at a given time $t$. The total number of listings increases linearly with time during the studied period, with no significant variations, showing that new listings are continuously added to the platform (see Fig. \ref{fig:active_adds}(a)). On the other hand, the number of active listings shows an increasing linear trend, which indicates that the number of listings available in the platform is growing over time (see Fig. \ref{fig:active_adds}(b)). Moreover, the number of active listings shows a weekly pattern, with a peak on Wednesday and a minimum on the weekends, consistent with human activity patterns in online platforms~\cite{goncalves-2008}. This weekly pattern is more evident when zooming into a 1-month period, as shown in the inset of Fig. \ref{fig:active_adds}(b).

\begin{figure}
    \vspace{0.2 cm}
    \centering
    \includegraphics[width = 0.8\textwidth]{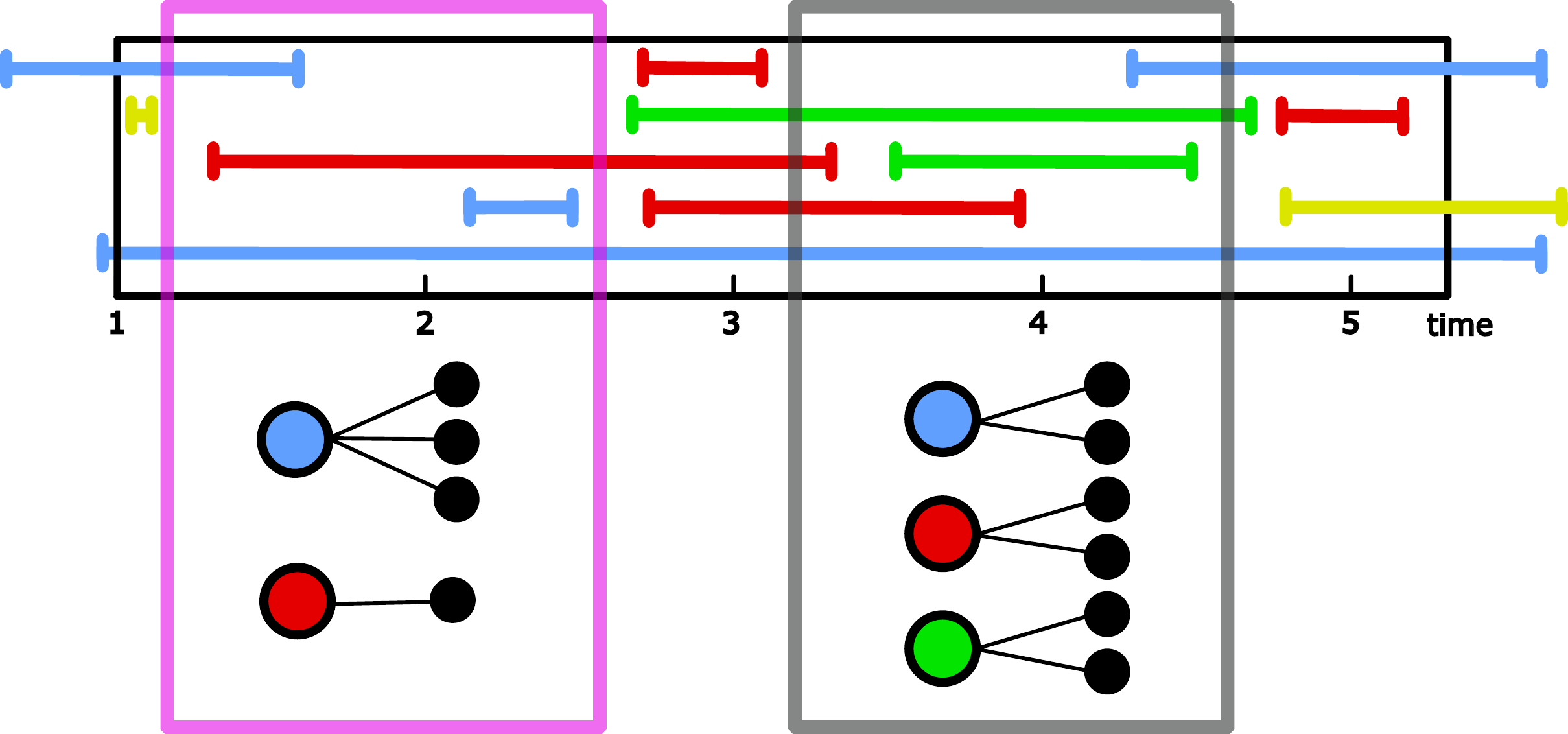}
	\caption[Housing market as a temporal bipartite network.]{Schematic representation of the temporal bipartite network between listings and agencies. On top, there is a temporal period, represented as a black box, in which the different listings are represented by colored horizontal lines, which represents the time span that a listing has been active in the platform. The color of the listings represent the real estate agency that posted them. For each time window, represented by purple and gray boxes (from $t_0$ to $t_f$), we build a bipartite network where listings are connected to the agency that posted it if the listing is active inside the temporal window. The colored big circles represent the agencies' nodes and the black small circles, the active listings. \label{fig:temporal_bipartite}}
\end{figure}

The differences between the total and active listings are driven by the listings' life $\tau$, defined as the number of days a certain listing has been posted in the platform. This measure is a proxy of a listings' attractiveness, because when a listing is sold/rented, the agency removes it from the platform. This is an obvious simplification of the reality, since it might be for other reasons, for example, a listing removed because a personal decision of the owner. Fig. \ref{fig:panel_time}(a-c) shows the listings' life distribution for the 3 regions studied. The purple line is the distribution for a time window including the whole period (2 years). In all cases, the listings' life distribution shows a fat tail distribution, indicating that, even though there are a lot of listings that are sold in a few days, there are a few present for long periods of time. This heterogeneous life distribution makes this system another example of the irregular human activity patterns mentioned in previous chapters (see section \ref{sec: Bursty Human Dynamics}).

However, bursty dynamics are usually characterized by fat tail inter-event time distributions, not life distributions. For the listings' dynamics, the inter-activations time distributions (time between new listings activations) shows a power-law distribution with exponent $\gamma \approx -2$ for all regions, as shown in Fig. \ref{fig:panel_time}(d). When we compare this distribution with the inter-event time distribution of human activity patterns, with an exponent around $\gamma \sim -1$, we find that the listings' inter-event times are less heterogeneous. A similar behavior is found in Fig. \ref{fig:panel_time}(e-f), where we show the bar sequence of the adds deactivations and activations for the Balearic Islands market. Even though we differentiate an irregular temporal dynamics on top of the weakly pattern, with days when there are more listings added or removed (see also sudden drop of active adds around 2019-08 in Fig.\ref{fig:active_adds}(b)), the overall activation behavior is not as bursty as in other human activities.

This process can be understood with the idea of ``aging'', introduced in Section \ref{sec:Aging mechanism}. Listings are added to the platform at some constant rate, evident by the linear increase of the total listings. But, the longer a listing is in the platform, the less likely it is to be sold/rented, leading to fat tail life distribution. In this particular scenario, aging does not reflect the attachment to a previous belief, as in the models in the previous part. Instead, aging acts as a temporal attractiveness of the listing. This stochastic process is known in the literature as \textit{delayed degradation}~\cite{lafuerza2013stochastic}, in which particles are added to a system at a constant rate, and get removed after a certain time $\tau$ after being created, where $\tau$ is usually randomly distributed. This process has been studied in many situations, such as gene regulation~\cite{lewis2003autoinhibition, barrio2006oscillatory, bratsun2005delay}, neuronal activity~\cite{flunkert2013dynamics}, and physiological processes~\cite{longtin1990noise}. In general, the delayed degradation stochastic process is solved in Ref.~\cite{lafuerza2013stochastic} but only for life distributions with a well-defined mean. In our case, the life distribution is fat-tailed, and the delayed degradation process needs to be treated with more detail.

\section{Agencies dynamics}

\begin{figure}
    \centering
    \includegraphics[width =\textwidth]{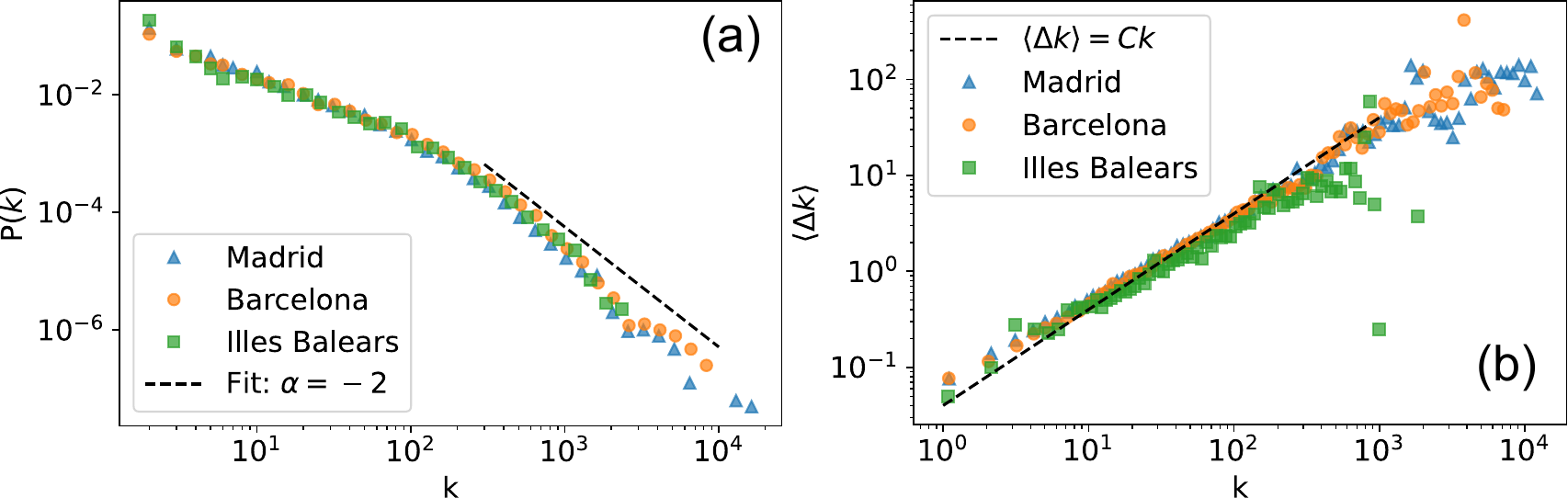}
	\caption[Preferential attachment to agencies.]{\label{fig:panel_degree}Preferential attachment to agencies. \textbf{(a)} Degree distribution of the agencies in the 3 regions. \textbf{(b)} Average degree increase of an agency as a function of the degree (listings posted) of the agency previously to the attachment. For both plots, different colors and markers indicate different regions or provinces. In \textbf{(a)}, the black dashed line shows a power-law distribution with exponent $\alpha  =-2$, and in \textbf{(b)} shows a linear increase with fitted slope $C = 0.04$.}
\end{figure}

In previous section, we have analyzed the listings' dynamics, just exploring how the total and active adds increase. However, the listings are not isolated entities, but they have a price, a location and are posted by a real estate agency. Now, we address the decision-making process that house owners follow when they decide to sell/rent their properties through a real estate agency, and how it is correlated with the pricing and the location of the listings.

To this end, we use the same approach as in the previous chapter: a multipartite network. In this case, we just need one part of the network, the bipartite network between listings and agencies. Another difference with the previous chapter is that we are interested in the temporal dynamics of the listings, so now we will build a temporal bipartite network. At each time window, defined by an initial date $t_0$ and a final date $t_f$, we construct a bipartite network where listings are connected to the agency that posted it if the listing is active in any time $t'$, inside the temporal window $t_0 \leq t' \leq t_f$. Does not matter if the listing was removed after $t_f$ or if it was added before $t_0$, we consider all active listings in the time window. Note that this network is very simple, as a listing cannot be connected to more than one agency. As it is observed in the schematic representation of the temporal bipartite network in Fig. \ref{fig:temporal_bipartite}, as the window moves in time, the bipartite network changes structure, allowing us to infer the agency-listings dynamics.

\subsection{Preferential attachment}

Considering the cumulated network of all the listings (setting as time window the whole 2-year period), we can analyze the degree distribution of the agencies. The degree of an agency in this context is the number of listings posted by the agency. As shown in Fig. \ref{fig:panel_degree}(a), the degree distribution of the agencies in the 3 regions follows a power-law distribution, highlighting the presence of very large agencies coexisting with several small agencies, in terms of the number of listings posted. This result suggests that agencies follow a preferential attachment mechanism, where the probability of a new listing being attached to an agency is proportional to the number of listings posted by the agency previously. This is a common mechanism in many real-world networks, such as the World Wide Web~\cite{barabasi1999emergence}, citation networks~\cite{redner1998popular}, transportation networks~\cite{barrat2004architecture}. In the case of the housing market, this mechanism is a signature of the popularity of the agencies, as the more listings an agency has, the more likely it is to have new listings attached to it.

To verify if the underlying mechanism is indeed preferential attachment, we analyze the average degree increase of an agency as a function of the degree of the agency previously to the attachment. For this purpose, we build a bipartite network for a time window of 1 month and compute the increase of degree of the agencies after a week. This methodology allows us to track the real estate agencies one by one to see if there is indeed a degree increase biased by the degree of the agency previously to the attachment. We repeat this process for all the time windows in the period, and we average the results. As shown in Fig. \ref{fig:panel_degree}(b), the average degree increase of an agency is linearly correlated with the degree of the agency previously to the attachment. This result confirms that the agencies follow a preferential attachment mechanism, where the probability of a new listing being attached to an agency is directly proportional to the number of listings posted by the agency previously. However, this agency popularity in terms of degree is not the only key factor in the attachment process. The price of the listings and the location of the listings are also important factors that influence the decision-making process of the house owners.

\begin{figure}
    \centering
    \includegraphics[width =\textwidth]{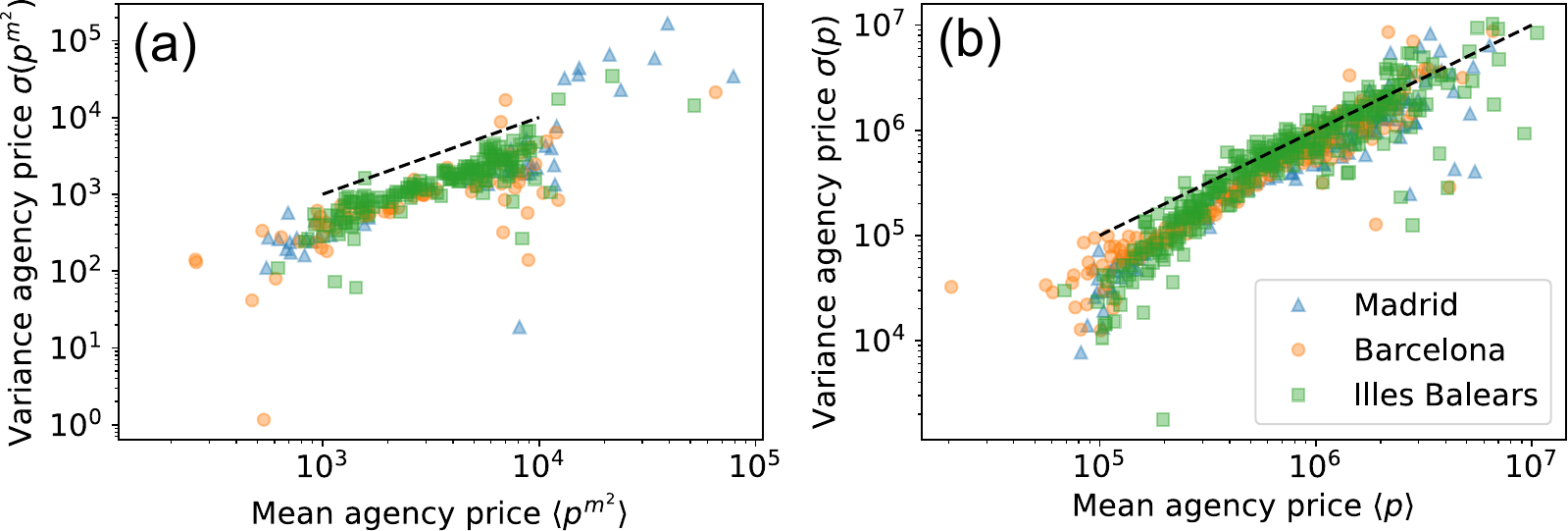}
	\caption[Variance of the agency price vs mean agency price.]{Variance of the listings price posted by an agency as a function of its mean price for the price per square meter \textbf{(a)} and the total price \textbf{(b)}. Different colors and markers indicate different regions or provinces. The black dashed line shows a linear fit with $\sigma(p) = \langle p \rangle$ for both plots. Here, $\langle \cdot \rangle$ stands for the average over the listings of an agency. \label{fig:sigma_price}}
\end{figure}

\subsection{Price correlations}

Since we have both the price and the surface area we compute the price per square meter of the listings, which is a common metric used in the real estate market. Both the price per square meter and the total price of the listings are distributed according to a log-normal distribution, an expected result in economic systems like ours~\cite{ibragimov2015heavy}. The price range of an agency is defined by the mean price of its listings and its variance, which are found to be correlated (see Fig. \ref{fig:sigma_price}). For both, the price per square meter and the total price, the variance of the listings price posted by an agency is proportional to the mean price of the agency, such that the higher the listings prices of an agency, the higher the variance. This proportionality has been observed for population count in ecological contexts~\cite{anderson1982variability, kilpatrick2003species}, and also recently in social systems~\cite{chen2020scaling}, and it is known as Taylor's law~\cite{taylor1961aggregation,eisler2008fluctuation}. In physics literature, this proportionality is known as fluctuation scaling~\cite{eisler2008fluctuation}, and it is shown to be related with heavy tailed data~\cite{brown2021taylor}, highlighting the heterogeneity in agencies pricing.

\begin{figure}
    \centering
    \includegraphics[width =\textwidth]{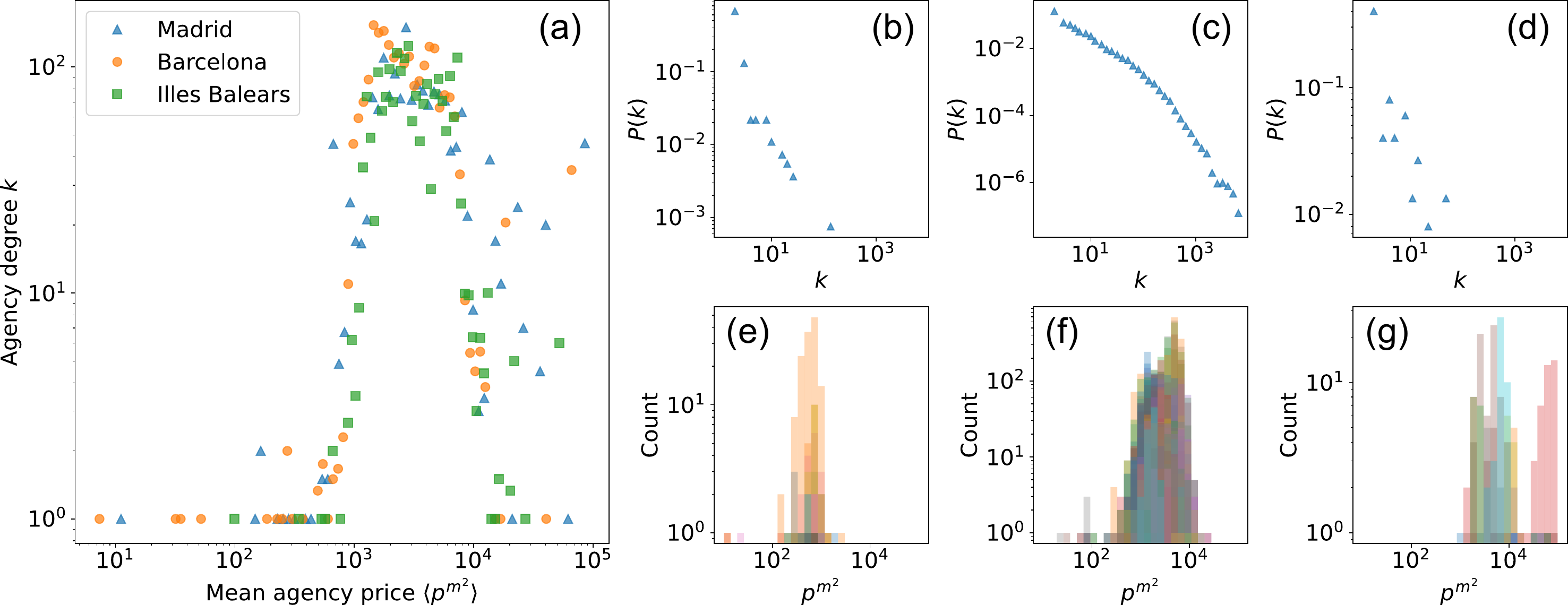}
	\caption[Price segmentation by the degree.]{\textbf{(a)} Average degree (number of listings) of an agency as a function of its mean price per square meter. Different colors and markers indicate different regions or provinces. Degree distribution among the agencies \textbf{(b)-(c)-(d)} and price histograms for 60 representative agencies \textbf{(e)-(f)-(g)} at the different price segments: \textbf{(b)} and \textbf{(e)} $p^{{m}^2} < 800 \, \textup{\euro} / m^2$, \textbf{(c)} and \textbf{(f)} $800 \, \textup{\euro}  / m^2 < p^{{m}^2} < 10^4 \, \textup{\euro}  / m^2$, and \textbf{(d)} and \textbf{(g)} $p^{{m}^2} > 10^4 \, \textup{\euro}  / m^2$. These plots correspond to the Madrid housing market. For the price distributions, the colors indicate histograms for different agencies. $\langle \cdot \rangle$ stands for the average over the listings of an agency. \label{fig:panel_price}}
\end{figure}

Besides price itself, we analyze how correlated is the price of the listings of an agency with its degree in the bipartite network. As shown in Fig. \ref{fig:panel_price}(a), the average degree of an agency shows a non-trivial dependence with the price per square meter. For low mean price, the degree values are very low. If we increase the price, we reach the typical prices in the market and degree increases very fast to a certain high value that keeps constant during a range of prices, but for very high prices, it decays again. This result suggests that agencies with low and high prices have few listings. The agencies with the highest number of listings are those with intermediate prices. This degree dependence allows us to segment the system in different price ranges according to how large is the mean degree. At the low prices segment (Fig. \ref{fig:panel_price}(b) and (e)), the degree distribution does not show a long tail anymore, and the agencies show a price distribution peaked at a certain price and with low fluctuations. For intermediate prices (Fig. \ref{fig:panel_price}(c) and  (f)), the degree distribution exhibits the power-law behavior observed in Fig. \ref{fig:panel_degree}(a). In this price segment, the agencies show price distributions similar to a log-normal distribution, with a peak at a certain price and fluctuations around it. For high prices (Fig. \ref{fig:panel_price}(d) and (g)), the degree distribution is rather homogeneous (do not have a fat tail) and the price distribution of the agencies shows higher variance, since the agencies include both high prices and middle prices listings. This segmentation is an effect of the coexistence of generalist agencies, that post agencies in a wide range of prices, and specialist agencies, which focus on a specific submarket segment. These results are consistent with Fig. \ref{fig:sigma_price}. 

\begin{figure}
    \centering
    \includegraphics[width =\textwidth]{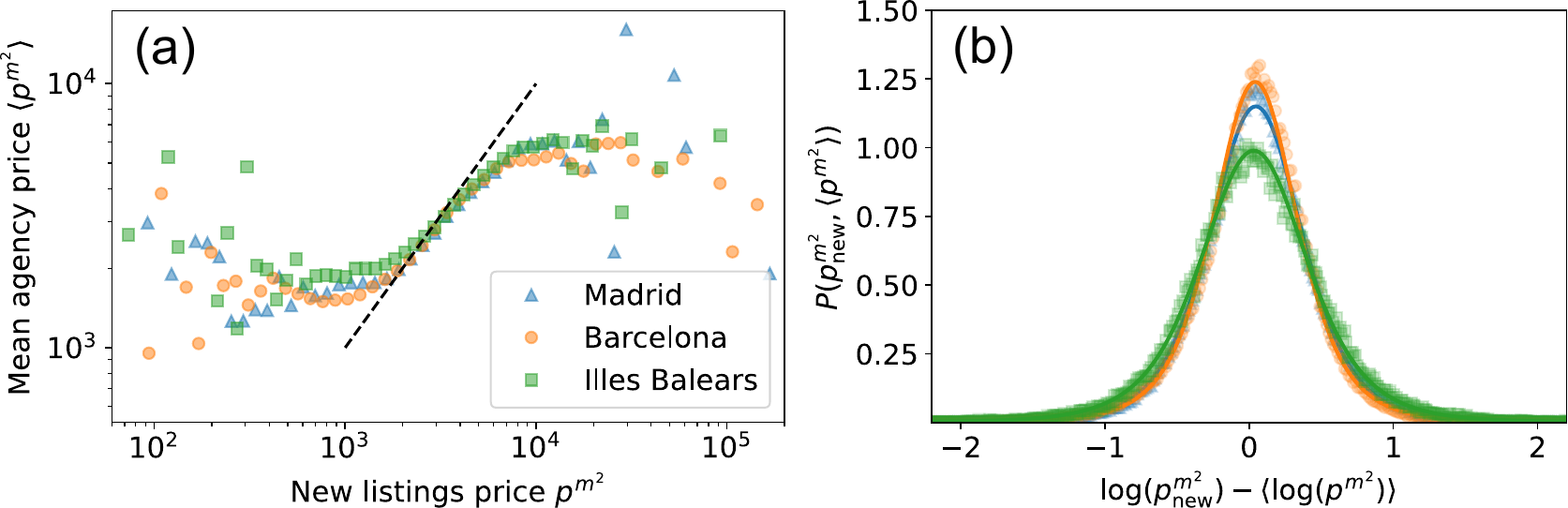}
	\caption[Attachment dynamics of new listings by price.]{Attachment of new listings by price. \textbf{(a)} Mean agency price per square meter as a function of the price of the new listing attached to the agency. Different colors and markers indicate different regions or provinces. The black dashed line shows a linear increase $\langle p^{{m}^2} \rangle = p^{{m}^2}_{\rm new}$. \textbf{(b)} Distribution of the logarithmic difference between the price of the new listing and the mean price of the agency. Different colors indicate different regions or provinces. $\langle \cdot \rangle$ stands for the average over the listings of an agency. Each solid colored line corresponds to a T-student fit for each region. The parameters are: $\nu = 5.47$, $\mu = 0.04$, $\sigma = 0.33$ for Madrid, $\nu = 4.62$, $\mu = 0.04$, $\sigma = 0.30$ for Barcelona, and $\nu = 4.49$, $\mu = 0.03$, $\sigma = 0.38$ for the Balearic Islands. $\nu$ is the degrees of freedom, $\mu$ is the location parameter, and $\sigma$ is the scale parameter. \label{fig:attach_price}}
\end{figure}

So far, the price correlations described have been analyzed from a static point of view. Regarding price dynamics, we explore how the attachment of new listings is correlated with the price per square meter. As we did for the degree preferential attachment, we compute the mean agency price during a time window of 1 month and price of the new listings attached to the agency during the next week. This process is repeated for all the time windows in the period. When we average over possible values of the new listings price, we observe a positive correlation, similar to linear, between the mean agency price and the new listings price (see Fig. \ref{fig:attach_price}(a)). This correlation seems to be robust across the different regions, highlighting how listings with a certain price attach to agencies that work in a similar price range.

For each new attachment, we can compute how the price of a new listing attached fluctuates around the mean price of the agency. Fig. \ref{fig:attach_price}(b) shows the distribution of the logarithmic difference between the price of the new listing and the mean price of the agency. This distribution is found to be centered around $\log( p^{{m}^2}_{\rm new}) - \langle \log ( p^{{m}^2} ) \rangle \approx 0$, as expected from the previous result, and is well-fitted by a T-student distribution. Therefore, besides the preferential attachment, the listing price proportionality is an important factor of the housing market dynamics. When a new house is advertised, is more likely to be listed by an agency with both, a high number of  active listings (degree) and a similar price range.

\subsection{Specialization in space}

\begin{figure}
    \centering
    \includegraphics[width =\textwidth]{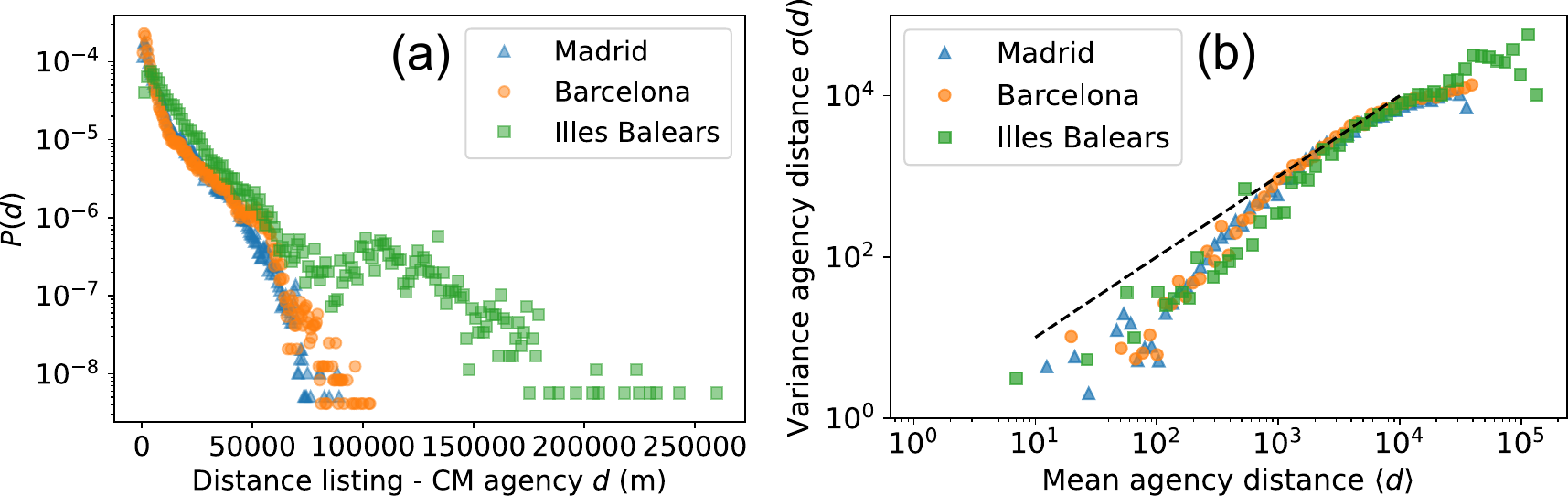}
	\caption[Distance correlations.]{\textbf{(a)} Distribution of the distance between listings of a certain agency and the center of mass of that same agency (mean location of the listings). \textbf{(b)} Variance of the listings distance to the agency center of mass of an agency as a function of its mean distance. The black dashed line shows a linear increase $\sigma(d) = \langle d \rangle$. Different colors indicate different regions or provinces. \label{fig:distance_panel}}
\end{figure}

It is said that in the real estate market, the three most important factors are location, location, and location~\cite{rosen1974hedonic}. In this section, we focus on the spatial correlations of the listings. The location of the listings is well-defined by its coordinates, but the location of an agency is not so clearly defined. We define as the agency location in a time window, the center of mass (CM) of the active listings posted by the agency in that time window. This definition is a simplification that allows us to quantify the spatial correlations of the listings via the distance from the listings and the agency center of mass. Fig. \ref{fig:distance_panel}(a) shows the distribution of the distance between listings and the agency center of mass for the 3 regions. This distribution is computed for all the listings in the period, is peaked around $10^2$ m, and follows an exponential decay. In fact, for Barcelona and Madrid, the largest distance is around $10^5$ m, but for the Balearic Islands, we observe a longer tail that reaches $2.5 \times 10^5$ m. This occurs due to the natural spatial segmentation of the Balearic Islands and the presence of generalist agencies that post listings in different islands.

As it occurred for the price of listings, the variance of the distance between the agency CM and the listings of an agency is proportional to the mean distance of the agency. This proportionality is shown in Fig. \ref{fig:distance_panel}(b) for the 3 regions studied, even though the relation is not as linear as for the price, specially for low mean agency distances. This result suggests that agencies with a higher effective radius $R$, defined as the mean distance $R = \langle d \rangle$, have a higher heterogeneity in the listings' location. Moreover, this fluctuation scaling process highlights the absence of a typical size of an agency in terms of  effective radius, going from large agencies that operate effectively everywhere, to localized agencies that focus on a specific region or neighborhood.

\begin{figure}
    \centering
    \includegraphics[width =\textwidth]{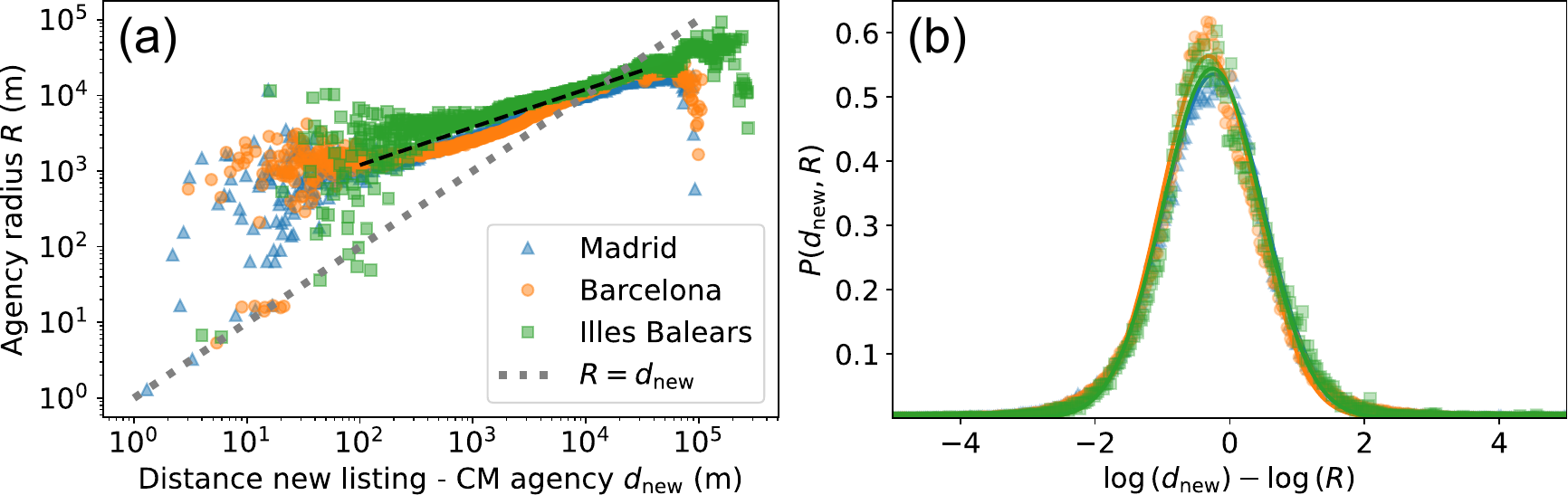}
	\caption[Attachment dynamics of new listings by distance.]{ Attachment dynamics of new listings by distance. \textbf{(a)} Agency effective radius (mean distance of the listings to the agency center of mass) as a function of the distance of the new listing (to the center of mass) attached to the agency. The black dotted gray line shows a linear increase $R = d_{\rm new}$, while the dashed black line shows a square root dependence $R = C \, d_{\rm new}^{1/2}$ being $C = 120$. \textbf{(b)} Distribution of the logarithmic difference between the distance of the new listing $d_{\rm new}$ and the effective radius of the agency $R$ (before the attachment). Different colors indicate different regions or provinces. Each solid colored line corresponds to a T-student fit for each region.  The parameters are: $\nu = 5.63$, $\mu = -0.25$, $\sigma = 0.69$ for Madrid, $\nu = 5.49$, $\mu = -0.30$, $\sigma = 0.67$ for Barcelona, and $\nu = 6.00$, $\mu = -0.25$, $\sigma = 0.67$ for the Balearic Islands. $\nu$ is the degrees of freedom, $\mu$ is the location parameter, and $\sigma$ is the scale parameter. \label{fig:distance_attach}}
\end{figure}

Regarding the dynamics, the location of a new listing also affects the attachment process to an agency. Fig. \ref{fig:distance_attach}(a) shows the effective radius as a function of the distance of the new listing attached to the agency. In this case, the scenario is very different to the price attachment. Here, the effective agency radius follows a proportional dependence to the distance listing - CM of the agency, but clearly sublinear (similar to $R \sim d_{\rm new}^{1/2}$). Thus, the majority of listings are attached to agencies with an effective radius larger or equal than the distance between the listing and the agency CM. This result is similar in all regions studied and consistent with the specialization of agencies in a certain area.

Moreover, as we did for the pricing and degree with a monthly window, we compute the distribution of the logarithmic difference between the distance of the new listing and the mean distance of the agency. Now, this distribution is not centered at $\log(d_{\rm new}) - \log(R) = 0$, but at negative values, as shown in Fig. \ref{fig:distance_attach}(b). Therefore, the new listings are attached to agencies where the effective radius is equal or larger than the distance between the new listing and the agency's CM but with present exceptions to this process. Notice that this distribution comes from an analysis of a dynamical process, and the agency CM and effective radius are changing in time and so the attachment process. From these results, we can conclude that when a new house is listed, is more likely to be listed by an agency with a high number of listings, with a similar price range to the house that also works in a similar area. These 3 mechanisms are the key factors that will compete and balance each other in the decision-making process of the house owners.

\section{Conclusions}

In this chapter, we have analyzed the dynamics of the housing market, using a comprehensive dataset of listings posted on an online platform for 3 provinces in Spain. We have shown that the listings' dynamics are non-stationary, with a weekly pattern in the number of active listings and a linear increase in the total number of listings. The listings' life distribution shows a fat tail, indicating irregular bursty dynamics, but the inter-activations time distribution not as heterogeneous as in other human activities. This process can be understood as a delayed degradation process, where listings are added to the platform at a constant rate, and they are removed after a certain time, similar to the attachment to a previous state in aging models. 

We have also shown that agencies follow a preferential attachment mechanism, where the probability of a new listing being attached to an agency is proportional to the number of listings posted by the agency previously, highlighting the presence of agency popularity in this market. Moreover, we found for both prices and distances (between add and agency center of mass), that the variance of the listings is proportional to the mean price or distance of the agency, showing a fluctuation scaling behavior. The presence of this relation in both magnitudes is a signature of heterogeneity and an underlying universal mechanism driving metadata in systems like this one to exhibit this Taylor's law behavior. The price of the listings is also correlated with the degree of the agency, showing a non-trivial dependence that allows to segment the market. 

For both, the distance and the price of a new house to be added into the offer system are key factors in the decision-making process of the house owners. So, in the housing market dynamics, there are 3 mechanisms competing directly: the popularity of the agency (preferential attachment), the price range of the listings of that agency (price proportionality), and the spatial specialization (distance lower than the agency radius). The balance of the importance on each of the mechanisms is what drives the housing market dynamics, and the results presented in this chapter serve as a basis to understand quantitatively emergent processes such as spatial specialization or price segmentation of the agencies.

The results presented in this chapter are based on a single dataset, and further work is needed to explore the generality of these results in other regions or countries. Nevertheless, it is important to remark that the results presented in this chapter were resilient to the different regions studied (Balearic Islands, Barcelona and Madrid), highlighting the robustness of the mechanisms found. Further work addressing the universality of these mechanisms need to be explored, as well as the exploration of other features, such as the price or distance correlation with the life of the listings in a certain agency, the spatial competition between agencies or the resilience of these mechanisms.

Further work will use this result to develop a model of the housing market dynamics, based on the idea similar to ``aging'' (explored in the first part of this thesis). This allows us to explore how the listings are posted on the online platform and how the agencies influence the decision-making process of the house owners. These models will be used to make a forecast of the market dynamics, which has important economic impact, as it can help to develop strategies to improve the housing market efficiency. Additionally, the understanding of the housing market dynamics could be useful for policymaking to predict market trends, and to develop strategies to reduce the housing market inequalities.


\part{Conclusions and future work}

\renewcommand{\thechapter}{9} 
\chapterimage{}
\chapterspaceabove{6.75cm}
\chapterspacebelow{7.25cm}

\chapter{Summary and main contributions}
In this concluding chapter, we offer a comprehensive overview of our most significant findings and their implications. We have explored in this thesis how aging and temporal dynamics take a relevant role in shaping the behavior in socio-economic complex systems. Through this thesis, we use a combination of theoretical and empirical approaches to study the emergent dynamics of complex social systems from a quantitative perspective, using tools from statistical physics, network theory, and data science. In particular, we have followed two complementary paths: the first one is the study of aging in threshold models, and the second one is the analysis of the temporal and spatial dynamics of the housing market from real online listings. We have found that aging and memory effects play a crucial role in the dynamics of these systems, leading to new insights and modeling challenges due to the non-Markovian nature of the dynamics.

\section{Remarks on Aging in threshold models \label{sec:aging_threshold_models}}

In the first part of the thesis, we delved into the analysis of aging in threshold models, in particular 3 different models: a segregation model (Sakoda-Schelling), a complex contagion model (Granovetter-Watts), and a consensus model (Symmetrical Threshold model). Aging in this context is introduced as a mechanism that modifies the microdynamics of the system, such that agents become increasingly resistant to change state or location the longer they remain in it. This mechanism is inspired by the empirical observation that individuals tend to stick to their decisions or locations over time, which can be seen as a form of inertia. This non-Markovian mechanism is found to be an important modification for both the stationary state and the dynamics of these models. In particular, in Chapter 3, we examined the aging effects in the Sakoda-Schelling segregation model. The study introduced aging as a factor where satisfied agents become increasingly resistant to changing their location the longer they remain satisfied in it. This modification led to the disappearance of the phase transition between segregated and mixed phases observed in the original model as the tolerance is increased, resulting in a phase diagram with segregated states even for high values of tolerance. This indicates that aging promotes segregation by creating a strong attachment to the current location, which counters the tendency to move based on satisfaction alone. Moreover, the stationary state of the model with aging present a higher global satisfaction than the original model. On the other hand, the coarsening process in the segregated phase was found to be slower in the presence of aging, as the interface density decayed with a lower exponent than in the original model. Furthermore, we found that the aging mechanism is able to break the time translational invariance of the model, leading to a different scaling of the autocorrelation function. For this model, the full study was conducted via numerical simulations, and a an analytical approach should be developed to understand the phase diagram of both the original and the version with aging.

Chapter 4 focused on extending the Approximate Master Equation (AME) for binary-state models to non-Markovian dynamics associated with aging. The original AME was developed to describe the dynamics of binary-state models in complex networks via a set of coupled differential equations where we treat sets of nodes with same degree and same number of neighbors in state $-1$ as compartments of a compartmental model. We extended this approach to include the aging mechanism, where age is treated as an additional variable of a node, as the state or the degree. With this framework, we were able to capture the non-Markovian dynamics associated with aging via two mechanisms: aging, when a node remains at its state and increases age, and resetting, when a node remains at its state and resets its age. With these two simple mechanisms, the AME is able to capture the non-Markovian dynamics for a large variety of binary state models with aging. The most important restriction of the AME is that it assumes an infinite large network with negligible levels of clustering. Nevertheless, this approach could be modified to include network size and stochastic effects related with aging, along the lines of Ref.~\cite{peralta-2020B}. In this thesis, Chapter 4 was very useful as a preamble for the analysis of the models presented in the following chapters.

In Chapter 5, we analyzed the impact of aging in the Granovetter-Watts model, a fundamental model for understanding complex contagion processes. In this context, we introduced two aging mechanisms: exogenous aging, where agents become increasingly resistant to adopting a behavior the longer they remain inactive, and endogenous aging, where agents become increasingly resistant to adopting a behavior the longer they remain in its state, even if the agents activate. For both mechanisms, aging was shown to slow down the cascade dynamics without altering the cascade condition. The model's behavior transitioned from exponential to stretched exponential or power-law adoption dynamics, depending on the aging mechanism (exogenous and endogenous). This chapter provided analytical expressions for the cascade conditions and exponents, derived from the AME, offering a comprehensive understanding of how aging influences the spread of information or behaviors in social networks. Furthermore, the results of this model in a Moore lattice were analyzed via numerical simulations, as the AME formalism is not valid for this network structure. In this regular network, the aging mechanism slows down the cascade dynamics dramatically, leading to very slow increasing adoption curves.

Chapter 6 was divided into two parts, addressing the phase diagram and the dynamics of the Symmetrical Threshold model, a consensus threshold model where both states are symmetric. The first part (Chapter 6A) discussed the ordering dynamics without aging, presenting a detailed analysis of the phase diagram and identifying three distinct phases: mixed, ordered, and frozen. The dynamical regimes of the model were characterized by a set of observables that allowed to distinguish the different phases, which were found to be in good agreement with the theoretical predictions of the AME. The second part (Chapter 6B) explored the implications of endogenous aging in this model. For sparse networks, aging introduced a new dynamical phase, at the mixed region of the phase diagram, characterized by an initial disordering followed by slow ordering. In this phase, it was found that consensus can be reached in the state of an initial minority due to aging. Moreover, aging significantly altered the interface density decay and persistence, showing that aging leads to a slower convergence towards the ordered state. The AME derived in Chapter 4 was used to analyze the model, providing a comprehensive understanding of the aging effects on consensus dynamics driven by group interactions. For both, the aging and non-aging models, the results were shown for 4 different network topologies (Complete graph, Erd\H{o}s-Rényi, random-regular and Moore lattice), showing how aging implications change for different network structures.

\section{Remarks on Assessing the housing market dynamics \label{sec:housing_market_dynamics}}

In the second part of the thesis, keeping this idea of burstiness and memory effects, we analyzed the static and dynamical properties of a real complex system: the housing market. We found that there are irregular temporal patterns in the listings data, which affect the strategic interactions between sellers and real estate agencies. Moreover, we identified that there is a spatial segmentation in the housing market, driven by the agencies influence and specialization.

Motivated by the spatial specialization of the real estate agencies, Chapter 7 explores the spatial segmentation of the housing market from online listings in 3 provinces in Spain and France. By developing a novel methodology to evaluate spatial segmentation using listings data, we identified robust submarkets defined by the influence of real estate agencies. This segmentation was found to be consistent across different spatial resolutions and community detection algorithms. When applied to census level data, we found that the identified communities differ from the ones using cells or municipalitiy level data. To avoid this particular scale-depencency of the detected submarkets, we proposed a stochastically aggregative method to obtain a coherent segmentation from the data at this level. This methodology was proven to be useful and allowed to identify communities from the French dataset, where the identified communities showed similar features with ones obtained from the Spanish dataset (submarkets larger than municipalites, with connected spatial structure). This suggests that the submarket partition is a result of an underlying robust mechanism and that market-based spatial divisions are crucial for effective policy-making and economical price modeling.

In Chapter 8, we extended the previous analysis to study the dynamics of real estate listings in 3 Spanish provinces, focusing on the listings temporal dynamics, the decision-making processes of house sellers and the role of real estate agencies. The analysis revealed that the temporal patterns in the listings show a persistence effect similar to aging, where the longer a listing remains available on the market, the less likely is to be sold. This effect was found to be consistent across different provinces and property types analyzed, suggesting that it is a general feature of the housing market. Additionally, via a temporal bipartite network analysis, we were able to identify via a quantitative approach the key factors influencing the decision-making process of house sellers when choosing a real estate agency. The results showed that there is a preferential attachment mechanism in the selection of agencies (where the most active agencies tend to attract more listings), a price similarity effect (where listings are posted by agencies with similar prices), and spatial specialization (where agencies tend to focus on specific areas). These findings provide a comprehensive understanding of the strategic interactions between sellers and agencies in the housing market from a quantitative perspective, with agreement with qualitative observations in the literature.

In conclusion, the findings of this thesis underscore the profound impact of aging and temporal dynamics on socio-economic systems. By incorporating aging into various models, we have demonstrated how memory effects alter system behaviors, leading to new insights into segregation dynamics, contagion processes, and consensus problems. Moreover, the analysis of real-world data from the housing market has revealed the importance of temporal (memory) and spatial (specialization) dynamics in shaping market structures and decision-making processes. These results not only advance theoretical understanding but also provide practical implications for policy and strategy in social and economic contexts.

\renewcommand{\thechapter}{10} 
\chapterimage{}
\chapterspaceabove{6.75cm}
\chapterspacebelow{7.25cm}

\chapter{General conclusions and outlook}
This thesis has delved into the intricate dynamics of aging and memory effects in social and economic systems, particularly focusing on their implications in various threshold models and housing market dynamics. Here, we provide a comprehensive discussion of our findings and propose future research directions, addressing potential advancements in the study of aging and its broader applications.

In the context of aging, our research has shown significant impacts on the dynamics of different threshold models. In agreement with the aging implications in the Voter model~\cite{artime-2018,artime2019herding,peralta-2020C,peralta-2018,peralta-2020A}, we found that aging promotes ordering in phases dynamically disordered in threshold models. In both, the Sakoda-Schelling and the Symmetrical Threshold model, aging is able to induce coarsening in a phase that would otherwise remain disordered. This effect occurs because aging introduces a persistence of small clusters, which can grow and eventually dominate the system, leading to segregation or consensus, depending on the model. On the other hand, the coarsening process is found to be slower in the presence of aging. This result is observed in the 3 models analyzed: a lower exponent at the interface decay of the Sakoda-Schelling model, an exponential cascade replaced by a power-law in the Granovetter-Watts model and an exponential interface decay replaced by a power-law decay in the Symmetrical Threshold model. Following these results, aging can be seen as a complex mechanism that prevents small clusters to dissolve, in a halfway between order and disorder, allowing for them to grow in a disordered regime and dominate the system (promote order), but at the same time preventing ordering dynamics to reach the segregated/fully adopted/consensus state (slowing down the coarsening process). 

Moreover, aging mechanism was found to change dramatically the dynamics in threshold models and its effects are found to be robust across different update rules. Nevertheless, typical agent-based Monte Carlo simulations are still using a Random Asynchronous Update, which misses the non-Markovian nature of human interactions. Through this thesis, we have shown how aging is an essential ingredient for the modeling of social and economic systems, and thus, it is crucial to incorporate it into future models. 

The next steps for studying aging involve extending this mechanism to different models of collective behavior, as the ones studied through this thesis. One promising direction is the inclusion of aging effects in the idiosyncratic behavior. The idiosyncratic behavior is a key feature of human decision-making, where individuals exhibit a certain degree of randomness or irrationality in their choices. By incorporating aging into these models, the independent behavior also decays with persistence time, motivated as an irrationality decay with the memory of their past decisions. Additionally, in the context of complex and group interactions, future modifications could explore the joint effect of aging and higher-order interactions in complex systems, considering the interplay between individual memory and collective behavior. This approach can help uncover the deeper, often nonlinear, relationships that govern aging processes in social and economic systems. 

Another exciting direction for future research is the application of aging models to new domains. For instance, aging effects in financial markets could be investigated to understand how traders' prolonged engagement with specific assets influences market volatility and price trends. In the context of game theory, aging could be used to model how players' memory of past strategies affects their decision-making and the evolution of strategies over time~\cite{samuele-ciardella-2023}. In fact, it is found that aging can promote the evolution of cooperation in the Prisoner's Dilemma game~\cite{attila-2009}, and this mechanism could be extended to other social dilemmas or multi-agent systems to explore the role of memory in shaping collective behavior. The applications in this last domain could adress questions like how memory to past strategies could help to avoid the tragedy of the commons~\cite{ostrom1990governing}, or if it could help to avoid the escalation of conflicts~\cite{axelrod1981evolution}.

Moreover, the validation of our results with empirical data could provide further insights into the real-world implications of aging in social and economic systems. By analyzing historical data on segregation patterns, consumer behavior, or adoption curves, we can test the predictions of our models and refine our understanding of aging effects in complex systems. However, the challenge lies in obtaining high-quality data that captures the temporal evolution of these systems accurately and the personal engagement or persistence of the individuals involved, as a proxy for aging. Collaborations with industry partners or government agencies could help in accessing such data and conducting empirical studies to validate our theoretical findings. Another interesting direction to validate the aging mechanism is to conduct experiments with human subjects, where the persistence of the individuals is measured and correlated with their decisions. This could provide valuable insights into how aging influences different tasks or decision-making processes.

Regarding the housing market dynamics, our research suggests several future directions. Firstly, the current findings are based on our dataset, and it is crucial to validate these results with larger and more diverse datasets. This would help to generalize our conclusions and ensure their robustness across different contexts. An interesting future work could involve developing an activity-driven model for the housing market, borrowing ideas from models with aging and for the 3 key factors that drive the decision-making of sellers (popularity, price and location). In this sense, Part 1 of the thesis, which focuses on aging and memory effects in abstract models, is closely related to Part 2, which applies these concepts to the real-world scenario of housing market dynamics. The listings in an online platform exhibit aging, as observed in the persistence of listings in the online platform. In this context, aging is not understood as a memory effect, but as a time decaying attractiveness of the listings. However, the same tools and mathematical concepts used in Part 1 can be applied to build an activity driven model to understand the dynamics of the housing market, which might be possible to describe via differential equations via the AME formalism described in Chapter 4.

Regarding segmentation, further investigation is needed to understand the dependence on data scale. For example, analyzing housing data for an entire country at a granular level, such as 1 $\times$ 1 km$^2$ cells, could reveal whether the observed communities persist or if larger communities emerge within which our identified communities reside. This scale-dependent analysis could help in identifying broader market trends and regional dependencies that are not apparent at smaller scales. Moreover, investigating temporal changes in market segmentation could provide insights into how economic cycles, policy changes, or significant events influence market structure over time.

Future work along these lines could involve more detailed analysis of other metadata and their impact on market segmentation. For example, investigating the role of demographic factors, economic indicators, or urban development plans on market dynamics could provide a richer understanding of the forces shaping the housing market. Integrating machine learning techniques with our segmentation methodology could also enhance the ability to predict and respond to market changes in real-time.

Finally, the analysis of real systems, like the housing market, has significant implications for policymaking and price analysis. By focusing on the dynamics of listings rather than the aggregated data, our thesis offers a more realistic perspective on how real estate agencies operate in the real world, where equilibrium is rarely achieved. Understanding these dynamics is crucial for developing effective policies and strategies that can adapt to the ever-changing nature of social and economic systems.

In conclusion, this thesis contributed to the understanding of aging and memory effects in social and economic dynamics. Future research should continue to refine these models, validate findings with broader datasets, and explore new applications in real-world systems. By addressing the temporal nature of these systems, our research opens up new avenues for innovation in modeling and understanding the complex interplay of factors that drive socio-economic behavior.

\fancyhead[LO]{\sffamily\normalsize\bfseries Chapter 10. General conclusions and outlook}

\fancyhead[LO]{\sffamily\normalsize\bfseries Chapter 10. General conclusions and outlook}

\stopcontents[part] 
\newpage
\thispagestyle{plain} 
\mbox{}

\chapterimage{} 
\chapterspaceabove{2.5cm} 
\chapterspacebelow{2cm} 

\fancyhead[LO]{\sffamily\normalsize\bfseries Bibliography}
\fancyhead[RE]{\sffamily\normalsize\bfseries Bibliography}
\printbibliography\addcontentsline{toc}{chapter}{\textcolor{ocre}{Bibliography}} 

@article{steinert-2022,
	title        = {{Emotions and Digital Well-Being: on Social Media's Emotional Affordances}},
	author       = {Steinert, Steffen and Dennis, Matthew},
	year         = {2022},
	month        = {4},
	journal      = {Philosophy and Technology},
	volume       = {35},
	number       = {2},
	doi          = {10.1007/s13347-022-00530-6},
	url          = {https://doi.org/10.1007/s13347-022-00530-6}
}

@article{yamir-2004,
  title={Dynamics of rumor spreading in complex networks},
  author={Moreno, Yamir and Nekovee, Maziar and Pacheco, Amalio F},
  journal={Physical review E},
  volume={69},
  number={6},
  pages={066130},
  year={2004},
  publisher={APS}
}

@article{borge2013cascading,
  title={Cascading behaviour in complex socio-technical networks},
  author={Borge-Holthoefer, Javier and Banos, Raquel A and Gonz{\'a}lez-Bail{\'o}n, Sandra and Moreno, Yamir},
  journal={Journal of Complex Networks},
  volume={1},
  number={1},
  pages={3--24},
  year={2013},
  publisher={OUP}
}

@article{gleeson2017temporal,
  title={Temporal profiles of avalanches on networks},
  author={Gleeson, James P and Durrett, Rick},
  journal={Nature communications},
  volume={8},
  number={1},
  pages={1227},
  year={2017},
  publisher={Nature Publishing Group UK London}
}

@incollection{cronin2006monte,
  title={Monte Carlo Simulation},
  author={Cronin, Kevin and Gleeson, James P},
  booktitle={Handbook of Food and Bioprocess Modeling Techniques},
  pages={501--532},
  year={2006},
  publisher={CRC Press}
}

@article{jkedrzejewski2018impact,
  title={Impact of memory on opinion dynamics},
  author={J{\k{e}}drzejewski, Arkadiusz and Sznajd-Weron, Katarzyna},
  journal={Physica A: Statistical Mechanics and its Applications},
  volume={505},
  pages={306--315},
  year={2018},
  publisher={Elsevier}
}

@article{przybyla2014diffusion,
  title={Diffusion of innovation within an agent-based model: Spinsons, independence and advertising},
  author={Przyby{\l}a, Piotr and Sznajd-Weron, Katarzyna and Weron, Rafa{\l}},
  journal={Advances in Complex Systems},
  volume={17},
  number={01},
  pages={1450004},
  year={2014},
  publisher={World Scientific}
}

@article{gonzalez2014assessing,
  title={Assessing the bias in samples of large online networks},
  author={Gonz{\'a}lez-Bail{\'o}n, Sandra and Wang, Ning and Rivero, Alejandro and Borge-Holthoefer, Javier and Moreno, Yamir},
  journal={Social Networks},
  volume={38},
  pages={16--27},
  year={2014},
  publisher={Elsevier}
}

@misc{online-platforms,
	title        = {{Online platforms: Economic and societal effects | Panel for the Future of Science and Technology (STOA) | European Parliament}},
	url          = {https://www.europarl.europa.eu/stoa/en/document/EPRS\_STU\%282021\%29656336}
}

@article{limburg2002complex,
	title        = {{Complex systems and valuation}},
	author       = {Karin E Limburg and Robert V O'Neill and Robert Costanza and Stephen Farber},
	year         = {2002},
	journal      = {Ecological Economics},
	volume       = {41},
	number       = {3},
	pages        = {409--420},
	doi          = {https://doi.org/10.1016/S0921-8009(02)00090-3},
	issn         = {0921-8009},
	url          = {https://www.sciencedirect.com/science/article/pii/S0921800902000903}
}

@article{aderem-2005,
	title        = {{Systems Biology: Its practice and Challenges}},
	author       = {Aderem, Alan},
	year         = {2005},
	month        = {5},
	journal      = {Cell},
	volume       = {121},
	number       = {4},
	pages        = {511--513},
	doi          = {10.1016/j.cell.2005.04.020},
	url          = {https://doi.org/10.1016/j.cell.2005.04.020}
}

@book{brown2000scaling,
	title        = {{Scaling in biology}},
	author       = {Brown, James H. and West, Geoffrey B.},
	year         = {2000},
	publisher    = {Oxford University Press, USA}
}

@article{boccaletti2006complex,
	title        = {{Complex networks: Structure and dynamics}},
	author       = {Boccaletti, Stefano and Latora, Vito and Moreno, Yamir and Chavez, Martin and Hwang, D-U},
	year         = {2006},
	journal      = {Physics reports},
	publisher    = {Elsevier},
	volume       = {424},
	number       = {4-5},
	pages        = {175--308}
}

@article{gao2011robustness,
	title        = {{Robustness of a network of networks}},
	author       = {Gao, Jianxi and Buldyrev, Sergey V and Havlin, Shlomo and Stanley, H. Eugene},
	year         = {2011},
	journal      = {Physical review letters},
	publisher    = {APS},
	volume       = {107},
	number       = {19},
	pages        = {195701}
}

@inproceedings{leskovec2010signed,
	title        = {{Signed networks in social media}},
	author       = {Leskovec, Jure and Huttenlocher, Daniel and Kleinberg, Jon},
	year         = {2010},
	booktitle    = {Proceedings of the SIGCHI conference on human factors in computing systems},
	pages        = {1361--1370}
}

@book{d2014networks,
	title        = {{Networks of networks: the last frontier of complexity}},
	author       = {D'Agostino, Gregorio and Scala, Antonio},
	year         = {2014},
	publisher    = {Springer},
	volume       = {340}
}

@article{gomez-2013,
	title        = {{Diffusion Dynamics on Multiplex Networks}},
	author       = {G\'omez, Sergio and D\'iaz-Guilera, Albert and G\'omez-Garde\~nes, Jesus and P\'erez-Vicente, C. J. and Moreno, Yamir and Arenas, Alex},
	year         = {2013},
	month        = {Jan},
	journal      = {Phys. Rev. Lett.},
	publisher    = {American Physical Society},
	volume       = {110},
	pages        = {028701},
	doi          = {10.1103/PhysRevLett.110.028701},
	url          = {https://link.aps.org/doi/10.1103/PhysRevLett.110.028701},
	issue        = {2},
	numpages     = {5}
}

@article{de2013mathematical,
	title        = {{Mathematical formulation of multilayer networks}},
	author       = {De Domenico, Manlio and Sol{\'e}-Ribalta, Albert and Cozzo, Emanuele and Kivel{\"a}, Mikko and Moreno, Yamir and Porter, Mason A and G{\'o}mez, Sergio and Arenas, Alex},
	year         = {2013},
	journal      = {Physical Review X},
	publisher    = {APS},
	volume       = {3},
	number       = {4},
	pages        = {041022}
}

@article{kivela2014multilayer,
	title        = {{Multilayer networks}},
	author       = {Kivel{\"a}, Mikko and Arenas, Alex and Barthelemy, Marc and Gleeson, James P and Moreno, Yamir and Porter, Mason A},
	year         = {2014},
	journal      = {Journal of complex networks},
	publisher    = {Oxford University Press},
	volume       = {2},
	number       = {3},
	pages        = {203--271}
}

@book{alon2019introduction,
	title        = {{An introduction to systems biology: design principles of biological circuits}},
	author       = {Alon, Uri},
	year         = {2019},
	publisher    = {Chapman and Hall/CRC}
}

@article{perc2013evolutionary,
	title        = {{Evolutionary dynamics of group interactions on structured populations: a review}},
	author       = {Perc, Matja{\v{z}} and G{\'o}mez-Gardenes, Jes{\'u}s and Szolnoki, Attila and Flor\'ia, Luis M and Moreno, Yamir},
	year         = {2013},
	journal      = {Journal of the royal society interface},
	publisher    = {The Royal Society},
	volume       = {10},
	number       = {80},
	pages        = {20120997}
}

@article{ferrara-2015,
	title        = {{Measuring emotional contagion in social media}},
	author       = {Ferrara, Emilio and Yang, Zeyao},
	year         = {2015},
	month        = {11},
	journal      = {PLOS ONE},
	volume       = {10},
	number       = {11},
	pages        = {e0142390},
	doi          = {10.1371/journal.pone.0142390},
	url          = {https://doi.org/10.1371/journal.pone.0142390}
}

@article{abella2023unraveling,
	title        = {{Unraveling higher-order dynamics in collaboration networks}},
	author       = {Abella, David and Birello, Piero and Di Gaetano, Leonardo and Ghivarello, Sara and Sabhahit, Narayan G and Sirocchi, Christel and Fern\'andez-Gracia, Juan},
	year         = {2023},
	journal      = {arXiv preprint arXiv:2306.17521}
}

@book{toral2014stochastic,
	title        = {{Stochastic numerical methods: an introduction for students and scientists}},
	author       = {Toral, Ra{{\'u}}l and Colet, Pere},
	year         = {2014},
	publisher    = {John Wiley \& Sons}
}

@article{jstor,
	title        = {{Creating social contagion through viral product design: a randomized trial of peer influence in networks on JSTOR}},
	journal      = {www.jstor.org},
	url          = {https://www.jstor.org/stable/41261920}
}

@article{jensen-2015,
	title        = {{Mechanisms and processes of peer contagion}},
	author       = {Jensen, Michaeline and Dishion, Thomas J.},
	year         = {2015},
	month        = {1},
	journal      = {Psychology},
	doi          = {10.1093/obo/9780199828340-0165},
	url          = {https://doi.org/10.1093/obo/9780199828340-0165}
}

@article{valente-1996,
	title        = {{Social network thresholds in the diffusion of innovations}},
	author       = {Thomas W. Valente},
	year         = {1996},
	journal      = {Social Networks},
	volume       = {18},
	number       = {1},
	pages        = {69--89},
	doi          = {https://doi.org/10.1016/0378-8733(95)00256-1},
	issn         = {0378-8733},
	url          = {https://www.sciencedirect.com/science/article/pii/0378873395002561}
}

@article{christakis2013social,
	title        = {{Social contagion theory: examining dynamic social networks and human behavior}},
	author       = {Christakis, Nicholas A and Fowler, James H},
	year         = {2013},
	journal      = {Statistics in medicine},
	publisher    = {Wiley Online Library},
	volume       = {32},
	number       = {4},
	pages        = {556--577}
}

@article{cugliandolo1993analytical,
	title        = {{Analytical solution of the off-equilibrium dynamics of a long-range spin-glass model}},
	author       = {Cugliandolo, Leticia F and Kurchan, Jorge},
	year         = {1993},
	journal      = {Physical Review Letters},
	publisher    = {APS},
	volume       = {71},
	number       = {1},
	pages        = {173}
}

@article{ravasz2004spreading,
	title        = {{Spreading of families in cyclic predator-prey models}},
	author       = {Ravasz, M{{\'a}}ria and Szab{{\'o}}, Gy{o}rgy and Szolnoki, Attila},
	year         = {2004},
	journal      = {Physical Review E},
	publisher    = {APS},
	volume       = {70},
	number       = {1},
	pages        = {012901}
}

@inproceedings{lejarraga2011let,
	title        = {{Let me handle this, I've done it before: Experience and self-involvement in superstitious learning}},
	author       = {Lejarraga, Jose and Lejarraga, Tomas and Gaston-Breton, Charlotte},
	year         = {2011},
	booktitle    = {Academy of Management Proceedings},
	volume       = {2011},
	number       = {1},
	pages        = {1--6},
	organization = {Academy of Management Briarcliff Manor, NY 10510}
}

@book{berge1984hypergraphs,
	title        = {{Hypergraphs: combinatorics of finite sets}},
	author       = {Berge, Claude},
	year         = {1984},
	publisher    = {Elsevier},
	volume       = {45}
}

@phdthesis{lucquiaud2022modeliser,
	title        = {{Modeliser l'emergence et detecter les structures de la segregation sociospatiale}},
	author       = {Lucquiaud, Antoine},
	year         = {2022},
	school       = {Universit{{'e}} Paris 1-Panth{{'e}}on Sorbonne}
}

@article{bollinger-2012,
	title        = {{Peer effects in the diffusion of solar photovoltaic panels}},
	author       = {Bollinger, Bryan and Gillingham, Kenneth},
	year         = {2012},
	month        = {11},
	journal      = {Marketing science (Providence, R.I.)},
	volume       = {31},
	number       = {6},
	pages        = {900--912},
	doi          = {10.1287/mksc.1120.0727},
	url          = {https://doi.org/10.1287/mksc.1120.0727}
}

@article{pastor2001epidemic,
	title        = {{Epidemic Spreading in Scale-Free Networks}},
	author       = {Pastor-Satorras, Romualdo and Vespignani, Alessandro},
	year         = {2001},
	month        = {Apr},
	journal      = {Phys. Rev. Lett.},
	publisher    = {American Physical Society},
	volume       = {86},
	pages        = {3200--3203},
	doi          = {10.1103/PhysRevLett.86.3200},
	url          = {https://link.aps.org/doi/10.1103/PhysRevLett.86.3200},
	issue        = {14},
	numpages     = {0}
}

@article{newman2002spread,
	title        = {{Spread of epidemic disease on networks}},
	author       = {Newman, Mark E. J.},
	year         = {2002},
	month        = {Jul},
	journal      = {Phys. Rev. E},
	publisher    = {American Physical Society},
	volume       = {66},
	pages        = {016128},
	doi          = {10.1103/PhysRevE.66.016128},
	url          = {https://link.aps.org/doi/10.1103/PhysRevE.66.016128},
	issue        = {1},
	numpages     = {11}
}

@article{christakis2007spread,
	title        = {{The Spread of Obesity in a Large Social Network over 32 Years}},
	author       = {Christakis, Nicholas A. and Fowler, James H.},
	year         = {2007},
	month        = {7},
	journal      = {The New England Journal of Medicine},
	volume       = {357},
	number       = {4},
	pages        = {370--379},
	doi          = {10.1056/nejmsa066082},
	url          = {https://doi.org/10.1056/nejmsa066082}
}

@article{fowler2009cooperative,
	title        = {{Cooperative behavior cascades in human social networks}},
	author       = {Fowler, James H. and Christakis, Nicholas A.},
	year         = {2010},
	month        = {3},
	journal      = {Proceedings of the National Academy of Sciences of the United States of America},
	volume       = {107},
	number       = {12},
	pages        = {5334--5338},
	doi          = {10.1073/pnas.0913149107},
	url          = {https://doi.org/10.1073/pnas.0913149107}
}

@misc{nobel-2021,
	title        = {{The Nobel Prize in Physics 2021}},
	url          = {https://www.nobelprize.org/prizes/physics/2021/press-release/}
}

@article{battiston-2021,
	title        = {{The physics of higher-order interactions in complex systems}},
	author       = {Battiston, Federico and Amico, Enrico and Barrat, Alain and Bianconi, Ginestra and De Arruda, Guilherme Ferraz and Franceschiello, Benedetta and Iacopini, Iacopo and K{\'e}fi, Sonia and Latora, Vito and Moreno, Yamir and Murray, Micah M. and Peixoto, Tiago P. and Vaccarino, Francesco and Petri, Giovanni},
	year         = {2021},
	month        = {10},
	journal      = {Nature Physics},
	volume       = {17},
	number       = {10},
	pages        = {1093--1098},
	doi          = {10.1038/s41567-021-01371-4},
	url          = {https://doi.org/10.1038/s41567-021-01371-4}
}

@article{cencetti-2023,
	title        = {{Distinguishing Simple and Complex Contagion Processes on Networks}},
	author       = {Cencetti, Giulia and Contreras, Diego Andr{'e}s and Mancastroppa, Marco and Barrat, Alain},
	year         = {2023},
	month        = {Jun},
	journal      = {Phys. Rev. Lett.},
	publisher    = {American Physical Society},
	volume       = {130},
	pages        = {247401},
	doi          = {10.1103/PhysRevLett.130.247401},
	url          = {https://link.aps.org/doi/10.1103/PhysRevLett.130.247401},
	issue        = {24},
	numpages     = {7}
}

@phdthesis{sakoda1949minidoka,
	title        = {{Minidoka: An analysis of changing patterns of social behavior}},
	author       = {Sakoda, James M.},
	year         = {1949},
	school       = {PhD thesis, University of California}
}

@article{sakoda1971checkerboard,
	title        = {{The checkerboard model of social interaction}},
	author       = {Sakoda, James M.},
	year         = {1971},
	journal      = {The Journal of Mathematical Sociology},
	publisher    = {Taylor \& Francis},
	volume       = {1},
	number       = {1},
	pages        = {119--132}
}

@article{Barabasi2005Bursts,
	title        = {{The origin of bursts and heavy tails in human dynamics}},
	author       = {Barab{\'a}si, Albert-L{\'a}szl{\'o}},
	year         = {2005},
	month        = {5},
	journal      = {Nature},
	volume       = {435},
	number       = {7039},
	pages        = {207--211},
	doi          = {10.1038/nature03459},
	url          = {https://nature.com/articles/nature03459}
}

@article{Vazquez2006Bursts,
	title        = {{Modeling bursts and heavy tails in human dynamics}},
	author       = {V{{\'a}}zquez, Alexei and Oliveira, J. G. and Dezső, Zolt{{\'a}}n and Goh, K.i. and Kondor, Imre and Barab{{\'a}}si, Albert L{{\'a}}szl{{\'o}}},
	year         = {2006},
	month        = {3},
	journal      = {Physical review. E, Statistical, nonlinear and soft matter physics},
	volume       = {73},
	number       = {3},
	doi          = {10.1103/physreve.73.036127},
	url          = {https://doi.org/10.1103/physreve.73.036127}
}

@article{karsai2012universal,
	title        = {{Universal features of correlated bursty behaviour}},
	author       = {Karsai, M{{\'a}}rton and Kaski, Kimmo and Barab{{\'a}}si, Albert-L{{\'a}}szl{{\'o}} and Kert{{\'e}}sz, J{{\'a}}nos},
	year         = {2012},
	month        = {5},
	journal      = {Scientific reports},
	volume       = {2},
	number       = {1},
	doi          = {10.1038/srep00397},
	url          = {https://www.nature.com/articles/srep00397}
}

@book{newman-book,
	title        = {{Networks: An Introduction}},
	author       = {Newman, Mark E. J.},
	year         = {2010},
	month        = {03},
	publisher    = {Oxford University Press},
	doi          = {10.1093/acprof:oso/9780199206650.001.0001},
	isbn         = {9780199206650},
	url          = {https://doi.org/10.1093/acprof:oso/9780199206650.001.0001}
}

@book{barrat-2008,
	title        = {{Dynamical processes on complex networks}},
	author       = {Barrat, Alain and Barthelemy, Marc and Vespignani, Alessandro},
	year         = {2008},
	publisher    = {Cambridge university press}
}

@article{Goh2008Burstiness,
	title        = {{Burstiness and memory in complex systems}},
	author       = {Goh, K.-i. and Barab{{'a}}si, Albert‐L{{'a}}szl{{'o}}},
	year         = {2008},
	month        = {1},
	journal      = {Europhysics letters},
	volume       = {81},
	number       = {4},
	pages        = {48002},
	doi          = {10.1209/0295-5075/81/48002},
	url          = {https://doi.org/10.1209/0295-5075/81/48002}
}

@article{Miritello2013Capacity,
	title        = {{Limited communication capacity unveils strategies for human interaction}},
	author       = {Miritello, Giovanna and Lara, Rub{{'e}}n and Cebri{{'a}}n, Manuel and Moro, Esteban},
	year         = {2013},
	month        = {6},
	journal      = {Scientific reports},
	volume       = {3},
	number       = {1},
	doi          = {10.1038/srep01950},
	url          = {https://www.nature.com/articles/srep01950}
}

@article{Eckmann2004Entropy,
	title        = {{Entropy of dialogues creates coherent structures in e-mail traffic}},
	author       = {Eckmann, Jean‐Pierre and Moses, Elisha and Sergi, Danilo},
	year         = {2004},
	month        = {9},
	journal      = {Proceedings of the National Academy of Sciences of the United States of America},
	volume       = {101},
	number       = {40},
	pages        = {14333--14337},
	doi          = {10.1073/pnas.0405728101},
	url          = {https://doi.org/10.1073/pnas.0405728101}
}

@article{Perra2012ActivityDriven,
	title        = {{Activity driven modeling of time varying networks}},
	author       = {Perra, Nicola and Gonçalves, Bruno and Pastor-Satorras, Romualdo and Vespignani, Alessandro},
	year         = {2012},
	month        = {6},
	journal      = {Scientific reports},
	volume       = {2},
	number       = {1},
	doi          = {10.1038/srep00469},
	url          = {https://www.nature.com/articles/srep00469}
}

@article{Jo2012Circadian,
	title        = {{Circadian pattern and burstiness in mobile phone communication}},
	author       = {Jo, Hang-Hyun and Karsai, M{{\'a}}rton and Kert{{\'e}}sz, J{{\'a}}nos and Kaski, Kimmo},
	year         = {2012},
	month        = {1},
	journal      = {New journal of physics},
	volume       = {14},
	number       = {1},
	pages        = {013055},
	doi          = {10.1088/1367-2630/14/1/013055},
	url          = {https://doi.org/10.1088/1367-2630/14/1/013055}
}

@article{Rocha2013Bursts,
	title        = {{Bursts of vertex activation and epidemics in evolving networks}},
	author       = {Rocha, Luis E. C. and Blondel, Vincent D.},
	year         = {2013},
	month        = {3},
	journal      = {PLOS computational biology/PLoS computational biology},
	volume       = {9},
	number       = {3},
	pages        = {e1002974},
	doi          = {10.1371/journal.pcbi.1002974},
	url          = {https://doi.org/10.1371/journal.pcbi.1002974}
}

@article{Wang2009Viruses,
	title        = {{Understanding the spreading patterns of mobile phone viruses}},
	author       = {Wang, Pu and Gonz{\'a}lez, Marta C. and Hidalgo, C{\'e}sar A. and Barab{\'a}si, Albert-L{\'a}szl{\'o}},
	year         = {2009},
	month        = {5},
	journal      = {Science},
	volume       = {324},
	number       = {5930},
	pages        = {1071--1076},
	doi          = {10.1126/science.1167053},
	url          = {https://doi.org/10.1126/science.1167053}
}

@article{Lazer2009CompSocSci,
	title        = {{Computational Social Science}},
	author       = {Lazer, David and Pentland, Alex and Adamic, Lada A. and Aral, Sinan and Barab{\'a}si, Albert-L{\'a}szl{\'o} and Brewer, Devon D. and Christakis, Nicholas A. and Contractor, Noshir and Fowler, James H. and Gutmann, Myron P. and Jebara, Tony and King, Gary and Macy, Michael W. and Roy, Deb and Van Alstyne, Marshall},
	year         = {2009},
	month        = {2},
	journal      = {Science},
	volume       = {323},
	number       = {5915},
	pages        = {721--723},
	doi          = {10.1126/science.1167742},
	url          = {https://pubmed.ncbi.nlm.nih.gov/19197046/}
}

@article{Eagle2006RealityMining,
	title        = {{Reality mining: sensing complex social systems}},
	author       = {Eagle, Nathan and Pentland, Alex},
	year         = {2005},
	month        = {11},
	journal      = {Personal and ubiquitous computing},
	volume       = {10},
	number       = {4},
	pages        = {255--268},
	doi          = {10.1007/s00779-005-0046-3},
	url          = {https://doi.org/10.1007/s00779-005-0046-3}
}

@article{Muchnik2013PowerLaw,
	title        = {{Origins of power-law degree distribution in the heterogeneity of human activity in social networks}},
	author       = {Muchnik, Lev and Pei, Sen and Parra, Lucas C. and Reis, Saulo D. S. and Andrade, Jos{\'e} S. and Havlin, Shlomo and Makse, Hern{\'a}n A.},
	year         = {2013},
	month        = {5},
	journal      = {Scientific reports},
	volume       = {3},
	number       = {1},
	doi          = {10.1038/srep01783},
	url          = {https://www.nature.com/articles/srep01783}
}

@misc{Clauset2007Proximity,
	title        = {{Persistence and periodicity in a dynamic proximity network}},
	author       = {Clauset, A. and Eagle, N.},
	year         = {2012},
	url          = {https://arxiv.org/abs/1211.7343}
}

@article{Radicchi2009WebActivity,
	title        = {{Human activity in the web}},
	author       = {Radicchi, Filippo},
	year         = {2009},
	month        = {8},
	journal      = {Physical review. E, Statistical, nonlinear and soft matter physics},
	volume       = {80},
	number       = {2},
	doi          = {10.1103/physreve.80.026118},
	url          = {https://doi.org/10.1103/physreve.80.026118}
}

@article{Holme2012Temporal,
	title        = {{Temporal networks}},
	author       = {Holme, Petter and Saramaki, Jari},
	year         = {2012},
	month        = {10},
	journal      = {Physics reports},
	volume       = {519},
	number       = {3},
	pages        = {97--125},
	doi          = {10.1016/j.physrep.2012.03.001},
	url          = {https://www.sciencedirect.com/science/article/abs/pii/S0370157312000841}
}

@book{stanley1971phase,
	title        = {{Phase transitions and critical phenomena}},
	author       = {Stanley, H. Eugene},
	year         = {1971},
	publisher    = {Clarendon Press, Oxford},
	volume       = {7}
}

@book{frisch1995turbulence,
	title        = {{Turbulence: the legacy of AN Kolmogorov}},
	author       = {Frisch, Uriel},
	year         = {1995},
	publisher    = {Cambridge university press}
}

@article{pikovsky2001universal,
	title        = {{A universal concept in nonlinear sciences}},
	author       = {Pikovsky, Arkady and Rosenblum, Michael and Kurths, Jurgen and Synchronization, A},
	year         = {2001},
	journal      = {Self},
	volume       = {2},
	pages        = {3}
}

@article{onsager-1944,
	title        = {{Crystal Statistics. I. A Two-Dimensional Model with an Order-Disorder Transition}},
	author       = {Onsager, Lars},
	year         = {1944},
	month        = {2},
	journal      = {Physical review},
	volume       = {65},
	number       = {3-4},
	pages        = {117--149},
	doi          = {10.1103/physrev.65.117},
	url          = {https://doi.org/10.1103/physrev.65.117}
}

@book{stauffer-1985,
	title        = {{Introduction to percolation Theory}},
	author       = {Stauffer, Dietrich and Aharony, Amnon},
	year         = {1985},
	month        = {1},
	doi          = {10.4324/9780203211595},
	url          = {https://doi.org/10.4324/9780203211595}
}

@book{may-2001,
	title        = {{Stability and complexity in model ecosystems}},
	author       = {May, Robert M.},
	year         = {2001},
	month        = {12},
	doi          = {10.1515/9780691206912},
	url          = {https://doi.org/10.1515/9780691206912}
}

@article{roche-1997,
	title        = {{The Ecology of Migrant Birds: A Neotropical Perspective.}},
	author       = {Roche, John P. and Rappole, John H.},
	year         = {1997},
	month        = {1},
	journal      = {Ecology},
	volume       = {78},
	number       = {1},
	pages        = {328},
	doi          = {10.2307/2266005},
	url          = {https://doi.org/10.2307/2266005}
}

@article{vosoughi-2018,
	title        = {{The spread of true and false news online}},
	author       = {Vosoughi, Soroush and Roy, Deb and Aral, Sinan},
	year         = {2018},
	month        = {3},
	journal      = {Science},
	volume       = {359},
	number       = {6380},
	pages        = {1146--1151},
	doi          = {10.1126/science.aap9559},
	url          = {https://doi.org/10.1126/science.aap9559}
}

@book{anderson1991infectious,
	title        = {{Infectious diseases of humans: dynamics and control}},
	author       = {Anderson, Roy M. and May, Robert M},
	year         = {1991},
	publisher    = {Oxford university press}
}

@article{anderson1972more,
	title        = {{More Is Different: Broken symmetry and the nature of the hierarchical structure of science.}},
	author       = {Anderson, Philip W},
	year         = {1972},
	journal      = {Science},
	publisher    = {American Association for the Advancement of Science},
	volume       = {177},
	number       = {4047},
	pages        = {393--396}
}

@article{anderson-2000,
	title        = {{Patterns of Democracy: Government Forms and Performance in Thirty-Six Countries}},
	author       = {Anderson, Jeffrey J. and Lijphart, Arend},
	year         = {2000},
	month        = {1},
	journal      = {CrossRef Listing of Deleted DOIs},
	volume       = {30},
	number       = {2},
	pages        = {117},
	doi          = {10.2307/3331092},
	url          = {https://doi.org/10.2307/3331092}
}

@article{ellickson-1999,
	title        = {{The Evolution of Social Norms: A Perspective from the Legal Academy}},
	author       = {Ellickson, Robert C.},
	year         = {1999},
	month        = {1},
	journal      = {Social Science Research Network},
	doi          = {10.2139/ssrn.191392},
	url          = {https://doi.org/10.2139/ssrn.191392}
}

@article{barabasi-2013,
	title        = {{Network science}},
	author       = {Barab{\'a}si, Albert-L{\'a}szl{\'o}},
	year         = {2013},
	month        = {3},
	journal      = {Philosophical transactions - Royal Society. Mathematical, Physical and engineering sciences/Philosophical transactions - Royal Society. Mathematical, physical and engineering sciences},
	volume       = {371},
	number       = {1987},
	pages        = {20120375},
	doi          = {10.1098/rsta.2012.0375},
	url          = {https://doi.org/10.1098/rsta.2012.0375}
}

@book{bryman-2010,
	title        = {{Social research methods}},
	author       = {Bryman, Alan},
	year         = {2010},
	month        = {2},
	booktitle    = {Taylor \& Francis},
	pages        = {157--184},
	doi          = {10.4324/9780203381175\{\_}chapter\{\_}

@article{karsai2019computational,
	title        = {{Computational human dynamics}},
	author       = {Karsai, M{{\'a}}rton},
	year         = {2019},
	journal      = {arXiv preprint arXiv:1907.07475}
}

@book{manyika-2011,
	title        = {{Big data: The next frontier for innovation, competition, and productivity}},
	author       = {Manyika, James},
	year         = {2011},
	month        = {5},
	url          = {http://abesit.in/wp-content/uploads/2014/07/big-data-frontier.pdf}
}

@article{boyd-2012,
	title        = {{Critical questions for big data: Provocations for a cultural, technological, and scholarly phenomenon}},
	author       = {Boyd, Danah and Crawford, Kate},
	year         = {2012},
	month        = {6},
	journal      = {Information, communication \& society},
	volume       = {15},
	number       = {5},
	pages        = {662--679},
	doi          = {10.1080/1369118x.2012.678878},
	url          = {https://doi.org/10.1080/1369118x.2012.678878}
}

@article{gracia-2011,
	title        = {{Selective advantage of tolerant cultural traits in the Axelrod-Schelling model}},
	author       = {Garcia-L{\'a}zaro, Carlos and Flor{{\'i}}a, L. Mario and Moreno, Yamir},
	year         = {2011},
	month        = {May},
	journal      = {Phys. Rev. E},
	publisher    = {American Physical Society},
	volume       = {83},
	pages        = {056103},
	doi          = {10.1103/PhysRevE.83.056103},
	url          = {https://link.aps.org/doi/10.1103/PhysRevE.83.056103},
	issue        = {5},
	numpages     = {8}
}

@article{watts-2007,
	title        = {{A twenty-first century science}},
	author       = {Watts, Duncan J.},
	year         = {2007},
	month        = {1},
	journal      = {Nature},
	volume       = {445},
	number       = {7127},
	pages        = {489},
	doi          = {10.1038/445489a},
	url          = {https://doi.org/10.1038/445489a}
}

@article{rudin-2019,
	title        = {{Stop explaining black box machine learning models for high stakes decisions and use interpretable models instead}},
	author       = {Rudin, Cynthia},
	year         = {2019},
	month        = {5},
	journal      = {Nature machine intelligence},
	volume       = {1},
	number       = {5},
	pages        = {206--215},
	doi          = {10.1038/s42256-019-0048-x},
	url          = {https://doi.org/10.1038/s42256-019-0048-x}
}

@book{goodfellow-2016,
	title        = {{Deep learning}},
	author       = {Goodfellow, Ian and Bengio, Yoshua and Courville, Aaron},
	year         = {2016},
	month        = {11},
	url          = {https://dl.acm.org/citation.cfm?id=3086952}
}

@book{murphy-2012,
	title        = {{Machine Learning : A Probabilistic Perspective}},
	author       = {Murphy, Kevin},
	year         = {2012},
	month        = {8},
	url          = {http://cds.cern.ch/record/1981503}
}

@article{baron2022analytical,
	title        = {{Analytical and numerical treatment of continuous ageing in the voter model}},
	author       = {Baron, Joseph W and Peralta, Antonio F. and Galla, Tobias and Toral, Ra{{\'u}}l},
	year         = {2022},
	journal      = {Entropy},
	publisher    = {MDPI},
	volume       = {24},
	number       = {10},
	pages        = {1331}
}

@article{artime2019herding,
	title        = {{Herding and idiosyncratic choices: Nonlinearity and aging-induced transitions in the noisy voter model}},
	author       = {Artime, Oriol and Carro, Adri\'an and Peralta, Antonio F. and Ramasco, Jos{{\'e}} J. and San Miguel, Maxi and Toral, Ra{{\'u}}l},
	year         = {2019},
	journal      = {Comptes Rendus. Physique},
	volume       = {20},
	number       = {4},
	pages        = {262--274}
}

@article{diaz2023network,
	title        = {{Network theory meets history. Local balance in global international relations}},
	author       = {Diaz-Diaz, Fernando and Bartesaghi, Paolo and Estrada, Ernesto},
	year         = {2023},
	journal      = {arXiv preprint arXiv:2303.03774}
}

@book{le2023crowd,
	title        = {{The Crowd a Study of the Popular Mind}},
	author       = {Le Bon, Gustave},
	year         = {2023},
	publisher    = {Beyond books hub}
}

@article{sporns-2004,
	title        = {{The small world of the cerebral cortex}},
	author       = {Sporns, Olaf and Zwi, Jonathan D.},
	year         = {2004},
	month        = {1},
	journal      = {Neuroinformatics},
	volume       = {2},
	number       = {2},
	pages        = {145--162},
	doi          = {10.1385/ni:2:2:145},
	url          = {https://doi.org/10.1385/ni:2:2:145}
}

@article{elith-2009,
	title        = {{Species Distribution Models: Ecological explanation and prediction across space and time}},
	author       = {Elith, Jane and Leathwick, John R.},
	year         = {2009},
	month        = {12},
	journal      = {Annual review of ecology, evolution, and systematics},
	volume       = {40},
	number       = {1},
	pages        = {677--697},
	doi          = {10.1146/annurev.ecolsys.110308.120159},
	url          = {https://doi.org/10.1146/annurev.ecolsys.110308.120159}
}

@article{bastolla-2009,
	title        = {{The architecture of mutualistic networks minimizes competition and increases biodiversity}},
	author       = {Bastolla, Ugo and Fortuna, Miguel A. and Pascual-Garc{{\'i}}a, Alberto and Ferrera, Antonio and Luque, Bartolo and Bascompte, Jordi},
	year         = {2009},
	month        = {4},
	journal      = {Nature},
	volume       = {458},
	number       = {7241},
	pages        = {1018--1020},
	doi          = {10.1038/nature07950},
	url          = {https://doi.org/10.1038/nature07950}
}

@article{ings-2008,
	title        = {{Review: Ecological networks - beyond food webs}},
	author       = {Ings, Thomas C. and Montoya, Jos{{\'e}} M. and Bascompte, Jordi and Bluthgen, Nico and Brown, Lee E. and Dormann, Carsten F. and Edwards, François and Figueroa, David and Jacob, Ute and Jones, J. Iwan and Lauridsen, Rasmus B. and Ledger, Mark E. and Lewis, Hannah M. and Olesen, Jes and Van Veen, F. J. Frank and Warren, Phil H. and Woodward, Guy},
	year         = {2008},
	month        = {12},
	journal      = {Journal of animal ecology},
	volume       = {78},
	number       = {1},
	pages        = {253--269},
	doi          = {10.1111/j.1365-2656.2008.01460.x},
	url          = {https://doi.org/10.1111/j.1365-2656.2008.01460.x}
}

@article{dunbar-2015,
	title        = {{The structure of online social networks mirrors those in the offline world}},
	author       = {Dunbar, Robin and Arnaboldi, Valerio and Conti, Marco and Passarella, Andrea},
	year         = {2015},
	month        = {10},
	journal      = {Social networks},
	volume       = {43},
	pages        = {39--47},
	doi          = {10.1016/j.socnet.2015.04.005},
	url          = {https://doi.org/10.1016/j.socnet.2015.04.005}
}

@article{newman-coll-2001,
	title        = {{The structure of scientific collaboration networks}},
	author       = {Newman, Mark E. J.},
	year         = {2001},
	month        = {1},
	journal      = {Proceedings of the National Academy of Sciences of the United States of America},
	publisher    = {National Acad Sciences},
	volume       = {98},
	number       = {2},
	pages        = {404--409},
	doi          = {10.1073/pnas.98.2.404},
	url          = {https://doi.org/10.1073/pnas.98.2.404}
}

@article{radicchi-2008,
	title        = {{Universality of citation distributions: Toward an objective measure of scientific impact}},
	author       = {Radicchi, Filippo and Fortunato, Santo and Castellano, Claudio},
	year         = {2008},
	month        = {11},
	journal      = {Proceedings of the National Academy of Sciences of the United States of America},
	volume       = {105},
	number       = {45},
	pages        = {17268--17272},
	doi          = {10.1073/pnas.0806977105},
	url          = {https://doi.org/10.1073/pnas.0806977105}
}

@article{hafnerburton-2009,
	title        = {{Network Analysis for International Relations}},
	author       = {Hafner-Burton, Emilie M. and Kahler, Miles and Montgomery, Alexander H.},
	year         = {2009},
	month        = {7},
	journal      = {International organization},
	volume       = {63},
	number       = {3},
	pages        = {559--592},
	doi          = {10.1017/s0020818309090195},
	url          = {https://doi.org/10.1017/s0020818309090195}
}

@article{kanter-1971,
	title        = {{Symbolic interactionism: perspective and method.}},
	author       = {Kanter, Rosabeth Moss and Blumer, Herbert},
	year         = {1971},
	month        = {4},
	journal      = {American sociological review},
	volume       = {36},
	number       = {2},
	pages        = {333},
	doi          = {10.2307/2094060},
	url          = {https://doi.org/10.2307/2094060}
}

@article{roenneberg-2013,
	title        = {{Light and the human circadian clock}},
	author       = {Roenneberg, Till and Kantermann, Thomas and Juda, Myriam and Vetter, C{\'e}line and Allebrandt, Karla V},
	year         = {2013},
	journal      = {Circadian clocks},
	publisher    = {Springer},
	pages        = {311--331}
}

@article{brown-1986,
	title        = {{Perceptions of peer pressure, peer conformity dispositions, and self-reported behavior among adolescents.}},
	author       = {Brown, B. Bradford and Clasen, Donna Rae and Eicher, Sue Ann},
	year         = {1986},
	month        = {7},
	journal      = {Developmental psychology},
	volume       = {22},
	number       = {4},
	pages        = {521--530},
	doi          = {10.1037/0012-1649.22.4.521},
	url          = {https://doi.org/10.1037/0012-1649.22.4.521}
}

@article{gonzalez2008understanding,
	title        = {{Understanding individual human mobility patterns}},
	author       = {Gonzalez, Marta C and Hidalgo, Cesar A and Barab{\'a}si, Albert-L{\'a}szl{\'o}},
	year         = {2008},
	journal      = {nature},
	publisher    = {Nature Publishing Group UK London},
	volume       = {453},
	number       = {7196},
	pages        = {779--782}
}

@article{dorogovtsev2002evolution,
	title        = {{Evolution of networks}},
	author       = {Dorogovtsev, Sergey N and Mendes, Jose FF},
	year         = {2002},
	journal      = {Advances in physics},
	publisher    = {Taylor \& Francis},
	volume       = {51},
	number       = {4},
	pages        = {1079--1187}
}

@article{chen-2014,
	title        = {{Big Data: a survey}},
	author       = {Chen, Min and Mao, Shiwen and Liu, Yunhao},
	year         = {2014},
	month        = {1},
	journal      = {Journal on special topics in mobile networks and applications Mobile networks and applications},
	volume       = {19},
	number       = {2},
	pages        = {171--209},
	doi          = {10.1007/s11036-013-0489-0},
	url          = {https://doi.org/10.1007/s11036-013-0489-0}
}

@article{blondel-2015,
	title        = {{A survey of results on mobile phone datasets analysis}},
	author       = {Blondel, Vincent D. and Decuyper, Adeline and Krings, Gautier},
	year         = {2015},
	month        = {8},
	journal      = {EPJ data science},
	volume       = {4},
	number       = {1},
	doi          = {10.1140/epjds/s13688-015-0046-0},
	url          = {https://doi.org/10.1140/epjds/s13688-015-0046-0}
}

@article{de-montjoye-2013,
	title        = {{Unique in the Crowd: The privacy bounds of human mobility}},
	author       = {De Montjoye, Yves Alexandre and Hidalgo, César A. and Verleysen, Michel and Blondel, Vincent D.},
	year         = {2013},
	month        = {3},
	journal      = {Scientific reports},
	volume       = {3},
	number       = {1},
	doi          = {10.1038/srep01376},
	url          = {https://doi.org/10.1038/srep01376}
}

@article{lazer-2014,
	title        = {{The parable of Google Flu: Traps in big data analysis}},
	author       = {Lazer, David and Kennedy, Ryan and King, Gary and Vespignani, Alessandro},
	year         = {2014},
	month        = {3},
	journal      = {Science},
	volume       = {343},
	number       = {6176},
	pages        = {1203--1205},
	doi          = {10.1126/science.1248506},
	url          = {https://doi.org/10.1126/science.1248506}
}

@article{zook-2017,
	title        = {{Ten simple rules for responsible big data research}},
	author       = {Zook, Matthew and Barocas, Solon and Boyd, Danah and Crawford, Kate and Keller, Emily and Gangadharan, Seeta Peña and Goodman, Alyssa A. and Hollander, Rachelle D. and Koenig, Barbara A. and Metcalf, Jacob and Narayanan, Arvind and Nelson, Alondra and Pasquale, Frank},
	year         = {2017},
	month        = {3},
	journal      = {PLOS computational biology},
	volume       = {13},
	number       = {3},
	pages        = {e1005399},
	doi          = {10.1371/journal.pcbi.1005399},
	url          = {https://doi.org/10.1371/journal.pcbi.1005399}
}

@misc{perry2012null,
	title        = {{Null models for network data}},
	author       = {Patrick O. Perry and Patrick J. Wolfe},
	year         = {2012},
	eprint       = {1201.5871},
	archiveprefix = {arXiv},
	primaryclass = {math.ST}
}

@book{gelman-2006,
	title        = {{Data analysis using regression and Multilevel/Hierarchical models}},
	author       = {Gelman, Andrew and Su, Yu-Sung},
	year         = {2006},
	month        = {12},
	doi          = {10.1017/cbo9780511790942},
	url          = {https://doi.org/10.1017/cbo9780511790942}
}

@book{hastie-2013,
	title        = {{The elements of statistical learning: data mining, inference, and prediction}},
	author       = {Hastie, Trevor and Tibshirani, Robert and Friedman, Jerome H.},
	year         = {2013},
	month        = {7},
	url          = {http://catalog.lib.kyushu-u.ac.jp/ja/recordID/1416361}
}

@book{witten-2005,
	title        = {{Data Mining: Practical Machine Learning Tools and Techniques, second edition (Morgan Kaufmann Series in Data Management Systems)}},
	author       = {Witten, Ian H. and Frank, Eibe},
	year         = {2005},
	month        = {6},
	url          = {https://dl.acm.org/citation.cfm?id=1205860}
}

@article{clauset-2008,
	title        = {{Hierarchical structure and the prediction of missing links in networks}},
	author       = {Clauset, Aaron and Moore, Cristopher and Newman, Mark E. J.},
	year         = {2008},
	month        = {5},
	journal      = {Nature},
	volume       = {453},
	number       = {7191},
	pages        = {98--101},
	doi          = {10.1038/nature06830},
	url          = {https://doi.org/10.1038/nature06830}
}

@book{stevens-2012,
	title        = {{Applied Multivariate Statistics for the Social Sciences}},
	author       = {Stevens, James P.},
	year         = {2012},
	month        = {11},
	doi          = {10.4324/9780203843130},
	url          = {https://doi.org/10.4324/9780203843130}
}

@article{gauvin-2022,
	title        = {{Randomized reference models for temporal networks}},
	author       = {Gauvin, Laetitia and Génois, Mathieu and Karsai, Márton and Kivela, Mikko and Takaguchi, Taro and Valdano, Eugenio and Vestergaard, Christian},
	year         = {2022},
	month        = {11},
	journal      = {SIAM review},
	volume       = {64},
	number       = {4},
	pages        = {763--830},
	doi          = {10.1137/19m1242252},
	url          = {https://doi.org/10.1137/19m1242252}
}

@book{gelman1995bayesian,
	title        = {{Bayesian data analysis}},
	author       = {Gelman, Andrew and Carlin, John B and Stern, Hal S and Rubin, Donald B},
	year         = {1995},
	publisher    = {Chapman and Hall/CRC}
}

@article{axelrod2006agent,
	title        = {{Agent-based modeling as a bridge between disciplines}},
	author       = {Axelrod, Robert},
	year         = {2006},
	journal      = {Handbook of computational economics},
	publisher    = {Elsevier},
	volume       = {2},
	pages        = {1565--1584}
}

@article{watts2004new,
	title        = {{The ''new'' science of networks}},
	author       = {Watts, Duncan J.},
	year         = {2004},
	journal      = {Annu. Rev. Sociol.},
	publisher    = {Annual Reviews},
	volume       = {30},
	pages        = {243--270}
}

@article{epstein1999agent,
	title        = {{Agent-based computational models and generative social science}},
	author       = {Epstein, Joshua M},
	year         = {1999},
	journal      = {Complexity},
	publisher    = {Wiley Online Library},
	volume       = {4},
	number       = {5},
	pages        = {41--60}
}

@article{vespignani2009predicting,
	title        = {{Predicting the behavior of techno-social systems}},
	author       = {Vespignani, Alessandro},
	year         = {2009},
	journal      = {Science},
	publisher    = {American Association for the Advancement of Science},
	volume       = {325},
	number       = {5939},
	pages        = {425--428}
}

@misc{committed-observatory-2023,
	title        = {{Observatorio de Fake News - Committed Observatory}},
	author       = {Committed Observatory},
	year         = {2023},
	month        = {5},
	url          = {https://committedobservatory.eu/observatorio/}
}

@misc{EDMO-observatory,
	title        = {{European Digital Media Observatory (EDMO)}},
	year         = {2024},
	month        = {5},
	url          = {https://digital-strategy.ec.europa.eu/en/policies/european-digital-media-observatory}
}

@misc{Polis-observatory,
	title        = {{Polis Fake News Observatory}},
	url          = {https://www.polisanalysis.com/fake-news-observatory}
}

@article{bianconi2023complex,
	title        = {{Complex systems in the spotlight: next steps after the 2021 Nobel Prize in Physics}},
	author       = {Bianconi, Ginestra and Arenas, Alex and Biamonte, Jacob and Carr, Lincoln D and Kahng, Byungnam and Kertesz, Janos and Kurths, Jurgen and Lu, Linyuan and Masoller, Cristina and Motter, Adilson E and others},
	year         = {2023},
	journal      = {Journal of Physics: Complexity},
	publisher    = {IOP Publishing},
	volume       = {4},
	number       = {1},
	pages        = {010201}
}

@book{posfai2016network,
	title        = {{Network science}},
	author       = {P{\'o}sfai, M{\'a}rton and Barab{\'a}si, Albert-L{\'a}szl{\'o}},
	year         = {2016},
	publisher    = {Citeseer}
}

@article{watts1998collective,
	title        = {{Collective dynamics of ''small-world'' networks}},
	author       = {Watts, Duncan J. and Strogatz, Steven H.},
	year         = {1998},
	journal      = {nature},
	publisher    = {Nature Publishing Group},
	volume       = {393},
	number       = {6684},
	pages        = {440--442}
}

@article{barabasi1999emergence,
	title        = {{Emergence of scaling in random networks}},
	author       = {Barab{\'a}si, Albert-L{\'a}szl{\'o} and Albert, R{\'e}ka},
	year         = {1999},
	journal      = {science},
	publisher    = {American Association for the Advancement of Science},
	volume       = {286},
	number       = {5439},
	pages        = {509--512}
}

@article{newman2003structure,
	title        = {{The structure and function of complex networks}},
	author       = {Newman, Mark E. J.},
	year         = {2003},
	journal      = {SIAM review},
	publisher    = {SIAM},
	volume       = {45},
	number       = {2},
	pages        = {167--256}
}

@article{barrat2004architecture,
	title        = {{The architecture of complex weighted networks}},
	author       = {Barrat, Alain and Barthelemy, Marc and Pastor-Satorras, Romualdo and Vespignani, Alessandro},
	year         = {2004},
	journal      = {Proceedings of the national academy of sciences},
	publisher    = {National Acad Sciences},
	volume       = {101},
	number       = {11},
	pages        = {3747--3752}
}

@article{onnela-2003,
	title        = {{Dynamics of market correlations: Taxonomy and portfolio analysis}},
	author       = {Onnela, Jukka-Pekka and Chakraborti, Anirban and Kaski, Kimmo and Kertesz, Janos and Kanto, Antti},
	year         = {2003},
	month        = {11},
	journal      = {Physical review. E, Statistical physics, plasmas, fluids, and related interdisciplinary topics},
	volume       = {68},
	number       = {5},
	doi          = {10.1103/physreve.68.056110},
	url          = {https://doi.org/10.1103/physreve.68.056110}
}

@article{latora-2002,
	title        = {{Is the Boston subway a small-world network?}},
	author       = {Latora, Vito and Marchiori, Massimo},
	year         = {2002},
	month        = {11},
	journal      = {Physica. A},
	volume       = {314},
	number       = {1-4},
	pages        = {109--113},
	doi          = {10.1016/s0378-4371(02)01089-0},
	url          = {https://doi.org/10.1016/s0378-4371(02)01089-0}
}

@article{tumminello-2005,
	title        = {{A tool for filtering information in complex systems}},
	author       = {Tumminello, M. and Aste, T. and Di Matteo, T. and Mantegna, R. N.},
	year         = {2005},
	month        = {7},
	journal      = {Proceedings of the National Academy of Sciences of the United States of America},
	volume       = {102},
	number       = {30},
	pages        = {10421--10426},
	doi          = {10.1073/pnas.0500298102},
	url          = {https://doi.org/10.1073/pnas.0500298102}
}

@article{girvan-2002,
	title        = {{Community structure in social and biological networks}},
	author       = {Girvan, Michelle and Newman, Mark E. J.},
	year         = {2002},
	month        = {6},
	journal      = {Proceedings of the National Academy of Sciences of the United States of America},
	publisher    = {National Acad Sciences},
	volume       = {99},
	number       = {12},
	pages        = {7821--7826},
	doi          = {10.1073/pnas.122653799},
	url          = {https://doi.org/10.1073/pnas.122653799}
}

@article{lancichinetti-2008,
	title        = {{Benchmark graphs for testing community detection algorithms}},
	author       = {Lancichinetti, Andrea and Fortunato, Santo and Radicchi, Filippo},
	year         = {2008},
	month        = {10},
	journal      = {Physical review. E, Statistical, nonlinear and soft matter physics},
	volume       = {78},
	number       = {4},
	doi          = {10.1103/physreve.78.046110},
	url          = {https://doi.org/10.1103/physreve.78.046110}
}

@article{blondel-2008,
	title        = {{Fast unfolding of communities in large networks}},
	author       = {Blondel, Vincent D. and Guillaume, Jean-Loup and Lambiotte, Renaud and Lefebvre, Etienne},
	year         = {2008},
	month        = {10},
	journal      = {Journal of statistical mechanics},
	publisher    = {IOP Publishing},
	volume       = {2008},
	number       = {10},
	pages        = {P10008},
	doi          = {10.1088/1742-5468/2008/10/p10008},
	url          = {https://doi.org/10.1088/1742-5468/2008/10/p10008}
}

@article{rosvall-2008,
	title        = {{Maps of random walks on complex networks reveal community structure}},
	author       = {Rosvall, Martin and Bergstrom, Carl T.},
	year         = {2008},
	month        = {1},
	journal      = {Proceedings of the National Academy of Sciences of the United States of America},
	publisher    = {National Acad Sciences},
	volume       = {105},
	number       = {4},
	pages        = {1118--1123},
	doi          = {10.1073/pnas.0706851105},
	url          = {https://doi.org/10.1073/pnas.0706851105}
}

@article{latapy-2008,
	title        = {{Basic notions for the analysis of large two-mode networks}},
	author       = {Latapy, Matthieu and Magnien, Clémence and Del Vecchio, Nathalie},
	year         = {2008},
	month        = {1},
	journal      = {Social networks},
	volume       = {30},
	number       = {1},
	pages        = {31--48},
	doi          = {10.1016/j.socnet.2007.04.006},
	url          = {https://doi.org/10.1016/j.socnet.2007.04.006}
}

@misc{netflix,
	title        = {{Netflix - Watch TV programmes online, watch films online}},
	url          = {https://www.netflix.com/}
}

@misc{HBO,
	title        = {{HBO Max}},
	url          = {https://play.hbomax.com/}
}

@book{ricci-2011,
	title        = {{Recommender Systems Handbook}},
	author       = {Ricci, Francesco and Rokach, Lior and Shapira, Bracha and Kantor, Paul B.},
	year         = {2011},
	month        = {1},
	doi          = {10.1007/978-0-387-85820-3},
	url          = {https://doi.org/10.1007/978-0-387-85820-3}
}

@article{newman-2001-collaboration,
	title        = {{Scientific collaboration networks. II. Shortest paths, weighted networks, and centrality}},
	author       = {Newman, Mark E. J.},
	year         = {2001},
	month        = {6},
	journal      = {Physical review. E, Statistical physics, plasmas, fluids, and related interdisciplinary topics},
	volume       = {64},
	number       = {1},
	doi          = {10.1103/physreve.64.016132},
	url          = {https://doi.org/10.1103/physreve.64.016132}
}

@article{bonabeau-2002,
	title        = {{Agent-based modeling: Methods and techniques for simulating human systems}},
	author       = {Bonabeau, Eric},
	year         = {2002},
	month        = {5},
	journal      = {Proceedings of the National Academy of Sciences of the United States of America},
	volume       = {99},
	number       = {3},
	pages        = {7280--7287},
	doi          = {10.1073/pnas.082080899},
	url          = {https://doi.org/10.1073/pnas.082080899}
}

@article{macal-2010,
	title        = {{Tutorial on agent-based modelling and simulation}},
	author       = {Macal, C. M. and North, M. J.},
	year         = {2010},
	month        = {9},
	journal      = {Journal of Simulation},
	volume       = {4},
	number       = {3},
	pages        = {151--162},
	doi          = {10.1057/jos.2010.3},
	url          = {https://doi.org/10.1057/jos.2010.3}
}

@book{railsback-2011,
	title        = {{Agent-Based and Individual-Based Modeling: A Practical Introduction}},
	author       = {Railsback, Steven F. and Grimm, Volker},
	year         = {2011},
	month        = {10},
	url          = {http://ci.nii.ac.jp/ncid/BB09226885}
}

@article{duffy-1998,
	title        = {{Growing Artificial Societies: Social Science from the Bottom Up}},
	author       = {Duffy, John and Epstein, Joshua M. and Axtell, Robert},
	year         = {1998},
	month        = {1},
	journal      = {Southern economic journal},
	volume       = {64},
	number       = {3},
	pages        = {791},
	doi          = {10.2307/1060800},
	url          = {https://doi.org/10.2307/1060800}
}

@article{metropolis-1949,
	title        = {{The Monte Carlo method}},
	author       = {Metropolis, Nicholas and Ulam, S.},
	year         = {1949},
	month        = {9},
	journal      = {Journal of the American Statistical Association},
	volume       = {44},
	number       = {247},
	pages        = {335--341},
	doi          = {10.1080/01621459.1949.10483310},
	url          = {https://doi.org/10.1080/01621459.1949.10483310}
}

@book{newman-1999,
	title        = {{Monte Carlo Methods in Statistical Physics}},
	author       = {Newman, Mark E. J. and Barkema, G T},
	year         = {1999},
	month        = {2},
	booktitle    = {Oxford University Press eBooks},
	doi          = {10.1093/oso/9780198517962.001.0001},
	url          = {https://doi.org/10.1093/oso/9780198517962.001.0001}
}

@book{landau-2014,
	title        = {{A guide to Monte Carlo simulations in Statistical Physics}},
	author       = {Landau, David P. and Binder, Kurt},
	year         = {2014},
	month        = {11},
	doi          = {10.1017/cbo9781139696463},
	url          = {https://doi.org/10.1017/cbo9781139696463}
}

@article{gillespie-1977,
	title        = {{Exact stochastic simulation of coupled chemical reactions}},
	author       = {Gillespie, Daniel T.},
	year         = {1977},
	month        = {12},
	journal      = {Journal of physical chemistry},
	volume       = {81},
	number       = {25},
	pages        = {2340--2361},
	doi          = {10.1021/j100540a008},
	url          = {https://doi.org/10.1021/j100540a008}
}

@article{gibson-2000,
	title        = {{Efficient Exact Stochastic Simulation of Chemical Systems with Many Species and Many Channels}},
	author       = {Gibson, Michael A. and Bruck, Jehoshua},
	year         = {2000},
	month        = {2},
	journal      = {The journal of physical chemistry. The journal of physical chemistry. A.},
	volume       = {104},
	number       = {9},
	pages        = {1876--1889},
	doi          = {10.1021/jp993732q},
	url          = {https://doi.org/10.1021/jp993732q}
}

@article{abella2024exploring,
	title        = {{Exploring the spatial segmentation of housing markets from online listings}},
	author       = {Abella, David and Mart\'inez, Johann H. and Mazzoli, Mattia and Corre, Thibault Le and Migozzi, Julien and Alonso-Paul\'i, Eduard and Cresp\'i-Cladera, Rafel and Louail, Thomas and Ramasco, Jos{\'e} J.},
	year         = {2024},
	journal      = {arXiv preprint arXiv:2405.08398},
	number       = {In submission}
}

@article{abella2024dynamics,
	title        = {{Housing market dynamics revealed through online listings}},
	author       = {Abella, David and Mart\'inez, Johann H. and Alonso-Paul\'i, Eduard and Cresp\'i-Cladera, Rafel and Ramasco, Jos{\'e} J.},
	year         = {2024},
	number       = {Currently in preparation}
}

@article{merton1968matthew,
	title        = {{The Matthew effect in science: The reward and communication systems of science are considered.}},
	author       = {Merton, Robert K},
	year         = {1968},
	journal      = {Science},
	publisher    = {American Association for the Advancement of Science},
	volume       = {159},
	number       = {3810},
	pages        = {56--63}
}

@book{goldenfeld-1992,
	title        = {{Lectures on phase transitions and the renormalization group}},
	author       = {Goldenfeld, Nigel},
	year         = {2018},
	publisher    = {CRC Press}
}

@article{axelrod1981evolution,
	title        = {{The evolution of cooperation}},
	author       = {Axelrod, Robert and Hamilton, William D},
	year         = {1981},
	journal      = {science},
	publisher    = {American Association for the Advancement of Science},
	volume       = {211},
	number       = {4489},
	pages        = {1390--1396}
}

@book{ostrom1990governing,
	title        = {{Governing the commons: The evolution of institutions for collective action}},
	author       = {Ostrom, Elinor},
	year         = {1990},
	publisher    = {Cambridge university press}
}

@article{attila-2009,
	title        = {{Impact of aging on the evolution of cooperation in the spatial prisoner's dilemma game}},
	author       = {Szolnoki, Attila and Perc, Matja\v{z} and Szab\'o, Gy\"orgy and Stark, Hans-Ulrich},
	year         = {2009},
	month        = {Aug},
	journal      = {Phys. Rev. E},
	publisher    = {American Physical Society},
	volume       = {80},
	pages        = {021901},
	doi          = {10.1103/PhysRevE.80.021901},
	url          = {https://link.aps.org/doi/10.1103/PhysRevE.80.021901},
	issue        = {2},
	numpages     = {7}
}

@misc{samuele-ciardella-2023,
	title        = {{Aging effects in Coordination games}},
	author       = {Samuele Ciardella},
	year         = {2023},
	url          = {https://ifisc.uib-csic.es/es/publications/aging-effects-in-coordination-games/}
}

@article{maria-2023,
	title        = {{Dynamical Model for Power Grid Frequency Fluctuations: Application to Islands With High Penetration of Wind Generation}},
	author       = {Martínez-Barbeito, María and Gomila, Damià and Colet, Pere},
	year         = {2023},
	journal      = {IEEE Transactions on Sustainable Energy},
	volume       = {14},
	number       = {3},
	pages        = {1436--1445},
	doi          = {10.1109/TSTE.2022.3231975}
}

@article{gleeson-2013,
	title        = {{Binary-State Dynamics on Complex Networks: Pair Approximation and Beyond}},
	author       = {Gleeson, James P.},
	year         = {2013},
	month        = {Apr},
	journal      = {Physical Review X},
	publisher    = {American Physical Society},
	volume       = {3},
	number       = {2},
	pages        = {021004},
	doi          = {10.1103/physrevx.3.021004},
	url          = {https://link.aps.org/doi/10.1103/PhysRevX.3.021004},
	issue        = {2},
	numpages     = {20}
}

@article{artime-2018,
	title        = {{Aging-induced continuous phase transition}},
	author       = {Artime, Oriol and Peralta, Antonio F. and Toral, Ra\'ul and Ramasco, Jos\'e J. and San Miguel, Maxi},
	year         = {2018},
	month        = {Sep},
	journal      = {Physical Review E},
	publisher    = {American Physical Society},
	volume       = {98},
	number       = {3},
	pages        = {032104},
	doi          = {10.1103/physreve.98.032104},
	url          = {https://link.aps.org/doi/10.1103/PhysRevE.98.032104},
	issue        = {3},
	numpages     = {7}
}

@article{centola-2007,
	title        = {{Cascade dynamics of complex propagation}},
	author       = {Centola, Damon and Egu\'iluz, V\'ictor M. and Macy, Michael W.},
	year         = {2007},
	journal      = {Physica A: Statistical Mechanics and its Applications},
	volume       = {374},
	number       = {1},
	pages        = {449--456},
	doi          = {10.1016/j.physa.2006.06.018}
}

@article{saeedian2017memory,
	title        = {{Memory effects on epidemic evolution: The susceptible-infected-recovered epidemic model}},
	author       = {Saeedian, Meghdad and Khalighi, Moein and Azimi-Tafreshi, Nahid and Jafari, GR and Ausloos, Marcel},
	year         = {2017},
	journal      = {Physical Review E},
	publisher    = {APS},
	volume       = {95},
	number       = {2},
	pages        = {022409}
}

@article{galam-2008,
	title        = {{Socio-physics: A review of Galam models}},
	author       = {given-i=S., given=Serge, family=Galam},
	year         = {2008},
	month        = {3},
	journal      = {International Journal of Modern Physics C},
	publisher    = {World Scientific Pub Co Pte Lt},
	volume       = {19},
	number       = {03},
	pages        = {409--440},
	doi          = {10.1142/s0129183108012297},
	url          = {http://dx.doi.org/10.1142/s0129183108012297}
}

@article{goncalves-2012,
	title        = {{Why, when, and how fast innovations are adopted}},
	author       = {given-i=S., given=S., family=Gonçalves and given-i={M. F.}, given={M. F.}, family=Laguna and given-i={J. R.}, given={J. R.}, family=Iglesias},
	year         = {2012},
	month        = {6},
	journal      = {The European Physical Journal B},
	publisher    = {Springer Science and Business Media LLC},
	volume       = {85},
	number       = {6},
	doi          = {10.1140/epjb/e2012-30082-6},
	url          = {http://dx.doi.org/10.1140/epjb/e2012-30082-6}
}

@article{bass1969,
	title        = {{A new product growth for model consumer durables}},
	author       = {Bass, Frank M},
	year         = {1969},
	journal      = {Management science},
	publisher    = {INFORMS},
	volume       = {15},
	number       = {5},
	pages        = {215--227}
}

@incollection{rogers2014,
	title        = {{Diffusion of innovations}},
	author       = {Rogers, Everett M and Singhal, Arvind and Quinlan, Margaret M},
	year         = {2014},
	booktitle    = {An integrated approach to communication theory and research},
	publisher    = {Routledge},
	pages        = {432--448}
}

@article{watts-2002,
	title        = {{A simple model of global cascades on random networks}},
	author       = {Watts, Duncan J.},
	year         = {2002},
	journal      = {Proceedings of the National Academy of Sciences},
	volume       = {99},
	number       = {9},
	pages        = {5766--5771},
	doi          = {10.1073/pnas.082090499}
}

@article{gleeson-2007,
	title        = {{Seed size strongly affects cascades on random networks}},
	author       = {Gleeson, James P. and Cahalane, Diarmuid J.},
	year         = {2007},
	month        = {May},
	journal      = {Physical Review E},
	publisher    = {American Physical Society},
	volume       = {75},
	number       = {5},
	pages        = {056103},
	doi          = {10.1103/physreve.75.056103},
	url          = {https://link.aps.org/doi/10.1103/PhysRevE.75.056103},
	issue        = {5},
	numpages     = {4}
}

@article{gleeson-2008,
	title        = {{Cascades on correlated and modular random networks}},
	author       = {Gleeson, James P.},
	year         = {2008},
	month        = {Apr},
	journal      = {Physical Review E},
	publisher    = {American Physical Society},
	volume       = {77},
	number       = {4},
	pages        = {046117},
	doi          = {10.1103/physreve.77.046117},
	url          = {https://link.aps.org/doi/10.1103/PhysRevE.77.046117},
	issue        = {4},
	numpages     = {10}
}

@article{singh-2013,
	title        = {{Threshold-limited spreading in social networks with multiple initiators}},
	author       = {Singh, P. and Sreenivasan, S. and Szymanski, B. K. and Korniss, G.},
	year         = {2013},
	month        = {7},
	journal      = {Scientific Reports},
	publisher    = {Nature Portfolio},
	volume       = {3},
	number       = {1},
	pages        = {2330},
	doi          = {10.1038/srep02330},
	url          = {https://doi.org/10.1038/srep02330}
}

@article{oh-2018,
	title        = {{Complex contagions with timers}},
	author       = {Oh, Se-Wook and Porter, Mason A.},
	year         = {2018},
	journal      = {Chaos: An Interdisciplinary Journal of Nonlinear Science},
	volume       = {28},
	number       = {3},
	pages        = {033101},
	doi          = {10.1063/1.4990038}
}

@article{chen-2020,
	title        = {{Non-Markovian majority-vote model}},
	author       = {Chen, Hanshuang and Wang, Shuang and Shen, Chuansheng and Zhang, Haifeng and Bianconi, Ginestra},
	year         = {2020},
	month        = {Dec},
	journal      = {Physical Review E},
	publisher    = {American Physical Society},
	volume       = {102},
	number       = {6},
	pages        = {062311},
	doi          = {10.1103/physreve.102.062311},
	url          = {https://link.aps.org/doi/10.1103/PhysRevE.102.062311},
	issue        = {6},
	numpages     = {11}
}

@article{karsai-2016,
	title        = {{Local cascades induced global contagion: How heterogeneous thresholds, exogenous effects, and unconcerned behaviour govern online adoption spreading}},
	author       = {Karsai, M\'arton and Iñiguez, Gerardo and Kikas, Riivo and Kaski, Kimmo and Kert\'esz, J\'anos},
	year         = {2016},
	journal      = {Scientific Reports},
	volume       = {6},
	number       = {1},
	pages        = {27178},
	doi          = {10.1038/srep27178}
}

@article{gleeson-2011,
	title        = {{High-Accuracy Approximation of Binary-State Dynamics on Networks}},
	author       = {Gleeson, James P.},
	year         = {2011},
	month        = {Aug},
	journal      = {Physical Review Letters},
	publisher    = {American Physical Society},
	volume       = {107},
	number       = {6},
	pages        = {068701},
	doi          = {10.1103/physrevlett.107.068701},
	url          = {https://link.aps.org/doi/10.1103/PhysRevLett.107.068701},
	issue        = {6},
	numpages     = {4}
}

@article{peralta-2020A,
	title        = {{Reduction from non-Markovian to Markovian dynamics: the case of aging in the noisy-Voter model}},
	author       = {Peralta, Antonio F. and Khalil, Nagi and Toral, Ra\'ul},
	year         = {2020},
	journal      = {Journal of Statistical Mechanics: Theory and Experiment},
	volume       = {2020},
	number       = {2},
	pages        = {024004},
	doi          = {10.1088/1742-5468/ab6847}
}

@article{peralta-2020B,
	title        = {{Binary-state dynamics on complex networks: Stochastic pair approximation and beyond}},
	author       = {Peralta, Antonio F. and Toral, Ra\'ul},
	year         = {2020},
	month        = {Dec},
	journal      = {Physical Review Research},
	publisher    = {American Physical Society},
	volume       = {2},
	number       = {4},
	pages        = {043370},
	doi          = {10.1103/physrevresearch.2.043370},
	url          = {https://link.aps.org/doi/10.1103/PhysRevResearch.2.043370},
	issue        = {4},
	numpages     = {25}
}

@article{peralta-2020C,
	title        = {{Ordering dynamics in the Voter model with aging}},
	author       = {Peralta, Antonio F. and Khalil, Nagi and Toral, Ra\'ul},
	year         = {2020},
	journal      = {Physica A: Statistical Mechanics and its Applications},
	volume       = {552},
	pages        = {122475},
	doi          = {10.1016/j.physa.2019.122475}
}

@article{granovetter-1978,
	title        = {{Threshold Models of Collective Behavior}},
	author       = {Granovetter, Mark},
	year         = {1978},
	journal      = {American Journal of Sociology},
	volume       = {83},
	number       = {6},
	pages        = {1420--1443},
	doi          = {10.1086/226707}
}

@article{fernandez-gracia-2013,
	title        = {{Timing Interactions in Social Simulations: The Voter Model}},
	author       = {Fern\'andez-Gracia, Juan and Egu\'iluz, V\'ictor M. and San Miguel, Maxi},
	year         = {2013},
	journal      = {Understanding Complex Systems},
	pages        = {331--352}
}

@article{artime-2017,
	title        = {{Dynamics on networks: competition of temporal and topological correlations}},
	author       = {Artime, Oriol and Ramasco, Jos\'e J. and San Miguel, Maxi},
	year         = {2017},
	journal      = {Scientific Reports},
	volume       = {7},
	number       = {1},
	pages        = {41627},
	doi          = {10.1038/srep41627}
}

@article{stark-2008,
	title        = {{Decelerating Microdynamics Can Accelerate Macrodynamics in the Voter Model}},
	author       = {Stark, Hans-Ulrich and Tessone, Claudio J. and Schweitzer, Frank},
	year         = {2008},
	month        = {Jun},
	journal      = {Physical Review Letters},
	publisher    = {American Physical Society},
	volume       = {101},
	number       = {1},
	pages        = {018701},
	doi          = {10.1103/physrevlett.101.018701},
	url          = {https://link.aps.org/doi/10.1103/PhysRevLett.101.018701},
	issue        = {1},
	numpages     = {4}
}

@article{rybski-2012,
	title        = {{Communication activity in a social network: relation between long-term correlations and inter-event clustering}},
	author       = {Rybski, Diego and Buldyrev, Sergey V. and Havlin, Shlomo and Liljeros, Fredrik and Makse, Hern\'an A.},
	year         = {2012},
	journal      = {Scientific Reports},
	volume       = {2},
	number       = {1},
	doi          = {10.1038/srep00560}
}

@article{zignani-2016,
	title        = {{Walls-in-one: usage and temporal patterns in a social media aggregator}},
	author       = {Zignani, Matteo and Esfandyari, Azadeh and Gaito, Sabrina and Rossi, Gian Paolo},
	year         = {2016},
	journal      = {Applied Network Science},
	volume       = {1},
	number       = {1},
	pages        = {5},
	doi          = {10.1007/s41109-016-0009-9}
}

@article{kumar-2020,
	title        = {{On interevent time distributions of avalanche dynamics}},
	author       = {Kumar, Pinaki and Korkolis, Evangelos and Benzi, Roberto and Denisov, Dmitry and Niemeijer, Andr\'e and Schall, Peter and Toschi, Federico and Trampert, Jeannot},
	year         = {2020},
	journal      = {Scientific Reports},
	volume       = {10},
	number       = {1},
	pages        = {626},
	doi          = {10.1038/s41598-019-56764-6}
}

@article{unicomb-2018,
	title        = {{Threshold driven contagion on weighted networks}},
	author       = {Unicomb, Samuel and Iñiguez, Gerardo and Karsai, M\'arton},
	year         = {2018},
	journal      = {Scientific Reports},
	volume       = {8},
	number       = {1},
	doi          = {10.1038/s41598-018-21261-9}
}

@article{iacopini-2019,
	title        = {{Simplicial models of social contagion}},
	author       = {Iacopini, Iacopo and Petri, Giovanni and Barrat, Alain and Latora, Vito},
	year         = {2019},
	journal      = {Nature Communications},
	volume       = {10},
	number       = {1},
	doi          = {10.1038/s41467-019-10431-6}
}

@article{karsai-2014,
	title        = {{Complex contagion process in spreading of online innovation}},
	author       = {Karsai, M\'arton and Iñiguez, Gerardo and Kaski, Kimmo and Kert\'esz, J\'anos},
	year         = {2014},
	journal      = {Journal of The Royal Society Interface},
	volume       = {11},
	number       = {101},
	pages        = {20140694},
	doi          = {10.1098/rsif.2014.0694}
}

@article{starnini-2017,
	title        = {{Equivalence between Non-Markovian and Markovian Dynamics in Epidemic Spreading Processes}},
	author       = {Starnini, Michele and Gleeson, James P. and Boguñ\'a, Mari\'an},
	year         = {2017},
	month        = {Mar},
	journal      = {Physical Review Letters},
	publisher    = {American Physical Society},
	volume       = {118},
	number       = {12},
	pages        = {128301},
	doi          = {10.1103/physrevlett.118.128301},
	url          = {https://link.aps.org/doi/10.1103/PhysRevLett.118.128301},
	issue        = {12},
	numpages     = {5}
}

@article{van-mieghem-2013,
	title        = {{Non-Markovian Infection Spread Dramatically Alters the Susceptible-Infected-Susceptible Epidemic Threshold in Networks}},
	author       = {Van Mieghem, P. and van de Bovenkamp, R.},
	year         = {2013},
	month        = {Mar},
	journal      = {Physical Review Letters},
	publisher    = {American Physical Society},
	volume       = {110},
	number       = {10},
	pages        = {108701},
	doi          = {10.1103/physrevlett.110.108701},
	url          = {https://link.aps.org/doi/10.1103/PhysRevLett.110.108701},
	issue        = {10},
	numpages     = {5}
}

@article{mnsted-2017,
	title        = {{Evidence of complex contagion of information in social media: An experiment using Twitter bots}},
	author       = {Mønsted, Bjarke and Sapieżyński, Piotr and Ferrara, Emilio and Lehmann, Sune},
	year         = {2017},
	journal      = {PLOS ONE},
	volume       = {12},
	number       = {9},
	pages        = {e0184148},
	doi          = {10.1371/journal.pone.0184148}
}

@article{liu-2018,
	title        = {{Impacts of opinion leaders on social contagions}},
	author       = {Liu, Quan-Hui and Lu, Feng-Mao and Zhang, Qian and Tang, Ming and Zhou, Tao},
	year         = {2018},
	journal      = {Chaos: An Interdisciplinary Journal of Nonlinear Science},
	volume       = {28},
	number       = {5},
	pages        = {053103},
	doi          = {10.1063/1.5017515}
}

@article{chen-2018,
	title        = {{Complex contagions with social reinforcement from different layers and neighbors}},
	author       = {Chen, Ling-Jiao and Chen, Xiao-Long and Cai, Meng and Wang, Wei},
	year         = {2018},
	journal      = {Physica A: Statistical Mechanics and its Applications},
	volume       = {503},
	pages        = {516--525},
	doi          = {10.1016/j.physa.2018.03.017}
}

@article{fernandez-gracia-2014,
	title        = {{Is the Voter Model a Model for Voters?}},
	author       = {Fern\'andez-Gracia, Juan and Suchecki, Krzysztof and Ramasco, Jos\'e J. and San Miguel, Maxi and Egu{\'i}luz, V{\'i}ctor M.},
	year         = {2014},
	month        = {Apr},
	journal      = {Physical Review Letters},
	publisher    = {American Physical Society},
	volume       = {112},
	number       = {15},
	pages        = {158701},
	doi          = {10.1103/physrevlett.112.158701},
	url          = {https://link.aps.org/doi/10.1103/PhysRevLett.112.158701},
	issue        = {15},
	numpages     = {5}
}

@article{sood-2005,
	title        = {{Voter Model on Heterogeneous Graphs}},
	author       = {Sood, V. and Redner, S.},
	year         = {2005},
	month        = {May},
	journal      = {Physical Review Letters},
	publisher    = {American Physical Society},
	volume       = {94},
	number       = {17},
	pages        = {178701},
	doi          = {10.1103/physrevlett.94.178701},
	url          = {https://link.aps.org/doi/10.1103/PhysRevLett.94.178701},
	issue        = {17},
	numpages     = {4}
}

@article{pastor-satorras-2015,
	title        = {{Epidemic processes in complex networks}},
	author       = {Pastor-Satorras, Romualdo and Castellano, Claudio and Van Mieghem, Piet and Vespignani, Alessandro},
	year         = {2015},
	journal      = {Reviews of Modern Physics},
	volume       = {87},
	number       = {3},
	pages        = {925--979},
	doi          = {10.1103/revmodphys.87.925}
}

@article{diakonova-2014,
	title        = {{Absorbing and shattered fragmentation transitions in multilayer coevolution}},
	author       = {Diakonova, Marina and San Miguel, Maxi and Egu\'iluz, V\'ictor M.},
	year         = {2014},
	month        = {Jun},
	journal      = {Physical Review E},
	publisher    = {American Physical Society},
	volume       = {89},
	number       = {6},
	pages        = {062818},
	doi          = {10.1103/physreve.89.062818},
	url          = {https://link.aps.org/doi/10.1103/PhysRevE.89.062818},
	issue        = {6},
	numpages     = {8}
}

@article{amato-2017,
	title        = {{Opinion competition dynamics on multiplex networks}},
	author       = {Amato, R and Kouvaris, N E and San Miguel, Maxi and D\'iaz-Guilera, A},
	year         = {2017},
	journal      = {New Journal of Physics},
	volume       = {19},
	number       = {12},
	pages        = {123019},
	doi          = {10.1088/1367-2630/aa936a}
}

@article{castellano-2009,
	title        = {{Nonlinearq-Voter model}},
	author       = {Castellano, Claudio and Muñoz, Miguel A. and Pastor-Satorras, Romualdo},
	year         = {2009},
	month        = {Oct},
	journal      = {Physical Review E},
	publisher    = {American Physical Society},
	volume       = {80},
	number       = {4},
	pages        = {041129},
	doi          = {10.1103/physreve.80.041129},
	url          = {https://link.aps.org/doi/10.1103/PhysRevE.80.041129},
	issue        = {4},
	numpages     = {8}
}

@article{peralta-2018,
	title        = {{Analytical and numerical study of the non-linear noisy Voter model on complex networks}},
	author       = {Peralta, Antonio F. and Carro, Adrian and San Miguel, Maxi and Toral, Ra\'ul},
	year         = {2018},
	journal      = {Chaos: An Interdisciplinary Journal of Nonlinear Science},
	volume       = {28},
	number       = {7},
	pages        = {075516},
	doi          = {10.1063/1.5030112}
}

@article{cencetti-2021,
	title        = {{Temporal properties of higher-order interactions in social networks}},
	author       = {Cencetti, Giulia and Battiston, Federico and Lepri, Bruno and Karsai, M\'arton},
	year         = {2021},
	journal      = {Scientific Reports},
	volume       = {11},
	number       = {1},
	doi          = {10.1038/s41598-021-86469-8}
}

@article{de-arruda-2020,
	title        = {{Social contagion models on hypergraphs}},
	author       = {de Arruda, Guilherme Ferraz and Petri, Giovanni and Moreno, Yamir},
	year         = {2020},
	month        = {Apr},
	journal      = {Physical Review Research},
	publisher    = {American Physical Society},
	volume       = {2},
	number       = {2},
	pages        = {023032},
	doi          = {10.1103/physrevresearch.2.023032},
	url          = {https://link.aps.org/doi/10.1103/PhysRevResearch.2.023032},
	issue        = {2},
	numpages     = {6}
}

@article{hackett-2011,
	title        = {{Cascades on a class of clustered random networks}},
	author       = {Hackett, Adam and Melnik, Sergey and Gleeson, James P.},
	year         = {2011},
	month        = {May},
	journal      = {Physical Review E},
	publisher    = {American Physical Society},
	volume       = {83},
	number       = {5},
	pages        = {056107},
	doi          = {10.1103/physreve.83.056107},
	url          = {https://link.aps.org/doi/10.1103/PhysRevE.83.056107},
	issue        = {5},
	numpages     = {9}
}

@article{hackett-2013,
	title        = {{Cascades on clique-based graphs}},
	author       = {Hackett, Adam and Gleeson, James P.},
	year         = {2013},
	month        = {Jun},
	journal      = {Physical Review E},
	publisher    = {American Physical Society},
	volume       = {87},
	number       = {6},
	pages        = {062801},
	doi          = {10.1103/physreve.87.062801},
	url          = {https://link.aps.org/doi/10.1103/PhysRevE.87.062801},
	issue        = {6},
	numpages     = {10}
}

@article{dodds-2013,
	title        = {{Limited Imitation Contagion on Random Networks: Chaos, Universality, and Unpredictability}},
	author       = {Dodds, Peter Sheridan and Harris, Kameron Decker and Danforth, Christopher M.},
	year         = {2013},
	month        = {Apr},
	journal      = {Physical Review Letters},
	publisher    = {American Physical Society},
	volume       = {110},
	number       = {15},
	pages        = {158701},
	doi          = {10.1103/physrevlett.110.158701},
	url          = {https://link.aps.org/doi/10.1103/PhysRevLett.110.158701},
	issue        = {15},
	numpages     = {5}
}

@article{dodds-2004,
	title        = {{Universal Behavior in a Generalized Model of Contagion}},
	author       = {Dodds, Peter Sheridan and Watts, Duncan J.},
	year         = {2004},
	month        = {May},
	journal      = {Physical Review Letters},
	publisher    = {American Physical Society},
	volume       = {92},
	number       = {21},
	pages        = {218701},
	doi          = {10.1103/physrevlett.92.218701},
	url          = {https://link.aps.org/doi/10.1103/PhysRevLett.92.218701},
	issue        = {21},
	numpages     = {4}
}

@article{shrestha-2014,
	title        = {{Message-passing approach for threshold models of behavior in networks}},
	author       = {Shrestha, Munik and Moore, Cristopher},
	year         = {2014},
	month        = {Feb},
	journal      = {Physical Review E},
	publisher    = {American Physical Society},
	volume       = {89},
	number       = {2},
	pages        = {022805},
	doi          = {10.1103/physreve.89.022805},
	url          = {https://link.aps.org/doi/10.1103/PhysRevE.89.022805},
	issue        = {2},
	numpages     = {9}
}

@article{iribarren-2009,
	title        = {{Impact of Human Activity Patterns on the Dynamics of Information Diffusion}},
	author       = {Iribarren, Jos\'e Luis and Moro, Esteban},
	year         = {2009},
	month        = {Jul},
	journal      = {Physical Review Letters},
	publisher    = {American Physical Society},
	volume       = {103},
	number       = {3},
	pages        = {038702},
	doi          = {10.1103/physrevlett.103.038702},
	url          = {https://link.aps.org/doi/10.1103/PhysRevLett.103.038702},
	issue        = {3},
	numpages     = {4}
}

@article{karimi-2013,
	title        = {{Threshold model of cascades in empirical temporal networks}},
	author       = {Karimi, Fariba and Holme, Petter},
	year         = {2013},
	journal      = {Physica A: Statistical Mechanics and its Applications},
	volume       = {392},
	number       = {16},
	pages        = {3476--3483},
	doi          = {10.1016/j.physa.2013.03.050}
}

@article{karsai-2011,
	title        = {{Small but slow world: How network topology and burstiness slow down spreading}},
	author       = {Karsai, M{\'a}rton and Kivela, M. and Pan, R. K. and Kaski, K. and Kert\'esz, J. and Barab\'asi, A.-L. and Saramaki, J.},
	year         = {2011},
	month        = {Feb},
	journal      = {Physical Review E},
	publisher    = {American Physical Society},
	volume       = {83},
	number       = {2},
	pages        = {025102},
	doi          = {10.1103/physreve.83.025102},
	url          = {https://link.aps.org/doi/10.1103/PhysRevE.83.025102},
	issue        = {2},
	numpages     = {4}
}

@article{perez-2016,
	title        = {{Competition in the presence of aging: dominance, coexistence, and alternation between states}},
	author       = {P\'erez, Toni and Klemm, Konstantin and Egu\'iluz, V\'ictor M.},
	year         = {2016},
	journal      = {Scientific Reports},
	volume       = {6},
	number       = {1},
	pages        = {21128},
	doi          = {10.1038/srep21128}
}

@article{boguna-2014,
	title        = {{Simulating non-Markovian stochastic processes}},
	author       = {Bogu\~n\'a, Marian and Lafuerza, Luis F. and Toral, Ra\'ul and Serrano, M. \'Angeles},
	year         = {2014},
	month        = {Oct},
	journal      = {Phys. Rev. E},
	publisher    = {American Physical Society},
	volume       = {90},
	number       = {4},
	pages        = {042108},
	doi          = {10.1103/PhysRevE.90.042108},
	url          = {https://link.aps.org/doi/10.1103/PhysRevE.90.042108},
	issue        = {4},
	numpages     = {9}
}

@article{fernandez-gracia-2011,
	title        = {{Update rules and interevent time distributions: Slow ordering versus no ordering in the Voter model}},
	author       = {Fern\'andez-Gracia, Juan and Egu\'iluz, V\'ictor M. and San Miguel, Maxi},
	year         = {2011},
	month        = {Jul},
	journal      = {Physical Review E},
	publisher    = {American Physical Society},
	volume       = {84},
	number       = {1},
	pages        = {015103},
	doi          = {10.1103/physreve.84.015103},
	url          = {https://link.aps.org/doi/10.1103/PhysRevE.84.015103},
	issue        = {1},
	numpages     = {4}
}

@article{gleeson-2016,
	title        = {{Effects of Network Structure, Competition and Memory Time on Social Spreading Phenomena}},
	author       = {Gleeson, James P. and O'Sullivan, Kevin P. and Baños, Raquel A. and Moreno, Yamir},
	year         = {2016},
	month        = {May},
	journal      = {Physical Review X},
	publisher    = {American Physical Society},
	volume       = {6},
	number       = {2},
	pages        = {021019},
	doi          = {10.1103/physrevx.6.021019},
	url          = {https://link.aps.org/doi/10.1103/PhysRevX.6.021019},
	issue        = {2},
	numpages     = {22}
}

@inbook{wormald_1999,
	title        = {{Models of Random Regular Graphs}},
	author       = {Wormald, N. C.},
	year         = {1999},
	journal      = {London Mathematical Society Lecture Note Series},
	booktitle    = {Surveys in Combinatorics, 1999},
	publisher    = {Cambridge University Press},
	series       = {London Mathematical Society Lecture Note Series},
	pages        = {239-298},
	doi          = {10.1017/CBO9780511721335.010},
	place        = {Cambridge},
	editor       = {Lamb, J. D. and Preece, D. A.Editors},
	collection   = {London Mathematical Society Lecture Note Series}
}

@article{erdos1960evolution,
	title        = {{On the evolution of random graphs}},
	author       = {Erd{\H{o}}s, Paul and R{\'e}nyi, Alfr{\'e}d and others},
	year         = {1960},
	journal      = {Publ. Math. Inst. Hung. Acad. Sci},
	volume       = {5},
	number       = {1},
	pages        = {17--60}
}

@article{barabasi2009scale,
	title        = {{Scale-free networks: a decade and beyond}},
	author       = {Barab{\'a}si, Albert-L{\'a}szl{\'o}},
	year         = {2009},
	journal      = {Science},
	publisher    = {American Association for the Advancement of Science},
	volume       = {325},
	number       = {5939},
	pages        = {412--413}
}

@article{molloy-1995,
	title        = {{A critical point for random graphs with a given degree sequence}},
	author       = {Molloy, Michael and Reed, Bruce},
	year         = {1995},
	journal      = {Random Structures and Algorithms},
	volume       = {6},
	number       = {2-3},
	pages        = {161--180},
	doi          = {10.1002/rsa.3240060204}
}

@article{newman-2001,
	title        = {{Random graphs with arbitrary degree distributions and their applications}},
	author       = {Newman, Mark E. J. and Strogatz, Stephen H. and Watts, Duncan J.},
	year         = {2001},
	journal      = {Physical Review E},
	volume       = {64},
	number       = {2},
	doi          = {10.1103/physreve.64.026118}
}

@article{Leah2022,
	title        = {{Multitype branching process method for modeling complex contagion on clustered networks}},
	author       = {Keating, Leah A. and Gleeson, James P. and O'Sullivan, David J. P.},
	year         = {2022},
	month        = {Mar},
	journal      = {Phys. Rev. E},
	publisher    = {American Physical Society},
	volume       = {105},
	pages        = {034306},
	doi          = {10.1103/PhysRevE.105.034306},
	url          = {https://link.aps.org/doi/10.1103/PhysRevE.105.034306},
	issue        = {3},
	numpages     = {12}
}

@misc{link_git,
	title        = {{GitHub repository}},
	howpublished = {\url{https://github.com/davidabbu/Aging-in-binary-state-models}}
}

@book{Voter-original,
	title        = {{Stochastic interacting systems: contact, Voter and exclusion processes}},
	author       = {Liggett, Thomas M and others},
	year         = {1999},
	publisher    = {springer science \& Business Media},
	volume       = {324}
}

@article{redner-2019,
	title        = {{Reality-inspired Voter models: A mini-review}},
	author       = {Redner, Sidney},
	year         = {2019},
	journal      = {Comptes Rendus Physique},
	volume       = {20},
	number       = {4},
	pages        = {275--292},
	doi          = {10.1016/j.crhy.2019.05.004}
}

@article{diakonova-2016,
	title        = {{Irreducibility of multilayer network dynamics: the case of the Voter model}},
	author       = {Diakonova, Marina and Nicosia, Vincenzo and Latora, Vito and San Miguel, Maxi},
	year         = {2016},
	journal      = {New Journal of Physics},
	volume       = {18},
	number       = {2},
	pages        = {023010},
	doi          = {10.1088/1367-2630/18/2/023010}
}

@article{carro-2016,
	title        = {{The noisy Voter model on complex networks}},
	author       = {Carro, Adri\'an and Toral, Ra\'ul and San Miguel, Maxi},
	year         = {2016},
	journal      = {Scientific Reports},
	volume       = {6},
	number       = {1},
	pages        = {24775},
	doi          = {10.1038/srep24775}
}

@article{vazquez-2008,
	title        = {{Generic Absorbing Transition in Coevolution Dynamics}},
	author       = {Vazquez, Federico and Egu\'iluz, V\'ictor M. and San Miguel, Maxi},
	year         = {2008},
	month        = {Mar},
	journal      = {Physical Review Letters},
	publisher    = {American Physical Society},
	volume       = {100},
	number       = {10},
	pages        = {108702},
	doi          = {10.1103/physrevlett.100.108702},
	url          = {https://link.aps.org/doi/10.1103/PhysRevLett.100.108702},
	issue        = {10},
	numpages     = {4}
}

@article{unknown-author-2018,
	title        = {{How Behavior Spreads: The Science of Complex Contagions How Behavior Spreads: The Science of Complex Contagions Damon Centola Princeton University Press, 2018. 308 pp.}},
	author       = {Centola, Damon},
	year         = {2018},
	journal      = {Science},
	volume       = {361},
	number       = {6409},
	pages        = {1320--1320},
	doi          = {10.1126/science.aav1974},
	url          = {https://www.science.org/doi/abs/10.1126/science.aav1974},
	eprint       = {https://www.science.org/doi/pdf/10.1126/science.aav1974}
}

@article{diaz-diaz-2022,
	title        = {{Echo chambers and information transmission biases in homophilic and heterophilic networks}},
	author       = {Diaz-Diaz, Fernando and San Miguel, Maxi and Meloni, Sandro},
	year         = {2022},
	journal      = {Scientific Reports},
	volume       = {12},
	number       = {1},
	pages        = {9350},
	doi          = {10.1038/s41598-022-13343-6}
}

@article{czaplicka-2016,
	title        = {{Competition of simple and complex adoption on interdependent networks}},
	author       = {Czaplicka, Agnieszka and Toral, Ra\'ul and San Miguel, Maxi},
	year         = {2016},
	month        = {Dec},
	journal      = {Physical Review E},
	publisher    = {American Physical Society},
	volume       = {94},
	number       = {6},
	pages        = {062301},
	doi          = {10.1103/physreve.94.062301},
	url          = {https://link.aps.org/doi/10.1103/PhysRevE.94.062301},
	issue        = {6},
	numpages     = {10}
}

@article{min-2018,
	title        = {{Competing contagion processes: Complex contagion triggered by simple contagion}},
	author       = {Min, Byungjoon and San Miguel, Maxi},
	year         = {2018},
	journal      = {Scientific Reports},
	volume       = {8},
	number       = {1},
	doi          = {10.1038/s41598-018-28615-3}
}

@article{min-2018-dual,
	title        = {{Competition and dual users in complex contagion processes}},
	author       = {Min, Byungjoon and Miguel, M. San},
	year         = {2018},
	month        = {10},
	journal      = {Scientific reports},
	volume       = {8},
	number       = {1},
	doi          = {10.1038/s41598-018-32643-4},
	url          = {https://doi.org/10.1038/s41598-018-32643-4}
}

@article{schelling-1969,
	title        = {{Models of Segregation}},
	author       = {Schelling, Thomas C.},
	year         = {1969},
	journal      = {American Economic Review},
	volume       = {59},
	number       = {2},
	pages        = {488}
}

@article{Schelling,
	title        = {{Dynamic models of segregation}},
	author       = {Schelling, Thomas C.},
	year         = {1971},
	journal      = {The Journal of Mathematical Sociology},
	publisher    = {Routledge},
	volume       = {1},
	number       = {2},
	pages        = {143--186},
	doi          = {10.1080/0022250X.1971.9989794},
	url          = {https://doi.org/10.1080/0022250X.1971.9989794}
}

@article{hegselmann-2017,
	title        = {{Thomas C. Schelling and James M. Sakoda: The Intellectual, Technical,and Social History of a Model}},
	author       = {Hegselmann, Rainer},
	year         = {2017},
	journal      = {Journal of Artificial Societies and Social Simulation},
	volume       = {20},
	number       = {3},
	pages        = {15},
	doi          = {10.18564/jasss.3511}
}

@article{granovetter,
	title        = {{Threshold models of collective behavior}},
	author       = {Granovetter, Mark},
	year         = {1978},
	journal      = {American Journal of Sociology},
	volume       = {83},
	pages        = {1420}
}

@article{Gauvin_2009,
	title        = {{Phase diagram of a Schelling segregation model}},
	author       = {Gauvin, Laetitia and Nadal, Jean-Pierre and Vannimenus, Jean},
	year         = {2009},
	month        = {Jul},
	journal      = {The European Physical Journal B},
	publisher    = {Springer Science and Business Media LLC},
	volume       = {70},
	number       = {2},
	pages        = {293-304},
	doi          = {10.1140/epjb/e2009-00234-0},
	issn         = {1434-6036},
	url          = {http://dx.doi.org/10.1140/epjb/e2009-00234-0}
}

@article{Dall_Asta_2008,
	title        = {{Statistical physics of the Schelling model of segregation}},
	author       = {Dall'Asta, L. and Castellano, C. and Marsili, M.},
	year         = {2008},
	month        = {Jul},
	journal      = {Journal of Statistical Mechanics: Theory and Experiment},
	publisher    = {IOP Publishing},
	volume       = {2008},
	number       = {07},
	pages        = {L07002},
	doi          = {10.1088/1742-5468/2008/07/l07002},
	issn         = {1742-5468},
	url          = {http://dx.doi.org/10.1088/1742-5468/2008/07/L07002}
}

@article{unified,
	title        = {{A unified framework for Schelling's model of segregation}},
	author       = {Rogers, Tim and McKane, Alan J},
	year         = {2011},
	month        = {Jul},
	journal      = {Journal of Statistical Mechanics: Theory and Experiment},
	publisher    = {IOP Publishing},
	volume       = {2011},
	number       = {07},
	pages        = {P07006},
	doi          = {10.1088/1742-5468/2011/07/p07006},
	issn         = {1742-5468},
	url          = {http://dx.doi.org/10.1088/1742-5468/2011/07/P07006}
}

@article{Gauvin_2010,
	title        = {{Schelling segregation in an open city: A kinetically constrained Blume-Emery-Griffiths spin-1 system}},
	author       = {Gauvin, Laetitia and Nadal, Jean-Pierre and Vannimenus, Jean},
	year         = {2010},
	month        = {Jun},
	journal      = {Physical Review E},
	publisher    = {American Physical Society (APS)},
	volume       = {81},
	number       = {6},
	doi          = {10.1103/physreve.81.066120},
	issn         = {1550-2376},
	url          = {http://dx.doi.org/10.1103/PhysRevE.81.066120}
}

@article{stauffer-2007,
	title        = {{Ising, Schelling and self-organising segregation}},
	author       = {Stauffer, Dietrich  and Solomon, S.},
	year         = {2007},
	journal      = {The European Physical Journal B},
	volume       = {57},
	number       = {4},
	pages        = {473--479},
	doi          = {10.1140/epjb/e2007-00181-8}
}

@article{grauwin-2009,
	title        = {{Competition between collective and individual dynamics}},
	author       = {Grauwin, S. and Bertin, E. and Lemoy, R. and Jensen, P.},
	year         = {2009},
	journal      = {Proceedings of the National Academy of Sciences},
	volume       = {106},
	number       = {49},
	pages        = {20622--20626},
	doi          = {10.1073/pnas.0906263106}
}

@article{BEG,
	title        = {{Ising Model for the Lambda Transition and Phase Separation in He3-He4 Mixtures}},
	author       = {Blume, M. and Emery, V.J. and Robert B. Griffiths},
	year         = {1971},
	journal      = {Phys. Rev. A},
	volume       = {4},
	pages        = {1071},
	doi          = {https://doi.org/10.1103/PhysRevA.4.1071}
}

@article{BlumeCapel,
	title        = {{Tricritical behavior of the Blume-Capel model}},
	author       = {Saul, D. M. and Wortis, Michael and Stauffer, Dietrich },
	year         = {1974},
	month        = {Jun},
	journal      = {Phys. Rev. B},
	publisher    = {American Physical Society},
	volume       = {9},
	pages        = {4964--4980},
	doi          = {10.1103/PhysRevB.9.4964},
	url          = {https://link.aps.org/doi/10.1103/PhysRevB.9.4964},
	issue        = {11},
	numpages     = {0}
}

@misc{randomwalks,
	title        = {{Quantifying ethnic segregation in cities through random walks}},
	author       = {Sandro Sousa and Vincenzo Nicosia},
	year         = {2020},
	eprint       = {2010.10462},
	archiveprefix = {arXiv},
	primaryclass = {physics.soc-ph}
}

@article{urban,
	title        = {{Quantifying segregation in an integrated urban physical-social space}},
	author       = {Xu, Yang and Belyi, Alexander and Santi, Paolo and Ratti, Carlo},
	year         = {2019},
	journal      = {Journal of the Royal Society Interface},
	volume       = {16},
	number       = {160},
	pages        = {20190536},
	doi          = {10.1098/rsif.2019.0536},
	issn         = {17425662},
	url          = {https://royalsocietypublishing.org/doi/abs/10.1098/rsif.2019.0536},
	pmid         = {31744420}
}

@book{silver2016scenescapes,
	title        = {{Scenescapes: How qualities of place shape social life}},
	author       = {Silver, Daniel Aaron and Clark, Terry Nichols},
	year         = {2016},
	publisher    = {University of Chicago Press}
}

@article{chetty-2016,
	title        = {{The Effects of Exposure to Better Neighborhoods on Children: New Evidence from the Moving to Opportunity Experiment}},
	author       = {given-i=R., given=Raj, family=Chetty and given-i=N., given=Nathaniel, family=Hendren and given-i={L. F.}, given={Lawrence F.}, family=Katz},
	year         = {2016},
	month        = {4},
	journal      = {American Economic Review},
	publisher    = {American Economic Association},
	volume       = {106},
	number       = {4},
	pages        = {855--902},
	doi          = {10.1257/aer.20150572},
	url          = {http://dx.doi.org/10.1257/aer.20150572}
}

@article{clark-2002,
	title        = {{Amenities Drive Urban Growth}},
	author       = {given-i={T. N.}, given={Terry Nichols}, family=Clark and given-i=R., given=Richard, family=Lloyd and given-i={K. K.}, given={Kenneth K.}, family=Wong and given-i=P., given=Pushpam, family=Jain},
	year         = {2002},
	month        = {12},
	journal      = {Journal of Urban Affairs},
	publisher    = {Informa UK Limited},
	volume       = {24},
	number       = {5},
	pages        = {493--515},
	doi          = {10.1111/1467-9906.00134},
	url          = {http://dx.doi.org/10.1111/1467-9906.00134}
}

@article{clark-2003,
	title        = {{Does commuting distance matter?}},
	author       = {given-i={W. A.}, given={William A.V.}, family=Clark and given-i=Y., given=Youqin, family=Huang and given-i=S., given=Suzanne, family=Withers},
	year         = {2003},
	month        = {3},
	journal      = {Regional Science and Urban Economics},
	publisher    = {Elsevier BV},
	volume       = {33},
	number       = {2},
	pages        = {199--221},
	doi          = {10.1016/s0166-0462(02)00012-1},
	url          = {http://dx.doi.org/10.1016/s0166-0462(02)00012-1}
}

@article{granovetter-1985,
	title        = {{Economic Action and Social Structure: The Problem of Embeddedness}},
	author       = {given=Mark, family=Granovetter},
	year         = {1985},
	month        = {11},
	journal      = {American Journal of Sociology},
	publisher    = {University of Chicago Press},
	volume       = {91},
	number       = {3},
	pages        = {481--510},
	doi          = {10.1086/228311},
	url          = {http://dx.doi.org/10.1086/228311}
}

@article{silver-2021,
	title        = {{Venues and segregation: A revised Schelling model}},
	author       = {given-i=D., given=Daniel, family=Silver and given-i=U., given=Ultan, family=Byrne and given-i=P., given=Patrick, family=Adler},
	year         = {2021},
	month        = {1},
	journal      = {PLOS ONE},
	publisher    = {Public Library of Science (PLoS)},
	volume       = {16},
	number       = {1},
	pages        = {e0242611},
	doi          = {10.1371/journal.pone.0242611},
	url          = {http://dx.doi.org/10.1371/journal.pone.0242611}
}

@article{wasserman-2001,
	title        = {{Segregation and the Provision of Spatially Defined Local Public Goods}},
	author       = {given-i=H., given=Henry, family=Wasserman and given-i=G., given=Gary, family=Yohe},
	year         = {2001},
	month        = {10},
	journal      = {The American Economist},
	publisher    = {SAGE Publications},
	volume       = {45},
	number       = {2},
	pages        = {13--24},
	doi          = {10.1177/056943450104500202},
	url          = {http://dx.doi.org/10.1177/056943450104500202}
}

@article{Heisemberg,
	title        = {{Aging dynamics of the Heisenberg spin glass}},
	author       = {Berthier, L. and Young, A. P.},
	year         = {2004},
	month        = {May},
	journal      = {Physical Review B},
	publisher    = {American Physical Society (APS)},
	volume       = {69},
	number       = {18},
	doi          = {10.1103/physrevb.69.184423},
	issn         = {1550-235X},
	url          = {http://dx.doi.org/10.1103/PhysRevB.69.184423}
}

@article{denton1995persistence,
	title        = {{The persistence of segregation: Links between residential segregation and school segregation}},
	author       = {Denton, Nancy A.},
	year         = {1995},
	journal      = {Minn. L. Rev.},
	publisher    = {HeinOnline},
	volume       = {80},
	pages        = {795}
}

@book{spinglassbook,
	title        = {{Spin glasses and random fields}},
	author       = {A P Young},
	year         = {1997},
	publisher    = {World Scientific},
	url          = {libgen.li/file.php?md5=5941f917f91576aa2f77de1c40712f2c}
}

@article{8Heisemberg,
	title        = {{Ordered Phase of Short-Range Ising Spin-Glasses}},
	author       = {Fisher, Daniel S. and Huse, David A.},
	year         = {1986},
	month        = {Apr},
	journal      = {Phys. Rev. Lett.},
	publisher    = {American Physical Society},
	volume       = {56},
	pages        = {1601--1604},
	doi          = {10.1103/PhysRevLett.56.1601},
	url          = {https://link.aps.org/doi/10.1103/PhysRevLett.56.1601},
	issue        = {15},
	numpages     = {0}
}

@article{Interfacial_roughening,
	title        = {{Interfacial roughening, segregation and dynamic behaviour in a generalized Schelling model}},
	author       = {Albano, Ezequiel},
	year         = {2012},
	month        = {03},
	journal      = {Journal of Statistical Mechanics-theory and Experiment - J STAT MECH-THEORY EXP},
	volume       = {2012},
	pages        = {},
	doi          = {10.1088/1742-5468/2012/03/P03013}
}

@article{Maxi,
	title        = {{The dynamics of first order phase transitions}},
	author       = {Gunton, D. and San Miguel, M. and Sahni, P.S.},
	year         = {1983},
	journal      = {Phase transitions and critical phenomena},
	publisher    = {Academic Press},
	volume       = {8},
	pages        = {267--466},
	annotate     = {Eds. C. Domb and J. Lebowitz}
}

@article{Abella_2024,
	title        = {{Ordering dynamics and aging in the symmetrical threshold model}},
	author       = {David Abella and Juan Carlos González-Avella and Maxi San Miguel and José J Ramasco},
	year         = {2024},
	month        = {Jan},
	journal      = {New Journal of Physics},
	publisher    = {IOP Publishing},
	volume       = {26},
	number       = {1},
	pages        = {013033},
	doi          = {10.1088/1367-2630/ad1ad4},
	url          = {https://dx.doi.org/10.1088/1367-2630/ad1ad4}
}

@article{abella2023many,
	title        = {{Many-Body Contributions in Water Nanoclusters}},
	author       = {Abella,  David and Franzese,  Giancarlo and Hernández-Rojas,  Javier},
	year         = {2023},
	month        = {Jan},
	journal      = {ACS Nano},
	publisher    = {American Chemical Society (ACS)},
	volume       = {17},
	number       = {3},
	pages        = {1959-1964},
	doi          = {10.1021/acsnano.2c06077},
	issn         = {1936-086X},
	url          = {http://dx.doi.org/10.1021/acsnano.2c06077}
}

@article{stauffer-2013,
	title        = {{A Biased Review of Sociophysics}},
	author       = {Stauffer, Dietrich},
	year         = {2013},
	journal      = {Journal of Statistical Physics},
	volume       = {151},
	number       = {1-2},
	pages        = {9--20},
	doi          = {10.1007/s10955-012-0604-9}
}

@article{henry-2011,
	title        = {{Emergence of segregation in evolving social networks}},
	author       = {Henry, A. D. and Pralat, P. and Zhang, C.-Q.},
	year         = {2011},
	journal      = {Proceedings of the National Academy of Sciences},
	volume       = {108},
	number       = {21},
	pages        = {8605--8610},
	doi          = {10.1073/pnas.1014486108}
}

@article{clark-1991,
	title        = {{Residential preferences and neighborhood racial segregation: A test of the schelling segregation model}},
	author       = {Clark, William A.V.},
	year         = {1991},
	journal      = {Demography},
	volume       = {28},
	number       = {1},
	pages        = {1--19},
	doi          = {10.2307/2061333}
}

@article{gracia-lazaro-2009,
	title        = {{Residential segregation and cultural dissemination: An Axelrod-Schelling model}},
	author       = {Gracia-L\'azaro, C. and Lafuerza, L. F. and Flor\'ia, L. M. and Moreno, Y.},
	year         = {2009},
	journal      = {Physical Review E},
	volume       = {80},
	number       = {4},
	doi          = {10.1103/physreve.80.046123}
}

@article{barabasi-2005,
	title        = {{The origin of bursts and heavy tails in human dynamics}},
	author       = {Barabasi, A. L.},
	year         = {2005},
	journal      = {Nature},
	volume       = {435},
	pages        = {207}
}

@article{HoKo,
	title        = {{Percolation and cluster distribution. I. Cluster multiple labeling technique and critical concentration algorithm}},
	author       = {Hoshen, J. and Kopelman, R.},
	year         = {1976},
	month        = {Oct},
	journal      = {Phys. Rev. B},
	publisher    = {American Physical Society},
	volume       = {14},
	pages        = {3438--3445},
	doi          = {10.1103/PhysRevB.14.3438},
	url          = {https://link.aps.org/doi/10.1103/PhysRevB.14.3438},
	issue        = {8},
	numpages     = {0}
}

@article{barmpalias-2018,
	title        = {{Minority Population in the One-Dimensional Schelling Model of Segregation}},
	author       = {Barmpalias, George and Elwes, Richard and Lewis-Pye, Andrew},
	year         = {2018},
	journal      = {Journal of Statistical Physics},
	volume       = {173},
	number       = {5},
	pages        = {1408--1458},
	doi          = {10.1007/s10955-018-2146-2}
}

@article{holden-2019,
	title        = {{Scaling limits of the Schelling model}},
	author       = {Holden, Nina and Sheffield, Scott},
	year         = {2019},
	journal      = {Probability Theory and Related Fields},
	volume       = {176},
	number       = {1-2},
	pages        = {219--292},
	doi          = {10.1007/s00440-019-00918-0}
}

@article{sert-2020,
	title        = {{Segregation dynamics with reinforcement learning and agent based modeling}},
	author       = {Sert, Egemen and Bar-Yam, Yaneer and Morales, Alfredo J.},
	year         = {2020},
	journal      = {Scientific Reports},
	volume       = {10},
	number       = {1},
	doi          = {10.1038/s41598-020-68447-8}
}

@article{ortega-2021,
	title        = {{Avalanches in an extended Schelling model: An explanation of urban gentrification}},
	author       = {Ortega, Diego and Rodr\'iguez-Laguna, Javier and Korutcheva, Elka},
	year         = {2021},
	journal      = {Physica A: Statistical Mechanics and its Applications},
	volume       = {573},
	pages        = {125943},
	doi          = {10.1016/j.physa.2021.125943}
}

@article{ortega-2021.2,
	title        = {{A Schelling model with a variable threshold in a closed city segregation model. Analysis of the universality classes}},
	author       = {Ortega, Diego and Rodr\'iguez-Laguna, Javier and Korutcheva, Elka},
	year         = {2021},
	journal      = {Physica A: Statistical Mechanics and its Applications},
	volume       = {574},
	pages        = {126010},
	doi          = {10.1016/j.physa.2021.126010}
}

@article{domic-2011,
	title        = {{Dynamics and complexity of the Schelling segregation model}},
	author       = {Domic, Nicol\'as Goles and Goles, Eric and Rica, Sergio},
	year         = {2011},
	journal      = {Physical Review E},
	volume       = {83},
	number       = {5},
	doi          = {10.1103/physreve.83.056111}
}

@article{vieira-2020,
	title        = {{Dynamics of extended Schelling models}},
	author       = {Vieira, Andr\'e P and Goles, Eric and Herrmann, Hans J},
	year         = {2020},
	journal      = {Journal of Statistical Mechanics: Theory and Experiment},
	volume       = {2020},
	number       = {1},
	pages        = {013212},
	doi          = {10.1088/1742-5468/ab5b8d}
}

@article{agarwal-2020,
	title        = {{Swap Stability in Schelling Games on Graphs}},
	author       = {Agarwal, Aishwarya and Elkind, Edith and Gan, Jiarui and Voudouris, Alexandros},
	year         = {2020},
	journal      = {Proceedings of the AAAI Conference on Artificial Intelligence},
	volume       = {34},
	number       = {02},
	pages        = {1758--1765},
	doi          = {10.1609/aaai.v34i02.5541}
}

@article{jensen-2018,
	title        = {{Giant Catalytic Effect of Altruists in Schelling’s Segregation Model}},
	author       = {Jensen, Pablo and Matreux, Thomas and Cambe, Jordan and Larralde, Hernan and Bertin, Eric},
	year         = {2018},
	journal      = {Physical Review Letters},
	volume       = {120},
	number       = {20},
	doi          = {10.1103/physrevlett.120.208301}
}

@article{lamanna-2018,
	title        = {{Immigrant community integration in world cities}},
	author       = {Lamanna, Fabio and Lenormand, Maxime and Salas-Olmedo, Mar\'ia Henar and Romanillos, Gustavo and Gonçalves, Bruno and Ramasco, Jos\'e J.},
	year         = {2018},
	journal      = {PLOS ONE},
	volume       = {13},
	number       = {3},
	pages        = {e0191612},
	doi          = {10.1371/journal.pone.0191612}
}

@article{Sassen,
	title        = {{The global city: introducing a concept}},
	author       = {S. Sassen},
	year         = {2005},
	journal      = {Brown Journal of World Affairs},
	volume       = {11},
	pages        = {27--43}
}

@article{Clark,
	title        = {{Understanding the social context of the Schelling segregation model}},
	author       = {Clark, William A.V. and Fossett, M.},
	year         = {2008},
	journal      = {Proceedings of the National Academy of Sciences},
	volume       = {105},
	number       = {11},
	pages        = {4109--4114},
	doi          = {10.1073/pnas.0708155105}
}

@article{lenormand-2015,
	title        = {{Comparing and modelling land use organization in cities}},
	author       = {Lenormand, Maxime and Picornell, Miguel and Cant\'u-Ros, Oliva G. and Louail, Thomas and Herranz, Ricardo and Barthelemy, Marc and Fr\'ias-Mart\'inez, Enrique and San Miguel, Maxi and Ramasco, Jos\'e J.},
	year         = {2015},
	journal      = {Royal Society Open Science},
	volume       = {2},
	number       = {12},
	pages        = {150449},
	doi          = {10.1098/rsos.150449}
}

@article{Vinkovic,
	title        = {{A physical analogue of the Schelling model}},
	author       = {Vinkovic, D. and Kirman, A.},
	year         = {2006},
	journal      = {Proceedings of the National Academy of Sciences},
	volume       = {103},
	number       = {51},
	pages        = {19261--19265},
	doi          = {10.1073/pnas.0609371103}
}

@article{Zhang,
	title        = {{Residential segregation in an all-integrationist world}},
	author       = {Junfu Zhang},
	year         = {2004},
	journal      = {Journal of Economic Behavior and Organization},
	volume       = {54},
	number       = {4},
	pages        = {533--550},
	doi          = {https://doi.org/10.1016/j.jebo.2003.03.005},
	issn         = {0167-2681},
	url          = {https://www.sciencedirect.com/science/article/pii/S0167268103001768}
}

@article{jusup2022social,
	title        = {{Social physics}},
	author       = {Jusup, Marko and Holme, Petter and Kanazawa, Kiyoshi and Takayasu, Misako and Romi{\'c}, Ivan and Wang, Zhen and Ge{\v{c}}ek, Sun{\v{c}}ana and Lipi{\'c}, Tomislav and Podobnik, Boris and Wang, Lin and others},
	year         = {2022},
	journal      = {Physics Reports},
	publisher    = {Elsevier},
	volume       = {948},
	pages        = {1--148}
}

@article{castellano2009statistical,
	title        = {{Statistical physics of social dynamics}},
	author       = {Castellano, Claudio and Fortunato, Santo and Loreto, Vittorio},
	year         = {2009},
	month        = {May},
	journal      = {Rev. Mod. Phys.},
	publisher    = {American Physical Society},
	volume       = {81},
	pages        = {591--646},
	doi          = {10.1103/RevModPhys.81.591},
	url          = {https://link.aps.org/doi/10.1103/RevModPhys.81.591},
	issue        = {2},
	numpages     = {0}
}

@article{min2023threshold,
	title        = {{Threshold Cascade Dynamics in Coevolving Networks}},
	author       = {Min, Byungjoon and San Miguel, Maxi},
	year         = {2023},
	journal      = {Entropy},
	publisher    = {MDPI},
	volume       = {25},
	number       = {6},
	pages        = {929}
}

@article{rybski-2009,
	title        = {{Scaling laws of human interaction activity}},
	author       = {Diego Rybski  and Sergey V. Buldyrev  and Shlomo Havlin  and Fredrik Liljeros  and Hern\'an A. Makse},
	year         = {2009},
	journal      = {Proceedings of the National Academy of Sciences},
	volume       = {106},
	number       = {31},
	pages        = {12640--12645},
	doi          = {10.1073/pnas.0902667106},
	url          = {https://www.pnas.org/doi/abs/10.1073/pnas.0902667106},
	eprint       = {https://www.pnas.org/doi/pdf/10.1073/pnas.0902667106}
}

@article{Zimmermann,
	title        = {{Coevolution of dynamical states and interactions in dynamic networks}},
	author       = {Zimmermann, Mart{\'i}n G. and Egu{\'i}luz, V{\'i}ctor M. and San Miguel, Maxi},
	year         = {2004},
	month        = {Jun},
	journal      = {Phys. Rev. E},
	publisher    = {American Physical Society},
	volume       = {69},
	pages        = {065102},
	doi          = {10.1103/PhysRevE.69.065102},
	url          = {https://link.aps.org/doi/10.1103/PhysRevE.69.065102},
	issue        = {6},
	numpages     = {4}
}

@article{glauber1963time,
	title        = {{Time-dependent statistics of the Ising model}},
	author       = {Glauber, Roy J},
	year         = {1963},
	journal      = {Journal of mathematical physics},
	publisher    = {American Institute of Physics},
	volume       = {4},
	number       = {2},
	pages        = {294--307}
}

@article{jewski-2017,
	title        = {{Pair approximation for the $q$-voter model with independence on complex networks}},
	author       = {Jedrzejewski, Arkadiusz},
	year         = {2017},
	month        = {Jan},
	journal      = {Phys. Rev. E},
	publisher    = {American Physical Society},
	volume       = {95},
	pages        = {012307},
	doi          = {10.1103/PhysRevE.95.012307},
	url          = {https://link.aps.org/doi/10.1103/PhysRevE.95.012307},
	issue        = {1},
	numpages     = {9}
}

@article{Min-2017,
	title        = {{Fragmentation transitions in a coevolving nonlinear voter model}},
	author       = {Min, Byungjoon and San Miguel, Maxi},
	year         = {2017},
	journal      = {Scientific Reports},
	volume       = {7},
	number       = {},
	pages        = {12864},
	doi          = {10.1038/s41598-017-13047-2}
}

@article{pereira2005majority,
	title        = {{Majority-vote model on random graphs}},
	author       = {Pereira, Luiz FC and Moreira, FG Brady},
	year         = {2005},
	journal      = {Physical Review E},
	publisher    = {APS},
	volume       = {71},
	number       = {1},
	pages        = {016123}
}

@article{de1992isotropic,
	title        = {{Isotropic majority-vote model on a square lattice}},
	author       = {de Oliveira, Mario J},
	year         = {1992},
	journal      = {Journal of Statistical Physics},
	publisher    = {Springer},
	volume       = {66},
	number       = {1},
	pages        = {273--281}
}

@article{campos2003small,
	title        = {{Small-world effects in the majority-vote model}},
	author       = {Campos, Paulo RA and de Oliveira, Viviane M and Moreira, FG Brady},
	year         = {2003},
	journal      = {Physical Review E},
	publisher    = {APS},
	volume       = {67},
	number       = {2},
	pages        = {026104}
}

@article{nowak2019homogeneous,
	title        = {{Homogeneous Symmetrical Threshold Model with Nonconformity: Independence versus Anticonformity}},
	author       = {Nowak, Bart{\l}omiej and Sznajd-Weron, Katarzyna},
	year         = {2019},
	month        = {4},
	journal      = {Complexity},
	publisher    = {Hindawi Publishing Corporation},
	volume       = {2019},
	pages        = {1--14},
	doi          = {10.1155/2019/5150825},
	url          = {https://doi.org/10.1155/2019/5150825}
}

@article{nowak2020symmetrical,
	title        = {{Symmetrical threshold model with independence on random graphs}},
	author       = {Nowak, Bart{\l}omiej and Sznajd-Weron, Katarzyna},
	year         = {2020},
	journal      = {Physical Review E},
	publisher    = {APS},
	volume       = {101},
	number       = {5},
	pages        = {052316}
}

@article{mobilia2015nonlinear,
	title        = {{Nonlinear q-voter model with inflexible zealots}},
	author       = {Mobilia, Mauro},
	year         = {2015},
	journal      = {Physical Review E},
	publisher    = {APS},
	volume       = {92},
	number       = {1},
	pages        = {012803}
}

@article{mellor2016characterization,
	title        = {{Characterization of the nonequilibrium steady state of a heterogeneous nonlinear q-voter model with zealotry}},
	author       = {Mellor, Andrew and Mobilia, Mauro and Zia, RKP},
	year         = {2016},
	journal      = {EPL (Europhysics Letters)},
	publisher    = {IOP Publishing},
	volume       = {113},
	number       = {4},
	pages        = {48001}
}

@misc{Pournaki-2022,
	title        = {{Order-disorder transition in the zero-temperature Ising model on random graphs}},
	author       = {Pournaki, Armin and Olbrich, Eckehard and Banisch, Sven and Klemm, Konstantin},
	year         = {2022},
	month        = {May},
	journal      = {Phys. Rev. E},
	publisher    = {arXiv},
	volume       = {107},
	pages        = {054112},
	doi          = {10.48550/ARXIV.2209.09325},
	url          = {https://arxiv.org/abs/2209.09325},
	copyright    = {arXiv.org perpetual, non-exclusive license},
	issue        = {5},
	numpages     = {5}
}

@article{Abella-2022,
	title        = {{Aging effects in Schelling segregation model}},
	author       = {Abella,  David and San Miguel,  Maxi and Ramasco,  José J.},
	year         = {2022},
	month        = {Nov},
	journal      = {Scientific Reports},
	publisher    = {Springer Science and Business Media LLC},
	volume       = {12},
	number       = {1},
	doi          = {10.1038/s41598-022-23224-7},
	issn         = {2045-2322},
	url          = {http://dx.doi.org/10.1038/s41598-022-23224-7}
}

@misc{supplementary_Abella_2022,
	title        = {{Supplementary Information Videos for: Aging effects in Schelling segregation model}},
	author       = {Author(s) of the paper},
	author       = {Abella,  David and San Miguel,  Maxi and Ramasco,  José J.},
	year         = {2022},
	month        = {Nov},
	journal      = {Scientific Reports},
	publisher    = {Springer Science and Business Media LLC},
	volume       = {12},
	number       = {1},
	url          = {http://dx.doi.org/10.1038/s41598-022-23224-7}
}

@article{Abella-2022-AME,
	title        = {{Aging in binary-state models: The Threshold model for complex contagion}},
	author       = {Abella, David and San Miguel, Maxi and Ramasco, Jos\'e J.},
	year         = {2023},
	month        = {Feb},
	journal      = {Phys. Rev. E},
	publisher    = {American Physical Society},
	volume       = {107},
	pages        = {024101},
	doi          = {10.1103/PhysRevE.107.024101},
	url          = {https://link.aps.org/doi/10.1103/PhysRevE.107.024101},
	issue        = {2},
	numpages     = {12}
}

@article{Suchecki-2005,
  title={Voter model dynamics in complex networks: Role of dimensionality, disorder, and degree distribution},
  author={Suchecki, Krzysztof and Egu{\'i}luz, V{\'i}ctor M and San Miguel, Maxi},
  journal={Physical Review E},
  volume={72},
  number={3},
  pages={036132},
  year={2005},
  publisher={APS}
}

@article{centola-2010,
	title        = {{The Spread of Behavior in an Online Social Network Experiment}},
	author       = {Centola, Damon},
	year         = {2010},
	journal      = {Science},
	volume       = {329},
	number       = {5996},
	pages        = {1194--1197},
	doi          = {10.1126/science.1185231}
}

@article{rosenthal-2015,
	title        = {{Revealing the hidden networks of interaction in mobile animal groups allows prediction of complex behavioral contagion}},
	author       = {Rosenthal, Sara Brin and Twomey, Colin R. and Hartnett, Andrew T. and Wu, Hai Shan and Couzin, Iain D.},
	year         = {2015},
	journal      = {Proceedings of the National Academy of Sciences},
	volume       = {112},
	number       = {15},
	pages        = {4690--4695},
	doi          = {10.1073/pnas.1420068112}
}






\cleardoublepage 

\fancyhead[LO]{\rightmark} 
\fancyhead[RE]{\leftmark} 

\chapterimage{} 
\chapterspaceabove{6.75cm} 
\chapterspacebelow{7.25cm} 

\begin{appendices}

\renewcommand{\chaptername}{Appendix} 


\chapterspaceabove{6.75cm}
\chapterspacebelow{7cm}

\chapter{\label{app:Vacancy density effect on the Schelling model dynamics} Vacancy density effect on the Schelling model dynamics}
\begin{figure}[ht]
    \centering
    \includegraphics[width=\linewidth]{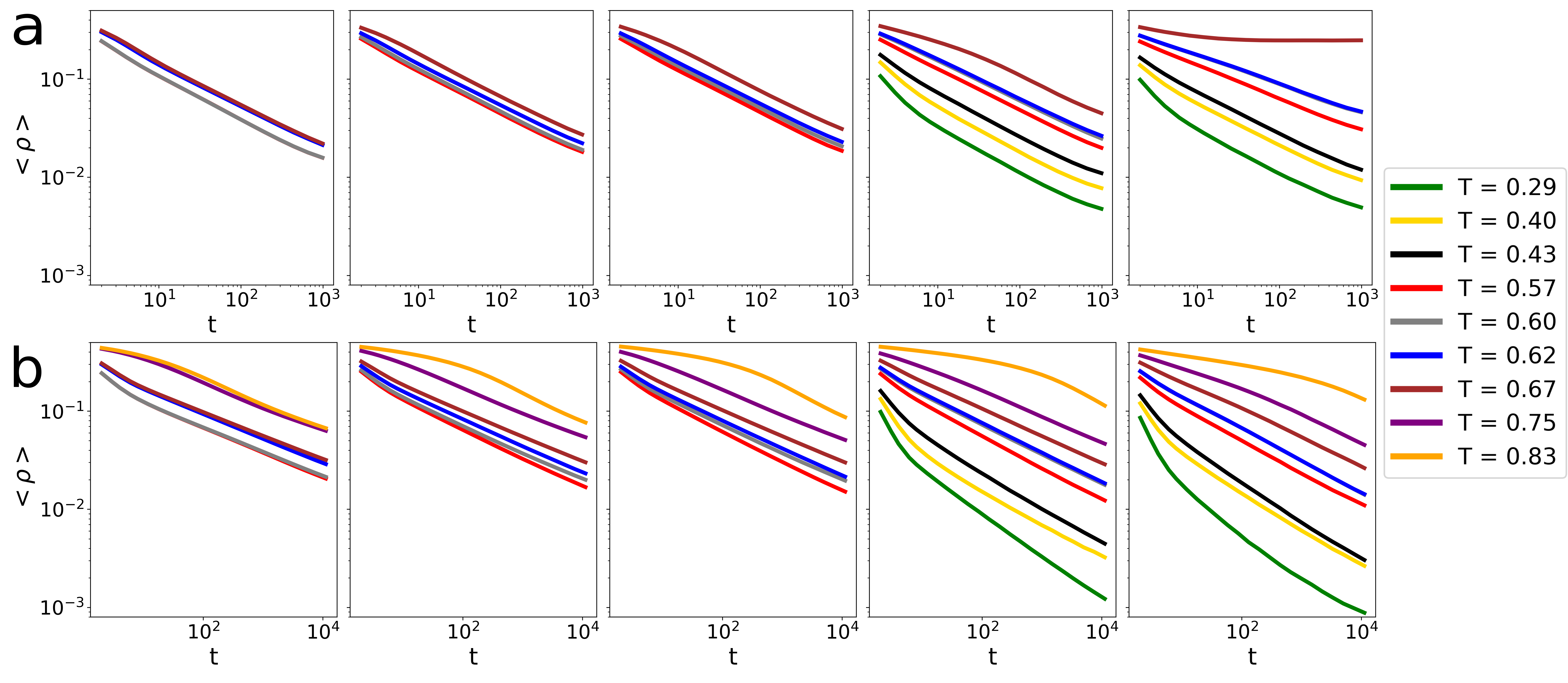} 
    \caption[Average interface density $ \langle \rho (t) \rangle$ for different $\rho_v$]{ Average interface density $ \langle \rho (t) \rangle$ as a function of time steps for different values of the tolerance parameter $T$ fr the Schelling model (\textbf{a}) and the version with aging (\textbf{b}). The different plots show a the evolution at a different value of the vacancy density, increasing from left to right $\rho_v = 0.005$, $0.15$, $0.2$, $0.3$ and $0.45$. Average performed over $10^{3}$ realisations with system size $100$ $x$ $100$.}
    \label{FigS1}
    \end{figure}
Since we restrain ourselves to the region $\rho_{v} < 0.5$, the increase/decrease of the number of vacancies does not change dramatically the behaviour. Above this value, we approach the segregated-dilute transition ($\rho_v \sim 0.62$). Nevertheless, it is worth to mention a few features we observe on the coarsening dynamics. Essentially, when we  set a higher vacancy density, the number of agents which see vacancies at their surroundings increases. This results in a family of similar power-law decays towards the segregated state for every meaningful value of $T$ (see Fig. \ref{FigS1}). 

\begin{figure}
\centering
\includegraphics[width=0.8\linewidth]{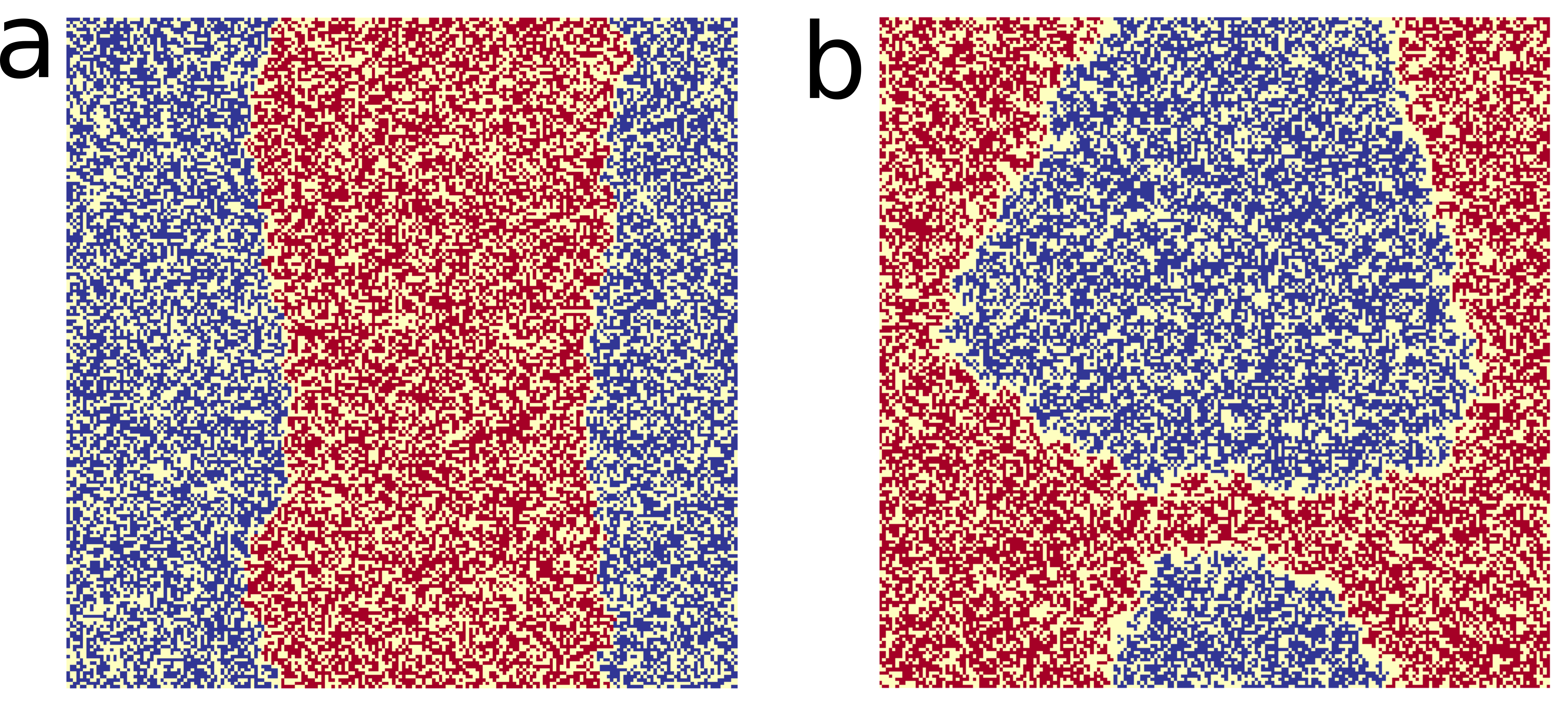} 
\caption[Snapshots of the system with large $\rho_{v}$]{Snapshots of the system at the final segregated state (after $10^{6}$ MC steps) for the Schelling model (\textbf{a}) and the version with (\textbf{b}). System size $200$ $x$ $200$ with $\rho_{v} = 0.45$ and $T = 0.29$.}
\label{FigS2}
\end{figure}

Moreover, a higher $\rho_v$ allows us to study the coarsening phenomena for lower values of $T$ according to the phase diagram for the original Schelling model. For those particular cases, when the aging is introduced, we observe a power law decay faster than without aging (Fig. \ref{FigS1}b). Therefore, the aging effect accelerates segregation in this region of the phase diagram, contrary as for lower values of $\rho_v$. This acceleration is not caused by reaching the 2-clusters state in less time. Since there is a large presence of vacancies, aging causes a formation of vacancy clusters at the interface. Fig. \ref{FigS2} shows the final segregated state with and without aging. This spontaneous behaviour is result of the low tolerance combined with the persistence of clusters (once formed) due to aging effect and the large number of vacancies that allows the possibility of the formation of clusters at the interface.

\begin{figure}[b!]
    \centering
    \includegraphics[width=0.85\linewidth]{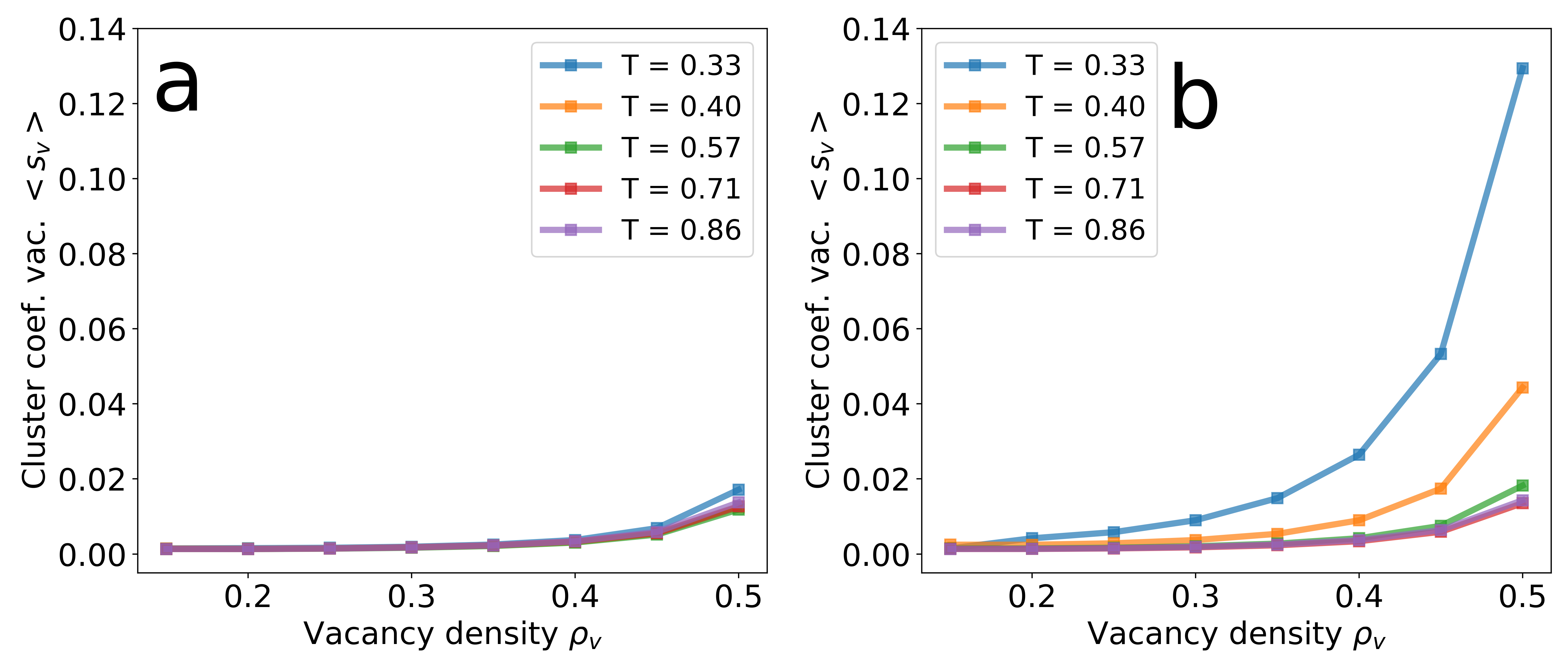} 
    \caption[Cluster coefficient of vacancies as a function of the vacancy density]{Cluster coefficient of vacancies as a function of the vacancy density $\rho_v$ for the Schelling model (\textbf{a}) and the version with (\textbf{b}) for different values of the tolerance $T$.}
    \label{FigS3}
    \end{figure}

In order to quantify this vacancy cluster formation, we define a measure inspired in the segregation coefficient:

\begin{equation}
    s_v = \frac{1}{\left(L^{2}\rho_v\right)^{2}} \sum_{\{c\}} n_{c}^{2}
\end{equation}

where $c$ is the size of a vacancy cluster and ${n_c}$ is the number of clusters with size $c$. The sample average of $s_{v}$ after reaching equilibrium is called the cluster coefficient of vacancies $\langle s_{v} \rangle$. 

The results of this measure as a function of $\rho_v$ for a few values of $T$ are represented in Fig.\ref{FigS3} for the Schelling model with and without aging. We observe an increasing dependence of $\langle s_v \rangle$ with $\rho_v$ for both models, but the effect reducing tolerance changes dramatically the behavior for the case with aging, highlighting the vacancy cluster formation.



\chapterspaceabove{6.75cm}
\chapterspacebelow{7cm}

\chapter{\label{appendix_HMFA} Heterogeneous mean-field taking into account aging (HMFA)}

Setting the time derivatives to 0 in Eqs. (\ref{eq:HMFaging2}), we obtain the relations for the stationary state:
\begin{equation}
    x^{\pm}_{k,0} = \sum_{j=0}^{\infty} x^{\mp}_{k,j} \,  \omega_{k,j}^{\mp}, \quad \quad x^{\pm}_{k,j} = x^{\pm}_{k,j-1} \, ( 1 - \omega_{k,j-1}^{\pm})  \qquad j > 0, \label{eq:SSaging4}
\end{equation}
from where we extract the stationary condition $x^{-}_{k,0} = x^{+}_{k,0}$, as in Ref. \cite{chen-2020}. Notice that by setting $p_A(j) = 1$ and summing over all ages $j$, we recover the HMF approximation (Eq. \ref{eq:HMF}) for the model without aging. Defining $x^{\pm}_{j}(t)$ as the fraction of agents in state $\pm 1$ with age $j$:
\begin{equation}
    x^{\pm}_{j} = \sum_k p_k \, x^{\pm}_{k,j},
\end{equation}
and using the degree distribution of a complete graph $p_k = \delta(k-N+1)$ (where $\delta(\cdot)$ is the Dirac delta), we sum over the variable $k$ and rewrite Eq. (\ref{eq:SSaging4}) in terms of $x^{\pm}_{j}$:
\begin{equation}
    x^{\pm}_{0} = \sum_{j=0}^{\infty} x^{\mp}_{j} \, \omega_{j}^{\mp},  \quad \quad x^{\pm}_{j} = x^{\pm}_{j-1} \, ( 1 - \omega_{j-1}^{\pm})  \qquad j > 0,   \label{eq:SSSaging2}
\end{equation}
where $\omega_{j}^{\pm} \equiv \omega_{N-1,j}^{\pm}$. Note that the stationary condition $x^{-}_{0} = x^{+}_{0}$ remains valid after summing over the degree variable. We compute the solution $x^{\pm}_{j}$ recursively as a function of $x^{\pm}_0$:
\begin{equation}
    x^{\pm}_{j} = x^{\pm}_0 \, F_j^{\pm} \qquad {\rm where} \qquad F_j^{\pm} = \prod_{a = 0}^{j-1} (1 - \omega_a^{\pm}),
\end{equation}
and summing all $j$,
\begin{equation}
    x^{\pm} = x^{\pm}_0 \, F^{\pm}  \qquad {\rm where} \qquad F^{\pm} = 1 + \sum_{j=1}^{\infty} F_j^{\pm}.
\end{equation}
Using the stationary condition $x^{-}_0 = x^{+}_0$, we reach:
\begin{equation}
    \frac{x^{+}}{x^{-}} = \frac{F^{+}}{F^{-}}.
\end{equation}
Notice that, for the complete graph, $\tilde{x}^{+} = x$, $\tilde{x}^{-} = 1 - x$. Therefore, $F^{\pm}$ is a function of the variable $x^{\mp}$ ($F^{+} = F(1 - x)$). Thus, we rewrite the previous expression just in terms of the variable $x$:
\begin{equation}
    \frac{x}{1- x} = \frac{F(1 - x)}{F(x)}.
\end{equation}



\chapterspaceabove{6.75cm}
\chapterspacebelow{7cm}

\chapter{\label{appendix_RR} Internal time recursive relation in Phase ${\rm {\bf I}}$/${\rm {\bf I}}^{*}$}

In Phase ${\rm I}$ and ${\rm I}^{*}$, the exceeding threshold condition ($m/k > T$) is full-filled for almost all agents in the system. Thus, agents will change their state and reset the internal time once activated. For the original model, all agents are activated once in a time step on average, but for the model with aging, the activation probability plays an important role. We consider here a set of $N$ agents that are activated randomly with an activation probability $p_A(j)$ and, once activated, they reset their internal time. Being $n_i(t)$ the fraction of agents with internal time $i$ at the time step $t$, we build a recursive relation for the previously described dynamics in terms of variables $i$ and $t$:

\begin{eqnarray}
    n_1(t) = \sum_{i=1}^{t-1} p_A(i) \, n_i(t-1) \quad \quad n_i(t) = (1 - p_A(i-1) ) \, n_{i-1}(t-1)  \qquad i > 1. \label{eq:RR1}
\end{eqnarray}

This recursion relation can be solved numerically from the initial condition ($n_1(0) = 1$, $n_i(0) = 0$ for $i > 1$). To obtain the mean internal time at time $t$, we just need to compute the following:

\begin{equation}
    \label{eq:RR}
    \bar{\tau}(t) = \sum_{i=1}^{t} i \, n_i(t).
\end{equation}

The solution from this recursive relation describes the mean internal time dynamics with great agreement with the numerical simulations performed at Phase I (for the complete graph) and Phase ${\rm  I}^{*}$ (for the Erd\H{o}s-R\'enyi and Moore lattice).

\newpage
\thispagestyle{plain} 
\mbox{}


\end{appendices}


\end{document}